\documentclass[]{aastex701}

\shortauthors{Couperus et al.}

\begin{document}

\title{The Solar Neighborhood. LV. M Dwarf Twin Binaries --- One in Five Twin Sibling Pairs Are Mismatched in Activity and/or Rotation}

\correspondingauthor{Andrew A. Couperus}
\author[0000-0001-9834-5792]{Andrew A. Couperus}
\affiliation{Five College Astronomy Department, Smith College, Northampton, MA 01063, USA}
\affiliation{RECONS Institute, Chambersburg, PA 17201, USA}
\email[show]{andcoup1@gmail.com}

\author[0000-0002-9061-2865]{Todd J. Henry}
\affiliation{RECONS Institute, Chambersburg, PA 17201, USA}
\email{thenry88@gsu.edu}

\author[0000-0002-9811-5521]{Aman Kar}
\affiliation{Department of Physics and Astronomy, Georgia State University, Atlanta, GA 30302, USA}
\affiliation{RECONS Institute, Chambersburg, PA 17201, USA}
\email{akar5@gsu.edu}

\author[0000-0003-0193-2187]{Wei-Chun Jao}
\affiliation{Department of Physics and Astronomy, Georgia State University, Atlanta, GA 30302, USA}
\email{wjao@gsu.edu}

\author[0000-0002-1864-6120]{Eliot Halley Vrijmoet}
\affiliation{Five College Astronomy Department, Smith College, Northampton, MA 01063, USA}
\affiliation{RECONS Institute, Chambersburg, PA 17201, USA}
\email{evrijmoet@smith.edu}

\author[0000-0001-5643-8421]{Rachel A. Osten}
\affiliation{Space Telescope Science Institute, Baltimore, MD 21218, USA}
\affiliation{Center for Astrophysical Sciences, Department of Physics and Astronomy, Johns Hopkins University, Baltimore, MD 21218, USA}
\email{osten@stsci.edu}

\begin{abstract}

\noindent
We report on a study of 36 pairs of `twin' M dwarfs in wide binaries and assess how similarly the stars behave. Stars in each twin pair have BP, RP, $J$, $H$, and $K_s$ differing by $<$\,0.10\,mag, mass estimates matching within $<$\,3\%, and presumably the same age and composition. We utilize short-- and long-term photometry, multi-epoch spectroscopy, and archival data to measure rotation periods, photometric activity levels, and H$\alpha$ equivalent widths for many systems. Speckle imaging, radial velocities, and long-term astrometry are used to identify unresolved companions, yielding three systems with unseen components. Among the 33 remaining twin systems, numerous remarkable pairs show nearly identical rotation rates and activity levels between their twin components, including cases throughout the lower main sequence and across a broad range of rotation-activity parameter space. In contrast, mismatches with $>$\,25\% differences exist in rotation period for $21\%_{-7\%}^{+14\%}$ of twin pairs, in rotation amplitude for $67\%_{-15\%}^{+10\%}$ of pairs, in multi-year photometric variability for $33\%_{-9\%}^{+12\%}$ of pairs, and in H$\alpha$ activity for $21\%_{-6\%}^{+9\%}$ of pairs, with fully convective systems generally mismatched more often. Thus, roughly one out of five M dwarf twin sets does not match in rotation and/or activity despite otherwise identical fundamental parameters. Furthermore, we compile three key systems showing larger relative active/inactive H$\alpha$ mismatches. We propose the various mismatches likely stem from factors such as dynamo stochasticity, activity cycles, formative disk aspects, and/or star-planet interactions, depending on the system. These well-vetted twins offer ripe targets for many future investigations.

\end{abstract}



\defcitealias{RTW_P1}{C25}
\defcitealias{Pass_2024_ApJ}{P24}
\defcitealias{Newton_2017}{N17}
\defcitealias{2024AJ....167..159L}{L24}
\defcitealias{Lu_2022}{L22}

\section{Introduction} \label{sec:intro}

Almost exactly 75\% of all stars in the solar neighborhood are M dwarfs \citep{Henry2024}. These enduring objects can live for trillions of years and are both smaller and lower mass than the Sun, spanning 0.08--0.62\,$\rm{M_\odot}$ \citep{Benedict_2016}. Those with masses $\gtrsim$\,0.35\,$\rm{M_\odot}$ are partially convective (PC) with an outer convection zone and inner radiative core, similar to the Sun. In contrast, M dwarfs at $\lesssim$\,0.35\,$\rm{M_\odot}$ --- which includes about half of all stars \citep{Henry2024} --- have fully convective (FC) interiors and lack a radiative core \citep{1997A&A...327.1039C}. The combination of stellar rotation and convection in M dwarfs drives a magnetic dynamo that generates magnetic fields \citep[e.g.,][]{Shulyak_2015}, subsequently producing activity behaviors as the magnetism influences material throughout and around the star. M dwarf stellar activity can include starspots, faculae, flares, heated chromospheres and coronae, and variable flux levels across a range of wavelengths and timescales \citep[e.g.,][and references therein]{2021isma.book.....B}.

M dwarfs have historically been observationally neglected due to their intrinsic faintness, but they are common targets of study today for three main reasons. First, their ubiquity makes them intrinsically connected to many other areas of astrophysics, such as measuring the initial mass function. Second, those with FC interiors allow us to explore magnetic dynamo behaviors in new regimes, enabling informative comparisons to better understand the solar dynamo. Third, their small masses and radii make them favorable targets for detecting rocky exoplanets in orbits where surface water may be  liquid \citep{2023AJ....165..265M}. The latter motivation is particularly important, but M dwarfs present a complication because their stellar activity could have complex impacts on exoplanet habitability. For example, strong flares and high energy emission due to activity could strip exoplanet atmospheres and dangerously irradiate planet surfaces \citep{Tarter_2007}. Alternately, activity-induced UV emission may trigger prebiotic chemistry conducive to the development of life \citep{2017ApJ...843..110R}.

The above factors motivate the desire for reliable predictions of M dwarfs' magnetism and activity throughout their lives, a task firmly linked with understanding their rotational evolution given that rotation helps drive activity via the dynamo. G/K/M stars nominally slow their rotation rates over time via magnetic braking, as demonstrated in the landmark work by \citet{1972ApJ...171..565S}, but modern observations indicate this is not a straightforward process. Starting with PC M dwarfs, similar stars at very young ages can show a scatter of rotation periods before finally converging to a consistent rotational sequence by ages of roughly half a billion years \citep[e.g.,][and references therein]{2023ApJ...947L...3B}. However, after this, the subsequent rotation spindown encounters the so-called intermediate period gap, a dearth of measured rotation periods near 20--30\,days first discovered using Kepler results \citep{McQuillan_2013,McQuillan_2014}. The cause of this rotation gap may relate to spot-faculae contrast balancing and/or core-envelope interactions and remains an active area of research \citep{2019A&A...621A..21R, 2020A&A...635A..43R, Spada_2020, Lu_2022} (\citet{Lu_2022} is hereafter referred to as \citetalias{Lu_2022}). \citetalias{Lu_2022} have shown observationally that the intermediate period gap spans the full range of PC M dwarfs, before disappearing somewhere near the transition to FC stars. However, once entering the FC regime, MEarth results from \citet{Newton_2016, Newton_2018} have demonstrated a strong bimodality in the distribution of field FC M dwarf rotation periods, a feature distinct from the aforementioned intermediate period gap in PC stars. This clustering of FC stars around short $\lesssim$\,10\,day and long $\gtrsim$\,70\,day periods is suspected to be caused by a relatively rapid phase of strong magnetic braking during a transition from fast to slow rotation, with the starting age of this fast braking phase varying among FC M dwarfs depending on their mass and possibly their formation histories \citep[][]{Pass2022, Medina_2022_spindown, Pass_2023, 2023MNRAS.526..870S, Pass_2024_ApJ, RTW_P1} (\citet{Pass_2024_ApJ} is hereafter referred to as \citetalias{Pass_2024_ApJ}, and \citet{RTW_P1} is hereafter referred to as \citetalias{RTW_P1}).

One might also expect M dwarfs with the same masses, compositions, ages, and rotation rates to have similar activity levels, but this is not always the case. For example, such ``twin" stars in the BL+UV~Ceti system \citep{Kervella_2016, Benedict_2016} show very different magnetic field strengths \citep{Kochukhov_2017}, spot activity \citep{Barnes_2017}, X-ray activity \citep{Audard_2003, Wolk-UVCet-CoolStars21}, and radio activity \citep{2024ApJ...970...56P}. Additional mismatches in H$\alpha$ activity levels are found in several nearly-twin M dwarf binaries studied by \citet{Gunning_2014}. Furthermore, M dwarfs with the same masses, compositions, and ages in young open clusters also sometimes show pronounced mismatches in rotation and/or activity \citep{2014ApJ...795..161D, Newton_2017, 2021AJ....161..277K, 2021ApJ...916...77P, Henry2024} (\citet{Newton_2017} is hereafter referred to as \citetalias{Newton_2017}).

This complex landscape has motivated our present project, which seeks to probe (1) how often such ``twin" M dwarfs display mis/matched rotation and activity behaviors, and (2) the extent to which any observed differences are a result of fundamental stellar parameters versus possibly stemming more from other factors. To this end, we examine the rotation and activity characteristics of 36 M dwarf twin wide binaries, carried out as part of the REsearch Consortium on Nearby Stars (RECONS; \href{www.recons.org}{www.recons.org}) effort. Our first paper in this series, \citetalias{RTW_P1}, reported our results for four such systems and showed that twin FC M dwarfs can still differ in rotation rate by a factor of $\sim$6 and in X-ray luminosity by a factor of $\gtrsim$\,40, and implicated disk formation factors and star-planet interactions as possible culprits. Here we expand the effort to another 32 twin systems.

This paper is structured as follows. We discuss our sample creation and vetting steps in Section~\ref{sec:Sample}, outline the observing campaigns in Section~\ref{sec:obs}, investigate possible blending and contamination factors in Section~\ref{sec:contam}, and detail the methods and results for our multiple observing campaigns in Sections \ref{sec:speckle}--\ref{sec:rot-activity}. We then discuss our results in Section~\ref{sec:discussion1}, assess specific systems of interest in Section~\ref{sec:sys-notes}, highlight worthwhile future work in Section~\ref{sec:future-work}, and summarize the key findings in Section~\ref{sec:conclusions}. Various additional details are included in Appendix~\ref{sec:appendix}.

\vfil\eject

\section{Sample} \label{sec:Sample}

Our investigation requires compiling pairs of M dwarfs with nearly identical underlying parameters, a process detailed in \citetalias{RTW_P1}. To summarize, we have selected 36 common-proper-motion wide binaries (the ``Full Sample") within 50\,pc from Gaia DR2 and DR3 \citep{2018A&A...616A...1G, GaiaDR3}, where each pair of M dwarfs has Gaia $BP$, $RP$, and 2MASS $J$, $H$, and $K_s$ apparent magnitudes differing by $<$\,0.10\,mag between components. We also note that all 36 component pairs have DR3 $\rm{M}_{BP}$/$\rm{M}_{RP}$/$\rm{M}_{G}$ and 2MASS $\rm{M}_J$/$\rm{M}_H$/$\rm{M}_{K_s}$ absolute magnitudes agreeing to within 0.10\,mag as well when using DR3 parallaxes for the individual stars, although this was not a selection criterion. For context, the median difference in DR3 apparent $G$ magnitude between components is 0.027\,mag. Component mass estimates differ by $<$\,3\% for all pairs, derived using the \citet{Benedict_2016} $V$-filter mass-luminosity relation (MLR) and Gaia DR2 $\rm{M}_{BP}$ values as described in \citetalias{RTW_P1}. We assume the pairs host co-eval stars with the same compositions, ages, and past environments, which combined with their virtually identical masses yields the desired twin status.

\begin{figure}[!t]
\centering
\includegraphics[width=0.79\textwidth]{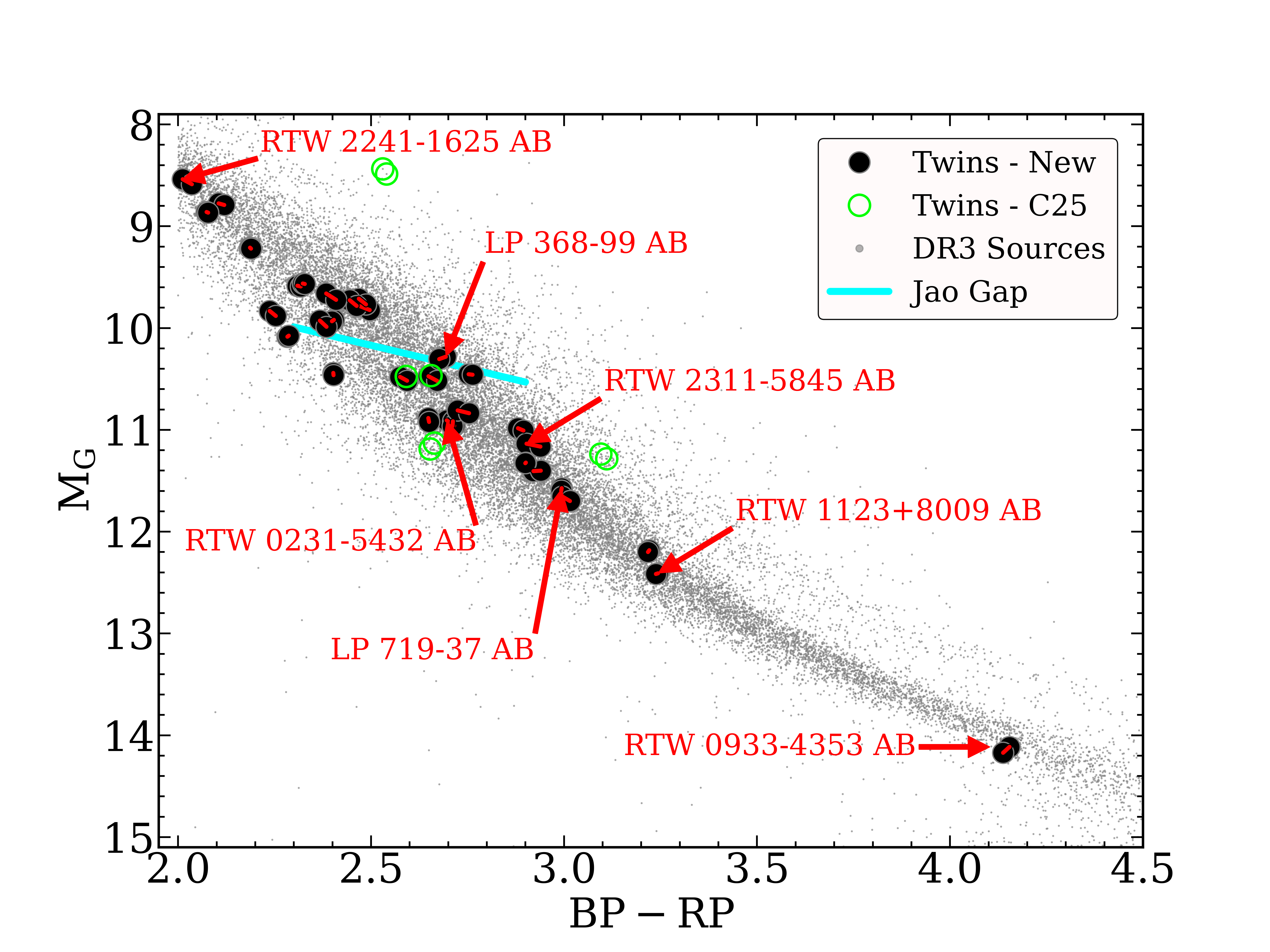}
\figcaption{An observational Hertzsprung–Russell diagram of our Full Sample using Gaia DR3 values. The 64 components comprising 32 pairs in this paper's New Systems are shown as black circles, with red lines connecting the components in each pair. Open green circles indicate the remaining four systems (eight components) previously discussed in our first paper in this series \citep[][\citetalias{RTW_P1}]{RTW_P1}. The main sequence is illustrated with gray points using a sample of Gaia DR3 sources within 50\,pc. A diagonal cyan line shows the Jao Gap near 0.35\,$\rm{M_\odot}$ \citep{2018ApJ...861L..11J, Jao_2023}, indicating the transition region for partially convective stars above it and fully convective stars below it --- we visually offset the gap upper edge definition from \citet{Jao_2023} downward by 0.05\,mag to better align with the center of the gap region. Various systems discussed throughout this paper are individually identified with red arrows and labels.\label{fig:HRD}}
\end{figure}

32 of our 36 systems are presented here for the first time, referred to as the 32 ``New Systems" throughout this work, while four other systems were explored in \citetalias{RTW_P1}. All of the checks and validation steps done for the New Systems here were also done for the four previously reported systems, as detailed in \citetalias{RTW_P1}. Three of the 36 total systems are found to be confirmed or suspected higher-order multiples (per \citetalias{RTW_P1}, and later in Section~\ref{subsec:multiplicity-checks}), and are therefore not twins so are excluded from all Figures here after Figure~\ref{fig:HRD}; they are included in some Tables of new measurements for completeness. The remaining 33 true twin pairs are referred to throughout this paper as the ``Results Sample", which includes systems and measurements reported from both \citetalias{RTW_P1} and this current work.

Astrometric and photometric properties of the Full Sample are presented in Table~\ref{tab:SampleTable-astr} and Table~\ref{tab:SampleTable-phot}, respectively. The parameters in Table~\ref{tab:SampleTable-astr} also include the parallax error, Renormalised Unit Weight Error \citep[RUWE;][]{RUWE_note}, and RV error from Gaia, all of which can be sensitive to possible unresolved companions and are explored further in Section~\ref{subsec:multiplicity-checks}. The photometric properties in Table~\ref{tab:SampleTable-phot} also include mass estimates and structural types, as well as a summary of observational coverage for each star, which are all referenced throughout this work. Stars are named using existing nearby star catalog designations when available; we otherwise use our own labeling scheme of RTW (RECONS TWins) with RA and Dec ICRS-2000 coordinate names. The designations of primary `A' components are based on Gaia DR2 $BP$ brightness, with the exception of GJ~745~AB that violates this pattern in favor of the commonly used GJ component labels. In this paper, B components always have the same base name as their primary, even if there may be existing alternate names available specific to the B star. We recommend researchers always carefully corroborate with precise coordinate matches, as different literature sources sometimes swap component labels, a largely irrelevant designation when considering such twin pairs.

Our Full Sample is shown on an observational Hertzsprung–Russell diagram (HRD) in Figure~\ref{fig:HRD}, highlighting the 32 New Systems reported here. The systems span nearly the entire M dwarf sequence, with our mass estimates ranging between 0.095\,$\rm{M_\odot}$ and 0.567\,$\rm{M_\odot}$. This includes both PC and FC M dwarfs, differentiated here and indicated in Table~\ref{tab:SampleTable-phot} using the observed PC/FC transition gap of \citet{2018ApJ...861L..11J, Jao_2023}. Of the 32 New Systems, 13 are PC above the gap, 17 are FC below the gap, and 2 reside within the gap.

%%%%%%%%%%%%%%% tab - Sample astrometry %%%%%%%%%%%%%%%
\startlongtable
\centerwidetable
\begin{deluxetable}{lccrrrrrrrc}
\tabletypesize{\footnotesize}
\tablewidth{0pt}
\tablecaption{Full RECONS Twin Binary Sample --- Astrometry\label{tab:SampleTable-astr}}
\tablehead{
\colhead{Name\tablenotemark{\rm{\dag}}} & \colhead{R.A.} & \colhead{Decl.} & \colhead{$\pi \pm \sigma$} & \colhead{$\mu_\alpha$} & \colhead{$\mu_\delta$} & \colhead{Ang. Sep.} & \colhead{2D Sep.} & \colhead{RUWE} & \colhead{$\rm{RV_{Gaia}} \pm \sigma$} \\
\colhead{} & \colhead{[ICRS-2016]} & \colhead{[ICRS-2016]} & \colhead{(mas)} & \colhead{($\mathrm{mas~yr^{-1}}$)} & \colhead{($\mathrm{mas~yr^{-1}}$)} & \colhead{(arcsec)} & \colhead{(au)} & \colhead{} & \colhead{($\mathrm{km~s^{-1}}$)}
}
\startdata
RTW 0143-0151 A   & 01 43 17.88 & $-$01 51 25.3 &  $21.25\pm0.03$ &    133.03 & $-$175.90 &   4.88 &  229.4 & 1.030 &    $14.09\pm2.41$  \\
RTW 0143-0151 B   & 01 43 17.76 & $-$01 51 29.8 &  $21.32\pm0.03$ &    133.33 & $-$181.04 &        &        & 1.094 &    $12.55\pm2.47$  \\
(2MA 0201+0117 A) & 02 01 47.00 & $+$01 17 05.1 &  $20.30\pm0.03$ &     74.77 &  $-$49.21 &  10.45 &  514.7 & 1.404 &     $5.18\pm0.95$  \\
(2MA 0201+0117 B) & 02 01 46.85 & $+$01 17 15.3 &  $20.31\pm0.03$ &     75.86 &  $-$46.73 &        &        & 1.501 &     $5.99\pm2.84$  \\
RTW 0231-5432 A   & 02 31 43.24 & $-$54 32 27.6 &  $43.49\pm0.02$ &  $-$15.69 & $-$175.03 &  10.23 &  235.4 & 1.432 &     $0.01\pm0.72$  \\
RTW 0231-5432 B   & 02 31 42.14 & $-$54 32 24.1 &  $43.43\pm0.02$ &  $-$15.28 & $-$173.44 &        &        & 1.395 &     $1.23\pm0.68$  \\
RTW 0231+4003 A   & 02 31 54.17 & $+$40 03 51.4 &  $22.34\pm0.02$ &    198.32 &   $-$1.10 &   5.82 &  260.1 & 1.320 &   $-20.98\pm1.71$  \\
RTW 0231+4003 B   & 02 31 54.17 & $+$40 03 45.5 &  $22.37\pm0.03$ &    198.33 &      1.05 &        &        & 1.316 &   $-17.97\pm2.32$  \\
RTW 0409+4623 A   & 04 09 43.08 & $+$46 23 35.2 &  $27.13\pm0.02$ &  $-$46.32 &      1.16 &   4.78 &  175.9 & 1.175 &    $-0.22\pm3.00$  \\
RTW 0409+4623 B   & 04 09 43.50 & $+$46 23 37.2 &  $27.17\pm0.02$ &  $-$44.51 &   $-$0.01 &        &        & 1.154 &     $0.16\pm1.34$  \\
KAR 0545+7254 A   & 05 45 50.02 & $+$72 54 09.0 &  $35.10\pm0.02$ &     80.06 &    114.13 &  81.37 & 2318.4 & 1.171 &    $-4.24\pm0.28$  \\
KAR 0545+7254 B   & 05 45 39.09 & $+$72 55 14.6 &  $35.10\pm0.01$ &     81.21 &    113.59 &        &        & 1.158 &    $-4.60\pm0.27$  \\
LP 719-37 A       & 06 14 12.72 & $-$14 36 07.5 &  $38.05\pm0.02$ &    212.49 & $-$298.46 &   6.18 &  162.5 & 1.186 &       \nodata      \\
LP 719-37 B       & 06 14 12.62 & $-$14 36 13.5 &  $38.06\pm0.02$ &    207.75 & $-$291.85 &        &        & 1.234 &     $7.03\pm1.81$  \\
G 103-63 A*       & 06 49 21.87 & $+$32 09 57.1 &  $32.76\pm0.02$ & $-$145.37 & $-$171.68 &   4.88 &  148.9 & 1.290 &    $31.34\pm17.21$ \\
G 103-63 B        & 06 49 21.94 & $+$32 09 52.3 &  $32.73\pm0.02$ & $-$150.79 & $-$167.55 &        &        & 1.288 &    $29.24\pm1.54$  \\
RTW 0824-3054 A   & 08 24 41.94 & $-$30 54 45.8 &  $32.50\pm0.02$ &  $-$74.11 &     42.23 &   7.63 &  234.5 & 1.312 &    $31.07\pm0.73$  \\
RTW 0824-3054 B*  & 08 24 41.81 & $-$30 54 53.2 &  $32.55\pm0.03$ &  $-$74.37 &     43.51 &        &        & 2.209 &    $30.21\pm0.80$  \\
LP 368-99 A       & 08 52 44.79 & $+$22 30 53.6 &  $30.64\pm0.03$ &  $-$40.55 & $-$153.15 &   4.61 &  150.5 & 1.221 &     $6.42\pm1.50$  \\
LP 368-99 B       & 08 52 44.62 & $+$22 30 49.7 &  $30.65\pm0.03$ &  $-$36.96 & $-$149.74 &        &        & 1.216 &       \nodata      \\
L 533-2 A         & 09 15 06.95 & $-$30 20 05.2 &  $33.90\pm0.02$ & $-$250.22 &    104.05 &   4.13 &  121.6 & 1.327 &    $19.94\pm0.23$  \\
L 533-2 B         & 09 15 07.03 & $-$30 20 01.2 &  $33.94\pm0.02$ & $-$244.83 &    108.12 &        &        & 1.216 &    $19.79\pm0.31$  \\
RTW 0933-4353 A   & 09 33 25.95 & $-$43 53 33.8 &  $56.98\pm0.04$ & $-$196.69 &    165.18 &  12.39 &  217.3 & 1.080 &       \nodata      \\
RTW 0933-4353 B   & 09 33 24.82 & $-$43 53 35.7 &  $57.04\pm0.04$ & $-$190.67 &    162.14 &        &        & 1.093 &       \nodata      \\
LP 551-62 A       & 11 06 16.91 & $+$07 50 17.2 &  $24.42\pm0.03$ & $-$178.56 &     41.26 &  25.14 & 1030.3 & 1.364 &    $28.56\pm1.22$  \\
LP 551-62 B       & 11 06 16.35 & $+$07 50 41.0 &  $24.38\pm0.03$ & $-$177.71 &     39.80 &        &        & 1.263 &    $27.69\pm1.38$  \\
RTW 1123+8009 A   & 11 23 12.63 & $+$80 09 02.1 &  $38.82\pm0.02$ &  $-$18.37 &  $-$24.04 &   9.96 &  256.5 & 1.128 &   $-15.02\pm2.61$  \\
RTW 1123+8009 B   & 11 23 16.46 & $+$80 09 03.8 &  $38.81\pm0.02$ &  $-$17.02 &  $-$27.34 &        &        & 1.167 &   $-21.90\pm1.91$  \\
RTW 1133-3447 A   & 11 33 15.39 & $-$34 47 01.9 &  $20.88\pm0.02$ & $-$244.08 &      2.99 & 106.07 & 5081.0 & 1.079 &       \nodata      \\
RTW 1133-3447 B   & 11 33 07.60 & $-$34 46 16.7 &  $20.87\pm0.02$ & $-$244.64 &      3.13 &        &        & 1.170 &     $5.98\pm0.72$  \\
(KX Com A)        & 12 56 52.80 & $+$23 29 50.7 &  $36.61\pm0.03$ &     70.27 &      4.10 &   7.72 &  211.2 & 1.234 &    $-7.89\pm0.84$  \\
(KX Com BC)       & 12 56 52.24 & $+$23 29 50.2 &  $36.52\pm0.03$ &     74.89 &      8.56 &        &        & 1.436 &   $-10.81\pm4.90$  \\
RTW 1336-3212 A   & 13 36 56.13 & $-$32 12 26.2 &  $41.75\pm0.02$ &    186.84 & $-$274.10 &   4.70 &  112.6 & 1.143 &    $41.94\pm1.38$  \\
RTW 1336-3212 B   & 13 36 56.34 & $-$32 12 30.0 &  $41.78\pm0.02$ &    193.52 & $-$268.48 &        &        & 1.197 &       \nodata      \\
L 197-165 A       & 14 04 00.31 & $-$59 24 03.3 &  $33.06\pm0.02$ &     18.49 & $-$494.51 &  78.09 & 2360.4 & 1.136 &   $-12.28\pm0.59$  \\
L 197-165 B       & 14 03 50.21 & $-$59 23 50.8 &  $33.10\pm0.02$ &     19.98 & $-$493.96 &        &        & 1.036 &   $-11.90\pm0.51$  \\
(GJ 1183 A)       & 14 27 55.69 & $-$00 22 30.5 &  $57.01\pm0.03$ & $-$361.15 &     41.70 &  13.07 &  229.3 & 1.510 &   $-11.72\pm3.27$  \\
(GJ 1183 B)       & 14 27 56.01 & $-$00 22 18.3 &  $57.03\pm0.03$ & $-$363.17 &     52.54 &        &        & 1.524 &       \nodata      \\
RTW 1433-6109 A   & 14 33 53.38 & $-$61 09 08.2 &  $32.64\pm0.02$ &  $-$54.09 &    155.46 &  10.20 &  312.5 & 1.042 &   $-28.44\pm0.37$  \\
RTW 1433-6109 B   & 14 33 54.74 & $-$61 09 11.0 &  $32.63\pm0.02$ &  $-$53.59 &    156.69 &        &        & 0.976 &   $-28.54\pm0.33$  \\
L 1197-68 A       & 14 36 34.22 & $+$11 02 41.8 &  $29.03\pm0.02$ &    264.47 & $-$155.02 &  11.26 &  387.7 & 1.472 &    $24.21\pm0.83$  \\
L 1197-68 B       & 14 36 34.04 & $+$11 02 30.8 &  $29.04\pm0.02$ &    267.02 & $-$151.38 &        &        & 1.430 &    $24.41\pm0.67$  \\
L 1198-23 A       & 14 56 46.97 & $+$12 45 21.4 &  $29.43\pm0.02$ & $-$281.44 &    187.17 &  44.79 & 1521.7 & 1.587 &   $-27.95\pm0.57$  \\
L 1198-23 B       & 14 56 49.91 & $+$12 45 09.1 &  $29.44\pm0.02$ & $-$277.44 &    187.06 &        &        & 1.604 &   $-27.79\pm0.88$  \\
RTW 1512-3941 A   & 15 12 08.28 & $-$39 41 57.6 &  $42.38\pm0.03$ &     77.65 &    108.07 &   6.06 &  143.0 & 1.207 &       \nodata      \\
RTW 1512-3941 B   & 15 12 08.03 & $-$39 41 52.3 &  $42.36\pm0.03$ &     65.62 &     98.13 &        &        & 1.184 &    $36.34\pm0.27$  \\
(NLTT 44989 A)    & 17 33 05.98 & $-$30 35 10.1 &  $54.68\pm0.02$ & $-$113.37 & $-$123.01 &   4.75 &   86.9 & 1.001 &    $40.09\pm0.68$  \\
(NLTT 44989 B)    & 17 33 05.62 & $-$30 35 11.3 &  $54.61\pm0.03$ & $-$121.23 & $-$122.86 &        &        & 1.167 &    $44.42\pm1.16$  \\
RTW 1812-4656 A   & 18 12 55.99 & $-$46 55 59.8 &  $22.16\pm0.02$ &     24.14 &     17.64 & 107.98 & 4868.5 & 1.119 &   $-18.05\pm0.92$  \\
RTW 1812-4656 B   & 18 12 48.73 & $-$46 54 41.6 &  $22.20\pm0.02$ &     24.36 &     18.38 &        &        & 1.152 &   $-18.93\pm0.81$  \\
GJ 745 A          & 19 07 05.02 & $+$20 53 11.4 & $113.25\pm0.03$ & $-$478.27 & $-$349.09 & 114.05 & 1007.2 & 0.807 &    $32.01\pm0.15$  \\
GJ 745 B          & 19 07 12.66 & $+$20 52 31.9 & $113.22\pm0.02$ & $-$480.75 & $-$332.50 &        &        & 1.015 &    $31.90\pm0.20$  \\
RTW 2011-3824 A   & 20 11 47.16 & $-$38 24 49.8 &  $31.94\pm0.03$ &  $-$29.81 &  $-$63.13 &   4.19 &  131.4 & 1.342 &    $-6.00\pm1.04$  \\
RTW 2011-3824 B   & 20 11 46.91 & $-$38 24 46.8 &  $31.83\pm0.02$ &  $-$39.45 &  $-$68.17 &        &        & 1.430 &    $-4.23\pm0.89$  \\
G 230-39 A        & 20 25 27.56 & $+$54 33 38.1 &  $31.08\pm0.01$ &    163.95 &    340.65 &  27.63 &  889.1 & 1.255 &   $-18.52\pm0.94$  \\
G 230-39 B        & 20 25 24.76 & $+$54 33 51.3 &  $31.07\pm0.01$ &    164.05 &    340.64 &        &        & 1.152 &   $-18.92\pm0.68$  \\
RTW 2202+5537 A   & 22 02 20.43 & $+$55 37 54.9 &  $38.81\pm0.02$ &    123.08 &   $-$4.73 &  23.41 &  603.6 & 1.356 &    $-7.47\pm2.57$  \\
RTW 2202+5537 B   & 22 02 20.18 & $+$55 38 18.3 &  $38.77\pm0.02$ &    133.90 &      0.05 &        &        & 1.127 &       \nodata      \\
RTW 2211+0058 A   & 22 11 32.55 & $+$00 58 47.6 &  $29.71\pm0.02$ &  $-$32.12 & $-$107.05 &   5.00 &  168.2 & 1.277 &     $2.35\pm1.16$  \\
RTW 2211+0058 B   & 22 11 32.26 & $+$00 58 49.9 &  $29.70\pm0.02$ &  $-$26.53 & $-$105.10 &        &        & 1.187 &     $3.81\pm1.04$  \\
RTW 2235+0032 A   & 22 35 36.48 & $+$00 32 32.6 &  $24.32\pm0.02$ &     12.58 &  $-$43.96 &   4.41 &  181.4 & 1.029 &     $9.94\pm2.30$  \\
RTW 2235+0032 B   & 22 35 36.49 & $+$00 32 37.0 &  $24.28\pm0.05$ &      8.69 &  $-$39.06 &        &        & 1.155 &     $9.25\pm3.79$  \\
RTW 2236+5923 A   & 22 36 32.41 & $+$59 23 33.0 &  $20.52\pm0.01$ &  $-$39.85 &  $-$14.81 &   5.56 &  270.5 & 1.174 &    $-1.55\pm0.65$  \\
RTW 2236+5923 B   & 22 36 31.72 & $+$59 23 31.3 &  $20.56\pm0.01$ &  $-$38.82 &  $-$10.70 &        &        & 1.303 &    $-0.50\pm0.68$  \\
RTW 2241-1625 A   & 22 41 10.44 & $-$16 25 10.7 &  $23.85\pm0.02$ &     64.97 &   $-$8.62 &   4.45 &  186.6 & 1.021 &     $1.83\pm0.62$  \\
RTW 2241-1625 B   & 22 41 10.72 & $-$16 25 08.8 &  $23.80\pm0.02$ &     62.12 &   $-$4.13 &        &        & 1.070 &     $2.87\pm0.74$  \\
RTW 2244+4030 A   & 22 44 06.16 & $+$40 29 58.1 &  $46.64\pm0.01$ &  $-$75.27 & $-$126.27 &  18.82 &  403.6 & 0.962 &    $12.06\pm0.16$  \\
RTW 2244+4030 B   & 22 44 04.51 & $+$40 29 58.5 &  $46.64\pm0.02$ &  $-$71.21 & $-$112.65 &        &        & 1.172 &    $10.61\pm0.19$  \\
L 718-71 A        & 23 00 33.63 & $-$23 57 15.8 &  $50.76\pm0.02$ &    191.29 & $-$345.50 &  74.58 & 1468.9 & 1.086 &     $0.80\pm0.25$  \\
L 718-71 B        & 23 00 36.82 & $-$23 58 16.2 &  $50.78\pm0.02$ &    194.13 & $-$347.39 &        &        & 1.041 &     $0.72\pm0.20$  \\
RTW 2311-5845 A   & 23 11 44.34 & $-$58 45 42.6 &  $22.77\pm0.02$ &     94.09 &     21.12 &   4.51 &  197.7 & 1.200 &    $-5.96\pm2.66$  \\
RTW 2311-5845 B   & 23 11 44.90 & $-$58 45 43.8 &  $22.84\pm0.02$ &     93.32 &     19.31 &        &        & 1.194 &       \nodata      \\
\enddata
\tablenotetext{\rm{\dag}}{Our B stars always adopt the same base name as their corresponding A star for clarity through this work, but the true common catalog names for several of the B components are as follows: LP~719-37~B $\rightarrow$ LP~719-38, G~103-63~B $\rightarrow$ G~103-64, LP~368-99~B $\rightarrow$ LP~368-98, L~533-2~B $\rightarrow$ L~533-3, LP~551-62~B $\rightarrow$ LP~551-61, L~1197-68~B $\rightarrow$ L~1197-67, L~1198-23~B $\rightarrow$ L~1198-22, G~230-39~B $\rightarrow$ G~230-38, L~718-71~B $\rightarrow$ L~718-70, and NLTT~44989~B $\rightarrow$ NLTT~44988. Likewise, the RTW~0143-0151~AB system also goes by SKF~116 or GIC~112 \citep{1980LowOB...8..157G}, GJ~745~A is Ross~730, and GJ~745~B is Ross~731 or HD~349726.}
\tablecomments{All 72 components from the 36 systems in our Full Sample are given, including the 32 New Systems reported in this paper and the four first detailed in \citetalias{RTW_P1}; the latter four are indicated with parentheses around their names. System entries are organized by increasing Right Ascension of the A components. Unresolved companions break a given system's twin nature, so the one confirmed case (KX~Com) uses additional component labels in the name while the two entries with suspected unseen companions are indicated with asterisks in their names. Measurements from Gaia DR3 are provided, including RUWE \citep[described in][]{RUWE_note}, and RV$_{\rm{Gaia}}$ where available. Angular separations are given for A-B pairs along with the corresponding projected physical separations based on their average distance. See Section~\ref{sec:Sample} for related details about the sample.}
\end{deluxetable}

%%%%%%%%%%%%%%% tab - Sample photometry %%%%%%%%%%%%%%%
\startlongtable
\centerwidetable
\begin{deluxetable}{lcccccc|ccc|cccc}
\tabletypesize{\footnotesize}
\tablewidth{0pt}
\tablecaption{Full RECONS Twin Binary Sample --- Photometry, Mass Estimates, and Outline of Observations\label{tab:SampleTable-phot}}
\tablehead{
\colhead{Name} & \colhead{\textit{G}} & \colhead{\textit{BP}} & \colhead{\textit{RP}} & \colhead{\textit{J}} & \colhead{\textit{H}} & \colhead{$K_s$} & \colhead{$M_G$} & \colhead{Mass} & \colhead{Type} & \colhead{0.9m-Long} & \colhead{0.9m-Rot} & \colhead{Speckle} & \colhead{CHIRON} \\
\colhead{} & \colhead{(mag)} & \colhead{(mag)} & \colhead{(mag)} & \colhead{(mag)} & \colhead{(mag)} & \colhead{(mag)} & \colhead{(mag)} & \colhead{($\rm{M_\odot}$)} & \colhead{} & \colhead{(years)} & \colhead{(N points)} & \colhead{} & \colhead{}
}
\startdata
RTW 0143-0151 A   & 14.77 & 16.45 & 13.53 & 11.74 & 11.18 & 10.90 & 11.41 &  0.214  & FC     & \nodata & \nodata & SOAR-u  & non-core \\
RTW 0143-0151 B   & 14.76 & 16.45 & 13.51 & 11.75 & 11.19 & 10.99 & 11.40 &  0.210  & FC     & \nodata & \nodata & SOAR-u  & non-core \\
(2MA 0201+0117 A) & 11.90 & 13.26 & 10.73 &  9.10 &  8.46 &  8.26 &  8.44 & (0.539) & PMS    & 4.0     & 230     & LDT     & core     \\
(2MA 0201+0117 B) & 11.95 & 13.33 & 10.79 &  9.15 &  8.53 &  8.27 &  8.49 & (0.534) & PMS    & 4.0     & 220     & LDT     & core     \\
RTW 0231-5432 A   & 12.71 & 14.21 & 11.52 &  9.91 &  9.35 &  9.11 & 10.91 &  0.271  & FC     & 4.2     & 263     & SOAR    & non-core \\
RTW 0231-5432 B   & 12.77 & 14.29 & 11.57 &  9.93 &  9.39 &  9.14 & 10.96 &  0.264  & FC     & 4.2     & 264     & SOAR    & non-core \\
RTW 0231+4003 A   & 13.71 & 15.25 & 12.50 & 10.80 & 10.22 &  9.95 & 10.45 &  0.310  & FC     & \nodata & \nodata & \nodata & \nodata  \\
RTW 0231+4003 B   & 13.71 & 15.26 & 12.50 & 10.79 & 10.20 &  9.95 & 10.46 &  0.309  & FC     & \nodata & \nodata & \nodata & \nodata  \\
RTW 0409+4623 A   & 13.75 & 15.26 & 12.55 & 10.93 & 10.36 & 10.10 & 10.92 &  0.267  & FC     & \nodata & \nodata & \nodata & \nodata  \\
RTW 0409+4623 B   & 13.79 & 15.30 & 12.59 & 10.99 & 10.40 & 10.16 & 10.96 &  0.264  & FC     & \nodata & \nodata & \nodata & \nodata  \\
KAR 0545+7254 A   & 11.99 & 13.31 & 10.85 &  9.34 &  8.70 &  8.45 &  9.71 &  0.401  & PC     & \nodata & \nodata & \nodata & \nodata  \\
KAR 0545+7254 B   & 12.04 & 13.38 & 10.90 &  9.40 &  8.76 &  8.51 &  9.77 &  0.394  & PC     & \nodata & \nodata & \nodata & \nodata  \\
LP 719-37 A       & 13.67 & 15.41 & 12.41 & 10.58 & 10.04 &  9.75 & 11.57 &  0.196  & FC     & 3.8     & 181     & SOAR    & core     \\
LP 719-37 B       & 13.69 & 15.43 & 12.44 & 10.63 & 10.07 &  9.78 & 11.60 &  0.194  & FC     & 3.8     & 181     & SOAR    & core     \\
G 103-63 A*       & 13.75 & 15.41 & 12.51 & 10.74 & 10.17 &  9.89 & 11.32 & (0.219) & FC     & \nodata & \nodata & \nodata & \nodata  \\
G 103-63 B        & 13.75 & 15.41 & 12.52 & 10.77 & 10.18 &  9.89 & 11.33 &  0.220  & FC     & \nodata & \nodata & \nodata & \nodata  \\
RTW 0824-3054 A   & 13.42 & 15.07 & 12.19 & 10.43 &  9.80 &  9.55 & 10.98 &  0.250  & FC     & 4.3     & \nodata & SOAR    & non-core \\
RTW 0824-3054 B*  & 13.44 & 15.10 & 12.21 & 10.42 &  9.83 &  9.56 & 11.01 & (0.247) & FC     & 4.3     & \nodata & SOAR    & non-core \\
LP 368-99 A       & 12.85 & 14.34 & 11.65 &  9.99 &  9.43 &  9.17 & 10.28 &  0.330  & Gap    & 2.9     & \nodata & LDT-X   & core     \\
LP 368-99 B       & 12.87 & 14.35 & 11.68 & 10.06 &  9.45 &  9.22 & 10.31 &  0.329  & Gap    & 2.9     & \nodata & LDT-X   & core     \\
L 533-2 A         & 12.14 & 13.46 & 10.99 &  9.52 &  8.46 &  8.65 &  9.79 &  0.394  & PC     & 3.4     & \nodata & SOAR    & non-core \\
L 533-2 B         & 12.17 & 13.51 & 11.02 &  9.49 &  8.47 &  8.63 &  9.82 &  0.387  & PC     & 3.4     & \nodata & SOAR    & non-core \\
RTW 0933-4353 A   & 15.34 & 18.05 & 13.90 & 11.48 & 10.83 & 10.43 & 14.11 &  0.096  & FC     & 4.1     & 77      & SOAR-u  & core     \\
RTW 0933-4353 B   & 15.39 & 18.08 & 13.94 & 11.54 & 10.88 & 10.52 & 14.17 &  0.095  & FC     & 4.1     & 73      & SOAR-u  & core     \\
LP 551-62 A       & 12.98 & 14.27 & 11.86 & 10.33 &  9.81 &  9.54 &  9.92 &  0.384  & PC     & \nodata & \nodata & LDT-X   & non-core \\
LP 551-62 B       & 13.00 & 14.28 & 11.88 & 10.39 &  9.84 &  9.58 &  9.93 &  0.379  & PC     & \nodata & \nodata & LDT-X   & non-core \\
RTW 1123+8009 A   & 14.46 & 16.40 & 13.16 & 11.23 & 10.59 & 10.33 & 12.41 &  0.142  & FC     & \nodata & \nodata & \nodata & \nodata  \\
RTW 1123+8009 B   & 14.47 & 16.41 & 13.17 & 11.24 & 10.62 & 10.35 & 12.42 &  0.142  & FC     & \nodata & \nodata & \nodata & \nodata  \\
RTW 1133-3447 A   & 13.23 & 14.40 & 12.16 & 10.75 & 10.16 &  9.94 &  9.83 &  0.406  & PC     & 4.3     & \nodata & SOAR    & non-core \\
RTW 1133-3447 B   & 13.28 & 14.46 & 12.21 & 10.78 & 10.23 &  9.97 &  9.88 &  0.399  & PC     & 4.3     & \nodata & SOAR    & non-core \\
(KX Com A)        & 12.66 & 14.07 & 11.48 &  9.86 &  9.33 &  9.09 & 10.47 &  0.319  & FC     & 2.2     & 214     & LDT     & core     \\
(KX Com BC)       & 12.65 & 14.12 & 11.46 &  9.83 &  9.29 &  9.04 & 10.47 & (0.316) & FC     & 2.2     & 223     & LDT     & core     \\
RTW 1336-3212 A   & 13.55 & 15.29 & 12.29 & 10.49 &  9.88 &  9.60 & 11.65 &  0.188  & FC     & 4.8     & \nodata & SOAR    & non-core \\
RTW 1336-3212 B   & 13.59 & 15.35 & 12.33 & 10.51 &  9.91 &  9.58 & 11.70 &  0.186  & FC     & 4.8     & \nodata & SOAR    & non-core \\
L 197-165 A       & 12.84 & 14.12 & 11.72 & 10.22 &  9.68 &  9.50 & 10.44 &  0.334  & FC     & 4.7     & \nodata & SOAR    & non-core \\
L 197-165 B       & 12.86 & 14.14 & 11.74 & 10.26 &  9.71 &  9.51 & 10.46 &  0.333  & FC     & 4.7     & \nodata & SOAR    & non-core \\
(GJ 1183 A)       & 12.46 & 14.27 & 11.17 &  9.31 &  8.70 &  8.40 & 11.24 &  0.215  & FC     & 10.2    & 281     & SOAR    & core     \\
(GJ 1183 B)       & 12.50 & 14.33 & 11.22 &  9.35 &  8.76 &  8.46 & 11.28 &  0.210  & FC     & 10.2    & 280     & SOAR    & core     \\
RTW 1433-6109 A   & 12.02 & 13.23 & 10.92 &  9.53 &  8.90 &  8.67 &  9.58 &  0.427  & PC     & 4.7     & \nodata & SOAR    & core     \\
RTW 1433-6109 B   & 12.02 & 13.24 & 10.92 &  9.50 &  8.93 &  8.63 &  9.59 &  0.424  & PC     & 4.7     & \nodata & SOAR    & core     \\
L 1197-68 A       & 12.61 & 13.87 & 11.50 & 10.04 &  9.46 &  9.21 &  9.93 &  0.387  & PC     & \nodata & \nodata & LDT-X   & non-core \\
L 1197-68 B       & 12.67 & 13.94 & 11.56 & 10.08 &  9.52 &  9.28 &  9.99 &  0.379  & PC     & \nodata & \nodata & LDT-X   & non-core \\
L 1198-23 A       & 12.38 & 13.70 & 11.25 &  9.74 &  9.16 &  8.92 &  9.73 &  0.401  & PC     & 3.1     & \nodata & LDT-X   & core     \\
L 1198-23 B       & 12.44 & 13.76 & 11.30 &  9.77 &  9.19 &  8.97 &  9.78 &  0.393  & PC     & 3.1     & \nodata & LDT-X   & core     \\
RTW 1512-3941 A   & 10.64 & 11.71 &  9.61 &  8.23 &  7.65 &  7.40 &  8.78 &  0.535  & PC     & 4.8     & \nodata & SOAR    & core     \\
RTW 1512-3941 B   & 10.66 & 11.74 &  9.62 &  8.26 &  7.62 &  7.39 &  8.79 &  0.532  & PC     & 4.8     & \nodata & SOAR    & core     \\
(NLTT 44989 A)    & 12.44 & 13.91 & 11.25 &  9.61 &  9.06 &  8.80 & 11.13 &  0.253  & FC     & 4.7     & 285     & SOAR    & core     \\
(NLTT 44989 B)    & 12.50 & 13.93 & 11.28 &  9.61 &  9.03 &  8.78 & 11.19 &  0.251  & FC     & 4.7     & 290     & SOAR    & core     \\
RTW 1812-4656 A   & 12.83 & 14.06 & 11.74 & 10.29 &  9.73 &  9.47 &  9.56 &  0.426  & PC     & 4.0     & \nodata & SOAR    & core     \\
RTW 1812-4656 B   & 12.84 & 14.07 & 11.74 & 10.32 &  9.71 &  9.50 &  9.57 &  0.426  & PC     & 4.0     & \nodata & SOAR    & core     \\
GJ 745 A          &  9.81 & 11.01 &  8.73 &  7.30 &  6.73 &  6.52 & 10.09 &  0.377  & Gap    & \nodata & \nodata & \nodata & non-core \\
GJ 745 B          &  9.81 & 11.01 &  8.72 &  7.28 &  6.75 &  6.52 & 10.07 &  0.378  & Gap    & \nodata & \nodata & \nodata & non-core \\
RTW 2011-3824 A   & 12.95 & 14.42 & 11.77 & 10.13 &  9.54 &  9.28 & 10.47 &  0.315  & FC     & \nodata & \nodata & SOAR    & non-core \\
RTW 2011-3824 B   & 13.01 & 14.48 & 11.81 & 10.18 &  9.60 &  9.33 & 10.52 &  0.309  & FC     & \nodata & \nodata & SOAR    & non-core \\
G 230-39 A        & 13.02 & 14.43 & 11.85 & 10.31 &  9.74 &  9.49 & 10.48 &  0.318  & FC     & \nodata & \nodata & \nodata & \nodata  \\
G 230-39 B        & 13.06 & 14.48 & 11.89 & 10.30 &  9.76 &  9.52 & 10.52 &  0.313  & FC     & \nodata & \nodata & \nodata & \nodata  \\
RTW 2202+5537 A   & 14.24 & 16.16 & 12.94 & 11.01 & 10.42 & 10.11 & 12.18 &  0.152  & FC     & \nodata & \nodata & \nodata & \nodata  \\
RTW 2202+5537 B   & 14.26 & 16.17 & 12.95 & 11.04 & 10.45 & 10.18 & 12.20 &  0.152  & FC     & \nodata & \nodata & \nodata & \nodata  \\
RTW 2211+0058 A   & 13.52 & 14.98 & 12.34 & 10.69 & 10.15 &  9.90 & 10.88 &  0.275  & FC     & \nodata & \nodata & LDT-X   & non-core \\
RTW 2211+0058 B   & 13.56 & 15.02 & 12.37 & 10.75 & 10.20 &  9.95 & 10.92 &  0.271  & FC     & \nodata & \nodata & LDT-X   & non-core \\
RTW 2235+0032 A   & 13.88 & 15.40 & 12.68 & 11.00 & 10.49 & 10.21 & 10.81 &  0.277  & FC     & 2.1     & 205     & LDT-X   & core     \\
RTW 2235+0032 B   & 13.91 & 15.46 & 12.70 & 10.98 & 10.46 & 10.21 & 10.84 &  0.272  & FC     & 2.1     & 206     & LDT-X   & core     \\
RTW 2236+5923 A   & 13.10 & 14.37 & 11.98 & 10.52 &  9.89 &  9.70 &  9.66 &  0.412  & PC     & \nodata & \nodata & \nodata & \nodata  \\
RTW 2236+5923 B   & 13.16 & 14.45 & 12.04 & 10.54 &  9.96 &  9.72 &  9.72 &  0.404  & PC     & \nodata & \nodata & \nodata & \nodata  \\
RTW 2241-1625 A   & 11.65 & 12.66 & 10.65 &  9.39 &  8.49 &  8.24 &  8.54 &  0.567  & PC     & 4.1     & 185     & SOAR    & core     \\
RTW 2241-1625 B   & 11.70 & 12.73 & 10.70 &  9.44 &  8.49 &  8.25 &  8.59 &  0.561  & PC     & 4.1     & 186     & SOAR    & core     \\
RTW 2244+4030 A   & 10.52 & 11.57 &  9.50 &  8.20 &  7.61 &  7.39 &  8.86 &  0.527  & PC     & \nodata & \nodata & \nodata & \nodata  \\
RTW 2244+4030 B   & 10.52 & 11.58 &  9.50 &  8.17 &  7.59 &  7.38 &  8.87 &  0.526  & PC     & \nodata & \nodata & \nodata & \nodata  \\
L 718-71 A        & 10.68 & 11.82 &  9.63 &  8.25 &  7.67 &  7.41 &  9.21 &  0.477  & PC     & 4.2     & \nodata & SOAR    & non-core \\
L 718-71 B        & 10.69 & 11.83 &  9.64 &  8.26 &  7.66 &  7.42 &  9.22 &  0.476  & PC     & 4.2     & \nodata & SOAR    & non-core \\
RTW 2311-5845 A   & 14.35 & 16.01 & 13.11 & 11.36 & 10.76 & 10.49 & 11.14 &  0.235  & FC     & 4.2     & 184     & SOAR-u  & core     \\
RTW 2311-5845 B   & 14.37 & 16.06 & 13.12 & 11.34 & 10.77 & 10.48 & 11.16 &  0.231  & FC     & 4.2     & 184     & SOAR-u  & core     \\
\enddata
\tablecomments{Photometric properties from Gaia DR3 and 2MASS are provided for all 72 components in our Full Sample. The four systems from \citetalias{RTW_P1} are indicated with parentheses around their names. Cases with known or suspected unresolved companions are indicated with component labels and asterisks the same as in Table~\ref{tab:SampleTable-astr}. Mass estimates were obtained using the \cite{Benedict_2016} MLR as detailed in \citetalias{RTW_P1} --- less reliable masses are indicated with parentheses, as discussed in Section~\ref{sec:Sample} and \citetalias{RTW_P1}. Stars are categorized as partially convective (PC), fully convective (FC), or within the transition gap between the two cases (Gap), based exclusively on their positions in the HRD relative to the Jao Gap \citep{2018ApJ...861L..11J, Jao_2023}. 2MA~0201+0117~AB is pre-main-sequence (PMS), as highlighted in \citetalias{RTW_P1}, but we often show it as FC for plotting purposes given its similar interior structure. For stars on the 0.9\,m long-term program, the time spans of their observations as of our most recent analysis for each are given. For stars observed at the 0.9\,m for rotation, the numbers of rotation data points used are given. Cases observed with speckle imaging indicate the instrument used (SOAR+HRCam or LDT+QWSSI), along with `u' flags for noisy low-quality observations and `X' flags for those with unusably bad signal-to-noise ratios (SNRs). Stars targeted by CHIRON for spectra indicate if they are in our ``core" sample with $\sim$5 visits or the ``non-core" group with only $\sim$1 spectral visit.}
\end{deluxetable}

%%%%%%%%%%%%%%% fig - RTWINS HRD %%%%%%%%%%%%%%%

\subsection{Young Association Membership} \label{subsec:youth-checks}

Of the four systems in \citetalias{RTW_P1}, 2MA~0201+0117~AB is a member of the young $\beta$ Pictoris association \citep{2015A&A...583A..85A, 2017A&A...607A...3M}\footnote{The recent $\beta$ Pictoris membership analysis by \citet{2024MNRAS.528.4760L} does not include 2MA~0201+0117~AB as a confirmed member, but we regard 2MA~0201+0117~AB as a member and pre-main-sequence star for the rest of this work based on the membership analyses by \citet{2015A&A...583A..85A} and \citet{2017A&A...607A...3M} and the stars' elevated nature in our HRD.} and the others are field stars. All 64 components in the 32 New Systems were also assessed for membership in nearby young associations using the BANYAN $\Sigma$ tool of \cite{banyan_2018}. We used Gaia DR3 astrometry, along with our own CHIRON radial velocities (RVs, Section~\ref{sec:chiron}) for observed systems; otherwise Gaia DR3 radial velocities were used when available. Our binaries may slightly deviate in RV compared to the systemic velocity owing to orbital motion, so we assessed BANYAN $\Sigma$ results both with and without RVs included. We found candidate memberships for RTW~0933-4353~AB and RTW~2202+5537~AB, as discussed next, but ultimately disregard these as not members of nearby young associations. All remaining New System targets present as high confidence field stars.

For RTW~0933-4353~AB, the BANYAN $\Sigma$ tool yields 70\% and 41\% probabilities of membership in the young 40--50\,Myr Argus association for A and B, respectively \citep{banyan_2018, Zuckerman2019}, although these drop to 18\% and 13\% when incorporating our mean CHIRON RVs. The extremely rapid $\sim$3\,hr rotation periods we find for the components (Section~\ref{sec:rot}) make it likely that very little spindown has occurred, also possibly supporting a young age. However, our mean CHIRON RVs at $19.70\pm2.46$\,$\mathrm{km\,s^{-1}}$ and $21.78\pm4.41$\,$\mathrm{km\,s^{-1}}$ for A and B, respectively, disagree with the RVs of $11.9\pm1.7$\,$\mathrm{km\,s^{-1}}$ predicted for both A and B by BANYAN $\Sigma$ if the Argus membership were real. In addition, the components do not appear elevated above the main sequence in Figure~\ref{fig:HRD}, as expected if the stars were truly only $\sim$50\,Myr old given their low masses of $\sim$0.1\,$\rm{M_\odot}$. We conclude that the system is most likely not a member of Argus.

For RTW~2202+5537~AB, the BANYAN $\Sigma$ tool yields 71\% and 72\% probabilities of membership in the young $\sim$150\,Myr AB Doradus moving group for A and B, respectively \citep{2015MNRAS.454..593B, banyan_2018}. This drops to 0\% for A when incorporating the Gaia DR3 RV of $-7.47\pm2.57$\,$\mathrm{km\,s^{-1}}$, while B does not have an RV available from Gaia, and neither component has an RV from our CHIRON work. The Gaia RV for A falls within roughly 1-sigma of the moving group's membership RVs given in \cite{banyan_2018} as $10^{+10}_{-20}$\,$\mathrm{km\,s^{-1}}$ but is discrepant from the RV value of $-26.6\pm0.9$\,$\mathrm{km\,s^{-1}}$ that BANYAN $\Sigma$ predicts for the A component if it were a true member. The pair is also not noticeably elevated above the main sequence in Figure~\ref{fig:HRD}. We conclude that the system is most likely not a member of AB Dor.

\subsection{Twin Status Validation Checks} \label{subsec:multiplicity-checks}

We made extensive efforts to confirm that the systems were true binaries that have no additional components. Crossmatching our systems to the SUPERWIDE catalog of wide binaries from \citet{2020ApJS..247...66H} reveals 30 of our 32 New Systems in the catalog, all deemed to be real binaries at $>$99\% probability, with most at $>$99.99\% probability. The remaining two systems, RTW~1123+8009~AB and RTW~1812-4656~AB, are not in SUPERWIDE owing to their proper motions falling below the 40\,$\rm{mas\,yr^{-1}}$ cutoff of SUPERWIDE. However, these two systems are present in the wide binary catalog of \cite{2018MNRAS.480.4884E}, have components with matching proper motions and parallaxes, display similar component RVs in Gaia DR3 (Table~\ref{tab:SampleTable-astr}), and have similar RVs from CHIRON in the case of RTW~1812-4656~AB (Section~\ref{sec:chiron}). We conclude that all 32 New Systems are reliable wide binaries.

More massive primaries or unresolved companions could invalidate our desired twin status for these systems. We have searched for additional components in the New Systems using largely the same checks as described in \citetalias{RTW_P1}, briefly summarized here. First, we found no astrometrically associated Gaia DR3 sources within a 10,000\,au 2D projected radius around each target star. We then matched system components against the Gaia DR3 non-single stars (NSS) catalogs \citep{2022yCat.1357....0G}, Washington Double Star Catalog \citep[WDS;][]{2001AJ....122.3466M}\footnote{WDS reports our twin pair LP~551-62~AB as the CD components of a larger ABCD system, but an investigation using archival DSS images \citep{1996ASPC..101...88L} and Gaia DR3 astrometry indicates the additional sources are astrometrically unaffiliated with the twin pair, leaving LP~551-62~AB as a twin system.}, SB9 spectroscopic binary catalog \citep{2004A&A...424..727P}, and Table 11 of \cite{Winters_2019}. We found only RTW~0824-3054~B to be likely hiding an additional companion based on Gaia NSS reporting an acceleration model compatible with its non-linear proper motions \citep{2023A&A...674A...9H}. The following cases were also observed with Robo-AO high-angular-resolution imaging in search of close-in companions \citep{2020AJ....159..139L}, but none were found: G~103-63~B, G~230-39~A and B, LP~551-62~A, L~1197-68~A, and L~1198-23~A and B. Finally, L~533-2~B was observed by the RAVE survey at five epochs over roughly 2 years \citep{2020AJ....160...82S}, with blending likely given the 6\farcs7 RAVE fiber size and 4\farcs13 AB separation, but no variable RV signal beyond the typical 1--2\,$\mathrm{km\,s^{-1}}$ uncertainties was found.

We also assessed the following relevant Gaia DR3 parameters --- which are detailed further in \citetalias{RTW_P1} --- and report most of them in Table~\ref{tab:SampleTable-astr}. \texttt{ipd\_frac\_multi\_peak} (IPDfmp) is $\leq$\,5\% for all 32 systems. RUWE is always $<$\,1.7, the ideal cutoff for nearby M dwarfs based on RECONS work by \cite{Vrijmoet_2020} and M.~R.~LeBlanc et al.~(2025, in preparation), except for the aforementioned RTW~0824-3054~B at $\rm{RUWE}=2.209$. The Gaia RV error is not markedly elevated above the expected value at each target's brightness based on comparison to many nearby single M dwarfs (M.~R.~LeBlanc et al.~2025, in preparation) --- and is $\lesssim$\,4\,$\mathrm{km\,s^{-1}}$ --- for all 32 systems, except G~103-63~A at $31.34\pm17.21$\,$\mathrm{km\,s^{-1}}$. Parallax errors are always $\lesssim$\,0.04\,mas, except for RTW~2235+0032~B at 0.0481\,mas, which is coupled with a slightly elevated $\sigma_{\rm{RV_{Gaia}}}$ of 3.79\,$\mathrm{km\,s^{-1}}$ compared to the A component's $\sigma_{\pi}$ of 0.0180\,mas and $\sigma_{\rm{RV_{Gaia}}}$ of 2.30\,$\mathrm{km\,s^{-1}}$. However, our own RVs measured at five epochs over 0.4\,years appear non-varying for both RTW~2235+0032~A and B down to the average single-visit noise of $\sim$0.22\,$\mathrm{km\,s^{-1}}$, and the Gaia RUWE and IPDfmp values are consistent with single stars for both components. We subsequently choose to treat the RTW~2235+0032~AB system as a true twin pair (i.e., it has no unresolved companions) in our results.

We further search for unseen companions to our New Systems in our observational analyses here using speckle imaging (Section~\ref{sec:speckle}), time series RVs (Section~\ref{sec:chiron}), and long-term ground-based astrometry (Section~\ref{sec:longterm}), but find no additional candidate or confirmed companions.

In summary, unresolved components are likely in the RTW~0824-3054 and G~103-63 systems, so we deem these to not be twin pairs. These are noted in Table~\ref{tab:SampleTable-astr} with asterisks --- their mass estimates in Table~\ref{tab:SampleTable-phot} are therefore likely erroneous as well. Combined with the four systems in \citetalias{RTW_P1}, in which only KX~Com is a triple, we conclude that 33 of the 36 systems in our twin study present as binaries with no additional components.

\section{Overview of Observations} \label{sec:obs}

The ultimate goals of our observations are to help rule out hidden companions and compare rotation rates and activity levels between twin components. This work utilizes largely the same observing campaigns and methods as outlined in \citetalias{RTW_P1}, with the exception of the Chandra X-ray observations that only occurred for the four targets in \citetalias{RTW_P1}. Given the breadth of our observations, we include both the methodology and results in unified sections focused on each observing campaign in turn to aid straightforward reading. To summarize, we have employed: (1) speckle imaging with HRCam on SOAR and QWSSI on LDT to identify potential additional companions in Section~\ref{sec:speckle}, (2) optical spectral observations with CHIRON on the CTIO/SMARTS 1.5\,m to check for RV companions and measure H$\alpha$ activity in Section~\ref{sec:chiron}, (3) multi-year photometry and astrometry via the RECONS long-term program on the CTIO/SMARTS 0.9\,m to search for astrometric perturbations from unseen companions and photometric variations from stellar activity cycles in Section~\ref{sec:longterm}, and (4) short-term photometry with the same 0.9\,m targeting rotation rates in Section~\ref{sec:rot}. The rotation investigation is further supported by substantial archival light curve data from Gaia DR3 \citep{GaiaDR3_Var}, ZTF \citep{ZTF, ZTF_data}, TESS \citep{TESS}, and ASAS-SN \citep{Shappee_2014, 2019MNRAS.486.1907J}.

The observations acquired, or lack thereof, for each star in the Full Sample are indicated in Table~\ref{tab:SampleTable-phot}. The various subsets targeted by each campaign are described in the observing programs' respective content across Sections~\ref{sec:speckle}--\ref{sec:rot}. All observational and analysis methodologies proceeded as described in \citetalias{RTW_P1}, with some factors expanded upon here where relevant or noted for exceptions. Details here refer to the 32 New Systems unless otherwise specified, as the remaining four systems were discussed extensively in \citetalias{RTW_P1}. Next, we discuss blending and contamination and then embark on a tour of the observational programs.

\section{Blending and Contamination} \label{sec:contam}

\begin{figure}[!t]
\centering
\includegraphics[width=0.49\textwidth]{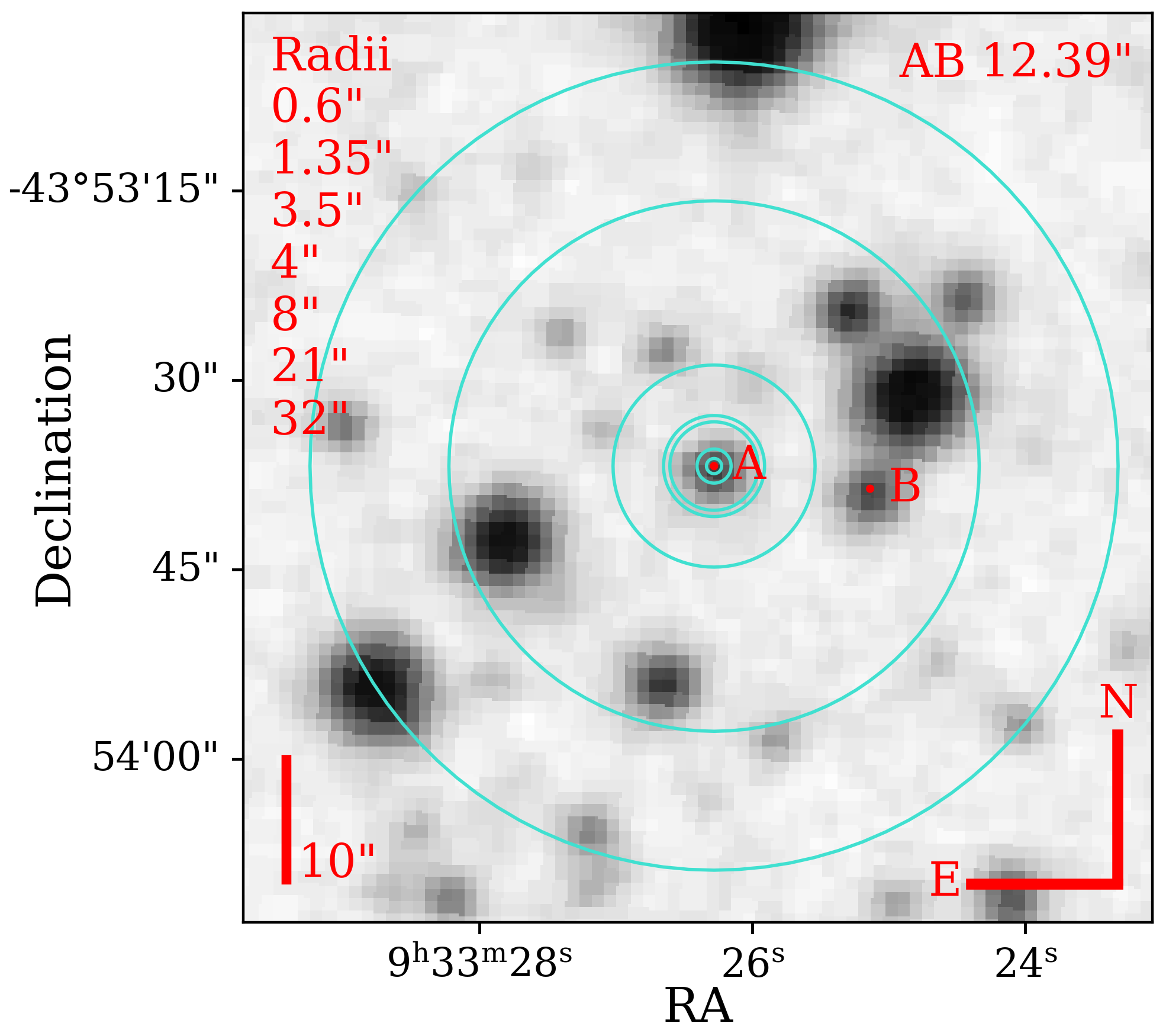}
\figcaption{An example archival image from the DSS2-Red survey \citep{1996ASPC..101...88L, 2004AJ....128.3082G} showing our twin stars RTW~0933-4353~A and B separated by 12$\arcsec$, with A and B shifted by their Gaia DR3 proper motions to 1998.0 for alignment with the image. In order of their listed increasing radii, the blue circles correspond to: (1) the typical 1\farcs2 seeing at CTIO for our SMARTS 0.9\,m and 1.5\,m observations, (2) the 2\farcs7 diameter CHIRON spectral fiber, (3) the 3\farcs5 Gaia BP and RP measurement region, (4) the 4$\arcsec$ minimum separation of our AB pairs and rough boundary for blending sources in our 0.9\,m observations depending on the seeing, (5) the 8$\arcsec$ boundary near which sources begin to blend in ZTF, (6) a single 21$\arcsec$ TESS pixel while keeping in mind that a common TESS aperture is a 3$\times$3 pixel grid, and (7) the 32$\arcsec$ boundary at which sources begin to blend in ASAS-SN data. All of our analyses and results here consider the variety of possible AB blending and background contamination impacts, as detailed in Section~\ref{sec:contam} and Section~\ref{subsec:rot-contam-syst}. \label{fig:field-radii}}
\end{figure}

Here we consider blending and contamination in the apparent magnitude measurements used for our sample construction (Section~\ref{subsec:contam-mags}) and then for the new data from our observing campaigns (Section~\ref{subsec:contam-new-obs}), ultimately concluding that the apparent magnitudes utilized and the derived results from our new observations are largely unaffected by AB blending or background contamination. Significant impacts due to blending and contamination in the additional archival data used as part of our rotation analysis are discussed later in Section~\ref{subsec:rot-contam-syst} as part of our rotation content. One target field and various measurement regions of interest are shown as examples in Figure~\ref{fig:field-radii}.

%%%%%%%%%%%%%%% fig - Example field for RTW0933-4353AB with various radii circles %%%%%%%%%%%%%%%

\subsection{Apparent Magnitudes} \label{subsec:contam-mags}

Using the same methodology described in \citetalias{RTW_P1}, we assessed the BP, RP, \textit{J}, \textit{H}, and $K_s$ magnitudes for possible contamination given their importance in our twin sample selection criteria. Starting with the Gaia measures, all 64 components in the 32 New Systems have (1) AB separations $>$\,4$\arcsec$ so do not blend A and B, (2) $<$\,1\% additional contaminating optical flux based on the $G$ mags of any nearby sources $<$\,3\farcs5 away, and (3) $\le$\,2.0\% excess flux in BP and RP based on the corrected\footnote{Factors were corrected to C* using the relations provided in \cite{2021A&A...649A...3R}.} \texttt{phot\_bp\_rp\_excess\_factor} parameter, altogether confirming no impactful contamination in the BP or RP magnitudes. For 2MASS magnitudes, mild blending between the wings of A and B sources is likely for our closer separation pairs $\lesssim$\,8$\arcsec$ apart, with nearby background sources possibly influencing the measures in some cases as well. By the same arguments given in \citetalias{RTW_P1} we do not expect this contamination in 2MASS to influence the results of our work, particularly because the higher resolution BP magnitudes used to derive our mass estimates already find mass differences $<3\%$ for each AB pair. 

\subsection{New Observations} \label{subsec:contam-new-obs}

In the new data from our four observing campaigns, A and B components are not blended in any systems at an impactful level. Some of the observed systems have $>$\,6$\arcsec$ separations, the point at which there are no AB blending concerns in any of the new observations, while the other pairs between 4--6$\arcsec$ were intentionally only observed in suitably good seeing to inhibit blending. For context, the 0.9\,m has 401\,mas pixels, the CHIRON spectral fiber is 2\farcs7 in diameter, and both observe from CTIO with 1--2$\arcsec$ seeing, as illustrated in Figure~\ref{fig:field-radii}. All 0.9\,m frames from both the long-term and rotation campaigns were manually reviewed to exclude any images with overlapping AB profile wings. We further tested the 0.9\,m rotation results with stricter seeing cuts on the data for a close-separation pair and confirmed our results remain consistent. For systems with multiple CHIRON visits, the H$\alpha$ equivalent width, RV, and $v\sin(i)$ results were assessed across the multiple epochs and showed no evident trends with changing seeing. Finally, the inherently high spatial resolution of speckle imaging by SOAR and LDT prevents any blending in those results.

Our new observations are also free from any impactful contamination from background sources near our target stars, but with some necessary notation. At the 2021.0 epoch, 47 of our 64 components have no background Gaia DR3 sources within 6$\arcsec$, leaving them entirely uncontaminated in any reasonable seeing conditions. Another 14 components do have background sources within 6$\arcsec$, but only with relative brightnesses of $\Delta G > 5$\,mag fainter, so the contamination is suitably faint to disregard. One of the three remaining cases is RTW~1123+8009~A with a source 5\farcs30 away at $\Delta G=2.19$\,mag fainter, but this twin component was not targeted by any of our new observing campaigns so the potential contamination is ignored. The other two remaining cases are RTW~0933-4353~A and B, where A is negligibly impacted by a single fainter $\Delta G=4.34$\,mag source 3\farcs50 away, but B has a point source 3\farcs66 away that is 1.35 $G$ mag brighter than the M dwarf target. We therefore only used 0.9\,m measurements for RTW~0933-4353~B with exquisite seeing $\lesssim$\,1\farcs2 along with a careful inspection of the frames for contamination, with a single contaminated CHIRON epoch thrown out for B. 

While the above rules out contamination at a single epoch near our observations, the long-term 0.9\,m campaign data series used here approach five years of coverage. This could result in a false variability signal if a star's proper motion carried it toward or away from background sources over time, especially with changing seeing. The largest proper motion shift over five years for a 0.9\,m target in our New Systems is $\sim$2\farcs5, so based on the resolution parameters mentioned above we conservatively selected any of the 0.9\,m targets with Gaia DR3 sources within 10$\arcsec$ at the 2021.0 epoch, yielding 13 components requiring review. Our careful inspection of the relative source locations, proper motion vectors, 0.9\,m images, $G$ magnitudes, and long-term 0.9\,m light curve results overall finds 4 of the 13 are entirely uncontaminated and the other 9 are only minimally contaminated by of order a few mmag depending on seeing and epoch. Impacts are thus well within the typical 0.9\,m noise level of $\sim$7--10\,mmag and do not markedly influence the takeaways of our results. A single exception, RTW~1133-3447~A, has no background sources within 10$\arcsec$ but sits on the large diffraction spike of a much brighter star about 25$\arcsec$ away --- A shows more photometric scatter than B in our long-term 0.9\,m results (Section~\ref{sec:longterm}), likely because of this diffraction spike contamination.

We also take advantage of the proper motions of our nearby stars along with archival DSS images from several decades ago to check for bright contaminating sources possibly hiding directly behind our targets today ($\lesssim$\,1$\arcsec$) and thereby not detected in Gaia as separate sources. For many targets in the New Systems we were able to entirely rule out any contaminating sources directly behind the stars at the epochs of our new observations, while a subset only have similar-brightness contaminating sources ruled out but can't entirely rule out fainter sources possibly lurking in the wings owing to insufficient proper motion shifts. We found no cases with background sources hiding directly behind our stars at present-day epochs within the brightness and separation limits that the proper motions and archival images allowed us to probe.

\section{Speckle Imaging --- SOAR and LDT} \label{sec:speckle}

The goal of our speckle work is to identify possible unseen stellar companions within $\lesssim$\,1$\arcsec$ of our targets that could invalidate the twin natures of our pairs.

\subsection{Speckle Observations and Data Processing} \label{subsec:speckle-methods}

All components in our 16 new southern systems at declination~$<$\,0\textdegree~were observed with the speckle imaging camera HRCam on SOAR and processed with the associated pipeline \citep{Tokovinin_2010, Tokovinin_2016, Tokovinin_2018}. The majority were newly observed through coauthor Vrijmoet's project \citep{Vrijmoet_2022}, while a few had already been captured by other SOAR speckle programs. All observations occurred during 2019--2022, with a range of 1--4 visits depending on the system. Three systems had notably worse signal-to-noise ratios (SNRs) than typical and so have more uncertain results, most often simply due to especially faint target stars; these cases are identified in Table~\ref{tab:SampleTable-phot} with a `u' symbol.

Of the 16 new northern pairs, components in six systems were observed with QWSSI on LDT \citep{QWSSI_2020} to cover the seven declination~$>$\,0\textdegree~pairs for which CHIRON spectra were also obtained. The one excluded case is GJ~745~AB, which was not observed with LDT because various efforts over the years have already directly or indirectly assessed both components for possible additional companions \citep[e.g.,][]{2010ApJ...725..875I, 2013MNRAS.429..859J, 2019A&A...623A..72K, 2020AJ....159..139L}, finding none except the AB wide binary. Unfortunately, all six systems observed with QWSSI ultimately had observations and results quite poor in SNR and were subsequently functionally useless (as indicated with X's in Table~\ref{tab:SampleTable-phot}), so we were unable to explore these six with speckle checks. However, we note that components in three of these six systems --- LP~368-99~AB, L~1198-23~AB, and RTW~2235+0032~AB --- have multi-epoch CHIRON spectra spanning 0.3--2.0\,yrs that found no RV variations (Section~\ref{sec:chiron}), thus providing some validation against unresolved companions in these cases despite the lack of speckle results.

\subsection{Speckle Results} \label{subsec:speckle-results}

No new sources were revealed among the 32 stars in 16 southern systems observed at SOAR, with average I-band detection limits of approximately $\Delta \rm{mag} \approx 2.3$ at 0\farcs15 and $\Delta \rm{mag} \approx 3.7$ at 1\farcs0. For context, a 0\farcs15 angular separation corresponds to a range of physical separations 1.3--7.3\,au for the range of distances 8.8--48.7\,pc across the 16 pairs. This provides useful validation against additional sources that could disrupt our twin comparisons. Intriguingly, the suspected unresolved companion to RTW~0824-3054~B noted in Section~\ref{subsec:multiplicity-checks} was {\it not} revealed by the speckle data, so it presumably exists below the speckle detection limits. The other component with a possible unresolved companion, G 103-63 A, was not observed with speckle imaging.

\section{Radial Velocities and H$\alpha$ Activity --- SMARTS 1.5\,m and CHIRON} \label{sec:chiron}

Here we discuss our spectral observations that seek to compare H$\alpha$ activity levels between components and search for possible RV variations from unresolved companions that would make the systems not true twins. Both efforts use spectra from the CHIRON high-resolution spectrograph at the SMARTS 1.5\,m in Chile \citep{Tokovinin_2013, Paredes_2021}.

\subsection{CHIRON Observations and Data Processing} \label{subsec:chiron-methods}

We acquired new CHIRON spectra at $R\approx27,000$ for the 23 New Systems observable from CTIO, using the same observational setup and data reduction routines as described in \citetalias{RTW_P1}. Targets were split into ``core" cases receiving $\sim$5 spectral visits spanning baselines of 0.3--2.0\,yrs and ``non-core" cases receiving only one visit, as indicated in the final column of Table~\ref{tab:SampleTable-phot}. Two of the non-core systems received additional second visits because of noisy or unclear H$\alpha$ results in their initial visits. The core systems were chosen to investigate each category of A/B component activity behaviors seen across our initial single-visit assessments --- both in emission, both in absorption, or mismatched H$\alpha$ signatures. Multiple spectral visits allow us to account for activity level changes, identify and remove flaring outlier cases, and search for possible RV variations. RVs, $v\sin(i)$ values, and H$\alpha$ EWs were measured using the same process as described in \citetalias{RTW_P1}.

For the faint targets RTW~0143-0151~B, RTW~0933-4353~A and B, and RTW~2311-5845~B, poor SNRs led to large and erroneous RV and $v\sin(i)$ measurements from certain spectral orders in certain epochs. These specific bad order cases were manually excluded such that our RV and $v\sin(i)$ values for these four stars were derived using only the remaining reliable measures. The CHIRON RVs for RTW~0143-0151~AB and RTW~2311-5845~AB are in good agreement between components in each system and with Gaia DR3 RVs. RTW~0933-4353~A and B have CHIRON RVs in agreement within their errors, but do not have Gaia DR3 RVs available to compare.

RTW~0933-4353~AB, the optically faintest system in our Full Sample, also displayed H$\alpha$ lines clearly and consistently offset from its rest frame wavelength over multiple epochs even after de-shifting by the initial measured stellar RVs. The deviations do not visually indicate SB2 behavior and they manifest similarly in both A and B, implying poor RV measurements due to the generally poor SNRs; no other stars had this noisy line offset issue. We thus added a corrective offset to each component's initial RVs based on the Gaussian fit central wavelength of H$\alpha$ at each epoch relative to the rest frame H$\alpha$ wavelength, and the resulting corrected RVs and H$\alpha$ positions are more consistent between epochs; our reported RVs are the values after this extra H$\alpha$-based correction, with the uncertainties left unchanged. The net impact is improved, but still poor, RV measures for RTW~0933-4353~A and B, only ruling out companions with RV amplitudes greater than the typical $\sim$10\,$\mathrm{km\,s^{-1}}$ single-visit uncertainties for this faint system.

Finally, we also considered the Li 6708\,\text{\AA} doublet as a possible age indicator, but there were no confident signals in any of our CHIRON spectra.

\subsection{CHIRON RV and $v\sin(i)$ Results} \label{subsec:chiron-rv-results}

The mean RV and $v\sin(i)$ results for our New Systems are reported in Table~\ref{tab:SpecTable}, with corresponding individual measurements reported in Table~\ref{tab:all-spec}. We reach typical single-visit RV uncertainties of $\sim$0.3\,$\mathrm{km\,s^{-1}}$ and single-visit $v\sin(i)$ uncertainties of 1.3\,$\mathrm{km\,s^{-1}}$, although our $v\sin(i)$ values $<$\,10\,$\mathrm{km\,s^{-1}}$ are less reliable, per \citet{Jao_2023}. None of our New Systems with multiple epochs of spectra show RV variations above the noise, nor do we see $v\sin(i)$ variations of significance. For reference, a 20\,$M_{\rm{Jup}}$ brown dwarf secondary in an edge-on, one-year, circular orbit around a 0.25\,$\rm{M_\odot}$ M dwarf would cause a velocity semi-amplitude of 1.5\,$\mathrm{km\,s^{-1}}$, well above our noise floor of $\sim$0.3 $\mathrm{km\,s^{-1}}$. For potential close-in tidally interacting companions, the $\sim$7-day orbital period tidal circularization timescale of M dwarfs \citep{Vrijmoet_thesis_2023} and above configuration would yield RV semi-amplitudes above our $\sim$0.3\,$\mathrm{km\,s^{-1}}$ noise floor for companions down to masses of nearly 1\,$M_{\rm{Jup}}$. Thus, we broadly rule out any stellar companions bright enough to affect any photometric or spectroscopic results reported here, or close enough to tidally enhance activity, for our targets with multiple epochs of CHIRON spectra.

\begin{figure}[!t]
\centering
\gridline{\fig{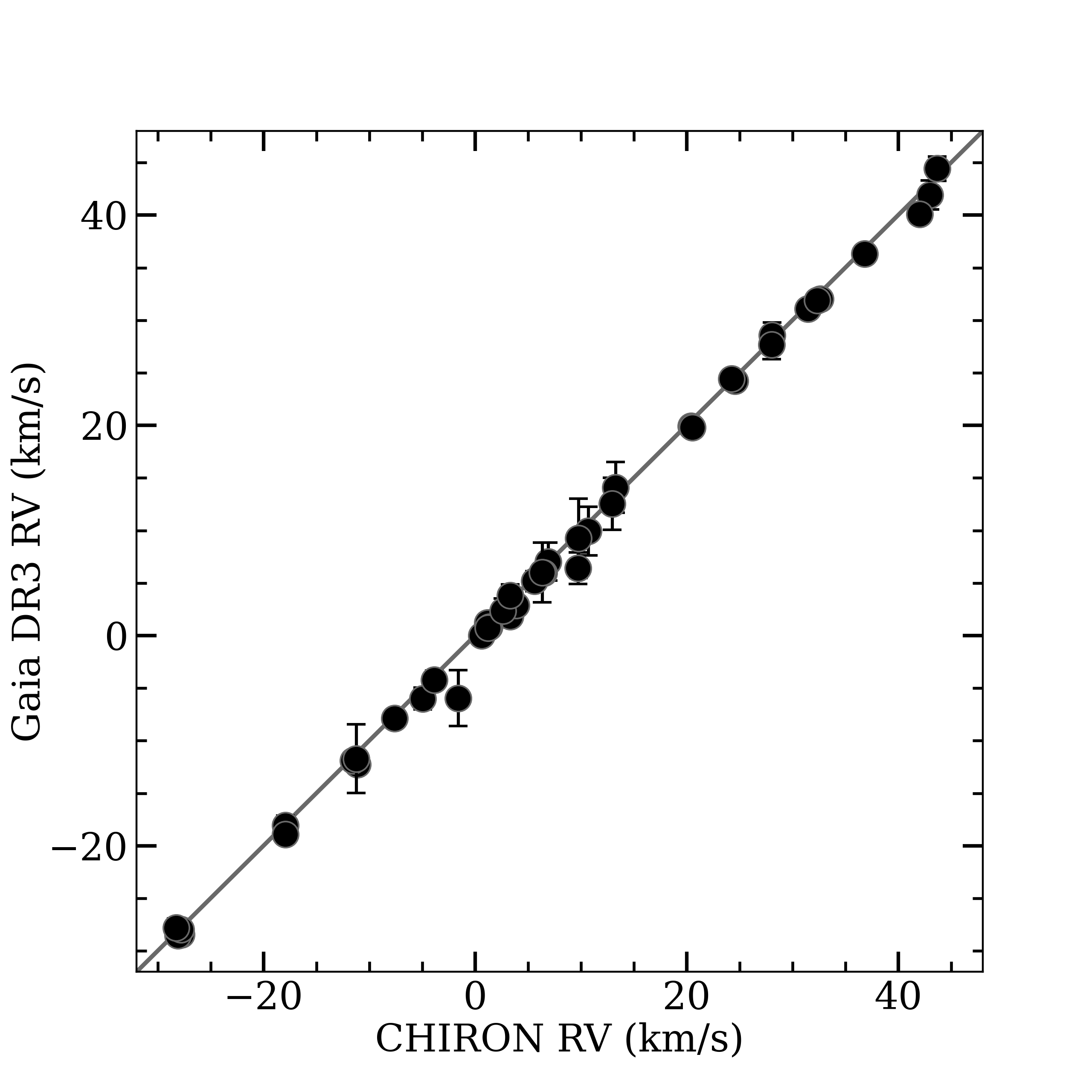}{0.495\textwidth}{(a)}
          \fig{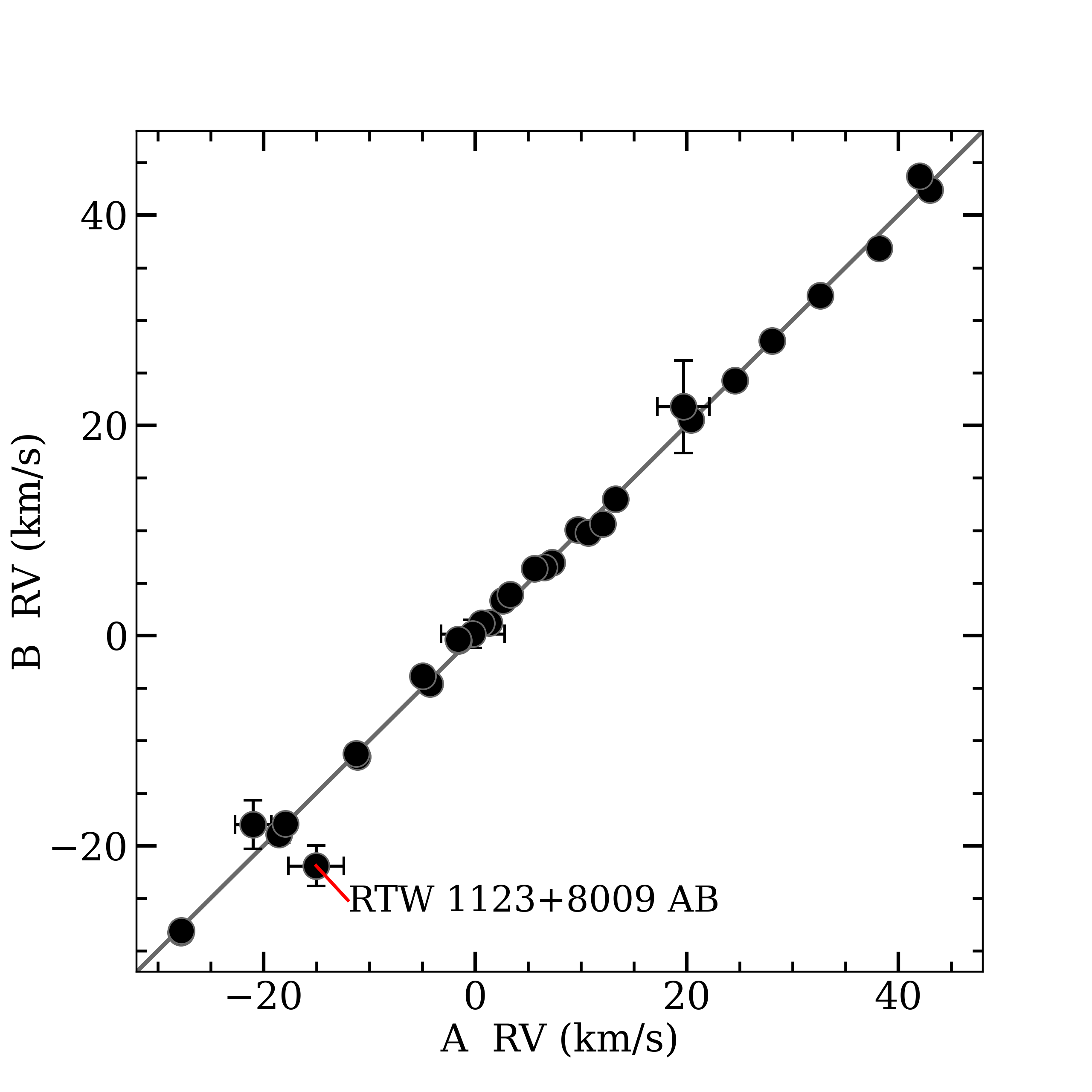}{0.495\textwidth}{(b)}}
\figcaption{Two equivalency plots assessing radial velocity measurements for the stars in our Results Sample. Gray one-to-one lines are underplotted. Error bars are always shown but often appear smaller than the points. (a) --- A comparison of our CHIRON RVs (Table~\ref{tab:SpecTable} and \citetalias{RTW_P1}) and the Gaia DR3 RVs (Table~\ref{tab:SampleTable-astr}) for the 43 individual stars with both measures available, using the weighted mean RVs for cases with multiple CHIRON epochs. (b) --- The RVs of our A and B components are compared for each system, except RTW~2202+5537~AB as B lacks an RV measure. We use CHIRON RVs if available, employing weighted mean RVs for cases with multiple epochs, and otherwise use Gaia DR3 RVs. Both panels demonstrate consistent agreement between the various measures, supporting the binary nature of our targets. \label{fig:RV-checks}}
\end{figure}

Our CHIRON RV measurements are compared to Gaia DR3 RVs in panel (a) of Figure~\ref{fig:RV-checks} for stars with both measurements available, showing consistent agreement. This supports the reliability of our overall CHIRON RV methodology and results. Panel (b) of Figure~\ref{fig:RV-checks} compares the RVs of the A and B components\footnote{A single star in our Results Sample, RTW~2202+5537~B, has neither CHIRON or Gaia RV measurements, so the system is not plotted.}, with the demonstrated agreement supporting their wide binary natures. These plots reveal no stars with significantly mismatched or changing RVs even for those without multiple epochs of CHIRON spectra --- excluding cases already known or suspected to be higher order multiples, which are not plotted --- again offering validation against close-in unresolved companions. The point in panel (b) that deviates modestly from the one-to-one line, RTW~1123+8009~AB, is plotted with Gaia DR3 RVs; more precise RV measures derived from TRES spectra by \citetalias{Pass_2024_ApJ} indicate that the pair has matching values of $-17.0 \pm 0.5$\,$\mathrm{km\,s^{-1}}$ for A and $-17.9 \pm 0.5$\,$\mathrm{km\,s^{-1}}$ for B.

%%%%%%%%%%%%%%% tab - mean spec measurements %%%%%%%%%%%%%%%
\startlongtable
\begin{deluxetable}{l|crcccccc|ccc}
\tabletypesize{\footnotesize}
\tablewidth{0pt}
\tablecaption{Mean CHIRON \& Literature Spectral Measurements for the 32 New Systems\label{tab:SpecTable}}
\tablehead{
\multicolumn{1}{c}{} & \multicolumn{8}{c}{\textbf{CHIRON}} & \multicolumn{3}{c}{\textbf{Literature}} \\
\colhead{Name} & \colhead{$N_{\rm{All}}$} & \colhead{$\rm\overline{RV}$} & \colhead{$\sigma_{\rm\overline{RV}}$} & \colhead{$\overline{v\sin(i)}$} & \colhead{$\sigma_{\overline{v\sin(i)}}$} & \colhead{$N_{\rm{H\alpha}}$} & \colhead{$\rm\overline{EW_{H\alpha}}$} & \colhead{$\rm{EW_{H\alpha}}$ Lo-Hi} & \colhead{$\rm{EW_{H\alpha}}$} & \colhead{$\sigma_{\rm{EW_{H\alpha}}}$} & \colhead{Ref.} \\
\colhead{} & \colhead{} & \colhead{($\mathrm{km~s^{-1}}$)} & \colhead{($\mathrm{km~s^{-1}}$)} & \colhead{($\mathrm{km~s^{-1}}$)} & \colhead{($\mathrm{km~s^{-1}}$)} & \colhead{} & \colhead{(\text{\AA})} & \colhead{(\text{\AA})} & \colhead{(\text{\AA})} & \colhead{(\text{\AA})} & \colhead{} \\
\colhead{(1)} & \colhead{(2)} & \colhead{(3)} & \colhead{(4)} & \colhead{(5)} & \colhead{(6)} & \colhead{(7)} & \colhead{(8)} & \colhead{(9)} & \colhead{(10)} & \colhead{(11)} & \colhead{(12)}
}
\startdata
RTW 0143-0151 A &  1      & $+$13.30 & 0.35    & (6.15)  & \nodata &  1      & $-$0.07 &     $\pm$0.02     & $-$0.388 & \nodata & [3]     \\
RTW 0143-0151 B &  1      & $+$12.97 & 0.35    & (7.90)  & \nodata &  1      & $-$0.06 &     $\pm$0.03     & $-$0.390 & \nodata & [3]     \\
RTW 0231-5432 A &  2      &  $+$0.63 & 0.06    & (1.66)  & \nodata &  2      & $+$0.16 & [$+$0.19,$+$0.13] & \nodata  & \nodata & \nodata \\
RTW 0231-5432 B &  2      &  $+$1.17 & 0.05    & (1.69)  & \nodata &  2      & $+$0.04 & [$+$0.08,$-$0.01] & \nodata  & \nodata & \nodata \\
RTW 0231+4003 A & \nodata & \nodata  & \nodata & \nodata & \nodata & \nodata & \nodata & \nodata           & \nodata  & \nodata & \nodata \\
RTW 0231+4003 B & \nodata & \nodata  & \nodata & \nodata & \nodata & \nodata & \nodata & \nodata           & \nodata  & \nodata & \nodata \\
RTW 0409+4623 A & \nodata & \nodata  & \nodata & \nodata & \nodata & \nodata & \nodata & \nodata           & $-$2.320 & 0.135   & [1]     \\
RTW 0409+4623 B & \nodata & \nodata  & \nodata & \nodata & \nodata & \nodata & \nodata & \nodata           & \nodata  & \nodata & \nodata \\
KAR 0545+7254 A & \nodata & \nodata  & \nodata & \nodata & \nodata & \nodata & \nodata & \nodata           & $+$0.0   & 0.3     & [2]     \\
KAR 0545+7254 B & \nodata & \nodata  & \nodata & \nodata & \nodata & \nodata & \nodata & \nodata           & $+$0.0   & 0.2     & [2]     \\
LP 719-37 A     &  5      &  $+$7.29 & 0.04    & (2.68)  & \nodata &  5      & $-$1.82 & [$-$1.16,$-$2.11] & $-$2.889 & 0.058   & [3]     \\
LP 719-37 B     &  5      &  $+$6.92 & 0.07    & (2.94)  & \nodata &  5      & $-$1.64 & [$-$1.09,$-$2.06] & $-$2.007 & 0.058   & [3]     \\
G 103-63 A*     & \nodata & \nodata  & \nodata & \nodata & \nodata & \nodata & \nodata & \nodata           & $-$0.008 & 0.038   & [3]     \\
G 103-63 B      & \nodata & \nodata  & \nodata & \nodata & \nodata & \nodata & \nodata & \nodata           & $+$0.050 & 0.041   & [3]     \\
RTW 0824-3054 A &  1      & $+$31.48 & 0.37    & (2.86)  & \nodata &  1      & $+$0.04 &     $\pm$0.03     & \nodata  & \nodata & \nodata \\
RTW 0824-3054 B*&  1      & $+$30.78 & 0.24    & (5.18)  & \nodata &  1      & $-$0.25 &     $\pm$0.02     & \nodata  & \nodata & \nodata \\
LP 368-99 A     &  5      &  $+$9.73 & 0.11    & 13.94   & 0.29    &  5      & $-$5.11 & [$-$4.39,$-$5.63] & \nodata  & \nodata & \nodata \\
LP 368-99 B     &  5      & $+$10.08 & 0.08    & (5.34)  & \nodata &  5      & $-$4.77 & [$-$3.93,$-$5.53] & \nodata  & \nodata & \nodata \\
L 533-2 A       &  2      & $+$20.41 & 0.11    & (1.82)  & \nodata &  2      & $+$0.23 & [$+$0.24,$+$0.22] & \nodata  & \nodata & \nodata \\
L 533-2 B       &  2      & $+$20.54 & 0.07    & (2.13)  & \nodata &  2      & $+$0.20 & [$+$0.24,$+$0.16] & \nodata  & \nodata & \nodata \\
RTW 0933-4353 A &  5      & $+$19.70 & 2.46    & 15.23   & 0.85    &  4      & $-$6.03 & [$-$3.56,$-$7.91] & \nodata  & \nodata & \nodata \\
RTW 0933-4353 B &  4      & $+$21.78 & 4.41    & 16.14   & 2.67    &  4      & $-$5.85 & [$-$4.66,$-$7.01] & \nodata  & \nodata & \nodata \\
LP 551-62 A     &  1      & $+$28.10 & 0.12    & (2.12)  & \nodata &  1      & $+$0.38 &     $\pm$0.02     & \nodata  & \nodata & \nodata \\
LP 551-62 B     &  1      & $+$28.03 & 0.16    & (3.17)  & \nodata &  1      & $+$0.27 &     $\pm$0.02     & \nodata  & \nodata & \nodata \\
RTW 1123+8009 A & \nodata & \nodata  & \nodata & \nodata & \nodata & \nodata & \nodata & \nodata           & $-$1.691 & \nodata & [3]     \\
RTW 1123+8009 B & \nodata & \nodata  & \nodata & \nodata & \nodata & \nodata & \nodata & \nodata           & $-$0.192 & \nodata & [3]     \\
RTW 1133-3447 A &  1      &  $+$6.54 & 0.36    & (4.22)  & \nodata &  1      & $+$0.30 &     $\pm$0.02     & \nodata  & \nodata & \nodata \\
RTW 1133-3447 B &  1      &  $+$6.50 & 0.17    & (5.98)  & \nodata &  1      & $+$0.24 &     $\pm$0.01     & \nodata  & \nodata & \nodata \\
RTW 1336-3212 A &  1      & $+$43.00 & 0.12    & (2.83)  & \nodata &  1      & $-$0.07 &     $\pm$0.02     & \nodata  & \nodata & \nodata \\
RTW 1336-3212 B &  1      & $+$42.39 & 0.26    & (1.91)  & \nodata &  1      & $+$0.03 &     $\pm$0.02     & \nodata  & \nodata & \nodata \\
L 197-165 A     &  1      & $-$11.11 & 0.22    & (2.88)  & \nodata &  1      & $+$0.07 &     $\pm$0.01     & \nodata  & \nodata & \nodata \\
L 197-165 B     &  1      & $-$11.52 & 0.11    & (3.21)  & \nodata &  1      & $+$0.19 &     $\pm$0.02     & \nodata  & \nodata & \nodata \\
RTW 1433-6109 A &  5      & $-$27.77 & 0.07    & (3.06)  & \nodata &  5      & $+$0.33 & [$+$0.34,$+$0.30] & \nodata  & \nodata & \nodata \\
RTW 1433-6109 B &  5      & $-$28.07 & 0.07    & (2.77)  & \nodata &  5      & $+$0.33 & [$+$0.35,$+$0.31] & \nodata  & \nodata & \nodata \\
L 1197-68 A     &  1      & $+$24.56 & 0.14    & (2.55)  & \nodata &  1      & $+$0.33 &     $\pm$0.01     & \nodata  & \nodata & \nodata \\
L 1197-68 B     &  1      & $+$24.26 & 0.13    & (2.56)  & \nodata &  1      & $+$0.34 &     $\pm$0.01     & \nodata  & \nodata & \nodata \\
L 1198-23 A     &  5      & $-$27.81 & 0.05    & (1.81)  & \nodata &  5      & $+$0.28 & [$+$0.32,$+$0.23] & \nodata  & \nodata & \nodata \\
L 1198-23 B     &  6      & $-$28.27 & 0.05    & (1.72)  & \nodata &  5      & $+$0.32 & [$+$0.34,$+$0.29] & \nodata  & \nodata & \nodata \\
RTW 1512-3941 A &  6      & $+$38.21 & 0.11    & (4.51)  & \nodata &  5      & $+$0.41 & [$+$0.43,$+$0.40] & \nodata  & \nodata & \nodata \\
RTW 1512-3941 B &  5      & $+$36.84 & 0.11    & (3.91)  & \nodata &  5      & $+$0.45 & [$+$0.45,$+$0.43] & \nodata  & \nodata & \nodata \\
RTW 1812-4656 A &  5      & $-$17.92 & 0.07    & (2.46)  & \nodata &  5      & $+$0.31 & [$+$0.35,$+$0.23] & \nodata  & \nodata & \nodata \\
RTW 1812-4656 B &  5      & $-$17.91 & 0.08    & (2.74)  & \nodata &  5      & $+$0.31 & [$+$0.32,$+$0.27] & \nodata  & \nodata & \nodata \\
GJ 745 A        &  1      & $+$32.62 & 0.16    & (3.71)  & \nodata &  1      & $+$0.19 &     $\pm$0.01     & $+$0.045 & 0.010   & [3]     \\
GJ 745 B        &  1      & $+$32.35 & 0.13    & (3.46)  & \nodata &  1      & $+$0.20 &     $\pm$0.01     & $+$0.037 & 0.010   & [3]     \\
RTW 2011-3824 A &  1      &  $-$4.93 & 0.32    & (1.72)  & \nodata &  1      & $+$0.15 &     $\pm$0.02     & \nodata  & \nodata & \nodata \\
RTW 2011-3824 B &  1      &  $-$3.85 & 0.21    & (1.38)  & \nodata &  1      & $+$0.17 &     $\pm$0.02     & \nodata  & \nodata & \nodata \\
G 230-39 A      & \nodata & \nodata  & \nodata & \nodata & \nodata & \nodata & \nodata & \nodata           & $+$0.077 & 0.038   & [3]     \\
G 230-39 B      & \nodata & \nodata  & \nodata & \nodata & \nodata & \nodata & \nodata & \nodata           & $+$0.114 & 0.038   & [3]     \\
RTW 2202+5537 A & \nodata & \nodata  & \nodata & \nodata & \nodata & \nodata & \nodata & \nodata           & \nodata  & \nodata & \nodata \\
RTW 2202+5537 B & \nodata & \nodata  & \nodata & \nodata & \nodata & \nodata & \nodata & \nodata           & \nodata  & \nodata & \nodata \\
RTW 2211+0058 A &  1      &  $+$2.62 & 0.12    & (1.84)  & \nodata &  1      & $+$0.22 &     $\pm$0.02     & $+$0.189 & 0.029   & [3]     \\
RTW 2211+0058 B &  1      &  $+$3.32 & 0.15    & (1.84)  & \nodata &  1      & $+$0.26 &     $\pm$0.02     & $+$0.127 & 0.030   & [3]     \\
RTW 2235+0032 A &  5      & $+$10.72 & 0.07    & (5.71)  & \nodata &  5      & $-$2.75 & [$-$2.44,$-$3.10] & $-$3.539 & 0.063   & [3]     \\
RTW 2235+0032 B &  5      &  $+$9.78 & 0.05    & (5.27)  & \nodata &  5      & $-$2.62 & [$-$2.52,$-$2.72] & $-$4.925 & 0.061   & [3]     \\
RTW 2236+5923 A & \nodata & \nodata  & \nodata & \nodata & \nodata & \nodata & \nodata & \nodata           & \nodata  & \nodata & \nodata \\
RTW 2236+5923 B & \nodata & \nodata  & \nodata & \nodata & \nodata & \nodata & \nodata & \nodata           & \nodata  & \nodata & \nodata \\
RTW 2241-1625 A &  5      &  $+$3.34 & 0.18    & (6.91)  & \nodata &  5      & $+$0.12 & [$+$0.15,$+$0.09] & \nodata  & \nodata & \nodata \\
RTW 2241-1625 B &  5      &  $+$3.90 & 0.17    & (6.46)  & \nodata &  5      & $+$0.32 & [$+$0.34,$+$0.31] & \nodata  & \nodata & \nodata \\
RTW 2244+4030 A & \nodata & \nodata  & \nodata & \nodata & \nodata & \nodata & \nodata & \nodata           & $+$0.53  & \nodata & [4]     \\
RTW 2244+4030 B & \nodata & \nodata  & \nodata & \nodata & \nodata & \nodata & \nodata & \nodata           & $+$0.54  & \nodata & [4]     \\
L 718-71 A      &  1      &  $+$1.34 & 0.22    & (3.60)  & \nodata &  1      & $+$0.41 &     $\pm$0.01     & \nodata  & \nodata & \nodata \\
L 718-71 B      &  1      &  $+$1.23 & 0.23    & (3.83)  & \nodata &  1      & $+$0.44 &     $\pm$0.01     & \nodata  & \nodata & \nodata \\
RTW 2311-5845 A &  5      &  $-$1.61 & 0.51    & 20.00   & 0.99    &  5      & $-$3.77 & [$-$3.22,$-$4.15] & \nodata  & \nodata & \nodata \\
RTW 2311-5845 B &  6      &  $-$0.35 & 0.30    & 16.89   & 0.48    &  5      & $-$6.04 & [$-$4.23,$-$8.02] & \nodata  & \nodata & \nodata \\
\enddata
\tablerefs{[1] \citet{2021ApJS..253...19Z}, [2] \citet{CARMENES_2015}, [3] \citetalias{Pass_2024_ApJ}, [4] \citet{2013AJ....145..102L}.}
\tablecomments{Of the 32 New Systems, 23 have spectral measurements from our new CHIRON observations reported in columns (2)--(9), while the remaining systems have literature H$\alpha$ EWs reported in columns (10)--(12) when available. Asterisks in their names indicate the two cases with suspected unresolved companions. \textbf{CHIRON:} 12 of the 23 CHIRON systems have multiple epochs of CHIRON spectra. For these systems, the values given are the weighted mean RV and $v\sin(i)$ measurements and uncertainties based on all available spectra ($N_{\rm{All}}$), along with the mean and range of the H$\alpha$ EWs based only on epochs with A and B successfully observed back-to-back and with no flares ($N_{\rm{H\alpha}}$). Parentheses indicate $v\sin(i)$ values $<$\,10\,$\mathrm{km\,s^{-1}}$ that are less reliable measurements with excluded unreliable errors. H$\alpha$ EW errors for single-epoch cases are given in the same column as the ranges for multi-epoch cases. The observed H$\alpha$ ranges are shown within square brackets, where the Lo-Hi values are numerically reversed to track activity strengths. \textbf{Literature:} Of the nine systems unobservable with CHIRON, five had H$\alpha$ EWs available in the literature for each component, one had only a single component measured, and we found no literature values for three. We include reported EW uncertainties if available from the literature sources. For \citetalias{Pass_2024_ApJ}, we adopted their higher resolution TRES spectra results if available and otherwise used their lower resolution FAST spectra results. For comparison, we include H$\alpha$ EWs from \citetalias{Pass_2024_ApJ} even if CHIRON measurements are available, but use our CHIRON measures for our plots and results analyses.}
\end{deluxetable}

\begin{figure}[!t]
\centering
\includegraphics[width=0.49\textwidth]{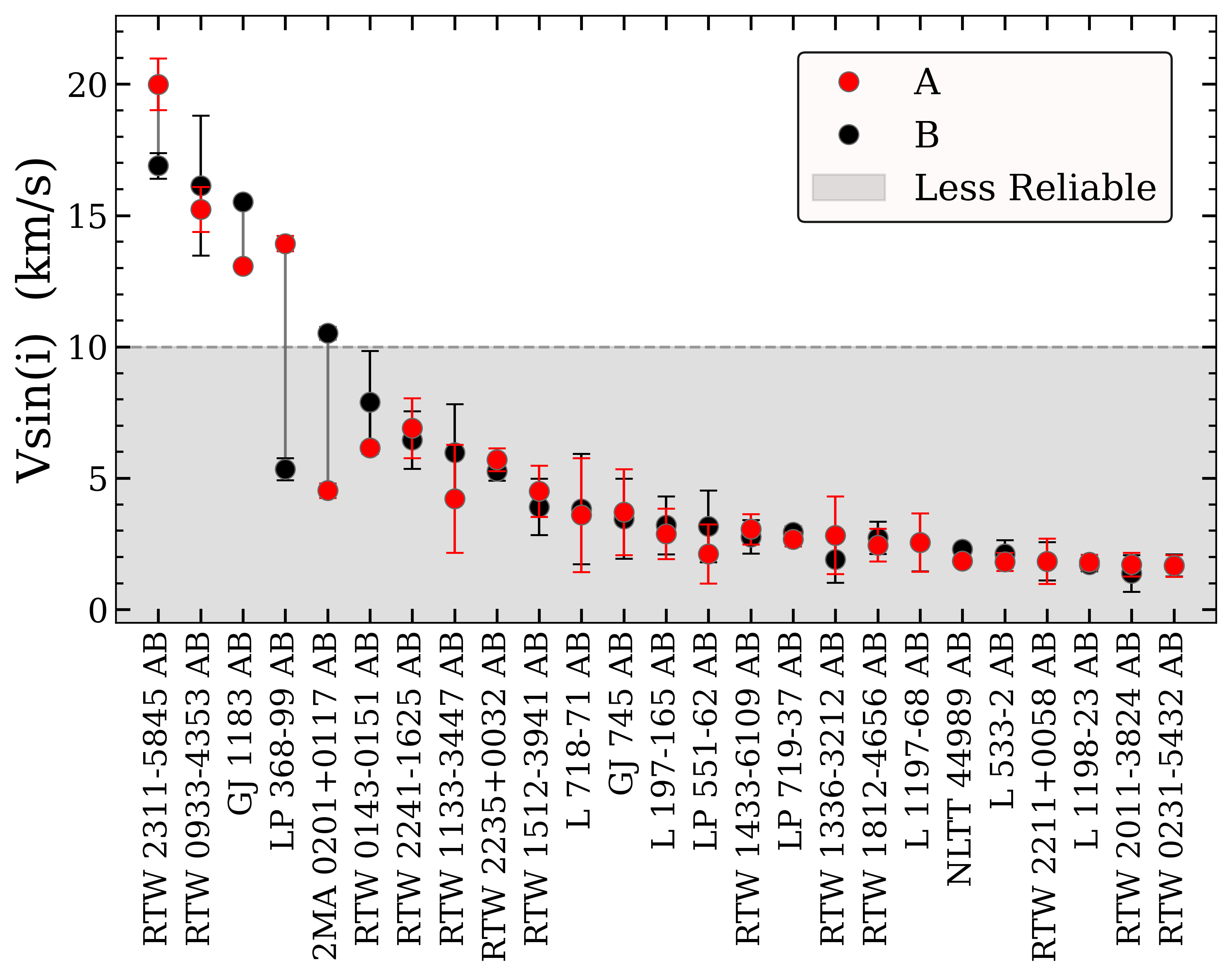}
\figcaption{Our CHIRON $v\sin(i)$ values compared between twin pair components from our Results Sample, using the measurements from Table~\ref{tab:SpecTable} and \citetalias{RTW_P1} and plotting error bars in all cases. Systems are ordered by descending $v\sin(i)$ value, with vertical gray lines connecting the component stars and a gray shaded region indicating less reliable measures at $<$\,10\,$\mathrm{km\,s^{-1}}$. The three systems with both stars above 10\,$\mathrm{km\,s^{-1}}$ show broadly similar values between components, while the remaining majority of systems also typically show well-aligned values despite residing in the region of less reliable measurements. The handful of exceptions are discussed in Section~\ref{subsec:chiron-rv-results} and align with their observed component rotation periods.
\label{fig:vsini}}
\end{figure}

Our CHIRON $v\sin(i)$ results are assessed in Figure~\ref{fig:vsini}. Despite the $10$\,$\mathrm{km\,s^{-1}}$ limit for reliable determinations, most of the binaries show well matched $v\sin(i)$ values. The fives systems hosting at least one component with $v\sin(i)>10$\,$\mathrm{km\,s^{-1}}$ have measured rotation periods for both stars (see Section~\ref{sec:rot}). The similar $v\sin(i)$ values and similar rotation periods between components in RTW~2311-5845~AB, RTW~0933-4353~AB, and GJ~1183~AB suggest that the components have similar inclinations, especially because the component in each pair with the slightly faster $v\sin(i)$ also has the slightly faster rotation period. Intriguingly, this further implies that the component rotation axes remain aligned for these three systems despite their component separations of $\sim$200\,au. The only two systems that show prominent $v\sin(i)$ mismatches are LP~368-99~AB and 2MA~0201+0117~AB, which agrees with each of these pairs hosting component rotation periods differing by $\sim$2$\times$ --- the component with a faster period has the larger $v\sin(i)$ for each case, as expected.

For systems lacking CHIRON spectra, a partial literature search for reliable $v\sin(i)$ measurements revealed no additional values. Similarly, Gaia DR3 reports spectral broadening with the Vbroad parameter, but of our systems lacking CHIRON spectra only RTW~2244+4030~A has a Vbroad measurement available --- its value at $5.0\pm3.1$\,$\mathrm{km\,s^{-1}}$ is $<$\,10\,$\mathrm{km\,s^{-1}}$ and therefore likely somewhat inaccurate per \citet{2023A&A...674A...8F}.

%%%%%%%%%%%%%% tab - all chiron spec results example table %%%%%%%%%%%%%%%
\begin{deluxetable}{lccccccccc}[!t]
\tablewidth{0pt}
\tablecaption{All CHIRON Spectral Measurements for the 32 New Systems\label{tab:all-spec}}
\tablehead{
\colhead{Name} & \colhead{J. Epoch} & \colhead{RV} & \colhead{$\sigma$--RV} & \colhead{$v\sin(i)$} & \colhead{$\sigma$--$v\sin(i)$} & \colhead{EW$_\mathrm{H\alpha}$} & \colhead{$\sigma$--EW$_\mathrm{H\alpha}$} & \colhead{SNR--$\mathrm{H\alpha}$} & \colhead{SNR--Continuum} \\
\colhead{} & \colhead{(YYYY.YYYY)} & \colhead{($\mathrm{km~s^{-1}}$)} & \colhead{($\mathrm{km~s^{-1}}$)} & \colhead{($\mathrm{km~s^{-1}}$)} & \colhead{($\mathrm{km~s^{-1}}$)} & \colhead{(\text{\AA})} & \colhead{(\text{\AA})} & \colhead{} & \colhead{}
}
\startdata
L 533-2 A   & 2020.1498 & $+$20.33 & 0.17 & 1.94 & 0.76 & $+$0.24 & 0.02 & 15.9 & 17.5 \\
L 533-2 A   & 2021.1873 & $+$20.46 & 0.14 & 1.79 & 0.39 & $+$0.22 & 0.02 & 26.1 & 28.4 \\
L 533-2 B   & 2020.1498 & $+$20.47 & 0.10 & 2.28 & 0.65 & $+$0.24 & 0.02 & 16.6 & 18.2 \\
L 533-2 B   & 2021.1873 & $+$20.63 & 0.11 & 1.86 & 0.85 & $+$0.16 & 0.02 & 14.8 & 15.8 \\
GJ 745 A    & 2021.3001 & $+$32.62 & 0.16 & 3.71 & 1.63 & $+$0.19 & 0.01 & 47.4 & 50.6 \\
GJ 745 B    & 2021.3002 & $+$32.35 & 0.13 & 3.46 & 1.52 & $+$0.20 & 0.01 & 50.7 & 54.4 \\
\enddata
\tablecomments{Example sets of CHIRON measurements for a random selection of our targets; all CHIRON measurements for the 32 New Systems are available in the machine-readable Table, including the candidate higher order multiple RTW~0824-3054~AB. None of the New Systems with multi-epoch spectra captured H$\alpha$ flares based on the criteria described in \citetalias{RTW_P1}, so we exclude flare indicators. Any $v\sin(i)$ measurements less than 10\,$\mathrm{km~s^{-1}}$ are less reliable. SNRs are provided for the H$\alpha$ line region (SNR--$\mathrm{H\alpha}$) as well as the average of the nearby continuum regions (SNR--Continuum), with details of these regions defined in \citetalias{RTW_P1}.}
\end{deluxetable}

%%%%%%%%%%%%%%% fig - CHIRON vs Gaia RVs, and A vs B RVs %%%%%%%%%%%%%%%

%%%%%%%%%%%%%%% fig - CHIRON vsini results %%%%%%%%%%%%%%%

\subsection{CHIRON H$\alpha$ Results} \label{subsec:chiron-ha-results}

Our mean H$\alpha$ EWs are reported in Table~\ref{tab:SpecTable} for the New Systems, with Table~\ref{tab:all-spec} containing all of the corresponding individual measurements. We ultimately encountered no H$\alpha$ flares for the New Systems with multi-epoch spectra, based on our flare criteria in \citetalias{RTW_P1}. For single epoch cases, we are unable to completely validate against flare snapshots --- however, observations are made with two back-to-back spectra on a single star before subsequent spectral co-adding during later processing, so we are able to somewhat rule out strong flares based on our manual review of the two sequential H$\alpha$ profiles. For systems without CHIRON spectra, we include H$\alpha$ EWs from several literature sources as given in Table~\ref{tab:SpecTable}, although four systems still ultimately lack H$\alpha$ EW measurements for at least one component in the pair. The literature sources used instruments with different resolutions to secure the spectra and used EW determination methods with different line and continuum windows, so the exact EW values should only be compared between A and B components from the same measurement source. For context, theoretical work by \citet{Cram_Mullan_1979} indicates that M dwarfs first shift to slightly deeper absorption as chromospheric heating and H$\alpha$ activity initially increase, before flipping to lessening absorption and eventually emission as activity and heating further increase. Here we assume our targets are sufficiently active to be in at least the maximal absorption state, similar to \citetalias{Newton_2017}, but because we compare between twin stars this assumption has minimal impact on our results.

\begin{figure}[!t]
\centering
\gridline{\fig{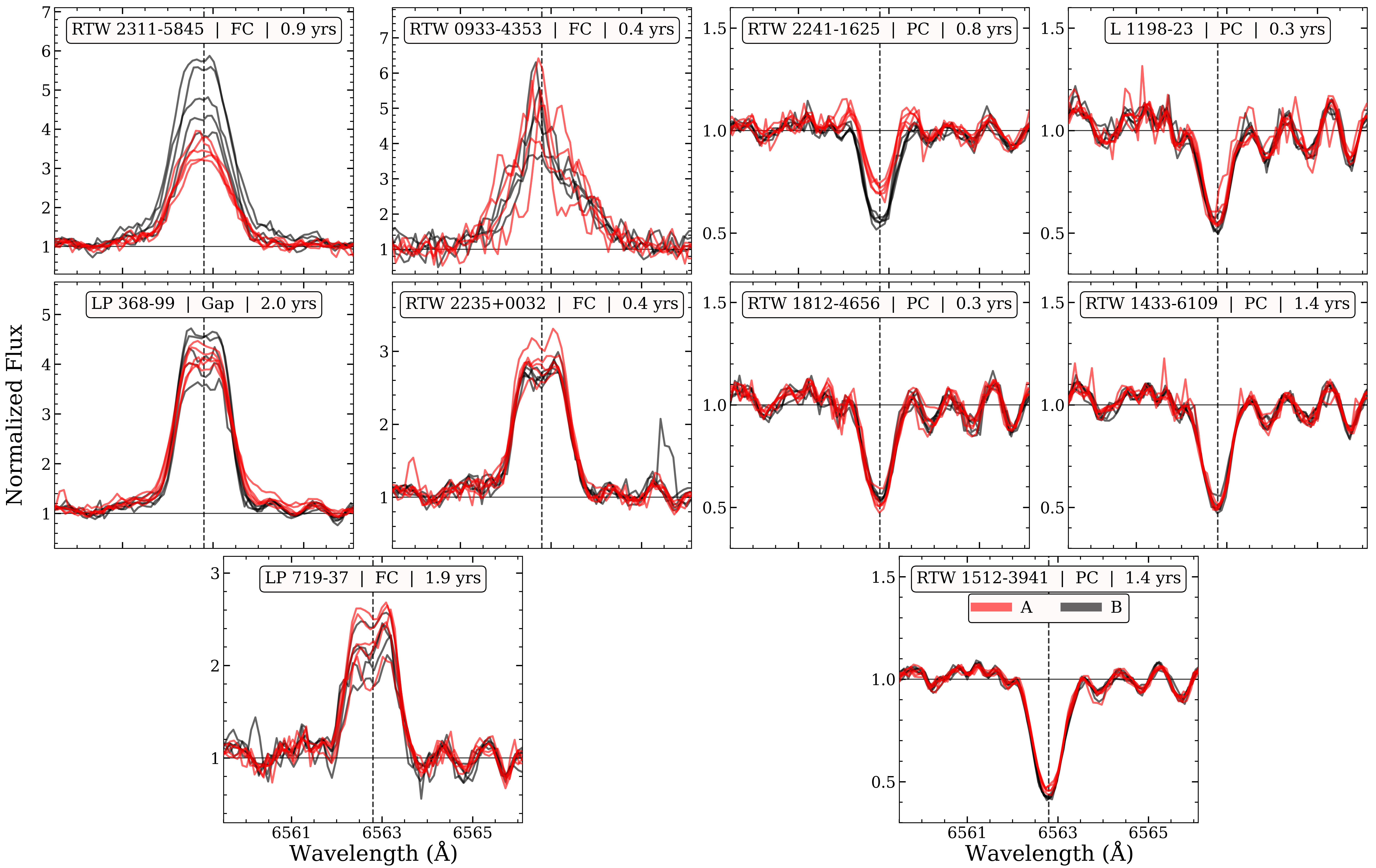}{0.99\textwidth}{}}
\vspace*{-15pt}
\gridline{\fig{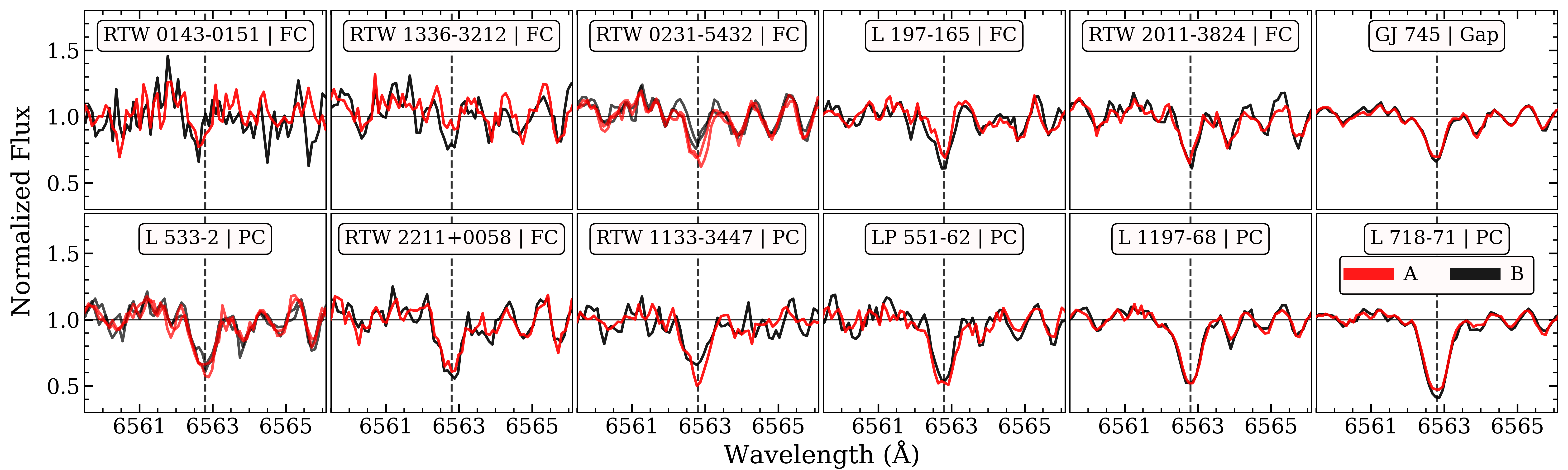}{0.99\textwidth}{}}
\vspace*{-15pt}
\figcaption{Spectra of the H$\alpha$ region for 22 New Systems observed by CHIRON, excluding the known or suspected non-twin higher order multiple systems. Each individual spectrum has been blaze corrected and shifted to the rest frame based on the measured stellar RVs. These plots only include epochs with both stars successfully observed back-to-back and with neither flaring. Each subplot compares the A star in red and its twin B star in black; cases with multiple epochs are shown stacked. The top grid of 10 pairs shows `core' targets where each system has five epochs, except RTW~0933-4353~AB with four epochs --- these multi-epoch cases are split into emission systems on the left (with changing y-axis scales) and absorption systems on the right. The bottom grid shows the remaining 12 non-core pairs observed at only one or two epochs. Titles in each subplot specify the binary system shown and their interior structure (PC/FC/Gap); those in the top grid also give the time baseline of their multiple epochs. It is remarkable how well-matched many pairs appear, even over multiple epochs of variability, but some key systems show sustained differences in H$\alpha$ strength. \label{fig:Ha-stacked_spec}}
\end{figure}

CHIRON H$\alpha$ line profiles are shown in Figure~\ref{fig:Ha-stacked_spec} for 22 pairs with A component spectra in red and B component B spectra in black, including both the core (10 pairs) and non-core (12 pairs) cases from our New Systems, and excluding higher-order multiples. The general categories of profile pairs are evident, including matching emission (e.g., RTW~2235+0032~AB), mismatched emission (e.g., RTW~2311-5845~AB), matching absorption (e.g., RTW~1512-3941~AB), and mismatched absorption (e.g., RTW~2241-1625~AB). We note that the H$\alpha$ strength for RTW~2311-5845~B is not steadily increasing or decreasing and instead grows both stronger and weaker across epochs. Mismatched features appear in both partially convective systems (RTW~2241-1625~AB) and fully convective systems (RTW~2311-5845~AB), the latter complementing the fully convective systems with pronounced activity differences in \citetalias{RTW_P1}. All of the systems here show extremely congruent overlap in the continuum features between components, emphasizing the twin natures of the pairs. The continuum and H$\alpha$ features in RTW~0933-4353~A and B show enhanced scatter due to the worse SNR of their spectra (see Section~\ref{subsec:chiron-methods}), but the averages of their visual profiles and EW measures indicate they are matching in H$\alpha$ strength; the somewhat triangular nature of their H$\alpha$ profiles is also notable, but broadening from their extremely rapid $\sim$3\,hr rotation periods (Section~\ref{sec:rot}) and the poor spectral SNRs ambiguate any obvious definite explanation. Higher SNR spectra of the RTW~0231-5432~AB and RTW~1336-3212~AB systems would also help elucidate how precisely similar or dissimilar their H$\alpha$ strengths are, given that our spectra show indications of possible slight differences in both cases but with obfuscating noise.

%%%%%%%%%%%%%%% fig - CHIRON Ha stacked spec %%%%%%%%%%%%%%%

\begin{figure}[!t]
\centering
\gridline{\fig{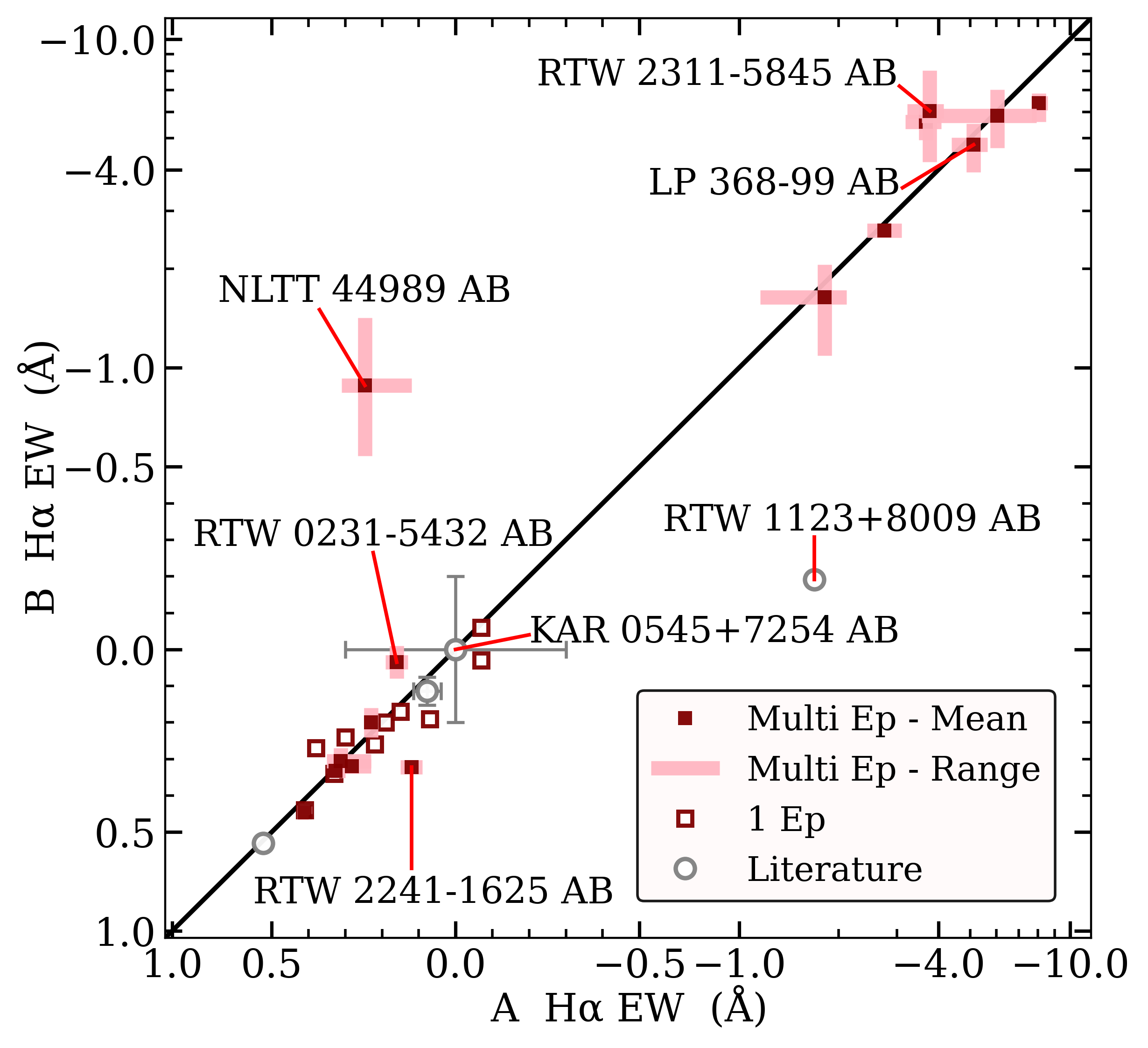}{0.462\textwidth}{(a)}
          \fig{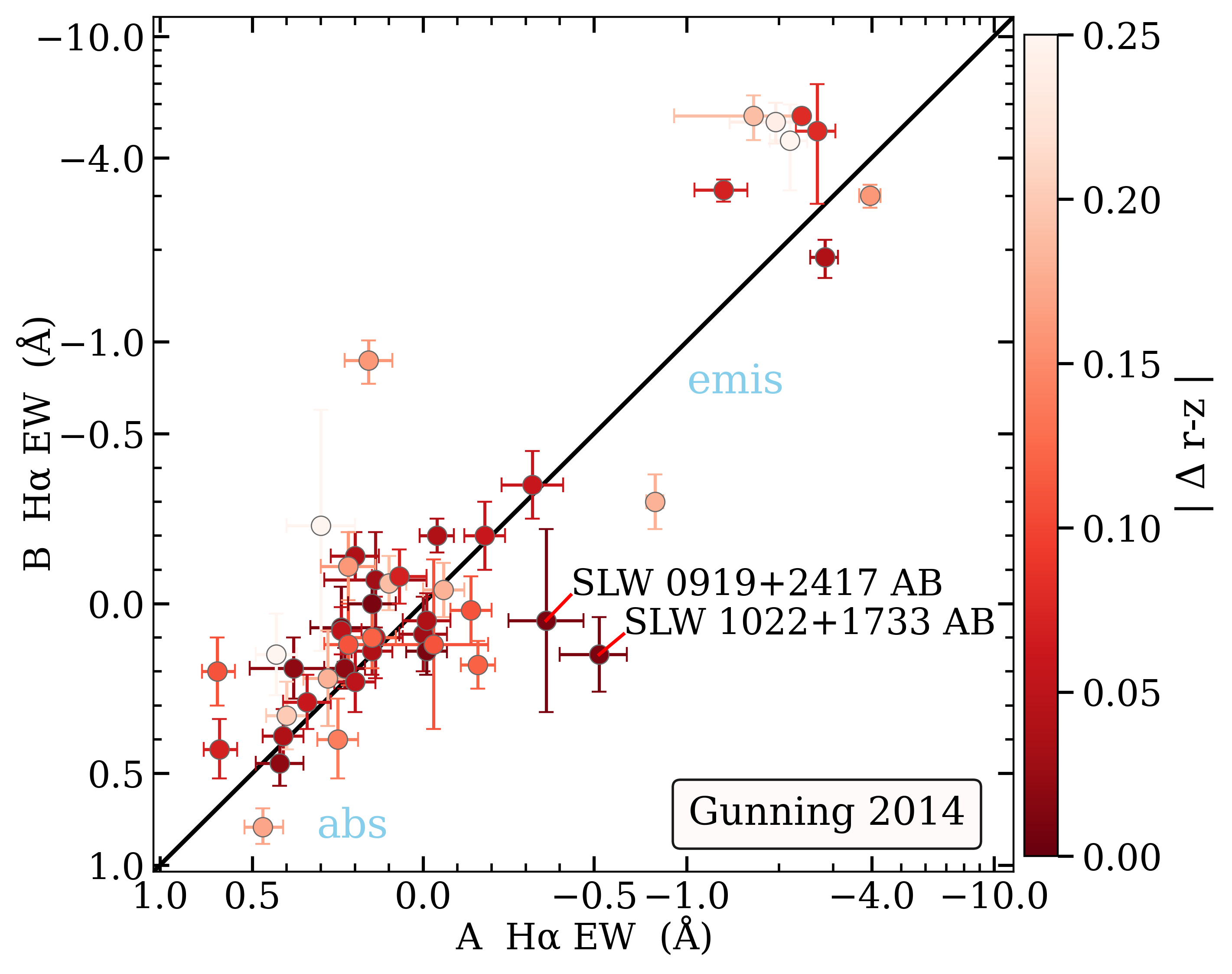}{0.534\textwidth}{(b)}}
\figcaption{H$\alpha$ EWs are compared between M dwarf binary components for two samples of twin or nearly-twin systems. Each plot is scaled linearly between $-$0.5 to $+$0.5 and logarithmically outside these bounds. Several systems discussed throughout the text are labeled. (a) --- The left panel shows our Results Sample, using measurements reported in Table~\ref{tab:SpecTable} and \citetalias{RTW_P1}. Cases with multiple CHIRON epochs show their observed ranges of H$\alpha$ EWs as shaded bars, while those with only a single CHIRON epoch have un-shown measurement uncertainties that are typically smaller than the points. The CHIRON H$\alpha$ results shown here only include epochs with both stars successfully observed back-to-back and with neither flaring. (b) --- A sample of 48 nearly-twin M dwarf wide binaries from \citet{Gunning_2014} is reproduced in the right panel using data from their Table 1, with their EWs inverted in sign to match our convention of negative values indicating emission. Their five systems with multiple epochs use the mean and standard deviation of observed activity. Points are color coded by the absolute difference in $r-z$ color between stars in a system, with more twin-like pairs having darker colors. Both panels show many well-matched systems alongside several key outliers hosting large activity mismatches. See Section~\ref{subsec:chiron-ha-results} for further discussion. \label{fig:Ha-equality}}
\end{figure}

We show the H$\alpha$ EW measurements from our Results Sample for A versus B in panel (a) of Figure~\ref{fig:Ha-equality}, including the literature cases. This reveals two contrasting results. First, the majority of pairs have remarkable agreement in their activity levels, especially when accounting for intrinsic variability in the multi-epoch cases. However, secondly, a few outliers show large differences despite their twin constituents. There are also relatively few systems falling between H$\alpha$ EWs of roughly $-$0.5\,\text{\AA} to $-$2\,\text{\AA} in both panels of Figure~\ref{fig:Ha-equality}, which we tentatively attribute to the dearth of intermediate-period intermediate-activity FC M dwarfs owing to their fast magnetic braking phase \citep[e.g.,][]{Newton_2017, Pass_2024_ApJ, RTW_P1}.

We consider six cases from the Results Sample to qualify as twin binaries meaningfully ``mismatched" in H$\alpha$ activity: RTW~1123+8009~AB, RTW~2241-1625~AB, RTW~2311-5845~AB, 2MA~0201+0117~AB, GJ~1183~AB, and NLTT~44989~AB, stemming from our manual inspection of all of the H$\alpha$ results in Figure~\ref{fig:Ha-stacked_spec}, Figure~\ref{fig:Ha-equality}, and \citetalias{RTW_P1}. We select these based on each system showing sustained activity differences beyond the spectral noise and beyond their intrinsic H$\alpha$ variability across multiple spectral epochs over months to years. These six mismatched systems are also identified if we were to instead choose mismatches by requiring such cases have a $>$\,25\% relative difference \textit{and} $\geq$\,0.2\,\text{\AA} absolute difference between the A and B H$\alpha$ EWs\footnote{GJ~1183~AB has a 23\% percent difference so technically does not meet the 25\% criteria to be considered mismatched. However, it has a large 1.68\,\text{\AA} absolute difference and visually evident mismatched H$\alpha$ behavior shown in \citetalias{RTW_P1}, with A also being 26\% stronger in H$\alpha$ EW relative to B on average, so we still designate the system as mismatched in H$\alpha$.}. The absolute difference criterion addresses cases with EWs near zero that have misleadingly large relative differences compared to their very small absolute differences. For example, the system RTW~0231-5432~AB is offset slightly above the one-to-one line in Figure~\ref{fig:Ha-equality}, but its two epochs of similar spectra visible in Figure~\ref{fig:Ha-stacked_spec} show how the weak absorption and noise somewhat inflate its perceived deviation in Figure~\ref{fig:Ha-equality}. With the six mismatch cases identified, we designate the remaining 23 twin systems in panel (a) of Figure~\ref{fig:Ha-equality} to be ``matching" in activity. {\bf This yields a cumulative mismatch rate of 21\% (6/29) for H$\alpha$ activity in otherwise twin stars.}  Of all 29 plotted systems, 16 have matching absorption, 4 have matching emission, 3 are matching with flat filled in absorption, 1 has mismatched absorption, 3 have mismatched emission, and the 2 cases RTW~1123+8009~AB and NLTT~44989~AB have significant active/inactive H$\alpha$ mismatches between their components.

We further investigated the chromospheric activity of our Full Sample based on the Ca II infrared triplet activity index reported in Gaia DR3 \citep{2023A&A...674A..30L}. Unfortunately, only six of our pairs have the corresponding \texttt{activityindex\_espcs} measurement available in DR3 for both components: one system is the suspected triple RTW~0824-3054~AB, three systems (GJ~1183~AB, KAR~0545+7254~AB, and RTW~2244+4030~AB) show similar values between components, RTW~2241-1625~AB shows a slightly more active A component consistent with our H$\alpha$ results, and NLTT~44989~AB shows an inactive A star and clearly active B star, which is consistent with the system's behavior discussed in \citetalias{RTW_P1}. This overall agreement is expected given that the Ca II infrared triplet and H$\alpha$ both probe chromospheric activity and generally track with each other as activity measures \citep[e.g.,][]{2017A&A...605A.113M}.

%%%%%%%%%%%%%%% fig - AB HaEW equal for all twins, including Gunning2014 comparison %%%%%%%%%%%%%%%

\subsubsection{Gunning et al. (2014) H$\alpha$ Results} \label{subsubsec:gunning-ha-results}

In panel (b) of Figure~\ref{fig:Ha-equality} we compare our results to those of \citet{Gunning_2014}, who report H$\alpha$ EWs for a sample of 48 nearly-twin M dwarf wide binaries in a similar effort to our own; there is no overlap between their sample and our Full Sample. We consider their sample to be comprised of ``nearly-twin" systems because they report using only a single color criterion of $\Delta(r-z) \le 0.25$\,mag to select pairs, in contrast to our own stricter ``twin" systems that required five magnitudes across the optical and IR to match within 0.1\,mag. For comparison, 27 of their systems have $|\Delta(r-z)| < 0.1$\,mag, and two of their systems are relatively un-twin-like with reported values of $|\Delta(r-z)| > 0.25$\,mag. Nonetheless, the \citet{Gunning_2014} results are consistent with ours, finding that the majority of systems broadly align in activity while key outlier systems can show large mismatches in activity. In comparison, our own higher precision measurements and stricter twin selection criteria yield less scatter with a tighter match to the one-to-one line for most systems. Some of their strongly mismatched pairs may also be the result of unresolved companions, as less data were used to vet for such cases in their work compared to our own stars. 

Two of the systems in \citet{Gunning_2014} --- SLW~0919+2417~AB and SLW~1022+1733~AB --- show intriguing activity mismatches alongside small $|\Delta(r-z)|$ values, and are labeled in panel (b) of Figure~\ref{fig:Ha-equality}. We further investigated these two systems given their standout behaviors. SLW~0919+2417~A and B are both FC while SLW~1022+1733~A and B are both PC, with both systems residing on the main sequence. For SLW~0919+2417~AB, the components have Gaia DR3 magnitude differences of 0.18, 0.13, and 0.14 in BP, RP, and $G$ respectively, making them not sufficiently twin-like by our criteria. The large H$\alpha$ EW uncertainties in the measurements for the pair also suggest their activity levels differ by roughly 1.5$\sigma$, so we disregard this system as a not true mismatch. For SLW~1022+1733~AB, the components have magnitude differences of 0.03, 0.01, 0.01, 0.02, 0.04, and 0.05 in BP, RP, $G$, $J$, $H$, and $K_s$ respectively, with a $\sim$10$\arcsec$ separation at their $\sim$200\,pc distance, making them non-interacting twins by our criteria. They have matching proper motions and parallaxes in DR3, with non-elevated IPDfmp and RUWE values consistent with a lack of unresolved companions. Their component H$\alpha$ EW values differ by $>$\,5$\sigma$, with A having mild emission ($\rm{EW} = -0.52 \pm 0.12$\,\text{\AA}) and B having absorption ($\rm{EW} = 0.15 \pm 0.11$\,\text{\AA}) in \citet{Gunning_2014}. While the scale of its activity mismatch is smaller than those seen in NLTT~44989~AB and RTW~1123+8009~AB, the SLW~1022+1733~AB system still demonstrates a pronounced difference, exceeds our 25\% and 0.2\,\text{\AA} mismatch criteria discussed above, and serves as a PC analog to our two FC mismatched systems. SLW~1022+1733~AB thus warrants study in future work, particularly for rotation period measurements and potential unseen massive companions. We do not see any rotation signals for A or B in the four sectors of available blended TESS data.

\section{Astrometry and Stellar Cycles --- SMARTS 0.9\,m Long-term Campaign} \label{sec:longterm}

We now consider the RECONS long-term program \citep{Henry_2018}, where our stars are targeted to probe their multi-year astrometric motion, photometric variability, and activity cycles.

\subsection{Long-Term Program Observations and Data Processing} \label{subsec:longterm-methods}

Our observations and analysis for data from the RECONS 25+\,yr 0.9\,m program at CTIO follow as described in \citetalias{RTW_P1}. Briefly, each target receives an average of two visits per year with five frames acquired per visit in a consistent filter of either \textit{V}, \textit{R}, or \textit{I}. Differential photometry and astrometry measurements are derived using the RECONS processing pipeline \citep[][]{2003AJ....125..332J, Jao_2011, Hosey_2015, Vrijmoet_2020}, where a consistent set of non-varying reference stars are used after manual selection and vetting. The resulting measures for our A and B components therefore derive from the same underlying images and calibrator stars, which supports the validity of any observed differences being truly astrophysical and not systematic. The systems targeted by our long-term program are specified in Table~\ref{tab:SampleTable-phot} and include most southern systems along with the northern core systems, chosen for a variety of logistical and astrophysical motivations. The 0.9\,m measurements used here and in \citetalias{RTW_P1} do not include any data acquired after 2024 April.

\subsection{Long-Term Astrometry Results} \label{subsec:longterm-astr-results}

\begin{figure}[!t]
\centering
\gridline{\fig{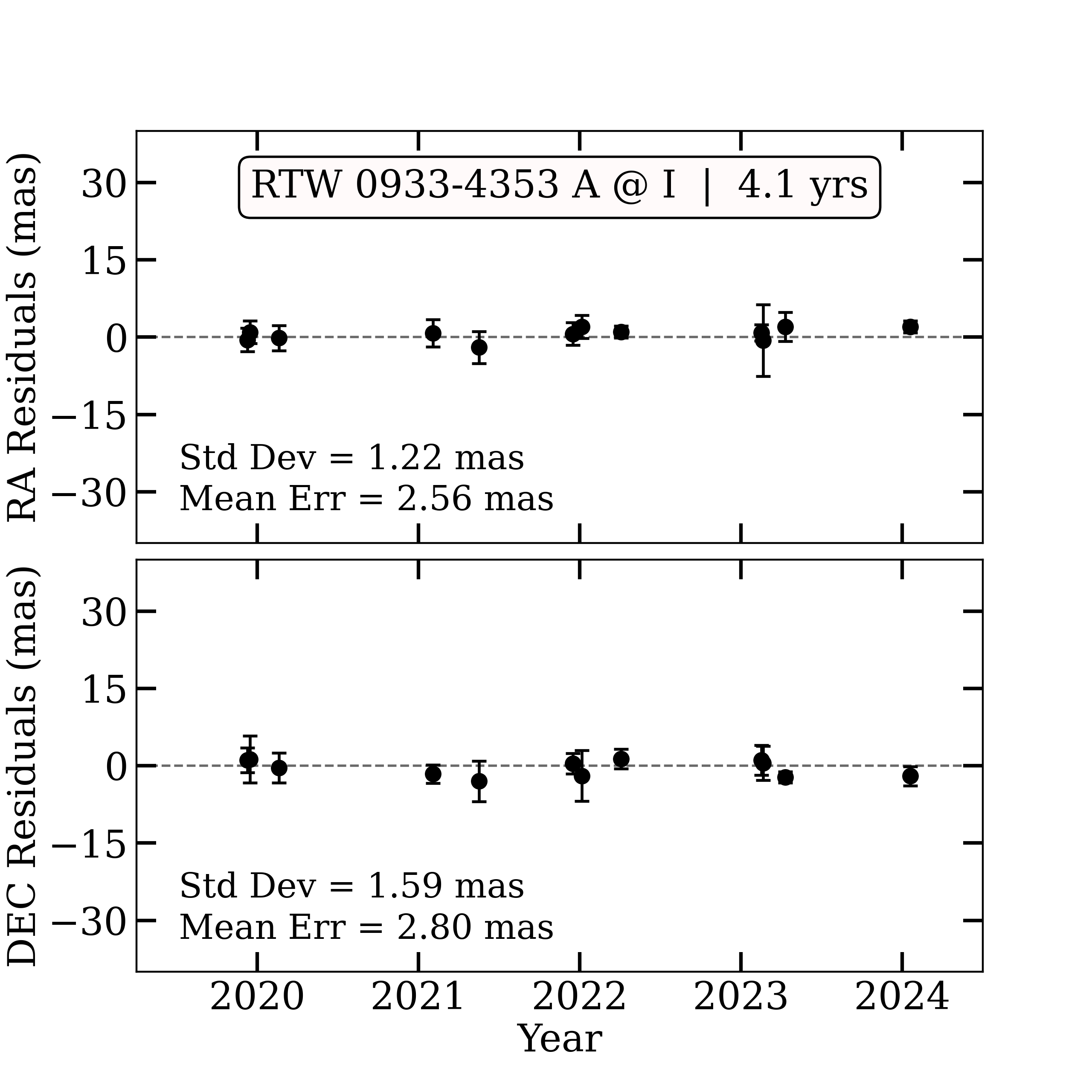}{0.49\textwidth}{(a)}
          \fig{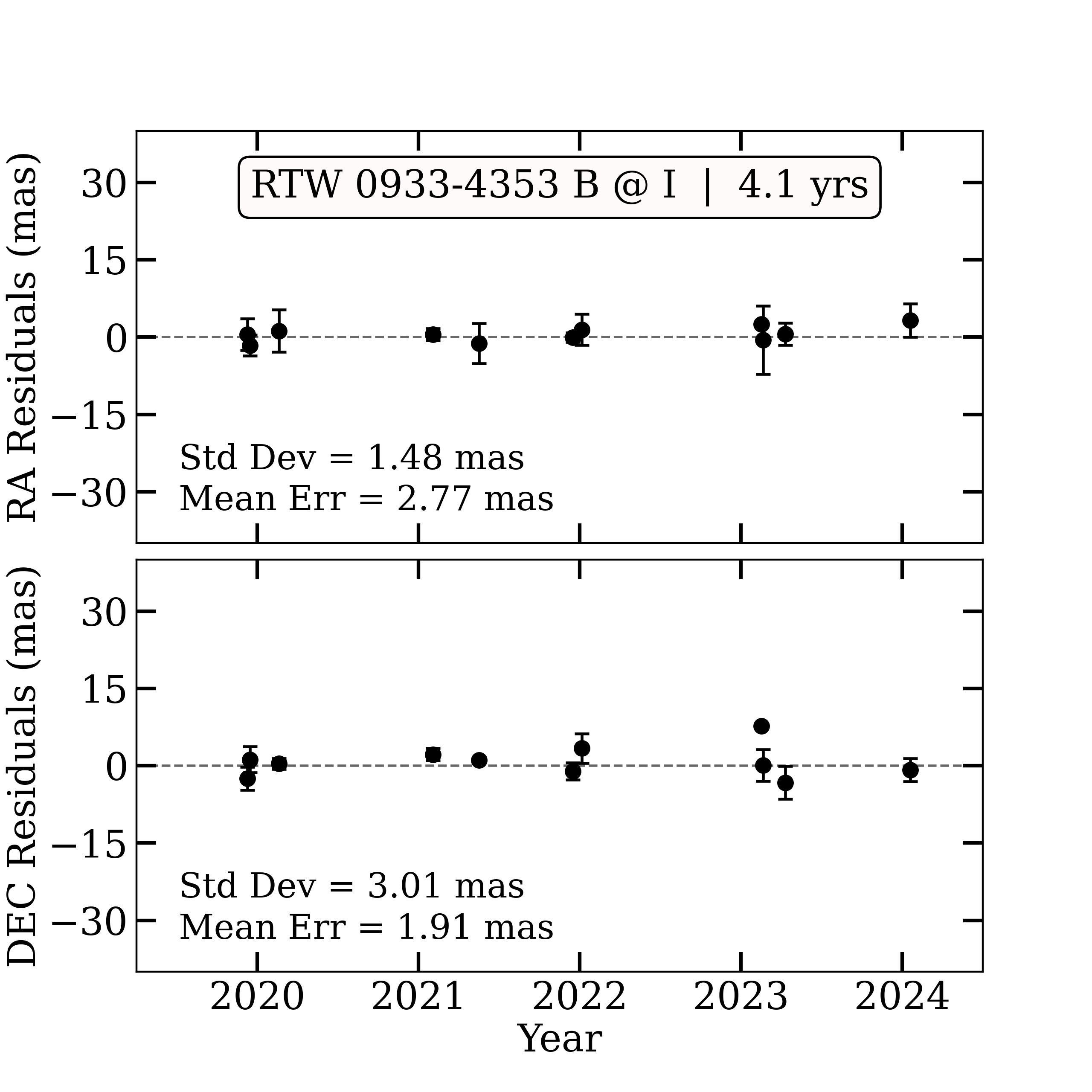}{0.49\textwidth}{(b)}}
\figcaption{An example set of astrometric results from the 0.9\,m long-term program for our twin stars RTW~0933-4353~A and B, using nightly mean points averaged from the multi-frame visits. The positional residuals in RA (top) and Declination (bottom) after fitting for parallactic motion and proper motion are shown, along with informative statistics and dashed lines at zero. The insets specify the observation filter and data time span. An additional unseen companion could present itself in these photocenter measurements as orbital motion if it existed. The absence of such motions in these data is therefore consistent with neither component hosting an unresolved companion down to the measurement limits, which typically reach to low mass brown dwarf companions in most orbits out to several times the duration of the time series. \label{fig:rtw0933-astrometry}}
\end{figure}

Of all stars in the 16 New Systems with long-term data available from the 0.9\,m, none show reliable astrometric perturbation signals beyond the noise after fitting for and removing their parallactic and proper motion movements. The same lack of perturbations was true for stars in the four systems in \citetalias{RTW_P1}. We show an example of 0.9\,m time series astrometry residuals in Figure~\ref{fig:rtw0933-astrometry} for RTW~0933-4353~A and B. Offsets in the astrometry from the dotted lines in RA and/or Dec are considered to be perturbations due to unseen orbiting companions, typically tens of milliarcseconds in size. Overall, the astrometric noise is typically $\sim$5\,mas for stars in the New Systems. The lack of perturbations is consistent with our components lacking additional unresolved stellar companions; we note that the known and suspected higher-order multiples also do not (yet) display orbital signatures in the available data.

%%%%%%%%%%%%%%% fig - 0.9m long-term astrometry example %%%%%%%%%%%%%%%

\subsection{Long-Term Photometry Results} \label{subsec:longterm-phot-results}

%%%%%%%%%%%%%%% fig - 0.9m long-term light curves %%%%%%%%%%%%%%%
\begin{figure}[!tp]
\centering
\includegraphics[width=0.906\textwidth]{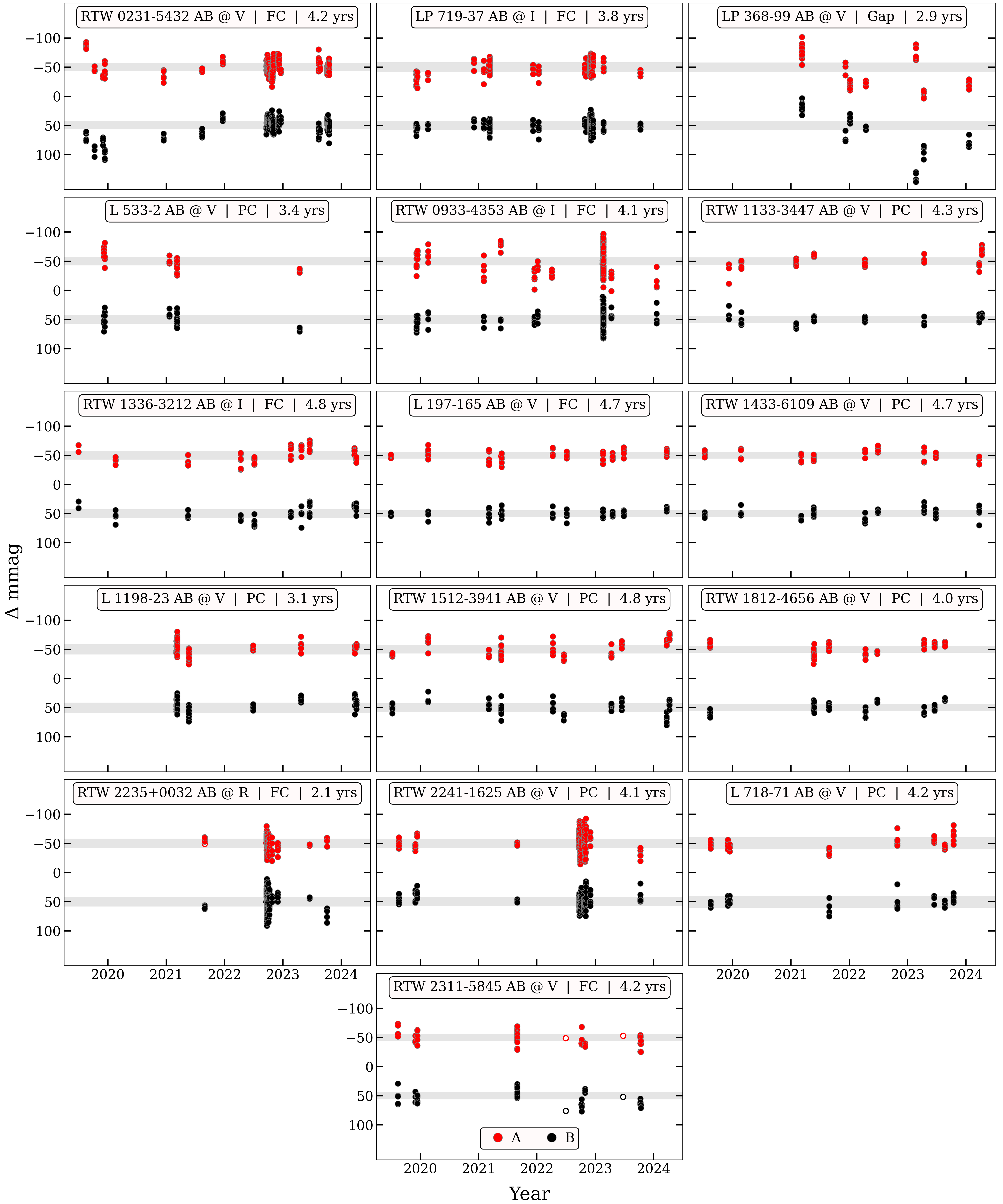}
\figcaption{Differential light curves from the 0.9\,m long-term program for 16 of our observed New Systems, ordered by ascending RA, with known or suspected non-twin higher order multiples excluded. A (red) and B (black) components are compared in each subplot, offset by $-50$\,mmag for A and $+50$\,mmag for B. Times are 2000 plus the number of Julian years since J2000. Grey shaded regions show the average noise of the non-varying reference stars in each field, spanning above and below each component's mean magnitude; typical noise levels are 5--10 mmag. Insets specify the filter, structure, and data time span. Corresponding long-term variability and noise measurements are given in Table~\ref{tab:long-term-MADs}. The large clusters of points for some stars are high-cadence rotation observations discussed in Section~\ref{sec:rot}, while open circles indicate visits with only one usable frame. Strong variations in LP~368-99~A and B are likely due to rotation, while RTW~1133-3447~A is contaminated by a nearby diffraction spike. These light curve data are available as Data behind the Figure (DbF) products. \label{fig:long-term-curves}}
\end{figure}

\begin{figure}[!t]
\centering
\includegraphics[width=0.49\textwidth]{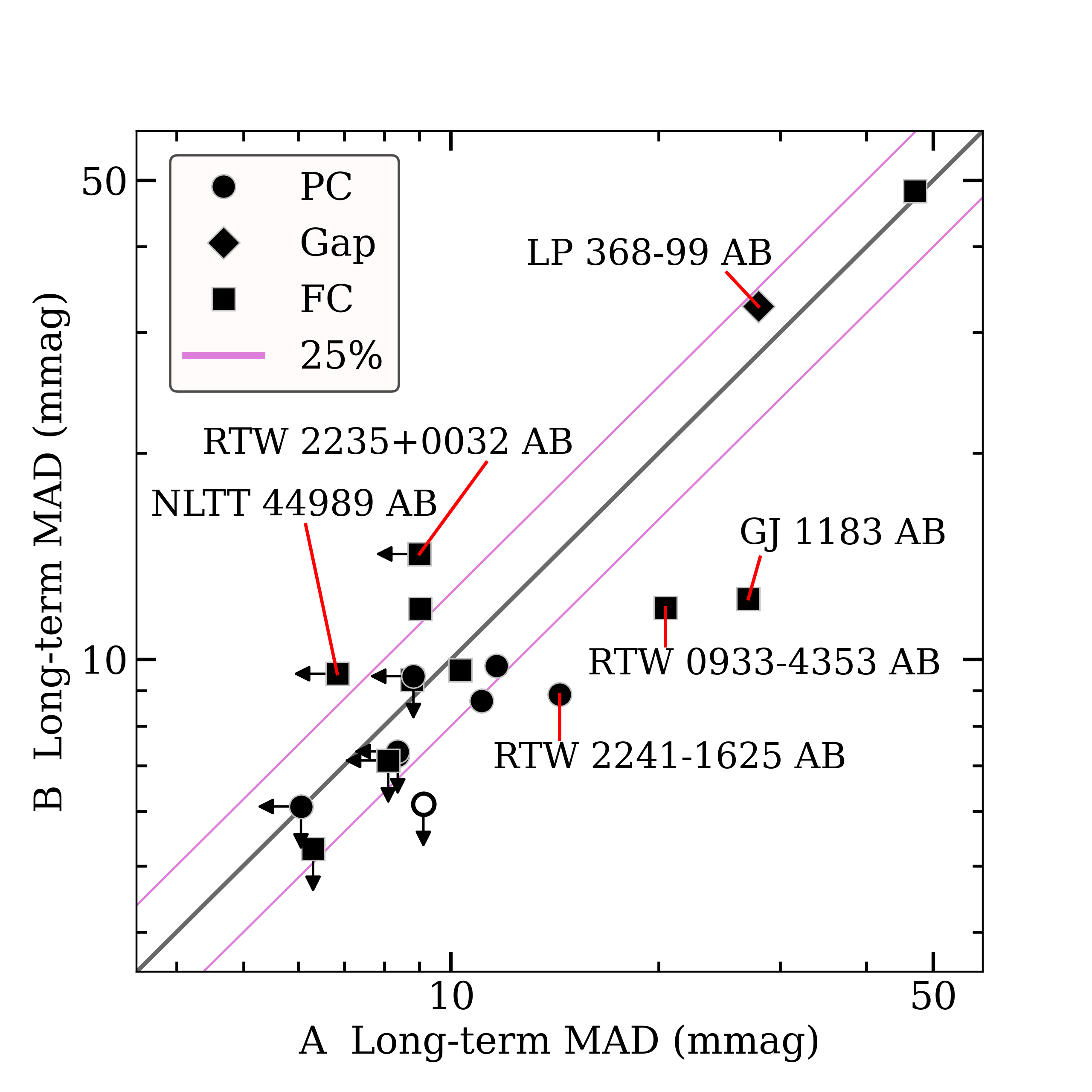}
\figcaption{A comparison is given of the 0.9\,m long-term photometric variability levels between twin pair components from our Results Sample, using values from Table~\ref{tab:long-term-MADs} and \citetalias{RTW_P1}. A one-to-one line and corresponding 25\% (1.25$\times$) bounding lines are shown. Point shapes correspond to the stars' interior structures as specified in the legend. A star with a MAD variability measure less than or within 10\% of its reference field MAD noise level is considered not varying beyond the noise and shows a corresponding upper limit arrow. The single open circle indicates the system RTW~1133-3447~AB, where the A component's variability measure is unreliable owing to a contaminating diffraction spike. Several key systems discussed throughout the text are labeled. While sometimes limited in baseline, cadence, and/or precision, these results begin to probe the predictability of long-term multi-year photometric variability in twin M dwarfs. \label{fig:long-term-MAD-equal}}
\end{figure}

We present the new long-term program light curves for 16 pairs in Figure~\ref{fig:long-term-curves}, visually comparing the A and B twin components in each binary. All components exhibit only minor variations in photometric variability, except for LP~368-99~A and B, for which the variability likely stems from spots dis/appearing over their 1-3\,d rotation periods, as explored in Section~\ref{sec:rot}. Long-term variability measurements are characterized using the mean absolute deviation (MAD) from the mean and are reported in Table~\ref{tab:long-term-MADs}, and shown for the A versus B components from our Results Sample in Figure~\ref{fig:long-term-MAD-equal}. The data are typically somewhat limited in baseline at 2--5\,yrs compared to the multi-year or even multi-decade timescales of M dwarf activity cycles \citep[e.g.,][A.~A.~Couperus et al.~2025 in preparation]{2016A&A...595A..12S, 2023ApJ...949...51I, Kar_2024, RTW_P1}, but preliminary assessments are possible. If we adopt a 25\% difference as the limit for A and B components ``matching" versus ``mismatching" in long-term photometric variability, then of the 19 cases in Figure~\ref{fig:long-term-MAD-equal} (16 from this paper and 3 from \citetalias{RTW_P1}) there are 12 systems with matching activity and 7 systems with mismatched activity\footnote{The RTW~1512-3941~AB system lands very close to the 25\% boundary line in Figure~\ref{fig:long-term-MAD-equal} but is just within 25\% for an actual percent difference calculation relative to the average component MAD.}. However, RTW~1133-3447~A resides on a contaminating diffraction spike in our 0.9\,m images (see Section~\ref{subsec:contam-new-obs}), very likely falsely inflating its perceived variability and thus necessitating this system's removal from our long-term variability assessment. {\bf Among the 18 systems with clean long-term observations at the 0.9\,m, 12 have matching and 6 have mismatching long-term variability, yielding a 33\% (6/18) activity mismatch rate across the full M sequence.}

%%%%%%%%%%%%%% tab - 0.9m long-term MAD values %%%%%%%%%%%%%%%
\begin{deluxetable}{lcccc}[!ht]
\tablewidth{0pt}
\tablecaption{Variability Measurements for 0.9\,m Long-term Program Data\label{tab:long-term-MADs}}
\tablehead{
\colhead{Name} & \colhead{Filter} & \colhead{$\rm{MAD_{long}}$} & \colhead{$\rm{MAD_{noise}}$} & \colhead{Ratio} \\
\colhead{} & \colhead{} & \colhead{(mmag)} & \colhead{(mmag)} & \colhead{}
}
\startdata
RTW 0231-5432 A  & V &  8.78  &  6.90 & 1.27  \\
RTW 0231-5432 B  &   &  9.33  &  6.91 & 1.35  \\
    LP 719-37 A  & I &  8.11  &  8.30 & 0.98  \\
    LP 719-37 B  &   &  7.12  &  8.30 & 0.86  \\
    LP 368-99 A  & V & 27.90  &  8.06 & 3.46  \\
    LP 368-99 B  &   & 32.74  &  8.06 & 4.06  \\
      L 533-2 A  & V & 11.64  &  7.38 & 1.58  \\
      L 533-2 B  &   &  9.78  &  7.38 & 1.33  \\
RTW 0933-4353 A  & I & 20.46  &  7.24 & 2.83  \\
RTW 0933-4353 B  &   & 11.89  &  7.54 & 1.58  \\
RTW 1133-3447 A  & V & (9.12) &  6.63 &(1.38) \\
RTW 1133-3447 B  &   &  6.15  &  6.62 & 0.93  \\
RTW 1336-3212 A  & I & 10.32  &  7.51 & 1.37  \\
RTW 1336-3212 B  &   &  9.64  &  7.51 & 1.28  \\
    L 197-165 A  & V &  6.31  &  5.72 & 1.10  \\
    L 197-165 B  &   &  5.29  &  5.72 & 0.92  \\
RTW 1433-6109 A  & V &  6.06  &  5.77 & 1.05  \\
RTW 1433-6109 B  &   &  6.10  &  5.77 & 1.06  \\
    L 1198-23 A  & V &  8.82  &  8.75 & 1.01  \\
    L 1198-23 B  &   &  9.45  &  8.74 & 1.08  \\
RTW 1512-3941 A  & V & 11.08  &  7.16 & 1.55  \\
RTW 1512-3941 B  &   &  8.70  &  7.19 & 1.21  \\
RTW 1812-4656 A  & V &  8.36  &  5.81 & 1.44  \\
RTW 1812-4656 B  &   &  7.24  &  5.81 & 1.25  \\
RTW 2235+0032 A  & R &  9.00  &  8.24 & 1.09  \\
RTW 2235+0032 B  &   & 14.25  &  8.25 & 1.73  \\
RTW 2241-1625 A  & V & 14.36  &  7.94 & 1.81  \\
RTW 2241-1625 B  &   &  8.89  &  7.93 & 1.12  \\
     L 718-71 A  & V &  8.37  & 10.41 & 0.80  \\
     L 718-71 B  &   &  7.34  & 10.41 & 0.71  \\
RTW 2311-5845 A  & V &  9.02  &  6.29 & 1.43  \\
RTW 2311-5845 B  &   & 11.85  &  6.29 & 1.88  \\
\enddata
\tablecomments{Given are the mean absolute deviation (MAD) values from the mean for our 0.9\,m long-term light curves presented in Figure~\ref{fig:long-term-curves}; single-frame epochs are excluded in the MADs. Each system's observation filter for the long-term program is specified in the A component's row. The $\rm{MAD_{long}}$ value measures variations in the stars' light curves, while the $\rm{MAD_{noise}}$ value gives the average MAD of the non-varying reference stars used in the differential photometry analysis. Values markedly above 1.0 in the Ratio column --- which divides the two MAD values --- indicate likely long-term variability; only LP~368-99~A and B stand out significantly, but those variations are likely due to rotation. Values for RTW~1133-3447~A are given in parentheses because its variability is likely inflated by a contaminating diffraction spike. A and B components use the same underlying images and reference fields so have very similar noise measures, but certain images are occasionally excluded for one star and not the other due to strong flares, cosmic rays, or seeing-dependent contamination, so slight differences in $\rm{MAD_{noise}}$ between A and B are possible.}
\end{deluxetable}

We caution that some of the systems have similar long-term MAD measures simply because both stars show no variations beyond the noise, and some have greater sampling in the short-term rotation versus long-term cycle regimes. For example, we see a candidate weak activity cycle signal for RTW~0231-5432~B and possibly A as well, analogous to the candidate cycle signals we reported for GJ~1183~A and B in \citetalias{RTW_P1}. For both systems, the presumed long-term cycle amplitudes appear slightly larger than the short-term rotation amplitudes (similar to the general findings of \citet{Kar_2024}), emphasizing the need for more long-term data with careful handling of short versus long variability signatures before conclusive claims can be made. There is also the bias that our long-term program did not look at every star observable --- unlike the H$\alpha$ activity effort --- with choices for 0.9\,m program target inclusion or exclusion sometimes based on the preliminary activity behaviors of our stars, thus introducing a small selection effect for a handful of systems. This does not impact our comparisons between A and B in a given observed pair or our overall probing for cycles, but could impact how representative our cumulative statistics are for the long-term activity mismatches.

Altogether, our ongoing long-term 0.9\,m program is regularly acquiring new data for the twin systems, and future work can continue this line of investigation. With additional data, using nightly means instead of all individual points may allow us to push down to 1--2\,mmag precision for some systems, enhancing the long-term variability measurements as long as careful binning of the shorter rotation signals is addressed simultaneously. The archival multi-year ZTF, ASAS-SN, and Gaia photometry data we employ later for our rotation analysis in Section~\ref{sec:rot} also sometimes show candidate long-term multi-year brightness trends\footnote{For example, long-term trends can be seen later in the rotation ZTF data for LP~368-99~A and B in Figure~\ref{fig:extra-rot}, but for these specific cases it is unclear if the trends are truly astrophysical or if they result from the moderate blending of A and B in the ZTF data.}, but we leave a deeper investigation of these signals to our future work alongside the growing 0.9\,m coverage.

%%%%%%%%%%%%%%% fig - 0.9m long-term MAD equality %%%%%%%%%%%%%%%

\section{Rotation Periods --- SMARTS 0.9\,m, Gaia, ZTF, TESS, and ASAS-SN} \label{sec:rot}

Our final observational program is the determination of stellar rotation periods via photometric brightness modulations as spots and faculae move in and out of view. We used the 0.9\,m and archival light curves from Gaia, ZTF, TESS, and ASAS-SN following the same general approaches from \citetalias{RTW_P1} and expanded upon next\footnote{We also crossmatched the New Systems against Kepler and K2 \citep{Kepler, K2}, but only had two matches, as detailed in the system-by-system rotation assessment notes. A handful of matches with MEarth are similarly discussed in the individual assessment notes \citep{Berta_2012, Newton_2016, Newton_2018}, although these cases were always rotation non-detections in the MEarth results.}. Because many of the binaries have separations less than 10$\arcsec$, leveraging multiple sources of light curve information is advantageous to tease out individual rotation periods. We outline the utilized data sources and methods in Section~\ref{subsec:rot-methods}, review relevant contamination and systematic factors for these data sources in Section~\ref{subsec:rot-contam-syst}, list the results in Table~\ref{tab:RotTable} (in which an ``X" indicates the star was observed by the particular program but no rotation period was derived), provide target-by-target rotation assessment notes in \ref{subsec:rot-notes}, and discuss the rotation results in Section~\ref{subsec:rot-results}.

\subsection{Rotation Observations and Data Processing} \label{subsec:rot-methods}

\textit{SMARTS 0.9\,m}: Of our New Systems on the long-term 0.9\,m program, six were observed with the 0.9\,m at a faster cadence via NOIRLab time (ID 2023A-549259; PI Couperus) to capture rotation signals, as indicated in Table~\ref{tab:SampleTable-phot}. We targeted systems visible from CTIO with activity signatures or archival blended rotation results that indicated rotation detections were feasible with the 0.9\,m. Systems not observed at the 0.9\,m often showed inactive H$\alpha$ absorption from CHIRON and/or non-detections of rotation in archival data, making it difficult or impossible for us to detect the likely very low-amplitude and/or long-duration periods in such cases even if they were observed. The 0.9\,m rotation observations and analysis procedures follow as detailed in \citetalias{RTW_P1}, with some stars here observed using the $R$ or $I$ filters instead of $V$.

\textit{Gaia}: Unlike \citetalias{RTW_P1}, in which none of the systems had Gaia DR3 light curves available, here we also employ Gaia DR3 epochal light curve photometry in $G$, BP, and RP \citep{GaiaDR3_Var}. DR3 only released these data for a subset of sources, inhibiting a comprehensive assessment, but 21 of our 64 components in the New Systems have the epochal photometry available for use. The exquisite spatial resolution of Gaia makes this particularly valuable as it resolves components in our pairs. We assessed the light curves in all three Gaia filters and excluded points rejected by the Gaia variability processing based on the \texttt{variability\_flag\_[G/BP/RP]\_reject} flags; the final curves typically had $\sim$50 points over the 34-month baseline of DR3.

\textit{ZTF}: For ZTF, we utilized PSF-fit light curves acquired from Data Release 20 for the New Systems \citep{ZTF_data}. We used data from both the $zg$ and $zr$ filters, the same as in \citetalias{RTW_P1}, but here we also assessed $zi$ results if $\gtrsim$\,10 points were available. These ZTF light curves for our stars have an average of roughly 400 points in $zg$ and 700 points in $zr$, but many of our southern targets lack coverage because ZTF is based in the northern hemisphere.

\textit{TESS}: The TESS data utilized were acquired from the Mikulski Archive for Space Telescopes (MAST) in April of 2024, with sector 75 being the latest sector of data employed. We used processed TESS time series data products from the TESS-SPOC pipeline for high-cadence and Full-Frame Image (FFI)-cadence light curves \citep{TESS, 2020RNAAS...4..201C}, but when these were unavailable we instead leveraged light curves from the MIT Quick-Look Pipeline \citep[QLP:][]{QLP1, QLP2, QLP3, QLP4}. For TESS-SPOC light curves, we used the Pre-search Data Conditioning Simple Aperture Photometry (PDCSAP) fluxes and errors. For QLP light curves, we used the provided KSPSAP fluxes and errors for sectors 1--55, while for later sectors we used the newly-provided \texttt{SYS\_RM\_FLUX} fluxes along with \texttt{DET\_FLUX\_ERR} errors, as this flux better preserves stellar variability. All of our New Systems were observed by TESS in at least one sector, averaging roughly three sectors per system, with some stars captured in as many as eleven sectors. 63 of the 64 components had light curves specific to their individual TESS Input Catalog (TIC) sources available from TESS-SPOC or QLP.

\textit{TESS-unpop}: We also used new TESS FFI light curves we derived from the \texttt{unpopular} package \citep{unpopular} to search for astrophysical signals beyond $\sim$13.5\,days \citep[][A.~Kar et al.~2025 in preparation]{2025AJ....170...15H}, referred to as ``TESS-unpop" data throughout this work. The preservation of long-term signals relies on the inclusion of a polynomial component in the unpopular processing, which we only included for our reported results if the raw Simple Aperture Photometry (SAP) fluxes also show a long-term signal, following \cite{Kar_2024}. Cases with ambiguous SAP flux trends are detailed in the target-by-target notes, and we always reviewed both the with-polynomial and without-polynomial unpopular data for thoroughness. The latest sector of data used for the unpopular results in this work was sector 76, one sector beyond the other TESS products above owing to the timing of our analyses. When multiple sectors were available with unpopular, we also generated a merged timeseries of all the sectors, wherein each individual sector's delta magnitude measurements where relative to that sector's mean magnitude. This simple merging does not robustly correct for sector-to-sector systematics, variations in the mean magnitude, or different sampling cadences across sectors, but still proved valuable in some cases to uncover preliminary rotation signals much longer than half a sector. All 64 components in the New Systems had FFI light curves available via unpopular with our manually chosen apertures, where a 3$\times$3 pixel grid was a commonly used aperture.

\textit{ASAS-SN}: For ASAS-SN, we assessed any pre-computed light curves available from the ASAS-SN photometry page \citep{Jayasinghe_2019}, but also generated full ASAS-SN Sky Patrol \citep[V1;][]{Kochanek_SkyPatrol} $V$-band and $g$-band aperture light curves for all components at their individual Gaia DR3 2016.0 positions (instead of only deriving these for cases without pre-computed curves available, as done in \citetalias{RTW_P1}). A few stars in the New Systems did not have the pre-computed curves available, and said curves also have far fewer data points than the up-to-date Sky Patrol service offers, motivating this change. For Sky Patrol data, we also removed any points with errors more than two standard deviations larger than the mean error, instead of simply removing points with errors $>$\,50\,mmag as done in \citetalias{RTW_P1}. The ASAS-SN Sky Patrol light curves for our stars typically have 900 points in $V$ and 3000 points in $g$.

Rotation analyses for all datasets excluded points using any provided quality flags and used the Lomb--Scargle periodogram methodology outlined in \citetalias{RTW_P1} \citep{Lomb_1976, Scargle_1982, Baluev_2008, Generalised_LS_2009, 2013A&A...558A..33A, Astropy2018}. Here we also sometimes used a modified periodogram range for certain cases to examine long-period signals with more precision and without erroneous 1-day sampling peaks. We report periods to 0.01 precision, except for uncertain TESS-unpop results using rounded estimates and the case of RTW~0933-4353~AB with extremely rapid $\sim$3\,hr periods precisely measured using multiple TESS sectors. We favored not adopting rotation periods if choices were uncertain or relied on only candidate detections, as indicated in Table~\ref{tab:RotTable}.

\subsection{Archival Rotation Data Blending, Contamination, and Systematics} \label{subsec:rot-contam-syst}

The archival light curve data from Gaia, ZTF, TESS, and ASAS-SN we use for rotation can have varying levels of AB blending and background contamination depending on the data source and binary system in question, as demonstrated in Figure~\ref{fig:field-radii}. Only six of our New Systems have separations larger than $60\arcsec$, roughly corresponding to a 3$\times$3 aperture of 21$\arcsec$ TESS pixels. Only seven systems are more widely separated than the 32$\arcsec$ diameter ASAS-SN apertures. ZTF utilizes robust PSF fitting with higher resolution 1\farcs01 pixels, so blending and contamination are much less of a concern for ZTF, e.g., we find that the 4\farcs61 separation binary LP~368-99~A and B have resolved rotation periods from ZTF. Gaia BP and RP are sensitive to contamination from sources within roughly 3\farcs5 of the target \citep{2023A&A...674A...2D}, while $G$ measures are better resolved using windows of roughly $1\farcs0\times2\farcs1$ \citep{2016A&A...595A...3F}, so our Gaia light curve results always resolve A and B in a pair but can potentially still have contamination from very close background sources.

These various factors are categorized with flags on the rotation results in Table~\ref{tab:RotTable}, using criteria similar to \cite{Kar_2024}, as follows. We identify non-target sources from the Gaia DR3 catalog falling within a radial distance around each target star --- ASAS-SN: 32\farcs0, ZTF: 8\farcs0, Gaia BP \& RP: 3\farcs5, Gaia $G$: 2\farcs1 --- or falling within the specific pixels used in the TESS SPOC and unpopular apertures\footnote{We do not utilize the contamination ratio provided in the TESS Input Catalog \citep{2018AJ....156..102S, 2019AJ....158..138S} because it uses a very large 10-pixel contaminant search region and also processes the other binary component as a contaminating source.} \citep[identified via \texttt{tpfplotter}\footnote{The Python package \texttt{tpfplotter} by J.~Lillo-Box is publicly available at \href{http://www.github.com/jlillo/tpfplotter}{www.github.com/jlillo/tpfplotter}.};][]{tpfplotter}. A flag of `C0' indicates negligible or no contamination, defined as no fainter contaminants in the measurement region with $\Delta G <$ 4\,mag compared to the target star $G$ mag. `C1' is mild contamination, where at least one background source exists with $\Delta G$ = 2--4\,mag. `C2' is major contamination, with at least one contaminant having $\Delta G <$ 2\,mag. We also visually assessed the TESS and TESS-unpop fields to identify any cases with very bright contaminating sources falling outside the chosen apertures but still heavily bleeding into them, which we consider in the subsequent target-by-target notes. Gaia results could also have different contamination flags in $G$ versus BP/RP for the same star, but this never occurred, so we report the consistent flags. The individual target stars will often have their other binary components nearby, but we do not count the pair members as contaminants and instead use the flag `bl' to indicate any data partially or fully blending the AB components; we use the `un' flag for unblended cases. We report all of these flags for cases that had data available, whether or not they showed a rotation signal. The new 0.9\,m rotation results are functionally free from impactful contamination and always suitably resolve A and B, as discussed in Section~\ref{sec:contam}.

%%%%%%%%%%%%%%% tab - Rotation Periods %%%%%%%%%%%%%%%
\begin{longrotatetable}
\movetabledown=10mm
\begin{deluxetable}{lllllll|cc}
\tabletypesize{\scriptsize}
\tablewidth{0pt}
\tablecaption{Compiled Rotation Periods and Amplitudes for the 32 New Systems\label{tab:RotTable}}
\tablehead{
\colhead{Name} & \colhead{0.9m $P_{\rm{rot}}$} & \colhead{Gaia $P_{\rm{rot}}$} & \colhead{ZTF $P_{\rm{rot}}$} & \colhead{TESS $P_{\rm{rot}}$} & \colhead{TESS-unpop $P_{\rm{rot}}$} & \colhead{ASAS-SN $P_{\rm{rot}}$} & \colhead{Final $P_{\rm{rot}}$} & \colhead{$\rm{amp}_{\rm{rot}}$}\\
\colhead{} & \colhead{(days)} & \colhead{(days)} & \colhead{(days)} & \colhead{(days)} & \colhead{(days)} & \colhead{(days)} & \colhead{(days)} & \colhead{(mmag)}
}
\startdata
RTW 0143-0151 A  &  \nodata   &  \nodata    &     bl-C0: X        &    \nodata                      &      bl-C1: X                 &     bl-C1: X           &   \nodata    &    \nodata       \\
RTW 0143-0151 B  &  \nodata   &  \nodata    &     bl-C0: X        &    bl-C0: X                     &      bl-C1: X                 &     bl-C1: X           &   \nodata    &    \nodata       \\
RTW 0231-5432 A  &un-C0: 69.78&  \nodata    &     \nodata         &    bl-C0: 7.22?                 &   bl-C0: $\sim$32?            & bl-C0: 66.14 \& 69.45  &    69.78     &   0.9m-V: 25.9   \\
RTW 0231-5432 B  &un-C0: 68.82&  \nodata    &     \nodata         &    bl-C0: 7.12?                 &   bl-C0: $\sim$30?            & bl-C0: 70.35 \& 66.27  &    68.82     &   0.9m-V: 18.8   \\
RTW 0231+4003 A  &  \nodata   &  \nodata    &     bl-C0: 29.47?   & bl-C1: 3.47? \& 4.19? \& 15.03? &   bl-C1: $\gtrsim$30?         &     bl-C2: X           &   \nodata    &    \nodata       \\
RTW 0231+4003 B  &  \nodata   &  \nodata    &     bl-C0: 29.65?   & bl-C0: 3.41? \& 4.69? \& 15.50? &   bl-C0: $\gtrsim$30?         &     bl-C1: X           &   \nodata    &    \nodata       \\
RTW 0409+4623 A  &  \nodata   &  \nodata    &     bl-C0: X        &    bl-C1: 2.24 \& 2.55          &   bl-C1: 2.26 \& 1.27?        &     bl-C1: X           & 2.24 or 2.55 & TESS: 5.9 or 7.1 \\
RTW 0409+4623 B  &  \nodata   &  \nodata    &     bl-C0: X        &    bl-C1: 2.29                  &   bl-C1: 2.25 \& 1.27?        &     bl-C1: X           & 2.24 or 2.55 & TESS: 5.9 or 7.1 \\
KAR 0545+7254 A  &  \nodata   &un-C0: X     &     un-C0: X        &    bl-C2: X                     &   bl-C2: 42.14?               &     un-C2: X           &   \nodata    &    \nodata       \\
KAR 0545+7254 B  &  \nodata   &  \nodata    &     un-C0: X        &    bl-C0: X                     &   bl-C0: 36.59?               &     un-C0: X           &   \nodata    &    \nodata       \\
LP 719-37 A      &un-C0: 5.08?&  \nodata    &     bl-C0: X        &    bl-C1: 5.18 \& 4.03          &   bl-C1: 5.19 \& 4.04         &     bl-C1: X           &     5.19     &   0.9m-I: 9.7    \\
LP 719-37 B      &un-C0: 4.07?&  \nodata    &     bl-C0: X        &    bl-C1: 4.04 \& 5.18          &   bl-C1: 4.04 \& 5.18         &     bl-C1: X           &     4.04     &   0.9m-I: 8.5    \\
G 103-63 A*      &  \nodata   &  \nodata    &     bl-C0: X        &    bl-C2: 6.36?                 &   bl-C2: $\sim$39?            &     bl-C2: X           &   [95.73]    &    \nodata       \\
G 103-63 B       &  \nodata   &  \nodata    &     bl-C0: X        &    bl-C2: X                     &   bl-C2: $\sim$39?            &     bl-C2: X           &   [85.68]    &    \nodata       \\
RTW 0824-3054 A  &  \nodata   &  \nodata    &     \nodata         &    bl-C2: 1.14                  &   bl-C1: 1.14 \& $\sim$39?    &     bl-C2: X           &   \nodata    &    \nodata       \\
RTW 0824-3054 B* &  \nodata   &  \nodata    &     \nodata         &    bl-C2: 1.15                  &   bl-C1: 1.14 \& $\sim$39?    &     bl-C2: X           &   \nodata    &    \nodata       \\
LP 368-99 A      &  \nodata   &un-C0: 1.19  &     bl-C0: 1.19     &    bl-C0: 1.19 \& 2.54          &   bl-C0: 1.19 \& 2.53         &  bl-C0: 1.19 \& 2.53   &     1.19     &  ZTF-zr: 53.0    \\
LP 368-99 B      &  \nodata   &  \nodata    &     bl-C0: 2.53     &    bl-C0: 1.19 \& 2.54          &   bl-C0: 1.19 \& 2.53         &  bl-C0: 1.19 \& 2.53   &     2.53     &  ZTF-zr: 100.7   \\
L 533-2 A        &  \nodata   &un-C0: 20--70?&    \nodata         &    bl-C1: X                     &   bl-C1: $\sim$30?            &     bl-C1: X           &   \nodata    &    \nodata       \\
L 533-2 B        &  \nodata   &un-C0: 10--80?&    \nodata         &    bl-C1: X                     &   bl-C1: $\sim$30?            &     bl-C1: X           &   \nodata    &    \nodata       \\
RTW 0933-4353 A  &un-C0: 0.131&un-C0: 0.134 &     \nodata         &    bl-C2: 0.134 \& 0.115        &   bl-C2: 0.134 \& 0.115       &     bl-C2: X           &     0.134    &   0.9m-I: 62.6   \\
RTW 0933-4353 B  &un-C0: 0.118&  \nodata    &     \nodata         &    bl-C2: 0.115 \& 0.134        &   bl-C2: 0.115 \& 0.134       &     bl-C2: X           &     0.115    &   0.9m-I: 37.3   \\
LP 551-62 A      &  \nodata   &un-C0: X     &     un-C0: X        &    bl-C0: X                     &   bl-C0: $\gtrsim$35?         &     bl-C0: X           &   \nodata    &    \nodata       \\
LP 551-62 B      &  \nodata   &  \nodata    &     un-C0: X        &    bl-C0: X                     &   bl-C0: $\gtrsim$35?         &     bl-C0: X           &   \nodata    &    \nodata       \\
RTW 1123+8009 A  &  \nodata   &  \nodata    &     un-C1: 92.20?   &    bl-C1: X                     &      bl-C1: 52.27?            &     bl-C1: X           &   [105.26]   &    \nodata       \\
RTW 1123+8009 B  &  \nodata   &  \nodata    &     un-C0: 101.65?  &    bl-C0: 7.56? \& 9.03?        &      bl-C1: 40.11?            &     bl-C1: X           &   [85.67]    &    \nodata       \\
RTW 1133-3447 A  &  \nodata   &un-C0: X     &     \nodata         &    un-C2: 0.90 \& 4.65?         &   un-C0: 0.90?                &     un-C2: X           &   \nodata    &    \nodata       \\
RTW 1133-3447 B  &  \nodata   &  \nodata    &     \nodata         &    un-C0: X                     &   un-C0: X                    &     un-C0: X           &   \nodata    &    \nodata       \\
RTW 1336-3212 A  &  \nodata   &  \nodata    &     \nodata         &    bl-C0: X                     &   bl-C0: $\gtrsim$30?         &     bl-C0: X           &   \nodata    &    \nodata       \\
RTW 1336-3212 B  &  \nodata   &  \nodata    &     \nodata         &    bl-C0: 4.53?                 &   bl-C0: $\gtrsim$30?         &     bl-C0: X           &   \nodata    &    \nodata       \\
L 197-165 A      &  \nodata   &  \nodata    &     \nodata         &    un-C1: X                     &   un-C1: $\sim$30?            &     un-C2: $\sim$30?   &   \nodata    &    \nodata       \\
L 197-165 B      &  \nodata   &un-C0: X     &     \nodata         &    un-C1: X                     &   un-C1: $\gtrsim$30?         &     un-C1: $\sim$30?   &   \nodata    &    \nodata       \\
RTW 1433-6109 A  &  \nodata   &un-C0: X     &     \nodata         &    bl-C2: 1.2--1.4 \& 2.8       &   bl-C2: $\sim$40?            &     bl-C2: X           &   \nodata    &    \nodata       \\
RTW 1433-6109 B  &  \nodata   &un-C0: X     &     \nodata         &    bl-C2: 1.2--1.4 \& 2.8       &   bl-C2: $\sim$40?            &     bl-C2: X           &   \nodata    &    \nodata       \\
L 1197-68 A      &  \nodata   &un-C0: X     &     un-C0: X        &    bl-C0: X                     &      bl-C0: X                 &     bl-C0: X           &   \nodata    &    \nodata       \\
L 1197-68 B      &  \nodata   &un-C0: X     &     un-C0: X        &    bl-C0: X                     &      bl-C0: X                 &     bl-C0: X           &   \nodata    &    \nodata       \\
L 1198-23 A      &  \nodata   &un-C0: X     &     un-C0: 53.13?   &    bl-C0: X                     &      bl-C0: X                 &     un-C0: 51.65?      &   \nodata    &    \nodata       \\
L 1198-23 B      &  \nodata   &  \nodata    &     un-C0: X        &    bl-C0: X                     &      bl-C0: $\gtrsim$30?      &     un-C0: 54.22?      &   \nodata    &    \nodata       \\
RTW 1512-3941 A  &  \nodata   &  \nodata    &     \nodata         &    bl-C0: X                     &   bl-C0: $\sim$27?            &     bl-C0: X           &   \nodata    &    \nodata       \\
RTW 1512-3941 B  &  \nodata   &  \nodata    &     \nodata         &    bl-C0: X                     &   bl-C0: $\sim$27?            &     bl-C0: X           &   \nodata    &    \nodata       \\
RTW 1812-4656 A  &  \nodata   &un-C0: X     &     \nodata         &    un-C2: 7.24? \& 12.30?       &   un-C2: $\sim$35?            &     un-C2: 29.43?      &   \nodata    &    \nodata       \\
RTW 1812-4656 B  &  \nodata   &un-C0: X     &     \nodata         &    un-C1: X                     &   un-C1: $\sim$28?            &     un-C1: 30.29?      &   \nodata    &    \nodata       \\
GJ 745 A         &  \nodata   &  \nodata    &     \nodata         &    un-C0: X                     &   un-C0: X                    &     un-C0: X           &   \nodata    &    \nodata       \\
GJ 745 B         &  \nodata   &  \nodata    &     \nodata         &    un-C0: X                     &   un-C0: X                    &     un-C0: X           &   \nodata    &    \nodata       \\
RTW 2011-3824 A  &  \nodata   &  \nodata    &     \nodata         &    bl-C0: X                     &      bl-C0: X                 &     bl-C0: X           &   \nodata    &    \nodata       \\
RTW 2011-3824 B  &  \nodata   &  \nodata    &     \nodata         &    bl-C0: X                     &      bl-C0: X                 &     bl-C0: X           &   \nodata    &    \nodata       \\
G 230-39 A       &  \nodata   &  \nodata    &     un-C0: 74.00?   &    bl-C0: X                     &      bl-C0: 68.82?            &     bl-C1: 72.01?      &   \nodata    &    \nodata       \\
G 230-39 B       &  \nodata   &  \nodata    &     un-C0: X        &    bl-C1: X                     &      bl-C0: 51.20?            &     bl-C1: 70.94?      &   \nodata    &    \nodata       \\
RTW 2202+5537 A  &  \nodata   &  \nodata    &     un-C0: 1.10?    &    bl-C1: 0.87 \& 0.97?         &   bl-C1: 0.87 \& 0.98         &     bl-C1: X           &     0.98     &    TESS: 5.0     \\
RTW 2202+5537 B  &  \nodata   &un-C0: 0.87  &     un-C0: 0.87     &    bl-C1: 0.87 \& 0.97?         &   bl-C1: 0.87 \& 0.98?        &     bl-C1: X           &     0.87     &    TESS: 6.2     \\
RTW 2211+0058 A  &  \nodata   &  \nodata    &     bl-C0: 73.35?   &    bl-C0: X                     &   bl-C0: $\gtrsim$25?         &     bl-C0: X           &   [77.98]    &    \nodata       \\
RTW 2211+0058 B  &  \nodata   &  \nodata    &     bl-C0: X        &    bl-C0: X                     &   bl-C0: $\gtrsim$25?         &     bl-C0: X           &   \nodata    &    \nodata       \\
RTW 2235+0032 A  &un-C0: 1.77 &  \nodata    &     bl-C0: 1.77     &    bl-C1: 1.78                  &      bl-C0: 1.77              &     bl-C1: X           &     1.77     &   0.9m-R: 16.8   \\
RTW 2235+0032 B  &un-C0: 1.80 &  \nodata    & bl-C0: 1.78 \& 2.27 &    bl-C1: 1.78                  &      bl-C0: 1.77              &     bl-C1: X           &     1.80     &   0.9m-R: 36.7   \\
RTW 2236+5923 A  &  \nodata   &  \nodata    &     bl-C0: 23.96?   &    bl-C0: X                     & bl-C0: 10.79? \& $\gtrsim$30? & bl-C0: 13.18? \& 27.55?&   [28.11]    &    \nodata       \\
RTW 2236+5923 B  &  \nodata   &  \nodata    &     bl-C0: 27.50?   &    bl-C0: X                     & bl-C0: 10.83? \& $\gtrsim$30? & bl-C0: 13.17? \& 27.58?&   [26.52]    &    \nodata       \\
RTW 2241-1625 A  &un-C0: 14.10&un-C0: X     &     bl-C0: 14.17?   &    bl-C0: 2.23? \& 4.70?        &   bl-C0: 13.87 \& 7.46?       & bl-C0: 13.91 \& 16.06? &     14.10    &   0.9m-V: 40.2   \\
RTW 2241-1625 B  &un-C0: 15.22&un-C0: 16.20?&     bl-C0: X        & bl-C0: 2.23? \& 4.72? \& 8.18?  &   bl-C0: 13.87 \& 7.46?       & bl-C0: 13.91 \& 16.06? &     15.22    &   0.9m-V: 19.7   \\
RTW 2244+4030 A  &  \nodata   &  \nodata    &     \nodata         &    bl-C1: 14.18?                &   bl-C1: $\gtrsim$30?         &     bl-C0: X           &   \nodata    &    \nodata       \\
RTW 2244+4030 B  &  \nodata   &  \nodata    &     \nodata         &    bl-C0: 12.18?                &   bl-C0: $\gtrsim$27?         &     bl-C0: X           &   \nodata    &    \nodata       \\
L 718-71 A       &  \nodata   &un-C0: X     &     un-C0: X        &    bl-C1: 6.47? \& 11.50?       &   bl-C1: $\sim$27?            &     un-C0: X           &   \nodata    &    \nodata       \\
L 718-71 B       &  \nodata   &  \nodata    &     un-C0: X        &    bl-C0: 6.41?                 &   bl-C0: $\gtrsim$25?         &     un-C0: X           &   \nodata    &    \nodata       \\
RTW 2311-5845 A  &un-C0: 0.88?&un-C0: 0.57  &     \nodata         &    bl-C0: 0.57 \& 0.73          &   bl-C0: 0.57 \& 0.73         &     bl-C1: 0.73?       &     0.57     &   0.9m-I: 10.1   \\
RTW 2311-5845 B  &un-C0: 0.73 &un-C0: 0.73  &     \nodata         &    bl-C0: 0.57 \& 0.73          &   bl-C0: 0.57 \& 0.73         &     bl-C1: 0.73?       &     0.73     &   0.9m-I: 25.6   \\
\enddata
\tablecomments{A compilation of rotation period measurements from multiple data sources for each component in the 32 New Systems. The two components that likely host additional unseen companions are indicated with asterisks in their names, as discussed in Section~\ref{subsec:multiplicity-checks}. TESS includes two treatments, with the TESS-unpop version better-preserving long-term signals beyond half a sector (13.7\,days) in duration if they exist. Entries of `\nodata' did not have data available, while entries noted as `X' did have light curve data available from the archival source but yielded no meaningful rotation signals. Leading ‘bl’ flags indicate results for which the underlying photometry partially or entirely blends the A and B components in a pair, while ‘un’ flags indicate A and B are unblended. Contamination from background sources can be different for each dataset and target star: C0 indicates negligible contamination, C1 is mild contamination, and C2 is major contamination, as defined in $\S$\ref{subsec:rot-contam-syst}. C2 cases are not trusted as reliably attributable to our target star given their contamination, unless there are other less-contaminated results that support the period measurement. Candidate but less confident periods that may not be reliable are noted with question marks and are not adopted into our set of final periods without other supporting measures. Final periods given in square brackets are values adopted from \citetalias{2024AJ....167..159L}. Peak-to-peak model amplitudes ($\rm{amp}_{\rm{rot}}$) are adopted from the 0.9\,m in $V$/$R$/$I$ if available and otherwise use an archival source consistent for A and B that also captured the adopted periods, prioritizing spatially resolved archival sources. Notes discussing the rotation assessment for each star are given in $\S$\ref{subsec:rot-notes} and the Appendix. Example light curves highlighting the rotation measurements are shown in Figure~\ref{fig:09m-rot} and Figure~\ref{fig:extra-rot} for each system with final rotation periods adopted here, excluding cases adopted from \citetalias{2024AJ....167..159L}. See $\S$\ref{sec:rot} for further details.}
\end{deluxetable}
\end{longrotatetable}

Our manual review also considered potential systematics in each case, including: (1) 1-day sampling and the related harmonics and aliases for ground-based data sources, (2) monthly lunar variations where relevant, (3) seasonal or annual sampling patterns for multi-year datasets, and (4) the various TESS systematics (see the TESS Instrument Handbook\footnote{The TESS Instrument Handbook V0.1 is available \href{https://archive.stsci.edu/missions/tess/doc/TESS_Instrument_Handbook_v0.1.pdf}{(here).}}, \citet{TESS}, and \citet{2024ApJ...962...47C}), including (a) 27\,day sector baselines, (b) 13.7\,day orbital scattered moonlight and earthshine, (c) 13.7\,day half-sector data downlink gaps, (d) 1\,day rotational earthshine in some sectors \citep{2019arXiv190312182L}, (e) 1.5--2\,day temperature changes, (f) $\sim$2.5--6.75\,day momentum dumps at different frequencies depending on the sector, and (g) spurious results from potentially imperfect detrending of long-term signals beyond $\sim$13\,days by the TESS-SPOC pipeline. Questionable signals related to any of these systematic timescales were regarded as less reliable and often outright ignored when obviously erroneous, based on manual inspection of each light curve. TESS systematics were particularly evident, especially in TESS-unpop data when merging multiple sectors to investigate long-term signals. We treated these uncertain TESS and TESS-unpop cases with enhanced scrutiny, considering factors such as how many sectors were available and showed a signal, if 2-minute-cadence and FFI-cadence data both showed a signal or not, if consecutive sectors aligned in their signals or not, if sectors captured minima/maxima or only showed linear trends, and if the signals aligned with any other available rotation information. For Gaia DR3, spurious periods can arise from various aspects of the Gaia systematics \citep{2023A&A...674A..25H}, but these are most often related to close separation sources within $<$\,0\farcs5 and as such are less of a concern given the vetting of our sample for close companions or contaminating sources. We generally did not encounter strong suspicious signals in our manual review of the Gaia light curves.

\subsection{Rotation Assessment Notes} \label{subsec:rot-notes}

Here we discuss the rotation results for the 12 systems in which both components have adopted final rotation periods in Table~\ref{tab:RotTable}; notes for the remaining cases are provided in the Appendix. The H$\alpha$ activity levels mentioned here are always from CHIRON (Section~\ref{sec:chiron}), unless otherwise specified. We often use the H$\alpha$ active/inactive nature of our stars and their estimated masses to compare to the results of \citetalias{Newton_2017} shown in their Figure 5, allowing us to estimate the approximate ranges of plausible rotation periods we might expect for our stars. We do not consider the CHIRON $v\sin(i)$ measurements when assessing the rotation unless a star's $v\sin(i)$ value is above 10\,$\mathrm{km\,s^{-1}}$.

\subsubsection{RTW 0231-5432 AB} \label{subsubsec:rot-notes_RTW-0231-5432}

\begin{figure}[!htbp]
\centering
\gridline{\fig{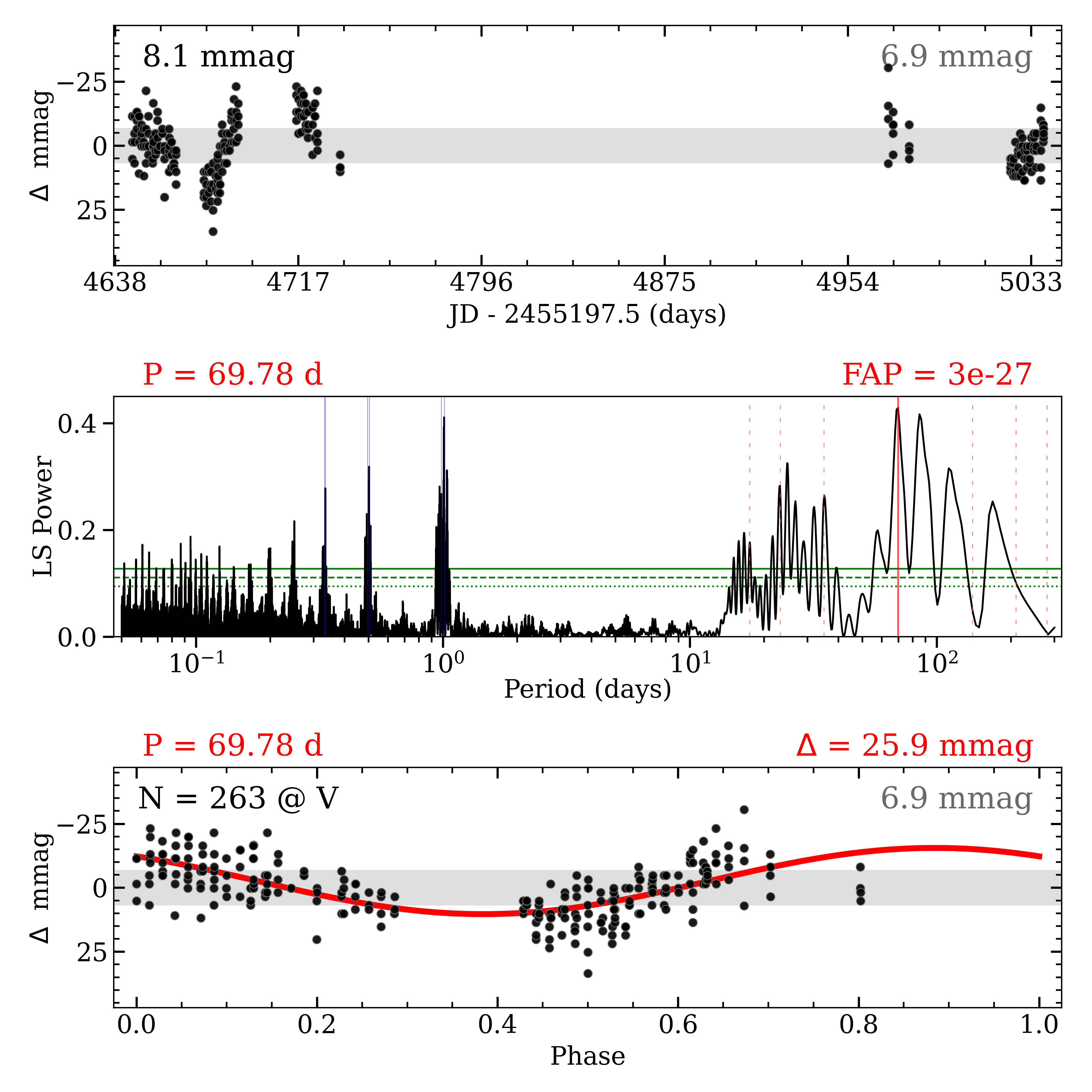}{0.47\textwidth}{$\uparrow$  RTW 0231-5432 A  $\uparrow$}
          \fig{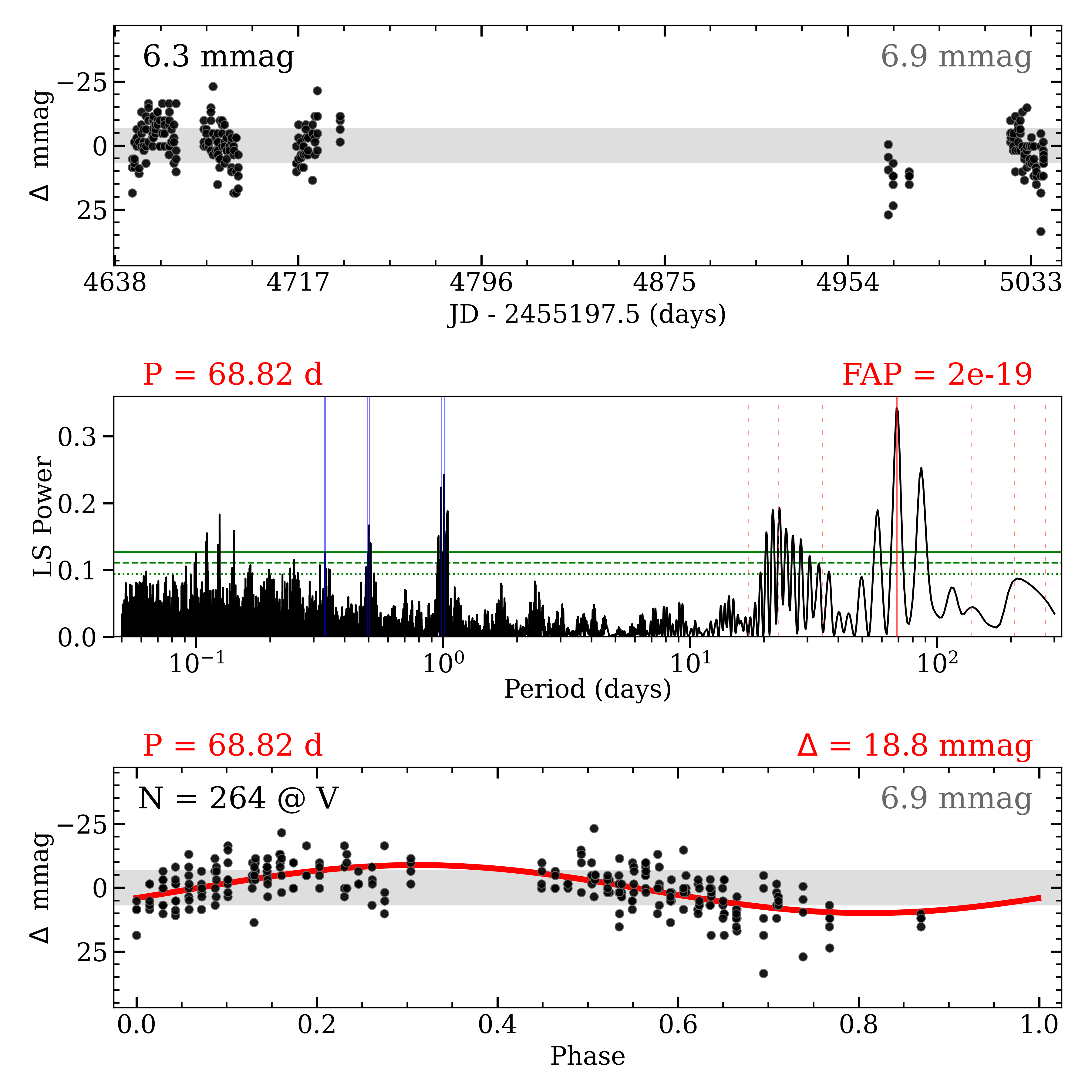}{0.47\textwidth}{$\uparrow$  RTW 0231-5432 B  $\uparrow$}}
\vspace*{-5.08pt}
\gridline{\fig{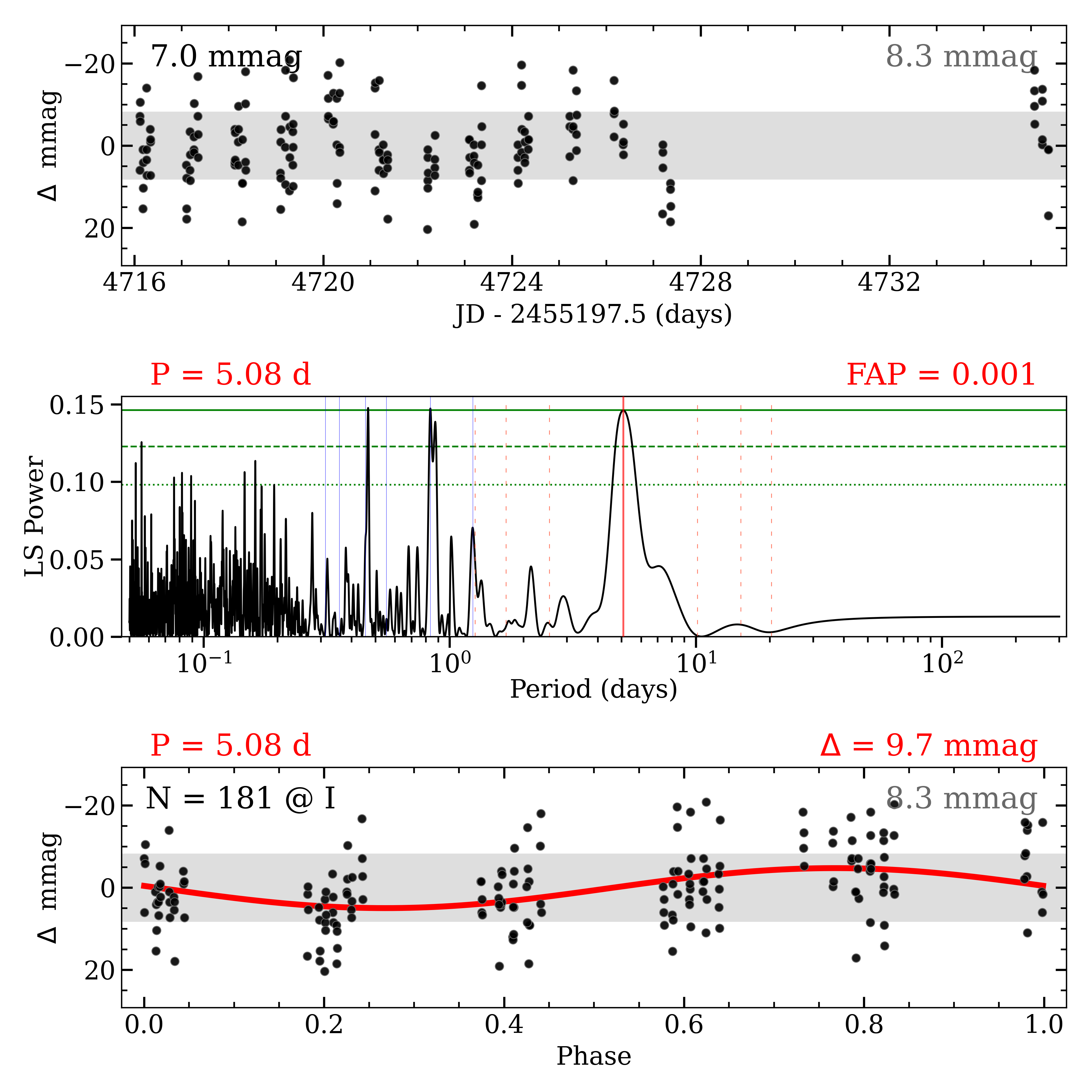}{0.47\textwidth}{$\uparrow$  LP 719-37 A  $\uparrow$}
          \fig{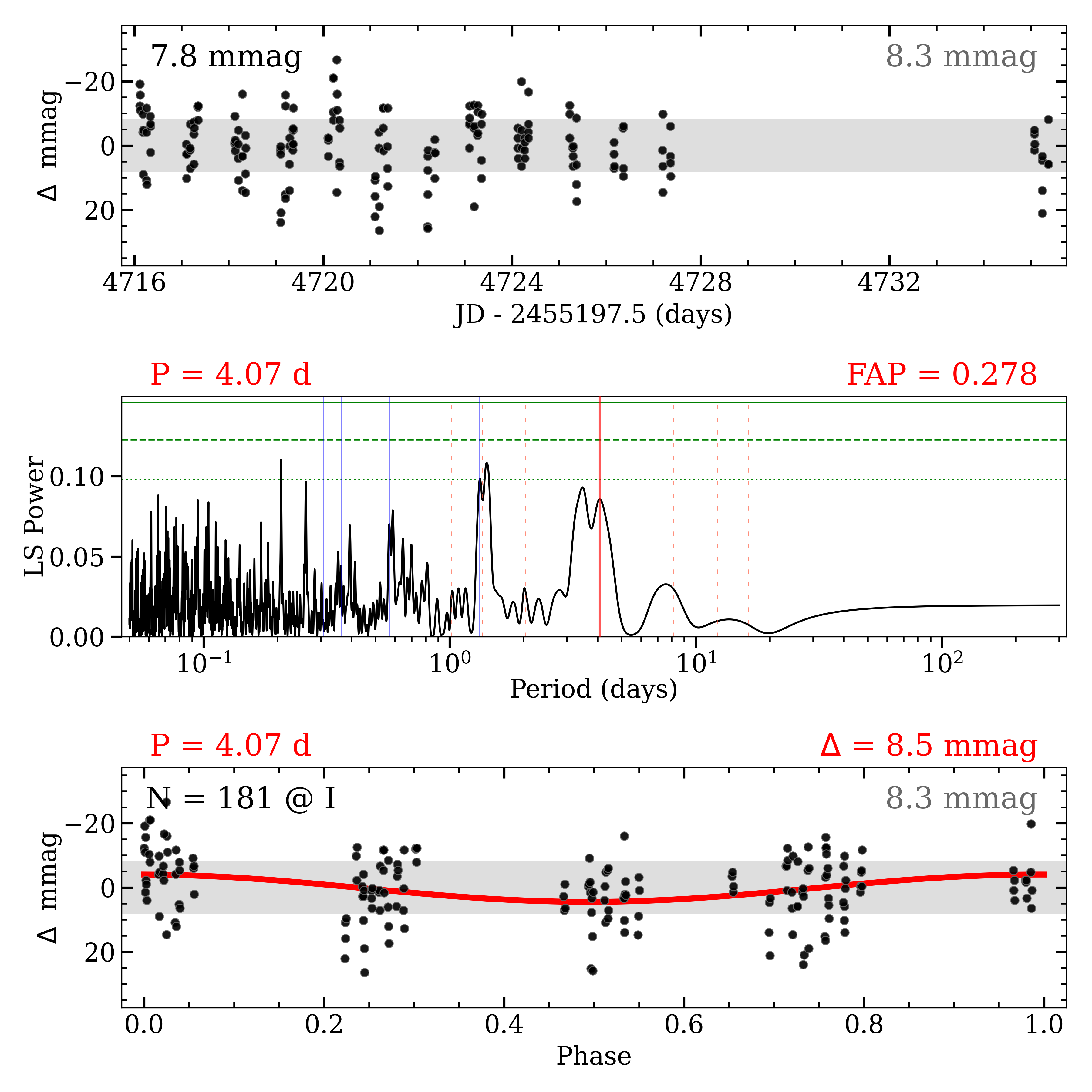}{0.47\textwidth}{$\uparrow$  LP 719-37 B  $\uparrow$}}
\vspace*{-5.08pt}
\figcaption{Rotation period results for the 6 New Systems with values in Table~\ref{tab:RotTable} using data from the CTIO/SMARTS 0.9\,m. Each vertical trio of panels assesses the data for a single star in a twin pair. The top panel is a differential light curve in a format similar to Figure~\ref{fig:long-term-curves}, with times relative to 2010.0 (JD 2455197.5). Black numbers in the top left of this panel indicate the MAD from the mean for the filled points, while grey numbers in the top right are the average MAD of the non-varying reference stars. These light curve data are available as Data behind the Figure (DbF) products. The middle panel shows a Lomb--Scargle periodogram of the data above, with horizontal green lines at False Alarm Probability (FAP) values of 10\% (dotted) / 1\% (dashed) / 0.1\% (solid), a vertical red line at the selected periodogram peak, vertical dashed orange lines at the 2/3/4 harmonic multiples, and vertical purple lines at the $n=[-3,...,3]$ 1-day sampling aliases. The bottom panel shows the corresponding phase-folded light curve based on the rotation period and the epoch of the first point, with the Lomb--Scargle sine wave, rotation period, FAP, and peak-to-peak amplitude ($\Delta$) all shown and given in red. The number of filled points (N) and observing filter are given in the top left of this panel; open points correspond to visits with only one usable frame. Our final adopted rotation periods are often supported by multiple results beyond just these, as discussed in Section~\ref{sec:rot}. \label{fig:09m-rot}}
\end{figure}

\textit{RTW~0231-5432} A and B both show rotation periods of about 70\,days in the resolved 0.9\,m data, as visible in Figure~\ref{fig:09m-rot}. A shows a similar but slightly weaker peak around $\sim$86\,days, but this peak further weakens relative to the stronger $\sim$70\,day peak when considering the epoch mean data (not shown here), so we select the $\sim$70\,day result for A. ASAS-SN shows two confident peaks around $\sim$70\,days and $\sim$66\,days in blended $g$-band data for apertures centered on either A or B, further supporting both stars having very similar but distinct long periods. Blended TESS-unpop data show long-term trends in each of five sectors, with sector-merged results finding candidate $\sim$30\,day periods --- we interpret these as likely half-period harmonic detections of the true longer periods, or possible sector-length systematics. TESS shows weak $\sim$7\,day signals in some, but not all, sectors, which we assume to be erroneous from the pipeline detrending of long-term signals. Our adopted long periods are commensurate with the stars' mild H$\alpha$ absorption at their masses of $\sim$0.27\,$\rm{M_\odot}$, based on comparison to \citetalias{Newton_2017}.

\subsubsection{RTW 0409+4623 AB} \label{subsubsec:rot-notes_RTW-0409+4623}

\textit{RTW~0409+4623} A and B are blended in TESS data that show two confident signals at 2.24\,days and 2.55\,days in one sector with an evident beat pattern in the light curve, as visible in Figure~\ref{fig:extra-rot}; a simulation of sinusoids at 2.24\,days and 2.55\,days closely matches the TESS beat pattern. A second available TESS sector shows two signals at 2.29\,days and 2.59\,days, but with much more asymmetry in their peak power. TESS-unpop shows the $\sim$2.25\,day signal, but the $\sim$2.55\,day signal has a less confident peak buried in the side-lobes of the stronger $\sim$2.25\,day peak. No unblended rotation measurements are available for further disentangling, so we cannot determine which period belongs to which star. We adopt measurements from TESS sector 19 as it shows both signals at roughly similar periodogram strengths, and in plots we simply use the 2.55\,day signal for A and 2.24\,day signal for B. Observations by \citet{2019ApJS..243...28L} and \citet{2021ApJS..253...19Z} via LAMOST find the A component to be H$\alpha$ active --- this supports a period of $\lesssim$~40\,days at A's 0.27\,$\rm{M_\odot}$ mass, consistent with either of our detected periods.

%%%%%%%%%%%%%%% fig - 0.9m rotation master plots %%%%%%%%%%%%%%%

\begin{figure}[!htbp]
\figurenum{10}
\centering
\gridline{\fig{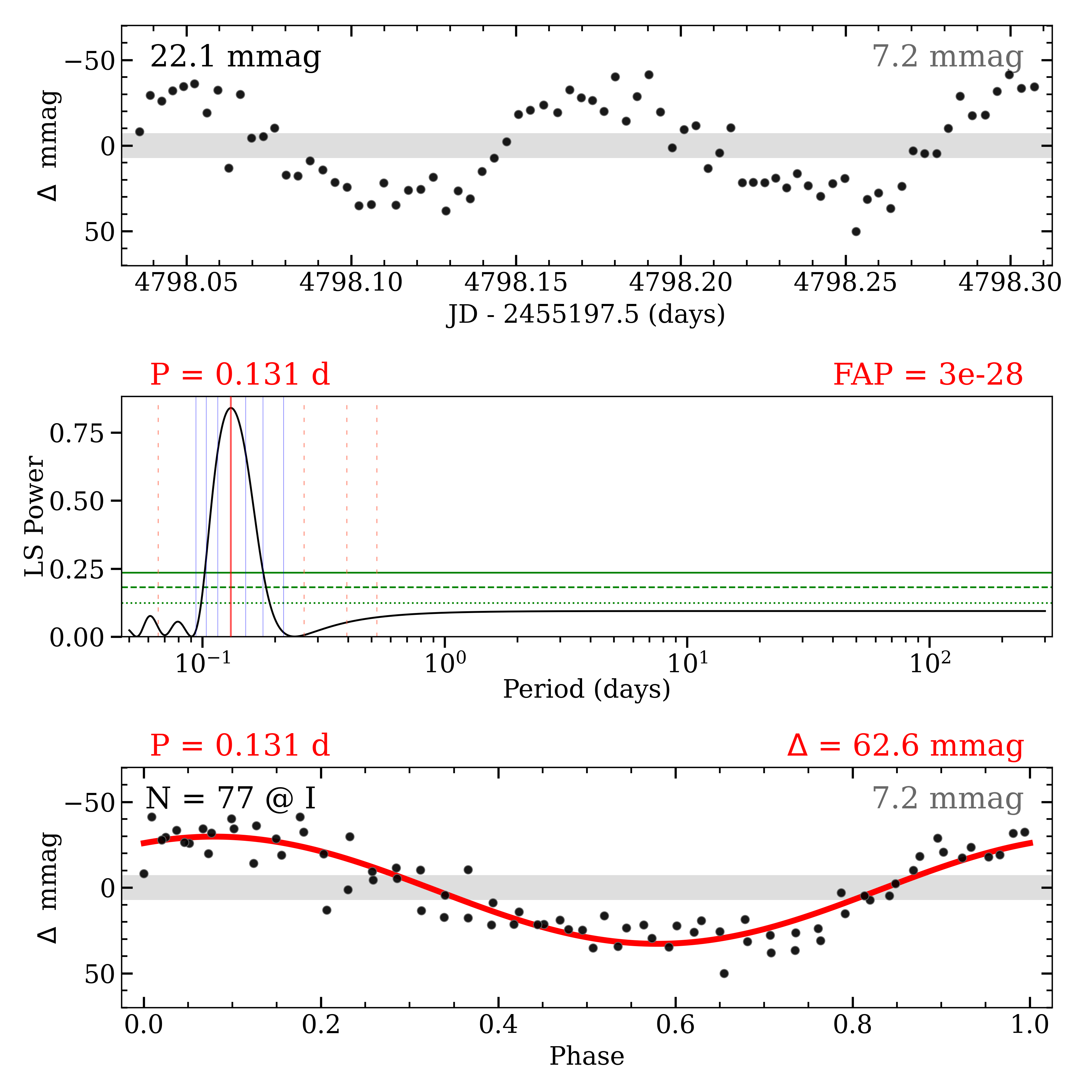}{0.47\textwidth}{$\uparrow$  RTW 0933-4353 A  $\uparrow$}
          \fig{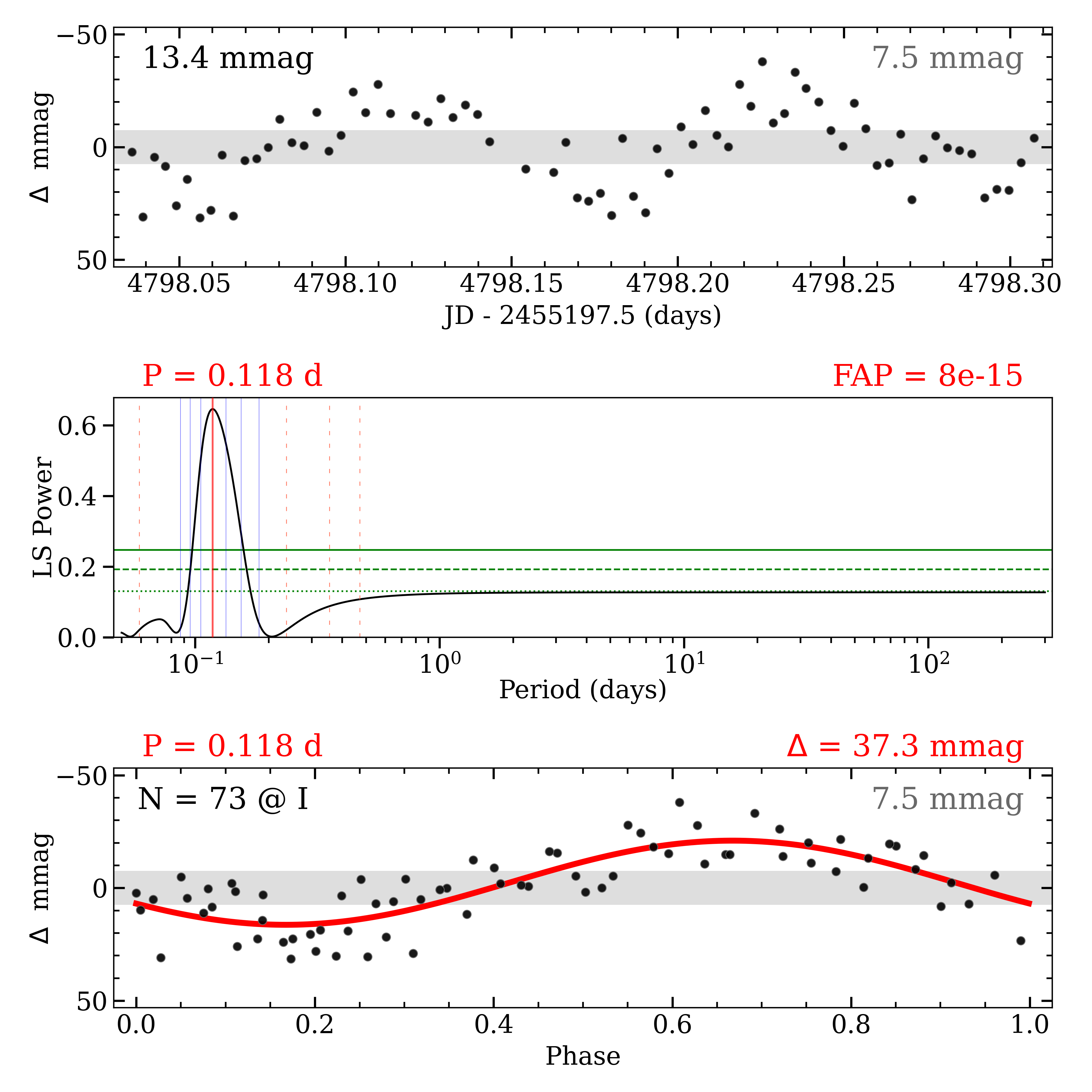}{0.47\textwidth}{$\uparrow$  RTW 0933-4353 B  $\uparrow$}}
\gridline{\fig{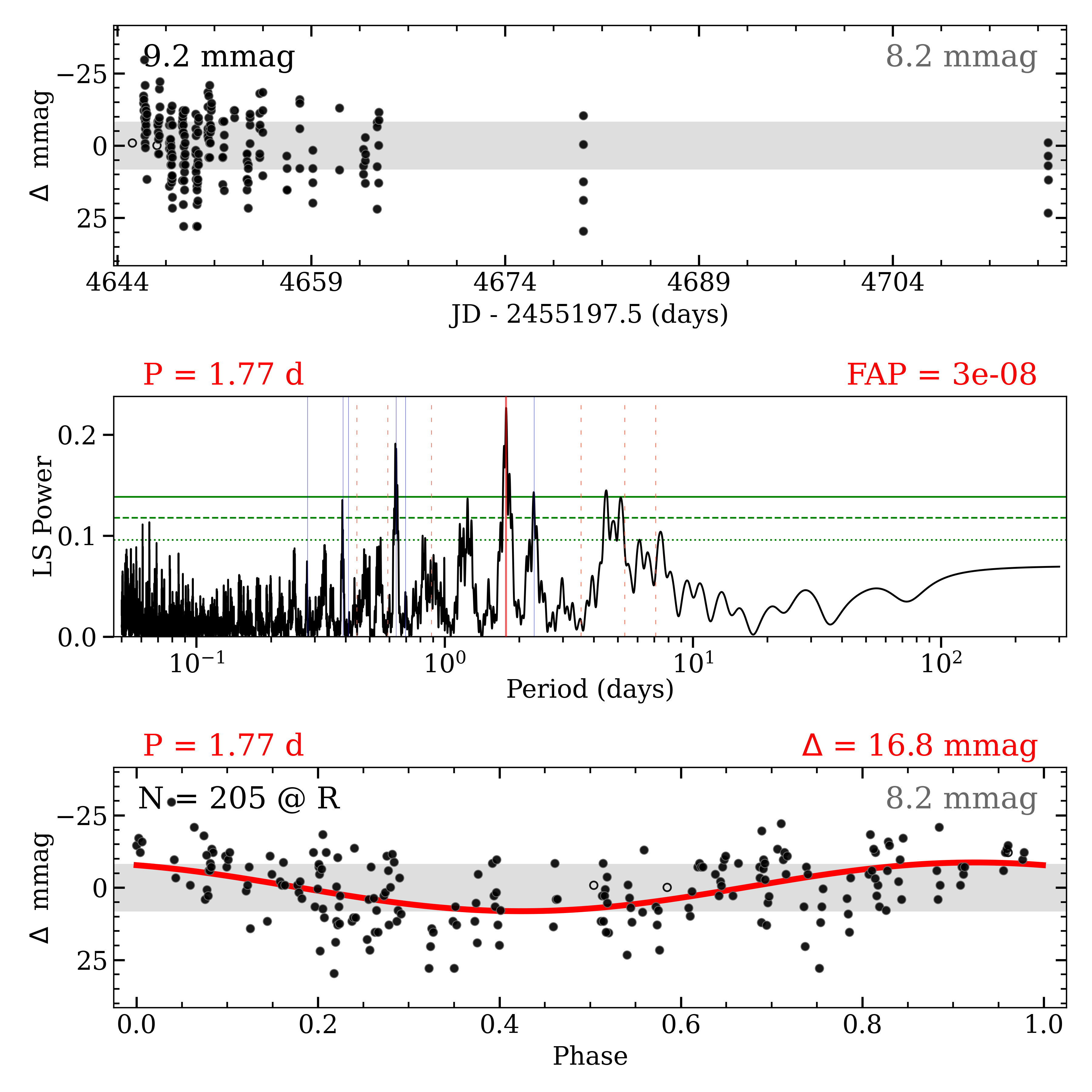}{0.47\textwidth}{$\uparrow$  RTW 2235+0032 A  $\uparrow$}
          \fig{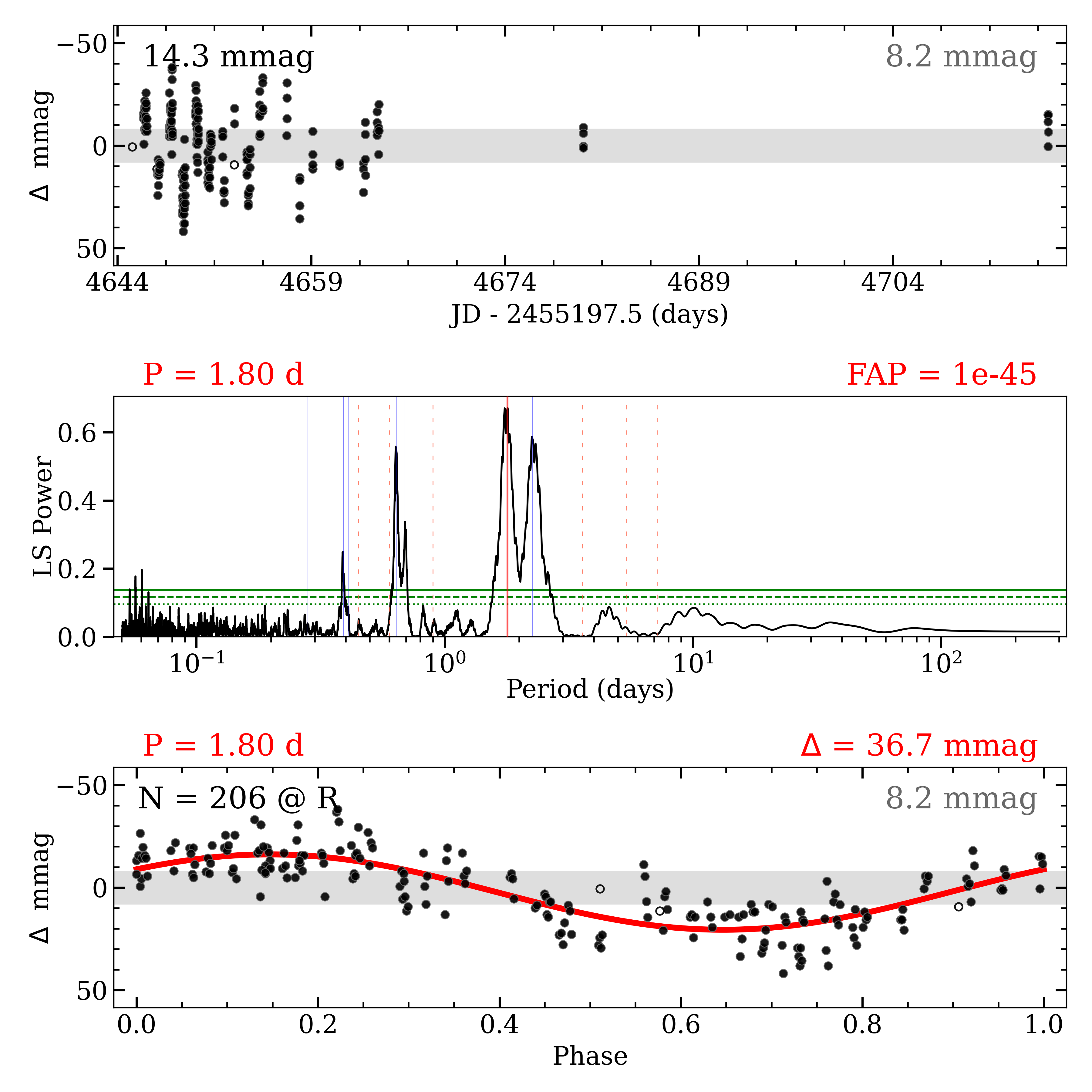}{0.47\textwidth}{$\uparrow$  RTW 2235+0032 B  $\uparrow$}}
\figcaption{(Continued.)}
\end{figure}

\begin{figure}[!htbp]
\figurenum{10}
\centering
\gridline{\fig{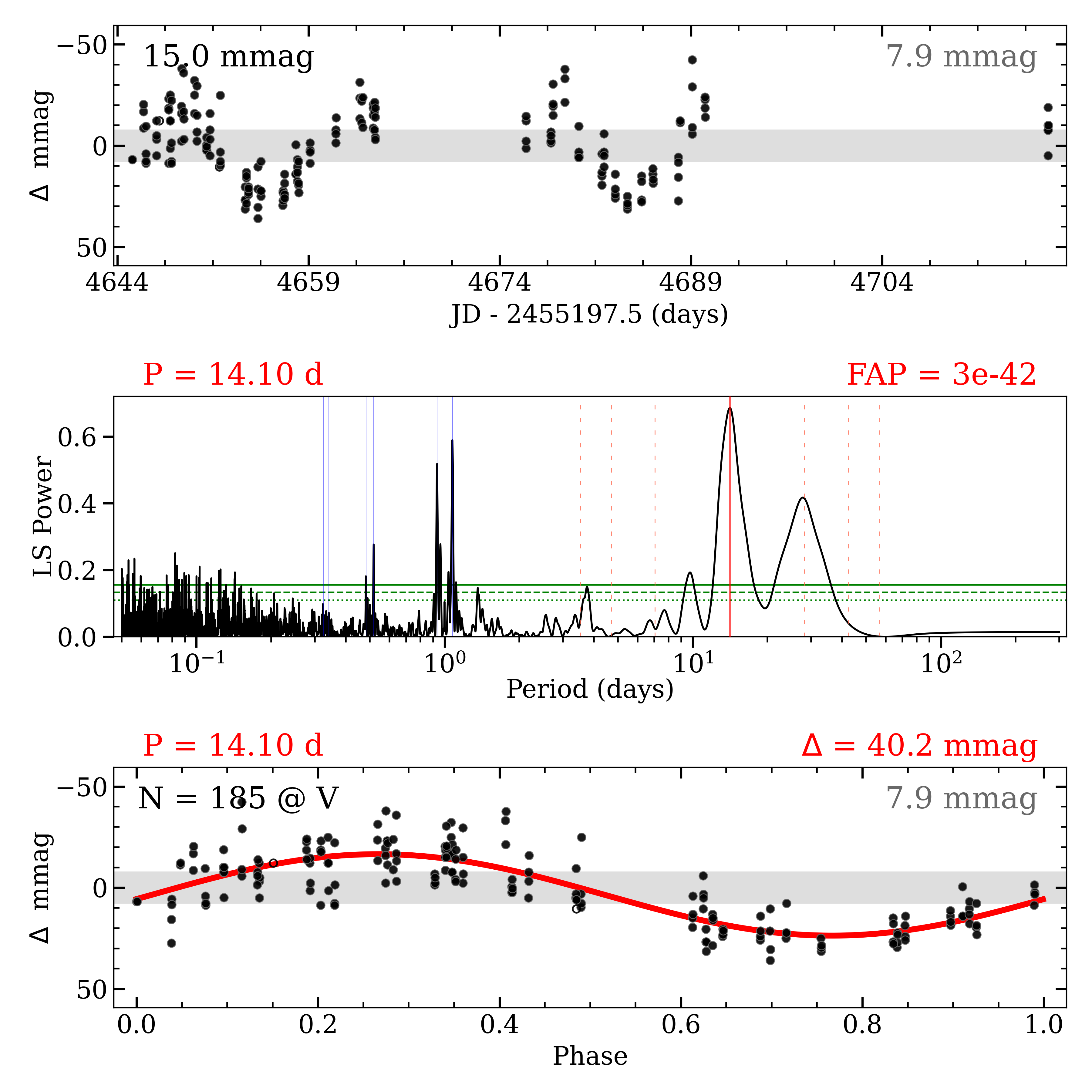}{0.47\textwidth}{$\uparrow$  RTW 2241-1625 A  $\uparrow$}
          \fig{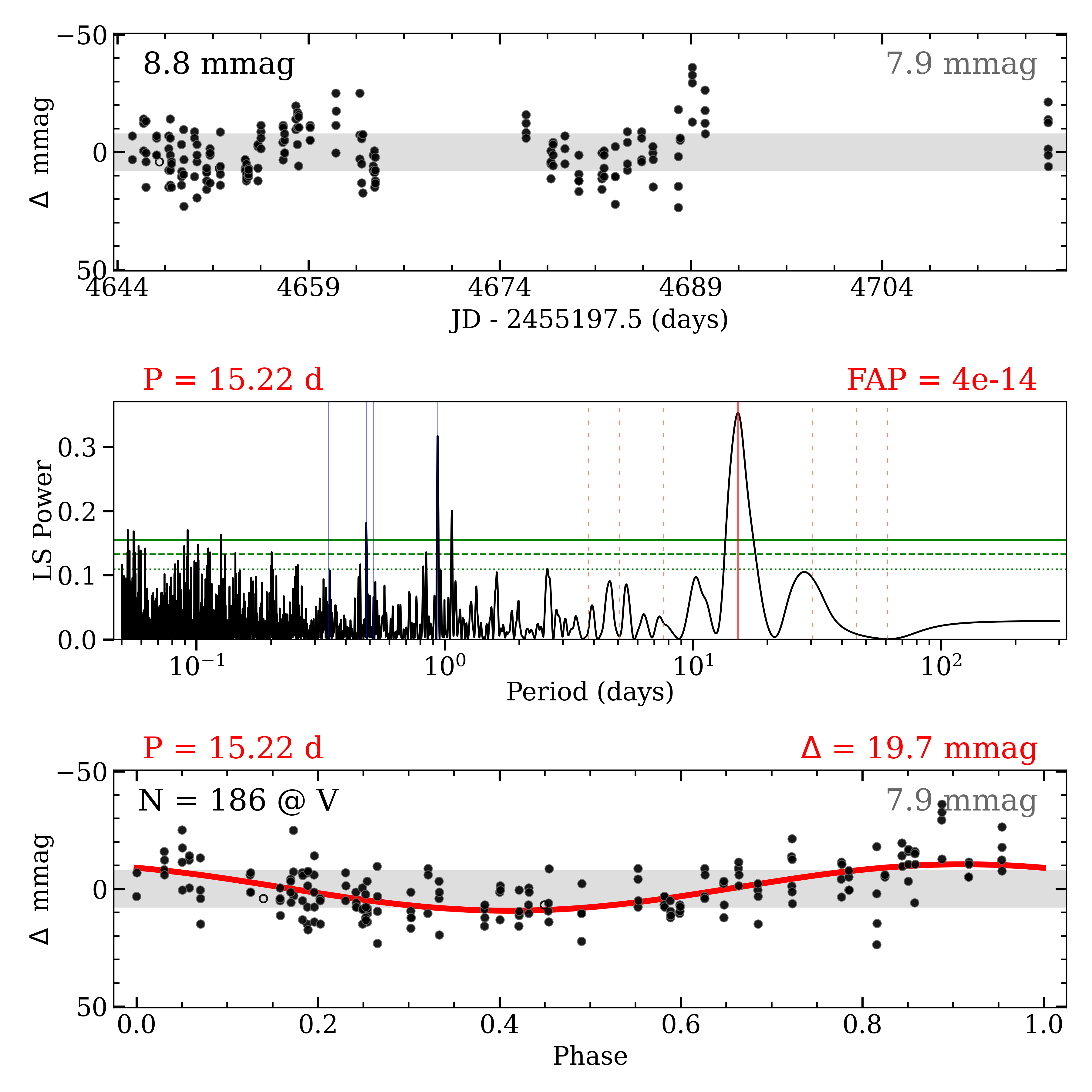}{0.47\textwidth}{$\uparrow$  RTW 2241-1625 B  $\uparrow$}}
\gridline{\fig{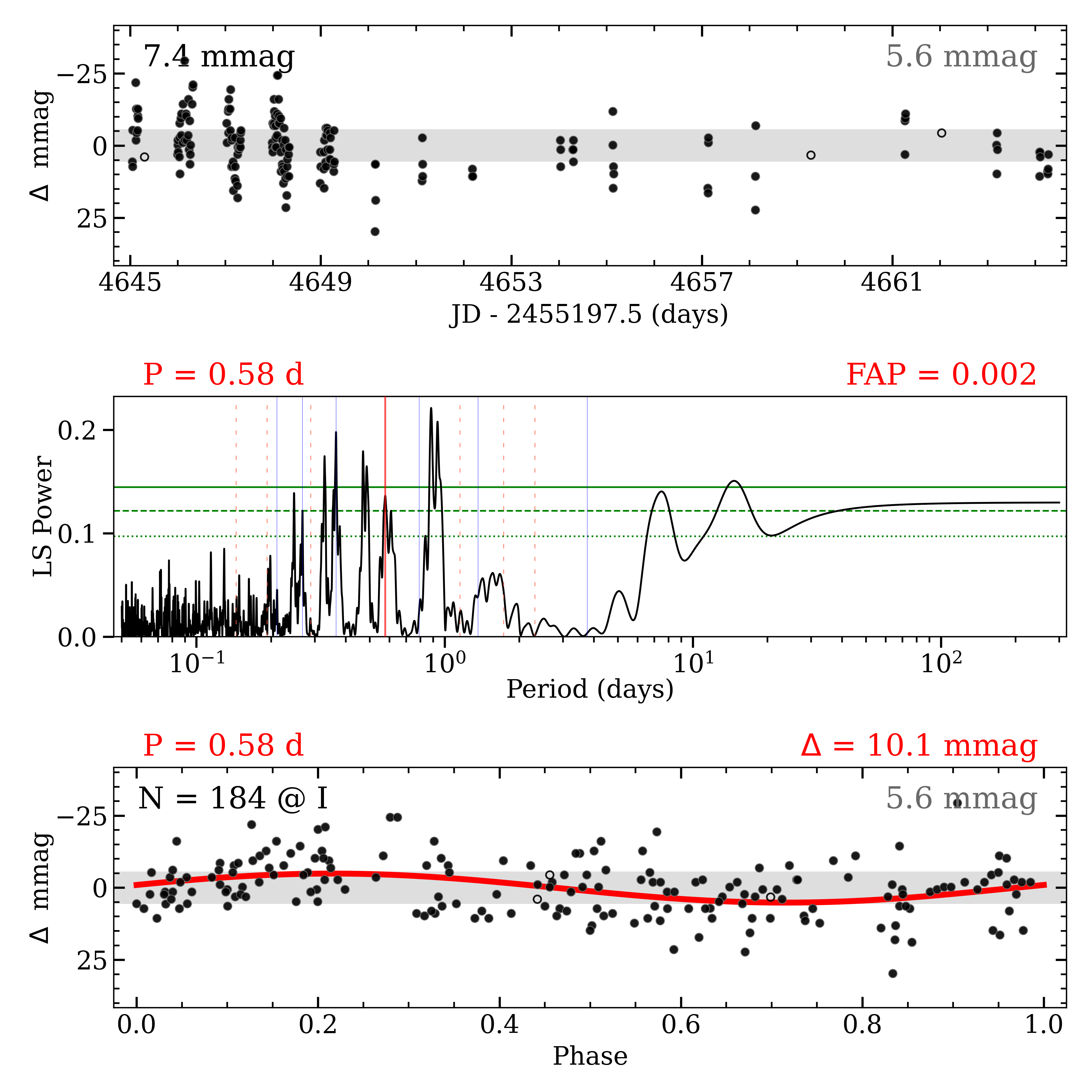}{0.47\textwidth}{$\uparrow$  RTW 2311-5845 A  $\uparrow$}
          \fig{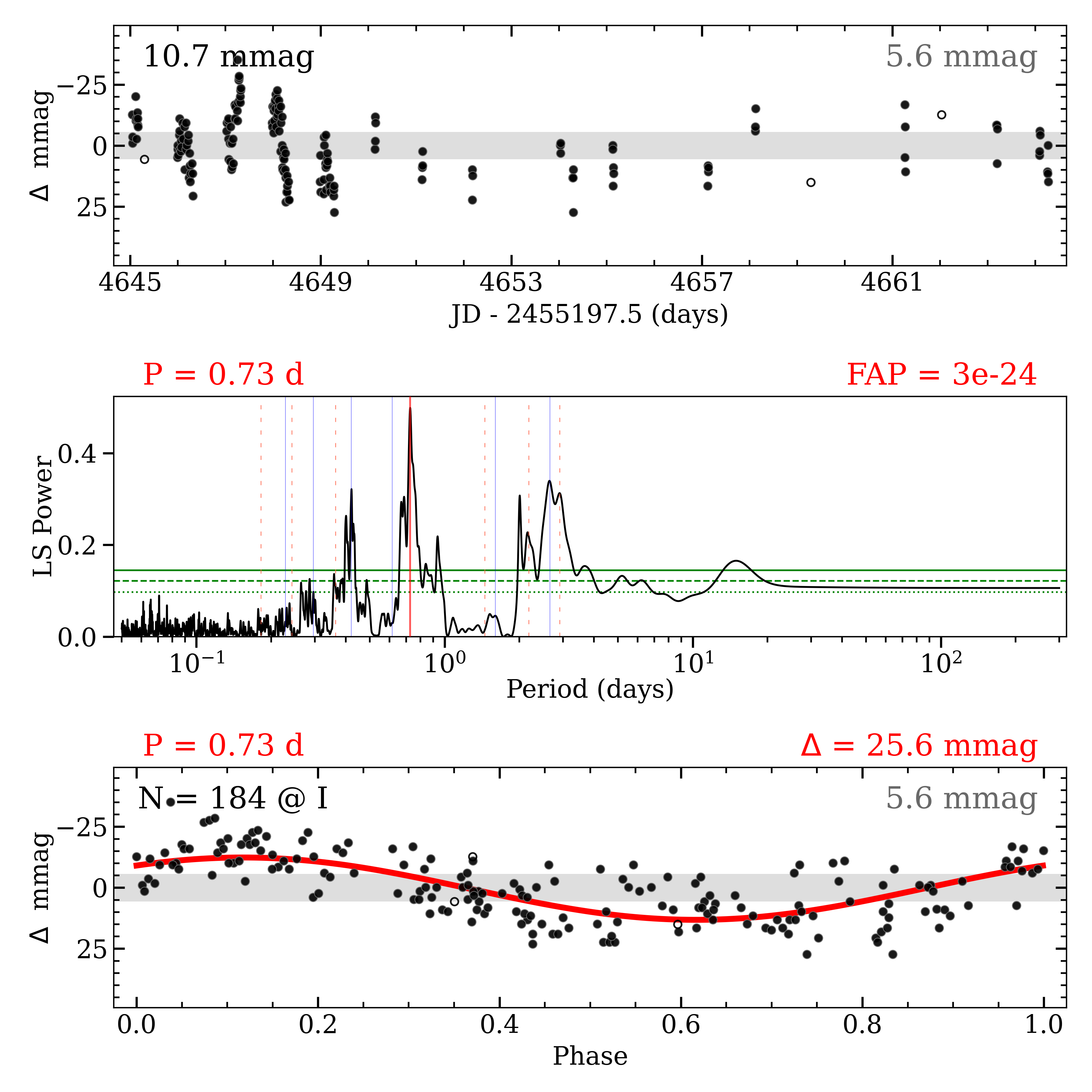}{0.47\textwidth}{$\uparrow$  RTW 2311-5845 B  $\uparrow$}}
\figcaption{(Continued.)}
\end{figure}

%%%%%%%%%%%%%%% fig - external extra rotation master plots %%%%%%%%%%%%%%%
\begin{figure}[!htbp]
\figurenum{11}
\centering
\gridline{\fig{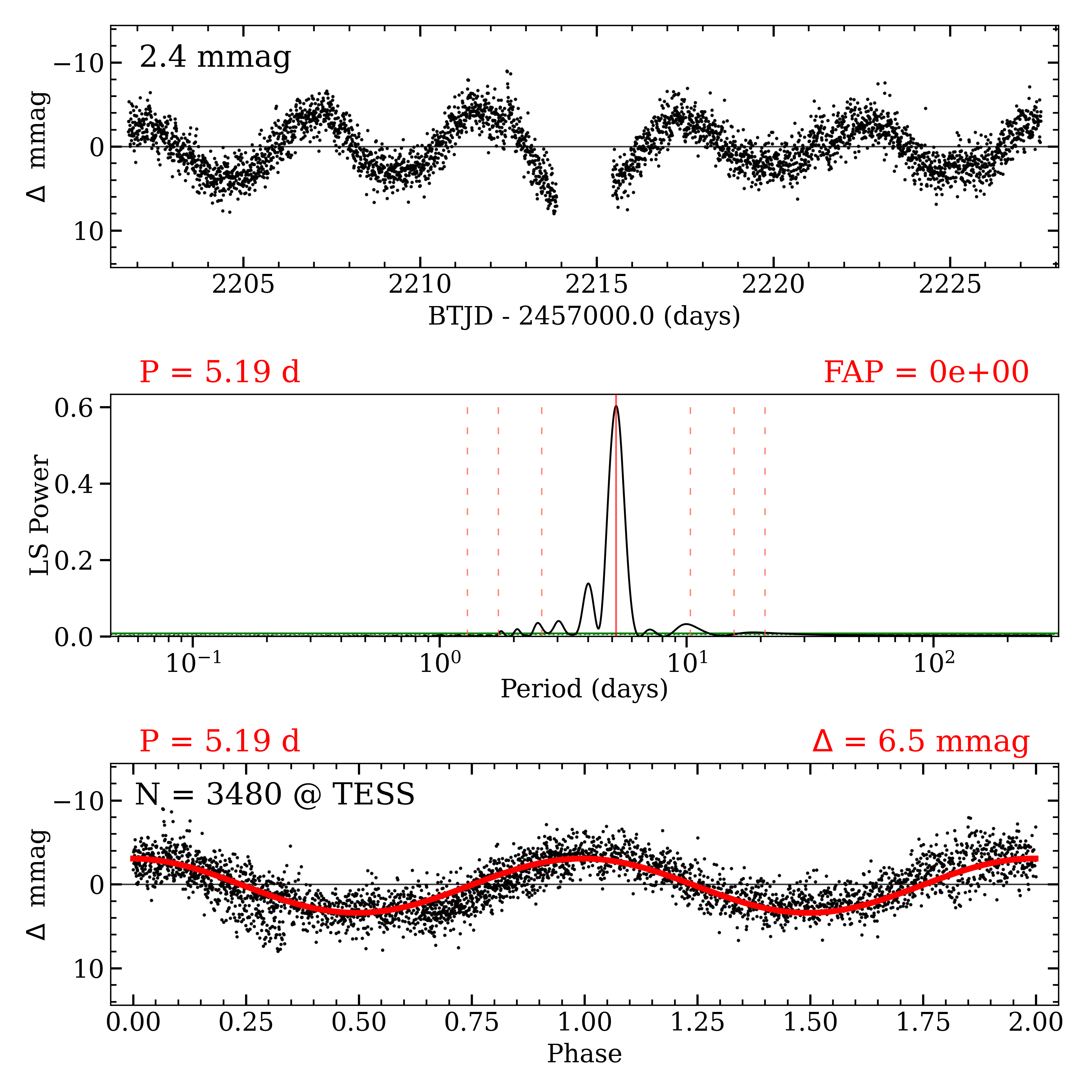}{0.47\textwidth}{$\uparrow$  LP 719-37 A : TESS-unpop : S33  $\uparrow$}
          \fig{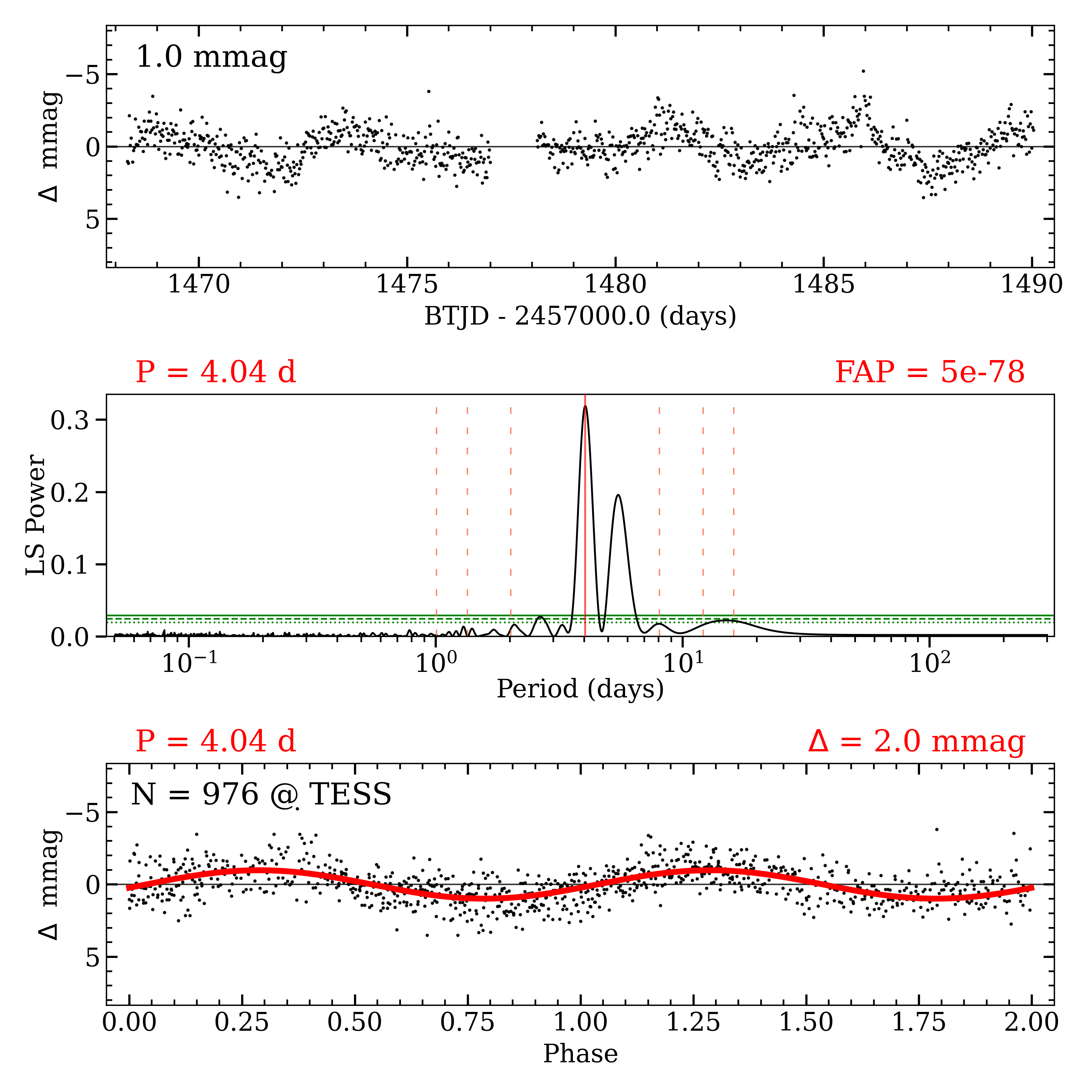}{0.47\textwidth}{$\uparrow$  LP 719-37 B : TESS-unpop : S6  $\uparrow$}}
\gridline{\fig{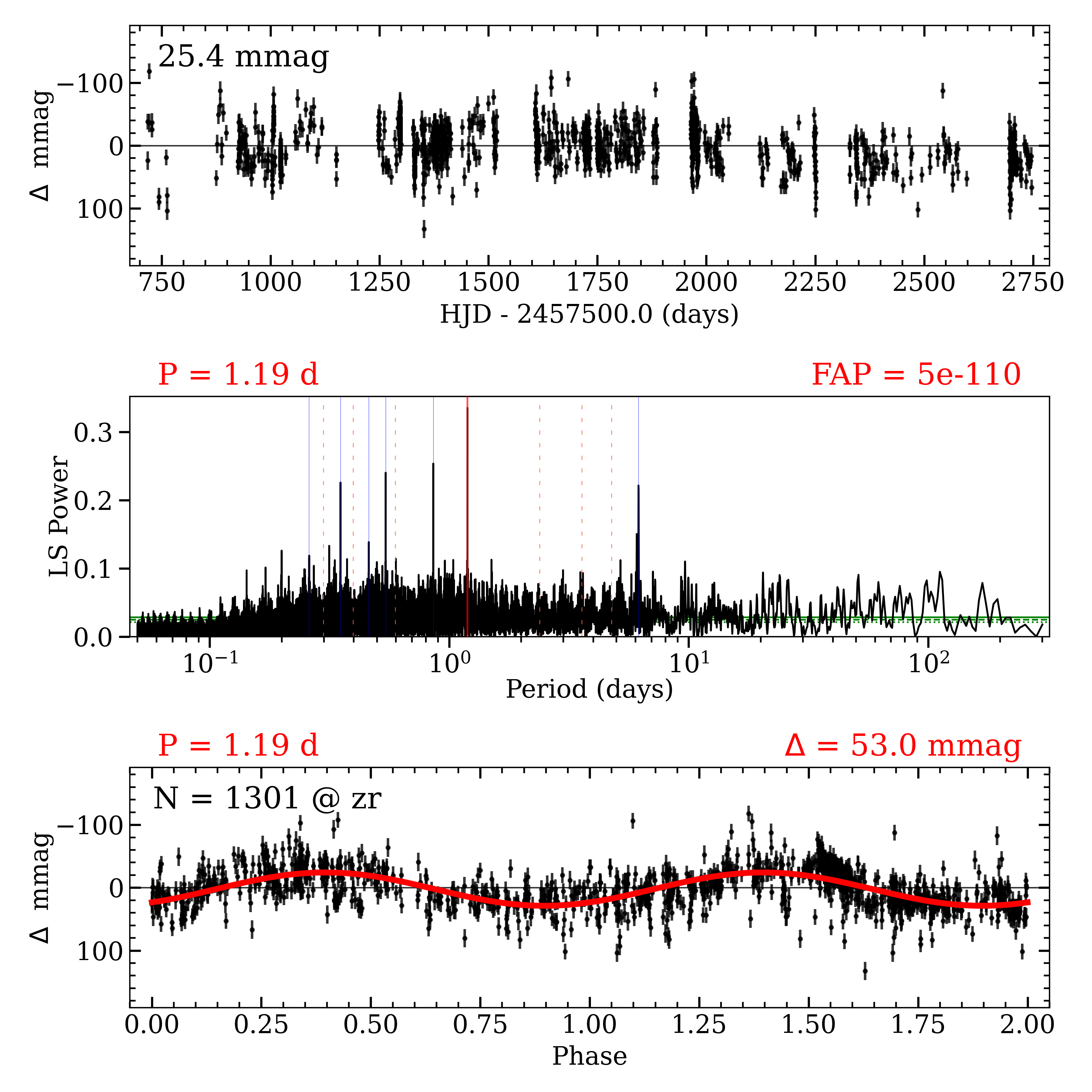}{0.47\textwidth}{$\uparrow$  LP 368-99 A : ZTF  $\uparrow$}
          \fig{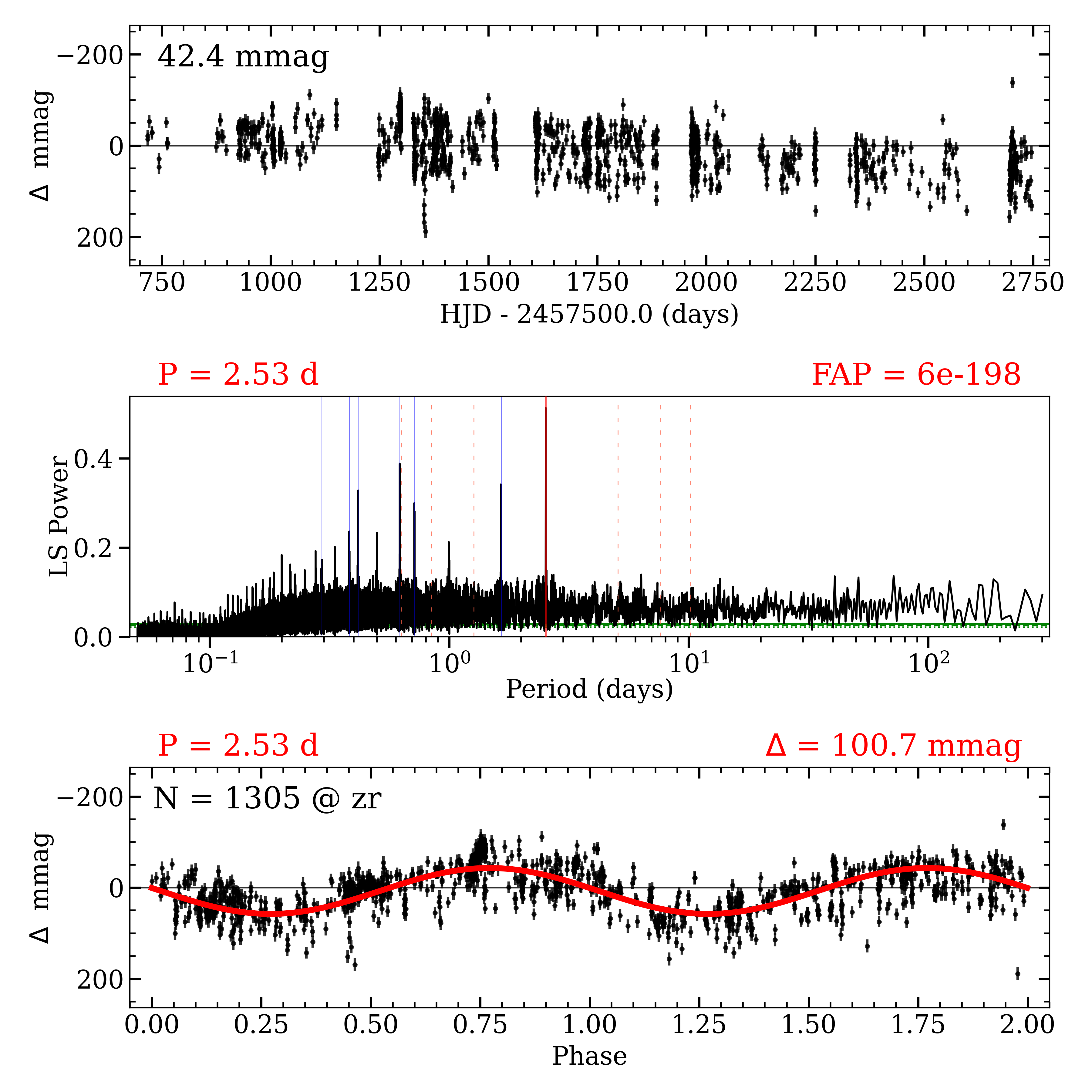}{0.47\textwidth}{$\uparrow$  LP 368-99 B : ZTF  $\uparrow$}}
\figcaption{Example useful rotation results from various external data sources are shown for stars in our New Systems, following the same general format as Figure~\ref{fig:09m-rot}, but without the gray shaded noise regions in the 0.9\,m plots. The TESS-unpop data for LP~719-37~AB are blended, but we show results from two different sectors that correspond to the disentangled periods for each star. LP~368-99~A and B are moderately blended in ZTF but their PSF-fit photometry yields resolved rotation results. RTW~2202+5537~A and B are blended in TESS-unpop but the resolved ZTF data for B allow us to disentangle the true rotation signal for each star. The TESS data for RTW~0409+4623~AB are blended, but the periodogram demonstrates the two peaks corresponding to the two distinct rotation signals we adopt. RTW~2311-5845~A shows an example result based on resolved Gaia DR3 data that helps confirm the system's rotation periods. \label{fig:extra-rot}}
\end{figure}

\begin{figure}[!htbp]
\figurenum{11}
\centering
\gridline{\fig{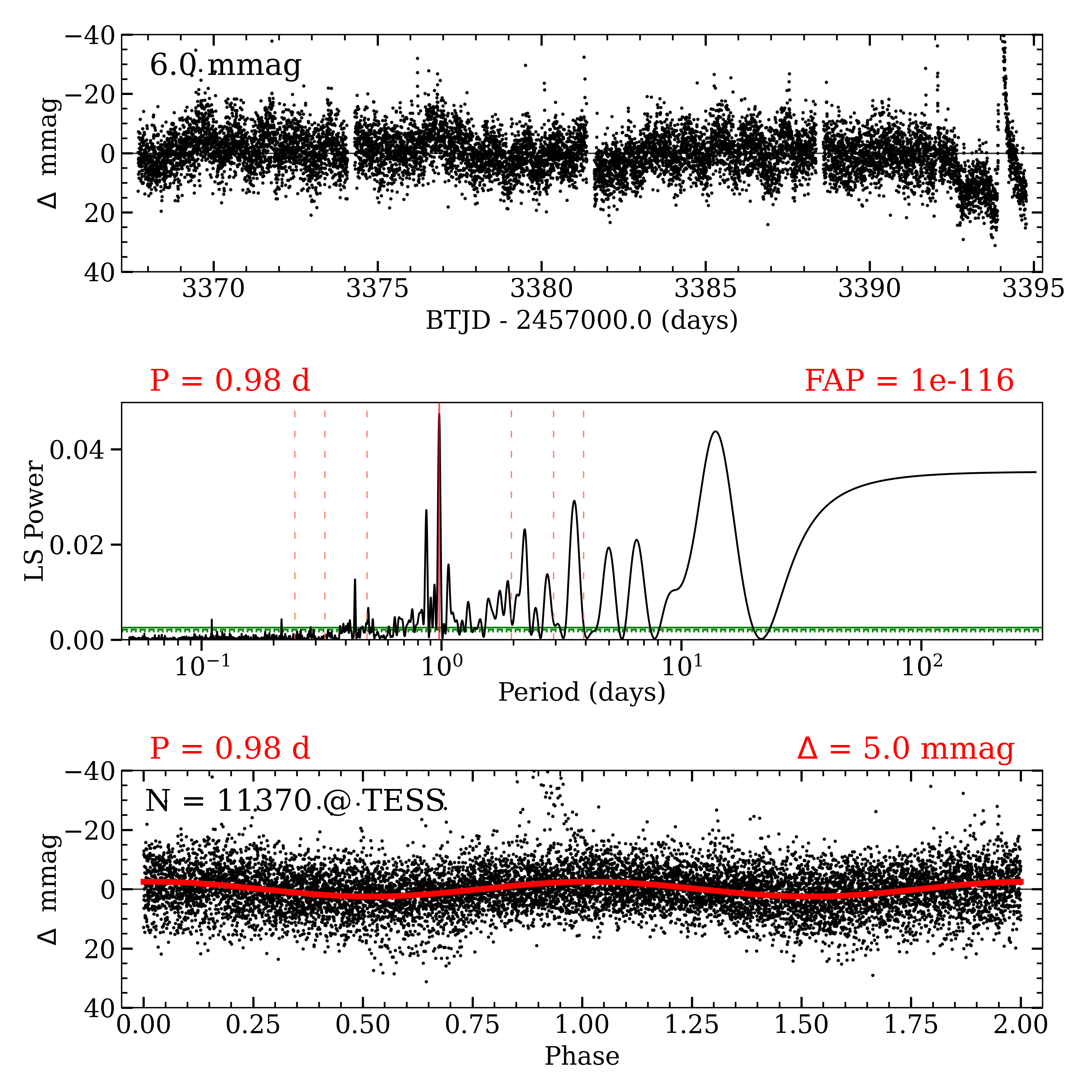}{0.47\textwidth}{$\uparrow$  RTW 2202+5537 A : TESS-unpop : S76  $\uparrow$}
          \fig{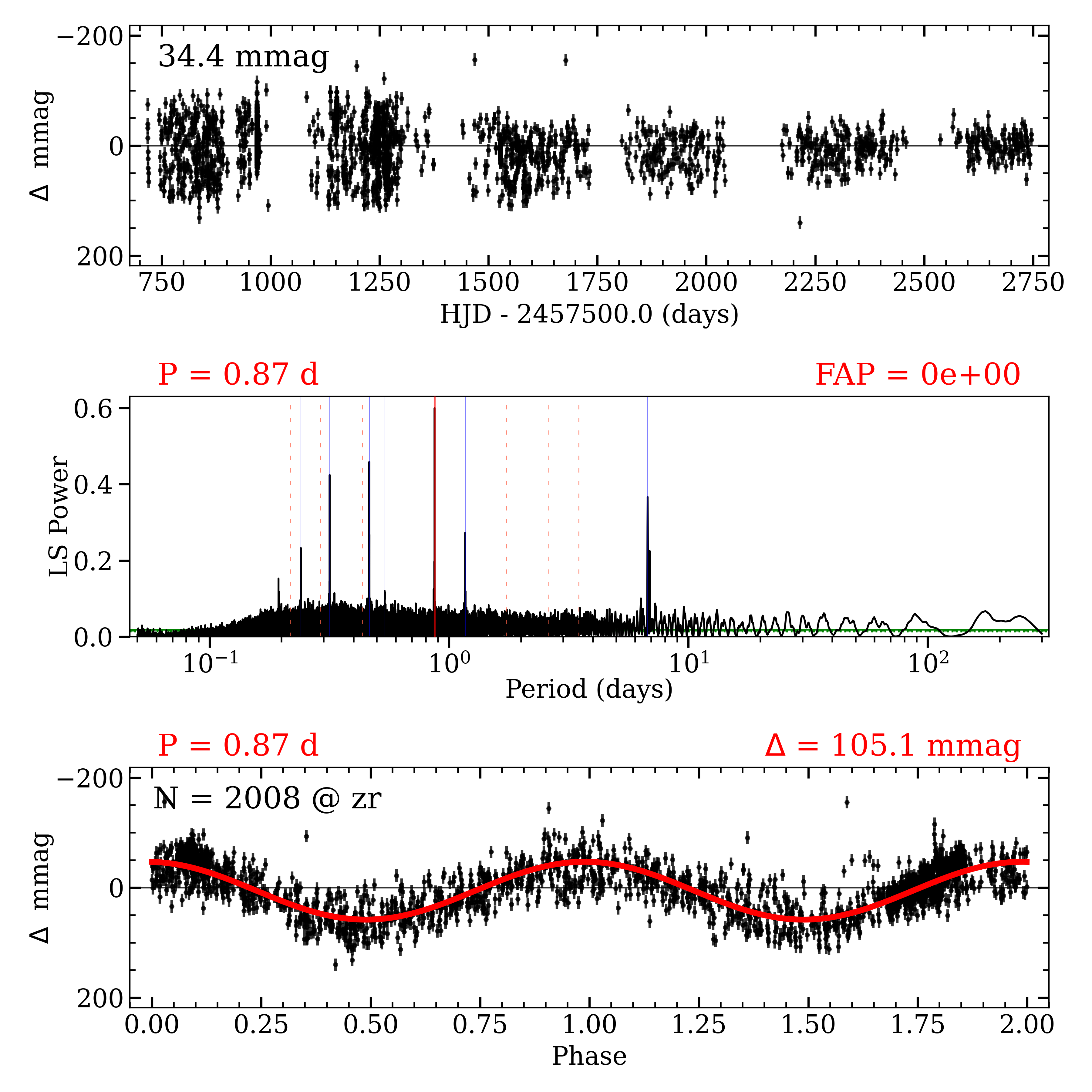}{0.47\textwidth}{$\uparrow$  RTW 2202+5537 B : ZTF  $\uparrow$}}
\gridline{\fig{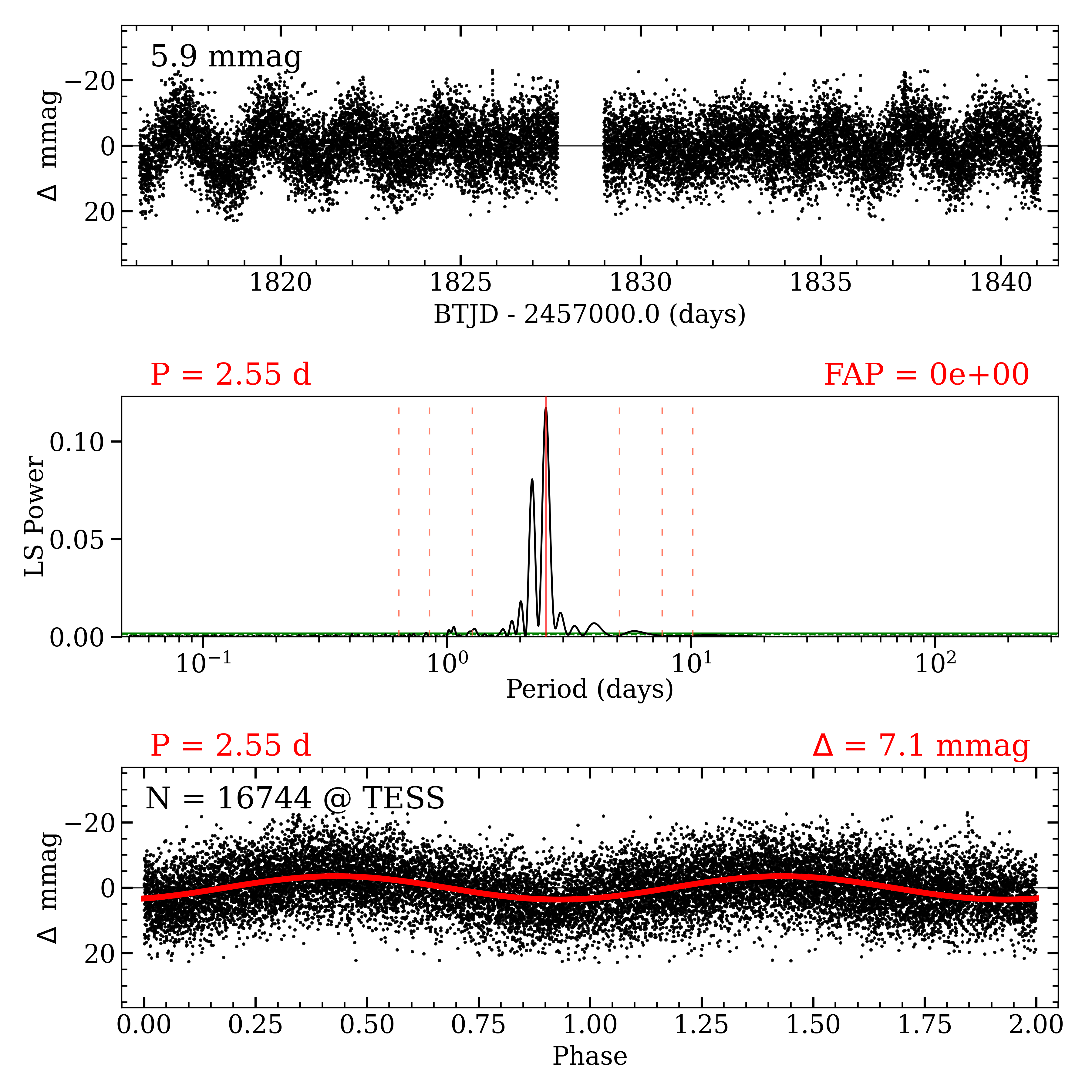}{0.47\textwidth}{$\uparrow$  RTW 0409+4623 AB : TESS-2min : S19  $\uparrow$}
          \fig{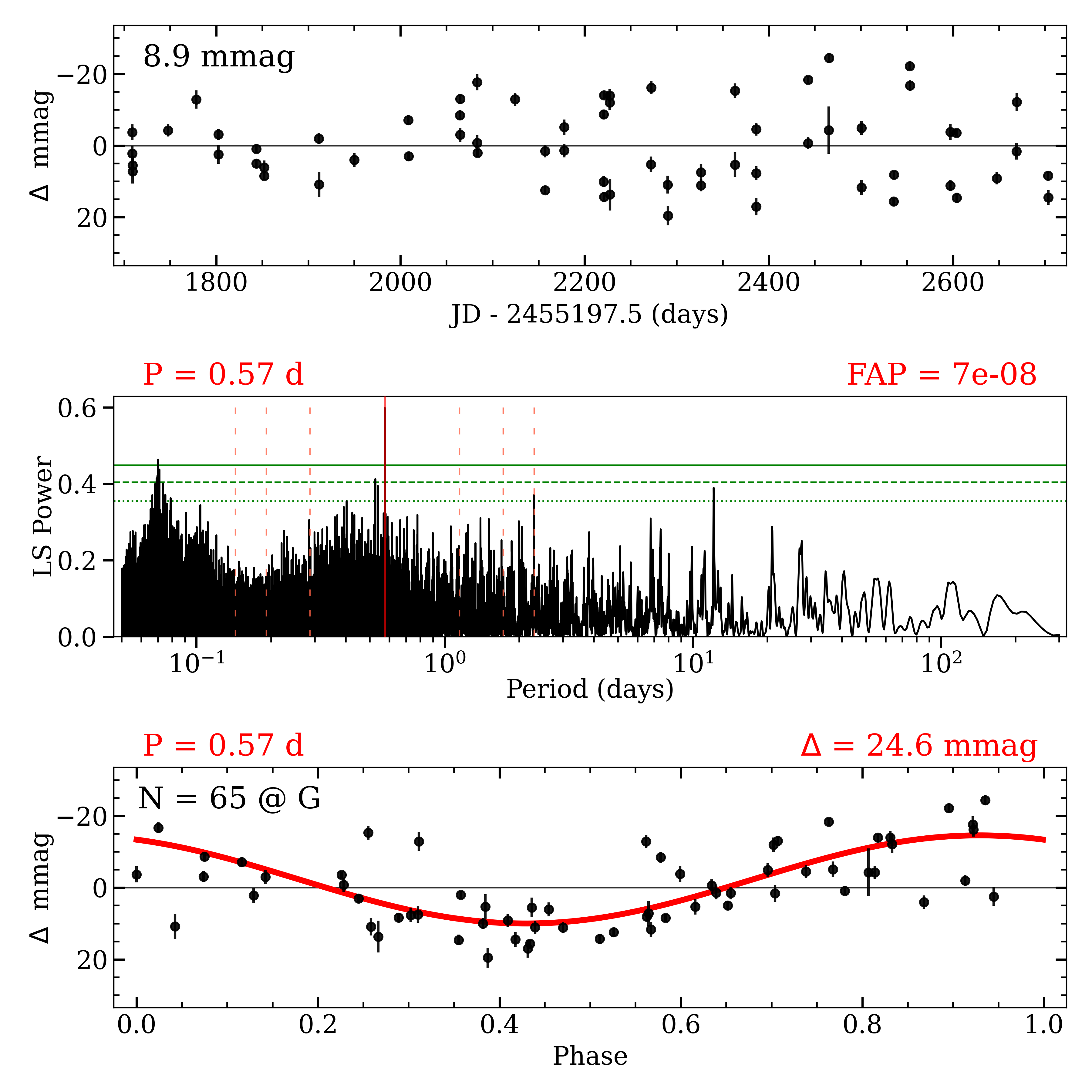}{0.47\textwidth}{$\uparrow$  RTW 2311-5845 A : Gaia  $\uparrow$}}
\figcaption{(Continued.)}
\end{figure}

\subsubsection{LP 719-37 AB} \label{subsubsec:rot-notes_LP-719-37}

\textit{LP~719-37} A and B have two sectors of blended TESS and TESS-unpop data that show a merging of two periodic signals around $\sim$4\,days and $\sim$5\,days, as visible in Figure~\ref{fig:extra-rot}. In the resolved 0.9\,m data we find weak peaks with enhanced periodogram power near $\sim$5\,days for A and $\sim$4\,days for B that we highlight in Figure~\ref{fig:09m-rot} --- while these are not the maximal power peaks, they remain present when folding in the extra long-term 0.9\,m data as well, supporting the features' legitimacy as corroboration for the TESS signals. We also note that in TESS-unpop, when using our custom blended apertures shifted slightly for the positions of each star, the $\sim$5\,day signal appears slightly stronger for A while the $\sim$4\,day signal conversely appears slightly stronger for B. We adopt our period measurements from the strongest signals from TESS-unpop: sector 33 for A at 5.19\,days and sector 6 for B at 4.04\,days. Both stars exhibit H$\alpha$ in emission, consistent with their masses at these rotation periods.

\subsubsection{G 103-63 A(C)-B} \label{subsubsec:rot-notes_G-103-63}

\textit{G~103-63} A is suspected to have an unresolved companion, per Section~\ref{subsec:multiplicity-checks}. Both A and B are inactive and flat in H$\alpha$ in the results of \citetalias{Pass_2024_ApJ}; at B's mass this implies a likely period of 40--130\,days via \citetalias{Newton_2017}, while A and its likely unresolved companion are both of lower mass and would likely have longer rotation periods. A and B have extra contamination leaking into their apertures from a nearby brighter source in TESS data, where A shows a low-confidence 6--7\,day signal in two sectors, but not in the remaining two sectors and not clearly in the heavily blended B data; thus, this is likely not A's rotation period. Blended TESS-unpop results merging six sectors show an uncertain $\sim$39\,day measurement from the combined long-term variations within each sector, along with several similar strength peaks between $\sim$30--100\,days, none of which is of high confidence. Separate from our own analysis, \cite{2024AJ....167..159L} (hereafter \citetalias{2024AJ....167..159L}) report rotation periods of 95.73\,days and 85.68\,days for A and B respectively based on their custom processing of ZTF light curve data in \citetalias{Lu_2022} that improves upon the standard ZTF pipeline light curve scatter. Our use of the standard ZTF pipeline data yields candidate peaks at similar periods but at relatively low confidence, hence why we do not note them in Table~\ref{tab:RotTable}. Our blended and contaminated TESS-unpop results may also be in alignment, where the uncertain $\sim$39\,day signal may be a noisy harmonic of the longer $\sim$86\,day or $\sim$96\,day signal. We choose to adopt the \citetalias{2024AJ....167..159L} periods (no amplitudes were given) but caution that these specific periods are possibly inaccurate or false positives given that (a) at their 4\farcs88 separation A and B are moderately blended in the ZTF data, and (b) the period results for A and B do not appear in the final \citetalias{Lu_2022} period catalog that employed stricter vetting criteria than \citetalias{2024AJ....167..159L}. Nonetheless, if 95.73\,days and 85.68\,days are indeed the accurate rotation periods, then this system's components display matching H$\alpha$ activity and rotation rates despite their likely non-twin triple nature.

\subsubsection{LP 368-99 AB} \label{subsubsec:rot-notes_LP-368-99}

\textit{LP~368-99} A and B consistently show strong intermixed signals at about 1.19\,days and 2.53\,days across several sectors of blended TESS and TESS-unpop data, in blended ASAS-SN data, and in blended K2 data \citep{K2, 2019ApJS..243...28L}. ZTF data show strong signals at 1.19\,days and 2.53\,days in A and B, respectively, congruent with the other results; the ZTF $zr$ data are visible in Figure~\ref{fig:extra-rot}. ZTF does moderately spatially blend A and B, but robust PSF fitting is employed in deriving the ZTF light curves we use, and these specific signals are also very strong in amplitude. Critically, we do not see evident periodogram peaks for each stars' ZTF data at the period of the other stars' signal, confirming the period measurements are functionally resolved for each star. A shows a confident 1.19\,day periodic signal in all three filters of resolved Gaia DR3 data as well, so multiple data sources suitably disentangle which signal belongs to which star. The fast rotation periods are consistent with the observed prominent H$\alpha$ emission from each star for their masses. The CHIRON $v\sin(i)$ measure is larger for A at 13.94\,$\mathrm{km\,s^{-1}}$ than for B at 5.34\,$\mathrm{km\,s^{-1}}$, consistent with A having a rotation period roughly half that of B.

\subsubsection{RTW 0933-4353 AB} \label{subsubsec:rot-notes_RTW-0933-4353}

\textit{RTW~0933-4353} A and B show confident rotation periods of 0.131\,days (3.14\,hrs) and 0.118\,days (2.83\,hrs) respectively in the spatially-resolved 0.9\,m data visible in Figure~\ref{fig:09m-rot}, further confirmed by Gaia DR3 showing a 0.134\,day periodic signal for A. Blended TESS and TESS-unpop data both show strong 0.134\,day and 0.115\,day signals as well but with major contamination from nearby bright sources; \citet{2024MNRAS.527.8290P} also found a period of 0.134\,days in the blended TESS data. These adopted rapid periods agree with the very strong H$\alpha$ emission and rapid $v\sin(i) \approx 15\,\mathrm{km\,s^{-1}}$ displayed by both components.

\subsubsection{RTW 1123+8009 AB} \label{subsubsec:rot-notes_RTW-1123+8009}

\begin{figure}[!t]
\figurenum{12}
\centering
\gridline{\fig{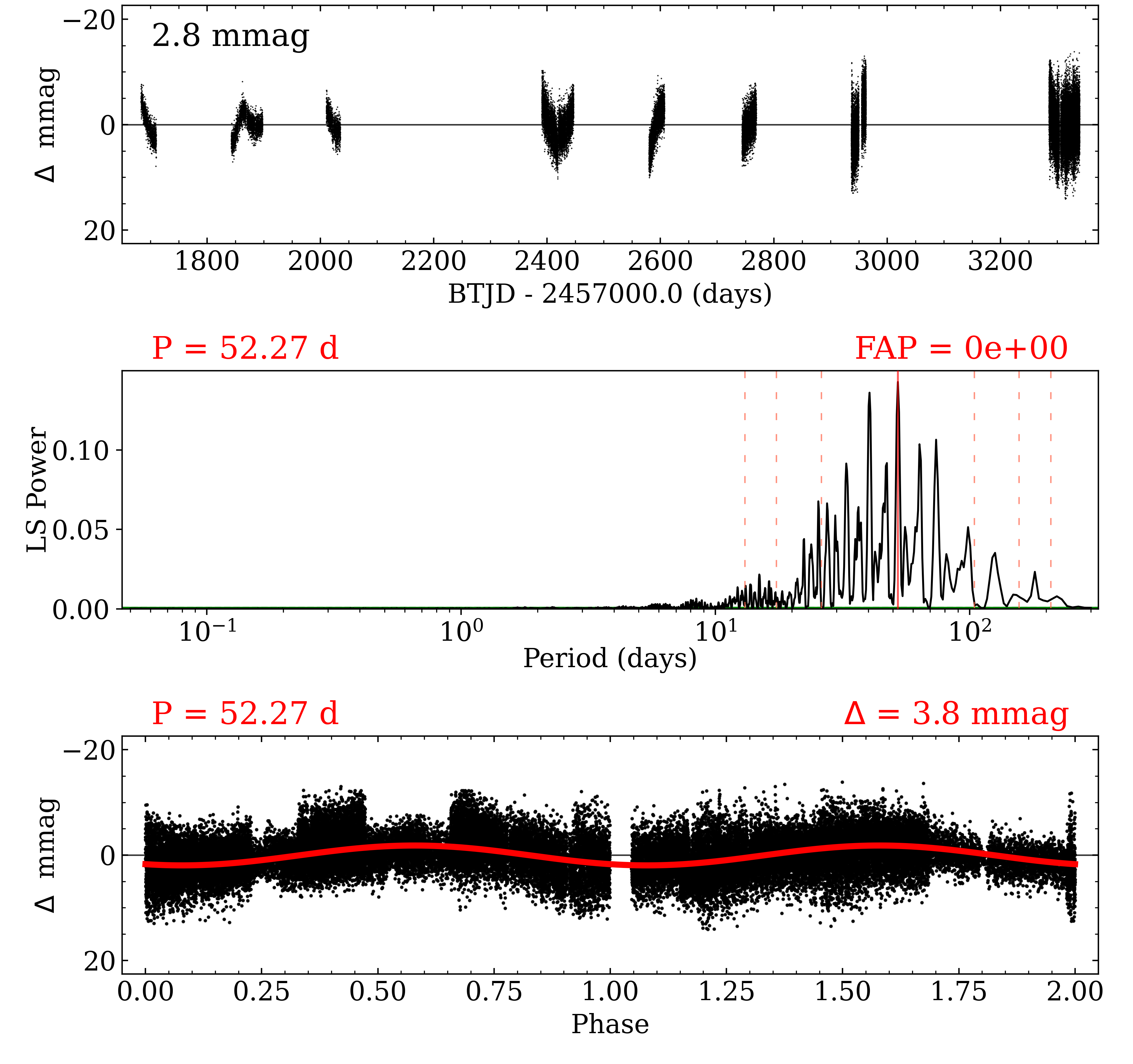}{0.47\textwidth}{$\uparrow$  RTW 1123+8009 A : TESS-unpop-merged  $\uparrow$}
          \fig{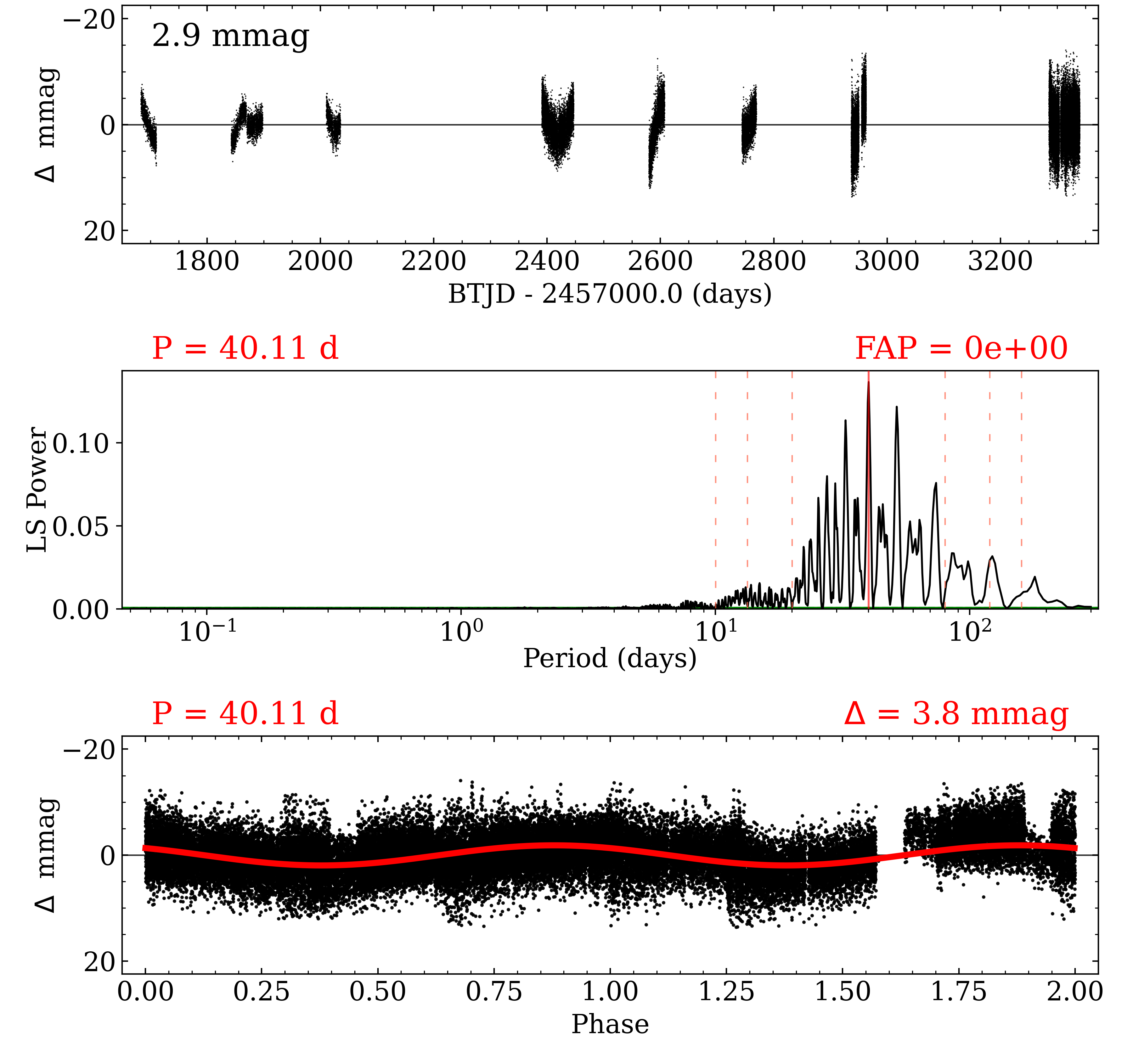}{0.47\textwidth}{$\uparrow$  RTW 1123+8009 B : TESS-unpop-merged  $\uparrow$}}
\figcaption{Rotation results from TESS-unpop for RTW~1123+8009~A and B, in the same general format as Figure~\ref{fig:extra-rot}. Data from sectors 14, 20, 21, 26, 40, 41, 47, 53, 60, 73, and 74 are merged in a joint analysis using the basic methods described in Section~\ref{subsec:rot-methods}. The TESS-unpop data are blended for the two stars given their 9\farcs96 separation, but our apertures were manually chosen to partially minimize the blending, so the results are very similar but slightly different between the two cases. These specific periods are possibly rough half-harmonics of the adopted longer periods from \citetalias{2024AJ....167..159L}, as discussed in Section~\ref{subsubsec:rot-notes_RTW-1123+8009}. \label{fig:rtw1123-unpop-rot}}
\end{figure}

\textit{RTW~1123+8009} A and B were found by \citetalias{Pass_2024_ApJ} to be H$\alpha$ active and inactive respectively, as shown in panel (a) of Figure~\ref{fig:Ha-equality}. Combining this information with our 0.14\,$\rm{M_\odot}$ mass estimates for each star implies likely periods of $\lesssim$\,100\,days for A and $\gtrsim$\,60\,days for B. There are many sectors of blended TESS data available, in which A shows no confident signal and B shows two uncertain signals at $\sim$7--9\,days in single sectors of TESS-SPOC and QLP data but not in other sectors. Merging the eleven sectors of blended TESS-unpop data uncovers candidate maximum power peaks near $\sim$52\,days if using A's aperture or $\sim$40\,days if using B's aperture, as shown in Figure~\ref{fig:rtw1123-unpop-rot}; both stars show those same two peaks given the blending. Unblended ZTF \textit{zr} data show uncertain candidate signals near 92\,days for A and 102\,days for B, with long-term rotational modulation weakly evident by eye in portions of the light curves. Similarly, \citetalias{2024AJ....167..159L} report rotation periods of 105.26\,days and 85.67\,days for A and B respectively based on their improved ZTF rotation analysis in \citetalias{Lu_2022}. Our \textit{zg} data for B do have an uncertain signal peak at 86\,days and a secondary peak around the same period in \textit{zr} data, so we consider our ZTF measures and those of \citetalias{2024AJ....167..159L} to be consistent. Our blended TESS-unpop results at $\sim$52\,days for A and $\sim$40\,days for B also align well as possible harmonic detections of the resolved 105.26\,day and 85.67\,day ZTF periods, given the shorter baselines of TESS sectors. The long ZTF periods are also generally consistent with each component's H$\alpha$ activity level, and we adopt the \citetalias{2024AJ....167..159L} results for the rotation periods, but do not adopt amplitudes because none were reported. However, these periods are reported in \citetalias{2024AJ....167..159L} but not \citetalias{Lu_2022} so may be inaccurate or false positives, as discussed earlier for G~103-63~AB in Section~\ref{subsubsec:rot-notes_G-103-63}; the fainter background source roughly 5$\arcsec$ away from RTW~1123+8009~A may also be influencing the period results. Erroneous or misselected periods are especially worth considering given the pair's pronounced H$\alpha$ activity mismatch. This system is ultimately quite valuable, being analogous to the strongly mismatched twin case of NLTT~44989~AB discussed in \citetalias{RTW_P1}, so measuring the rotation rate for each star more confidently in the future is key.

%%%%%%%%%%%%%%% fig - rtw1123+8009ab unpop sector merged rotation master plots %%%%%%%%%%%%%%%

\subsubsection{RTW 2202+5537 AB} \label{subsubsec:rot-notes_RTW-2202+5537}

\textit{RTW~2202+5537} B shows a confident 0.87\,day rotation period result across multiple datasets, including resolved space-based Gaia data that rule out alternate alias peaks. This result for B is shown in the ZTF \textit{zr} data in Figure~\ref{fig:extra-rot}, and matches the 0.87\,day period reported for B in ZTF data by \citet{Chen2020} and in TESS data by \cite{2024AJ....167..189C}. Blended TESS and TESS-unpop data also show a second signal at 0.98\,days, weak in sector 56 but strong in sector 76, as shown in Figure~\ref{fig:extra-rot}. The 0.98\,day result from TESS-unpop sector 76 appears significantly stronger in A than B when using our custom apertures to minimize their blending, supporting a 0.98\,day period for A and 0.87\,day period for B. Ascribing 0.98\,days to A also aligns with the weak candidate 1.10\,day signal for A from resolved ZTF data, where the true period's proximity to 1.0\,days might be the reason ZTF did not confidently find the 0.98\,day signal. We adopt periods of 0.98\,days for A and 0.87\,days for B, and amplitudes from sector 76 of the blended TESS-unpop data using the stars' respective apertures because this dataset is the only one showing both stars' signals reliably. We note that the amplitudes of rotational modulation for A and B are clearly changing relative to each other over time across the various datasets, with B typically showing significantly larger amplitudes than A. H$\alpha$ EWs are not available for either star but would likely show strong emission given the two components' rapid rotation and 0.15\,$\rm{M_\odot}$ masses.

\subsubsection{RTW 2235+0032 AB} \label{subsubsec:rot-notes_RTW-2235+0032}

\textit{RTW~2235+0032} has period detections at 1.77\,days and 1.80\,days for A and B respectively from our resolved 0.9\,m data, as shown in Figure~\ref{fig:09m-rot}. The 0.9\,m periodograms have a few moderately strong alternative alias peaks, but the blended space-based TESS and TESS-unpop data have strong $\sim$1.77\,day signals and do not show prominent detections at the various alias periods. Blended ZTF data show a mix of evident signals at the true $\sim$1.8\,day periods and near 2.27\,days as well --- which are sampling aliases of each other --- congruent with the other results. The adopted 0.9\,m rotation periods and estimated stellar masses are consistent with the active H$\alpha$ emission we observe in each component.

\subsubsection{RTW 2236+5923 AB} \label{subsubsec:rot-notes_RTW-2236+5923}

\textit{RTW~2236+5923} A and B have no H$\alpha$ EW measurements available to inform rotation. Both components have unclear SAP flux trends for deciding if the TESS-unpop analysis should use an included polynomial or not for possible long-term trends. With a polynomial, we see candidate long-term variations $\gtrsim$\,30\,days based on merging the two available sectors of blended data. Without a polynomial, we see a candidate $\sim$11\,day signal in a single sector. Blended ASAS-SN data show weak peaks near $\sim$13\,days and $\sim$28\,days in \textit{g}-band but not in \textit{V}-band, which we only note for their possible congruence with other results and otherwise would consider them unreliable. We see weak candidate signals near $\sim$24\,days and $\sim$28\,days for A and B respectively in ZTF, but at relatively low confidence. However, \citetalias{2024AJ....167..159L} report rotation periods of 28.11\,days and 26.52\,days for A and B respectively based on their analysis of improved ZTF data. Thus, there is a cumulative congruence of possible roughly month-long signals in TESS-unpop, ASAS-SN, and ZTF, and we adopt the reported literature periods from \citetalias{2024AJ....167..159L} based on their alternate ZTF analysis with reduced photometric scatter, with no reported amplitudes to adopt. We note that the \citetalias{2024AJ....167..159L} periods may be inaccurate or false positives given they are not present in \citetalias{Lu_2022}, as discussed for G~103-63~AB earlier in Section~\ref{subsubsec:rot-notes_G-103-63}.

\subsubsection{RTW 2241-1625 AB} \label{subsubsec:rot-notes_RTW-2241-1625}

\textit{RTW~2241-1625} A and B have resolved periods of about 14\,days and 15\,days respectively from the 0.9\,m data, as shown in Figure~\ref{fig:09m-rot}. The 0.9\,m data for A also show a moderate-strength harmonic peak near 28\,days, but this longer peak becomes statistically insignificant compared to the 14\,day signal when using visit-averaged 0.9\,m data (not shown here), and the 28\,day signal does not appear prominently in the other datasets. Blended ASAS-SN data in both $g$-band and $V$-band show matching signals near $\sim$14\,days and $\sim$16\,days, supporting two distinct but similar periods. Gaia doesn't show a clear result for A, but does have a weak signal for B at $\sim$16\,days in two filters. Similarly, ZTF shows a weak $\sim$14\,day signal for A but nothing reliable for B. Blended TESS-unpop data show a confident signal at $\sim$14\,days in one sector, along with likely harmonics around $\sim$7.5\,days in the other two sectors. Blended TESS data show a complex mix of uncertain faster signals $<$\,10\,days, but given the other results we ascribe all of these TESS cases to be due to a combination of detrending out the true longer periods, harmonics of the true periods, and/or systematics. Beyond Table~\ref{tab:RotTable}, this twin pair was also observed by K2 \citep{K2}, although it was blended given the 4$\arcsec$ Kepler pixels and 4\farcs45 AB separation --- the K2 data yielded periods of 15.54\,days as given in \cite{2016MNRAS.456.2260A} and 15.83\,days as given in \cite{2020A&A...635A..43R}, congruent with the other results here. We ultimately adopt the 0.9\,m-based periods given that they are the only resolved reliable measurements for each star. The adopted rotation rates are also consistent with the two stars' H$\alpha$ inactivity at their masses.

\subsubsection{RTW 2311-5845 AB} \label{subsubsec:rot-notes_RTW-2311-5845}

\textit{RTW~2311-5845} A and B are blended in TESS and TESS-unpop but still show consistent confident signals at 0.57\,days and 0.73\,days across three sectors, with evident beat patterns. Resolved Gaia data show the 0.57\,day and 0.73\,day signals in A and B respectively (see A in Figure~\ref{fig:extra-rot}), in two different filters for each case, confirming which period belongs to which star. Our resolved 0.9\,m data also reveal a rotation period of 0.73\,days for B, but the signal for A is weaker and has several plausible peaks between about 0.2--1\,days --- the strongest peak for A is 0.88\,days, with an alternative one at 0.58\,days, which we select for our plot in Figure~\ref{fig:09m-rot} given that it is the true signal. The adopted rapid periods are consistent with the two stars' very strong H$\alpha$ emission at their masses. The large $v\sin(i)$ values of 20.00\,$\mathrm{km\,s^{-1}}$ and 16.89\,$\mathrm{km\,s^{-1}}$ for A and B, respectively, align with their short rotation periods as well, and with A having a slightly faster rotation.

\subsection{Rotation Results} \label{subsec:rot-results}

Our adopted rotation periods and amplitudes for the New Systems are given in Table~\ref{tab:RotTable}. We adopt rotation periods for both components in only 12 of our 32 New Systems, or 16 of all 36 systems when including those in \citetalias{RTW_P1}. Most of the targets we are unable to determine periods for likely have relatively slow rotation, low activity, and/or small photometric changes, as supported by our observations of prominent H$\alpha$ absorption in many of these cases.

For the rotation photometric amplitudes in Table~\ref{tab:RotTable} not from the 0.9\,m, RTW~0409+4623~AB and RTW~2202+5537~AB entirely or significantly blend their equal-brightness companion stars together in TESS, thereby diluting the fluxes. Their resulting amplitude measurements should thus be doubled to yield the more accurate amplitudes of variation for the individual stars. We therefore leave the amplitudes for RTW~0409+4623~AB and RTW~2202+5537~AB as-measured and uncorrected in Table~\ref{tab:RotTable} but double the individual amplitude values when plotted. We consider each case's `C1' mild background contamination in TESS suitably negligible for our purposes here and do not attempt to correct for it. The only other non-0.9\,m case, LP~368-99~AB, has amplitudes from ZTF where the two stars are moderately blended. However, ZTF uses PSF-fit photometry, and we ultimately consider the stars suitably resolved for the measurements of their large amplitudes following the evidence used earlier indicating their periods are well-resolved.

%%%%%%%%%%%%%%% fig - Prot equality and amp_rot equality %%%%%%%%%%%%%%%
\begin{figure}[!t]
\figurenum{13}
\centering
\gridline{\fig{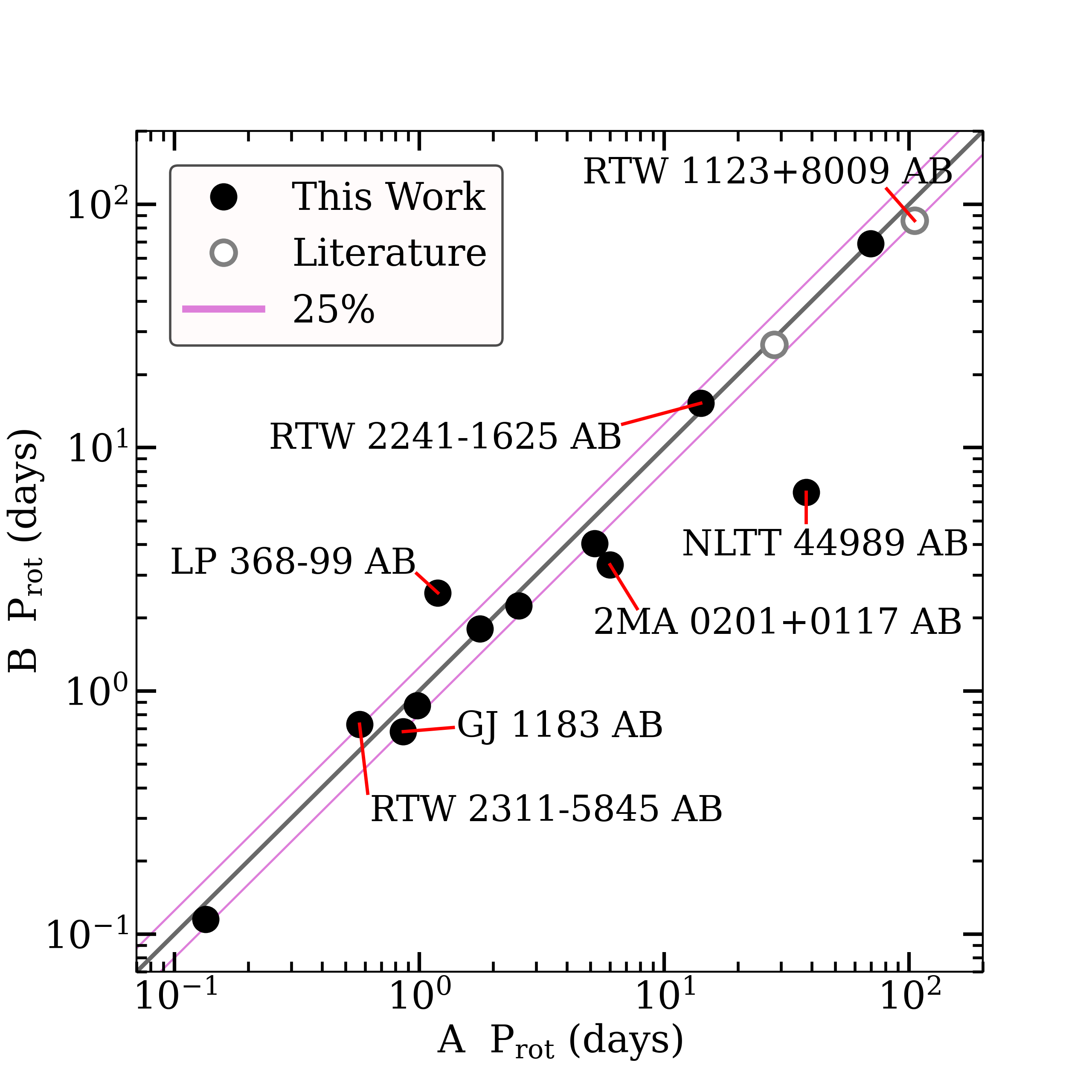}{0.49\textwidth}{(a)}
          \fig{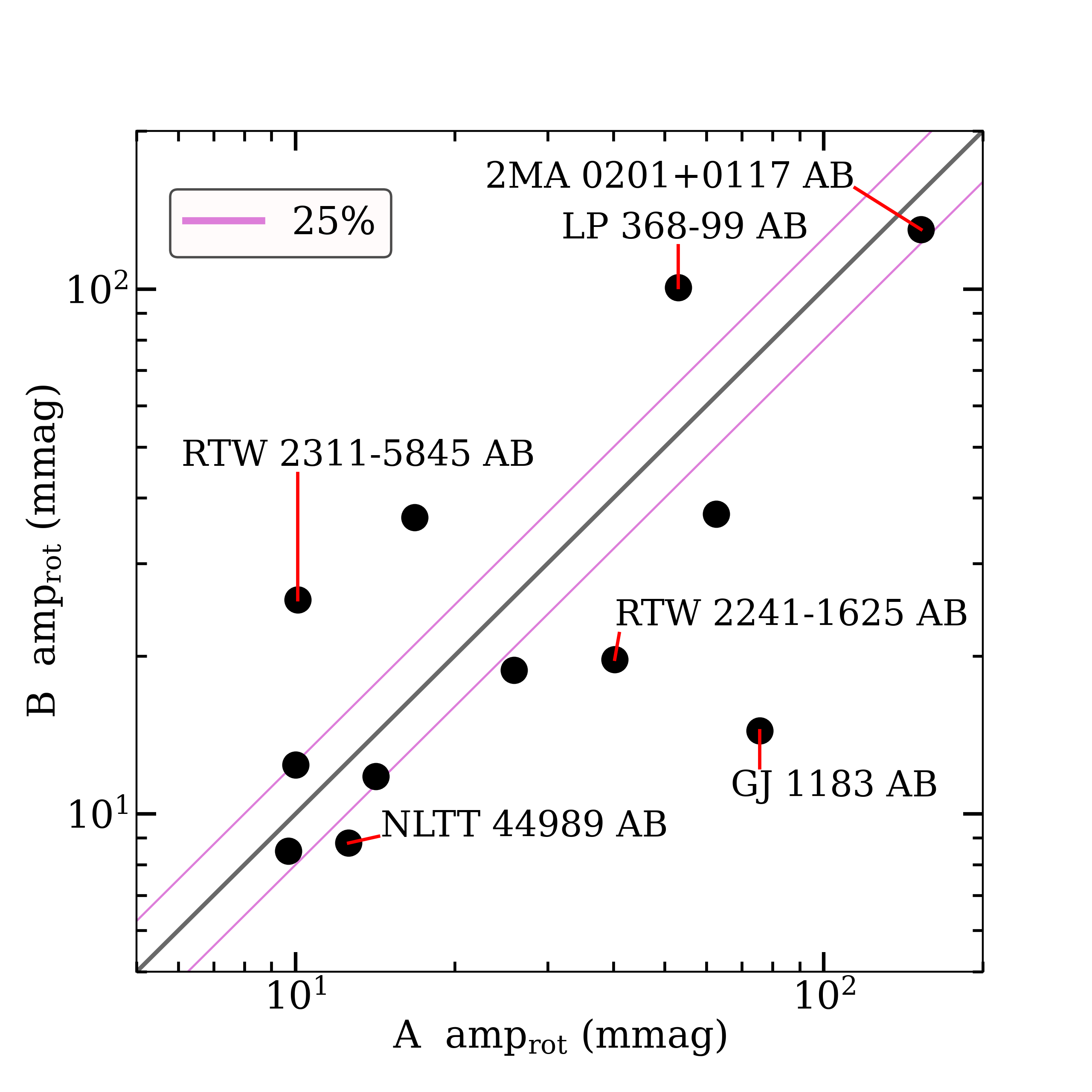}{0.49\textwidth}{(b)}}
\figcaption{Comparisons of rotation periods and peak-to-peak rotation amplitudes ($\rm{amp}_{\rm{rot}}$) between the A and B components in our twin binary Results Sample are shown, using values from Table~\ref{tab:RotTable} and \citetalias{RTW_P1}. One-to-one lines and corresponding 25\% (1.25\,$\times$) bounding lines are shown. For plotting purposes we assign RTW~0409+4623~A the 2.55\,day period and corresponding 7.1\,mmag amplitude, RTW~0409+4623~B the 2.24\,day period and corresponding 5.9\,mmag amplitude, and then double the amplitudes for RTW~0409+4623~AB and RTW~2202+5537~AB, as discussed in Section~\ref{sec:rot}. The measurements for different systems are derived from photometry in different filters, so care should be taken if comparing amplitudes in panel (b) across systems and not just between twin components. Several systems discussed in the text are labeled. Many systems show stunning alignment in their components' resolved rotation periods, including cases spanning rapid to slow rotation and across a large range of the M dwarf sequence, but key mismatched systems appear as well. The corresponding rotational photometric variability amplitudes show larger scatter, as expected given the potentially changing nature of photometric contrast effects. \label{fig:rot-equal}}
\end{figure}

We compare the A and B twin component rotation periods in panel (a) of Figure~\ref{fig:rot-equal} for systems in our Results Sample with final periods adopted for both stars. We see remarkable consistency, with 11 of the 14 plotted systems having component periods matching within 25\%, including cases with periods ranging from $\sim$0.1\,days up to $\sim$100\,days, and for both PC and FC systems. The three mismatched systems are LP~368-99~AB from this work as well as 2MA~0201+0117~AB and NLTT~44989~AB from \citetalias{RTW_P1}. LP~368-99~AB resides directly in the PC/FC transition gap region and is discussed further in Section~\ref{subsec:LP368}, 2MA~0201+0117~AB is a pre-main-sequence system, and NLTT~44989~AB is a FC system. There are only two PC systems with rotation periods measured for both stars, RTW~2236+5923~AB and RTW~2241-1625~AB, both of which have periods differing by $<$10\%. \textbf{Considering all M structural types, these results indicate a 21\% (3/14) mismatch rate for rotation periods.} We caution that we have not recovered rotation periods for \textit{all} twin pairs in our sample or within a volume complete set, so the cumulative statistics may not be accurately representative of the entire M dwarf population. For example, we are biased towards stars active enough to have detectable photometric amplitudes of variation, whereas very slowly rotating inactive systems with small amplitudes are underrepresented here.

Panel (b) of Figure~\ref{fig:rot-equal} compares the corresponding rotation photometric amplitudes for the twin systems, and exhibits significant scatter. This is perhaps expected given the varying nature of spot and faculae net brightness modulations. That said, we do still see the evident pattern of stronger (or weaker) amplitudes in one star generally tracking with stronger (or weaker) amplitudes in its twin companion. Using a 25\% percent difference boundary for matched versus mismatched rotation amplitudes --- the same limit used for the rotation periods and long-term variability assessment --- \textbf{we find that only 4 of the 12 systems match in rotation amplitude, corresponding to a mismatch rate of 67\% (8/12) when considering all structural types}. There is only one PC system with amplitudes available for both components, RTW~2241-1625~AB, which is mismatched. We again suggest caution when considering these population-level rates given the aforementioned rotation activity detection biases, relatively small number of twin systems with measurements, and the lack of a volume complete sample.

\section{Rotation--Activity Comparisons} \label{sec:rot-activity}

%%%%%%%%%%%%%%% fig - Prot versus Ha %%%%%%%%%%%%%%%
\begin{figure}[!t]
\figurenum{14}
\centering
\includegraphics[width=0.79\textwidth]{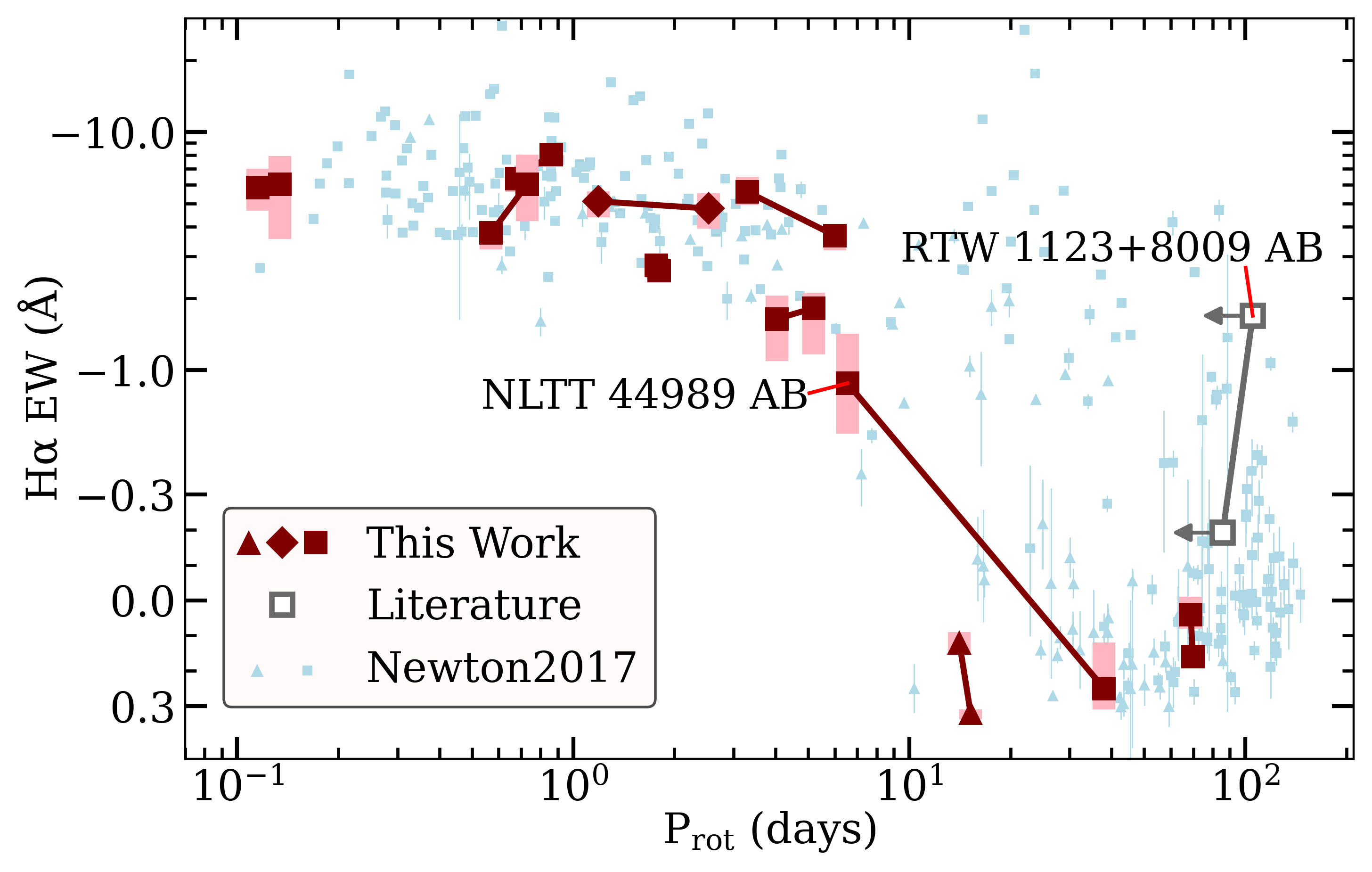}
\figcaption{Twin binaries from our Results Sample shown in the rotation-activity plane for M dwarfs, with lines connecting the two twin components in each binary. The H$\alpha$ EW scale is linear between 0.3 to $-$0.3 and logarithmic beyond this range. Mean measurements are from Table~\ref{tab:SpecTable}, Table~\ref{tab:RotTable}, and \citetalias{RTW_P1}. The CHIRON H$\alpha$ EWs include only non-flaring epochs with both stars successfully observed back-to-back, and shaded bars show the ranges of observed EWs. Square symbols represent fully convective stars, triangles are partially convective, and diamonds specify the LP~368-99~AB system residing within the gap between these structural types (see Section~\ref{subsec:LP368}). Comparison data for a large sample of field M dwarfs from \citetalias{Newton_2017} are underplotted in blue, and these points use the same PC/FC symbol scheme based on their estimated masses relative to 0.35\,$\rm{M_\odot}$. The two significantly mismatched twin pairs are (1) NLTT~44989~AB from \citetalias{RTW_P1}, and (2) RTW~1123+8009~AB using H$\alpha$ EWs from \citetalias{Pass_2024_ApJ} and rotation periods from \citetalias{2024AJ....167..159L}. It is notable that all of the stars generally follow the rotation-activity relationship, even for NLTT~44989~AB where the slower rotator is correspondingly less active; the single exception is RTW~1123+8009~AB, which has rotation period measurements that are less confident and possibly somewhat inaccurate hence the arrows plotted on its points (see Section~\ref{subsubsec:rot-notes_RTW-1123+8009}). \label{fig:prot-Ha}}
\end{figure}

%%%%%%%%%%%%%%% fig - rot amplitude versus prot and versus HaEW %%%%%%%%%%%%%%%
\begin{figure}[!t]
\figurenum{15}
\centering
\gridline{\fig{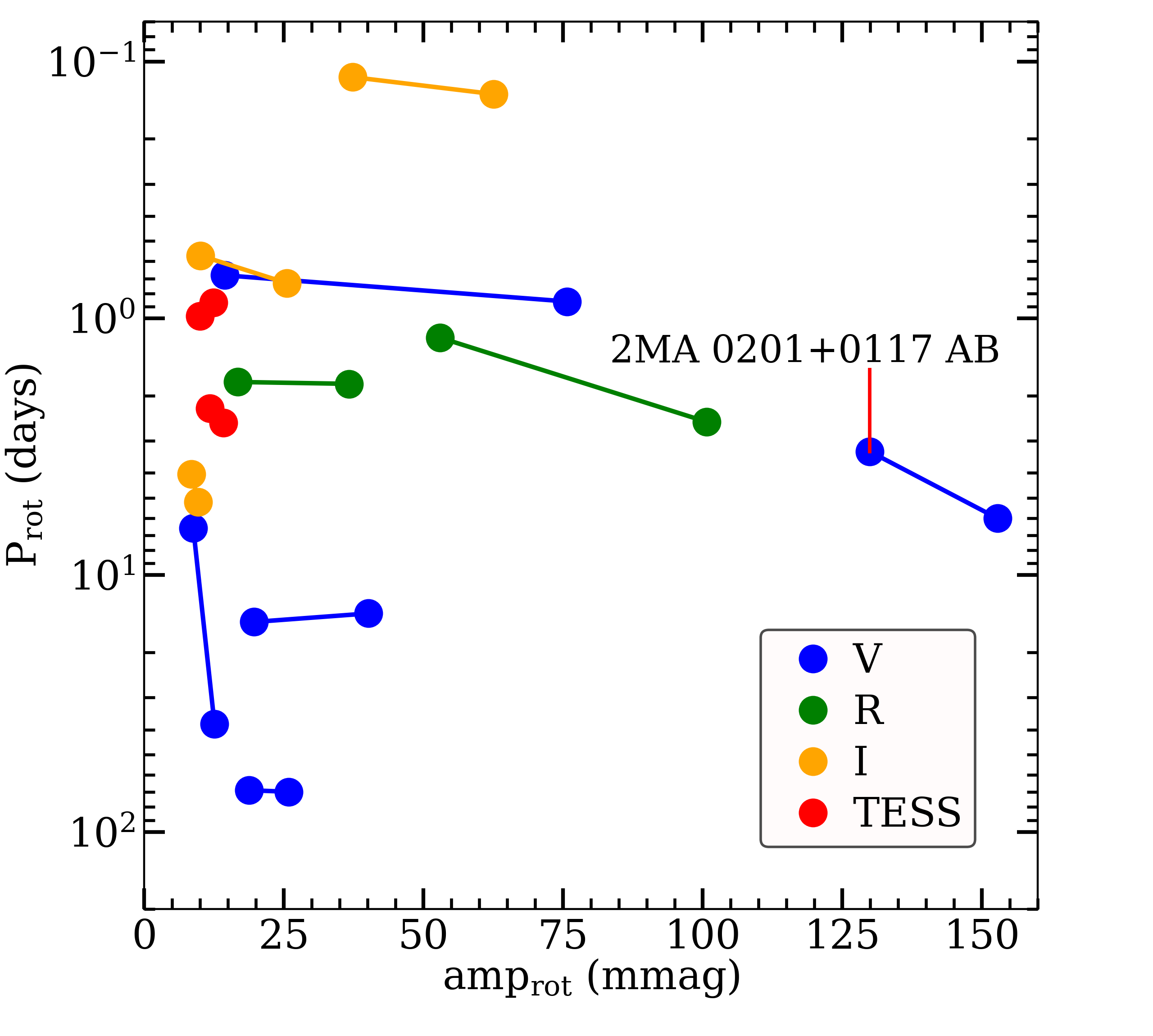}{0.49\textwidth}{(a)}
          \fig{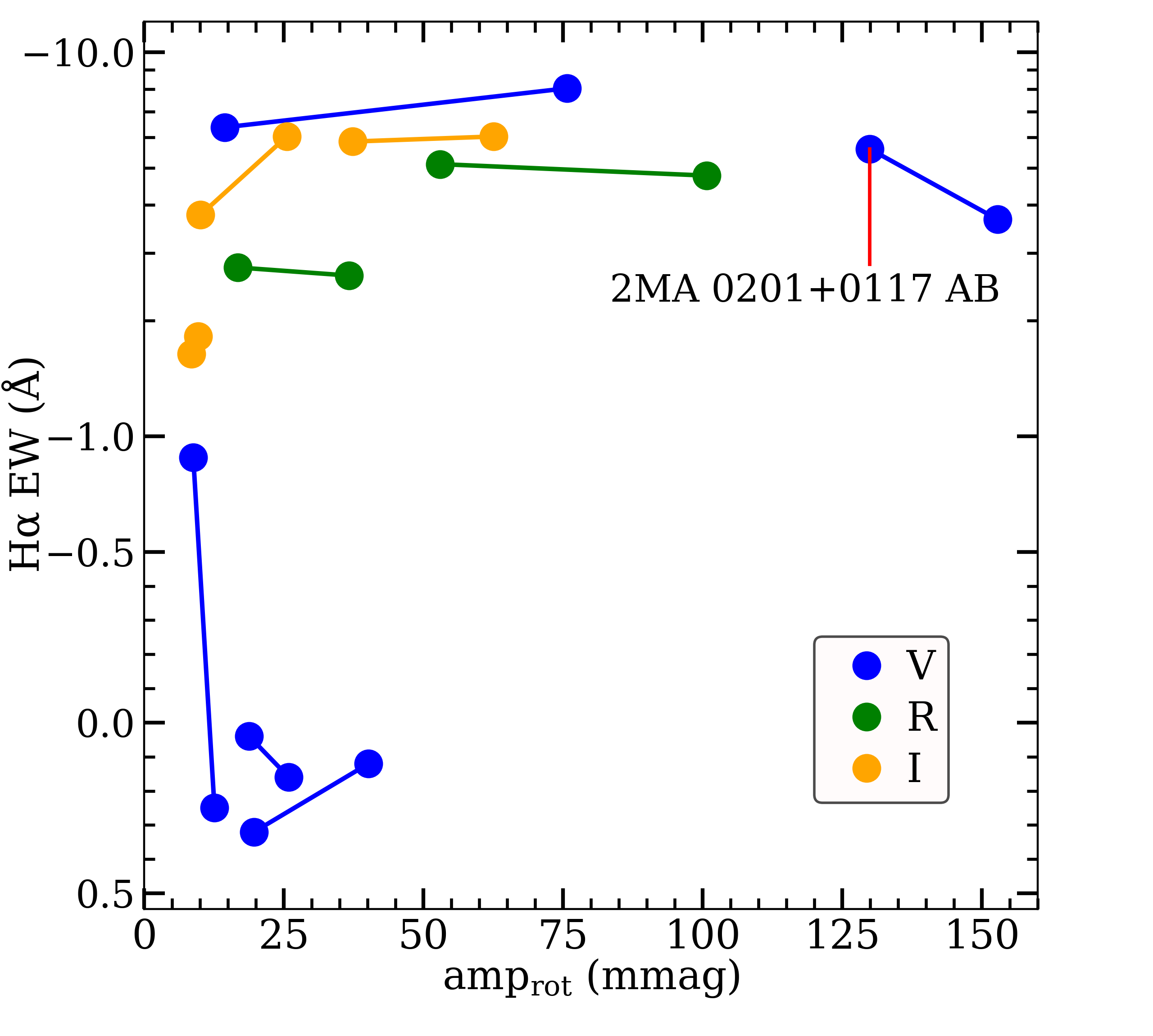}{0.49\textwidth}{(b)}}
\figcaption{Peak-to-peak rotation photometric amplitudes ($\rm{amp}_{\rm{rot}}$) compared with rotation periods in panel (a) and with mean H$\alpha$ EWs in panel (b) for twin stars from our Results Sample, with lines connecting the components in each binary. Values are from Table~\ref{tab:SpecTable}, Table~\ref{tab:RotTable}, and \citetalias{RTW_P1}, with most systems shown here being FC cases. We do not show the range of observed H$\alpha$ EWs in panel (b) for visual clarity, and only use non-flaring CHIRON epochs with both stars observed back-to-back. The y-axes are set such that a star with more emission or more rapid rotation moves upward in each plot. Points are color coded by their respective observation filters, which is important considering photometric variability amplitudes will generally appear larger in bluer filters \citep[][]{Hosey_2015, Kar_2024}. We include LP~368-99~AB under `R' given the similarity between R and the ZTF $zr$ filter actually used for its amplitude measurements. RTW~0409+4623~AB and RTW~2202+5537~AB are handled as described in Figure~\ref{fig:rot-equal}. The pair with the largest photometric amplitudes is the pre-main sequence system 2MA~0201+0117~AB described in \citetalias{RTW_P1}. Stars with the largest amplitudes are generally more likely to be active in H$\alpha$ and relatively rapidly rotating, as expected. \label{fig:amp_prot_ha}}
\end{figure}

Here we use our measurements directly as $P_{\rm{rot}}$ and H$\alpha$ EW instead of converting to the oft-used Rossby number ($R_o = P_{\rm{rot}}/\tau_{\rm{conv}}$) and $L_{\rm{H\alpha}} / L_{\rm{bol}}$ parameters. This is done for the same reasons presented in \citetalias{RTW_P1}, namely that the twin nature of our pairs allows us to avoid conversion relations and the possible unreliability of the resulting Rossby numbers \citep[e.g.,][]{2014ApJ...794..144R, Jao2022_rossby}. Care should thus be taken to only compare results between A and B pairs unless keeping in mind the mass differences across systems.

The rotation periods and H$\alpha$ EWs are shown in Figure~\ref{fig:prot-Ha} for our Results Sample twins, with most plotted pairs being FC and only one being PC and one Gap. Most of our twin pairs in the rotation--activity plane are broadly similar in their component behaviors throughout the saturated and unsaturated regimes --- at least compared to the general field scatter --- while the two outlier cases of NLTT~44989~AB and RTW~1123+8009~AB significantly violate this trend. These two active/inactive mismatched cases demonstrate in Figure~\ref{fig:prot-Ha} how their components span the saturated and unsaturated regimes despite their underlying twin natures. RTW~1123+8009~AB hosting longer rotation periods but similar H$\alpha$ activity to NLTT~44989~AB is somewhat expected when considering the strongly mass-dependent nature of FC M dwarf spindown and H$\alpha$ activity (e.g., \citetalias{Newton_2017, Pass_2024_ApJ}), given the former system has smaller component masses of 0.14\,$\rm{M_\odot}$ compared to the latter system's component masses of 0.25\,$\rm{M_\odot}$. Overall, it is also remarkable that every case generally follows the rotation-activity relationship except for RTW~1123+8009~AB, and we suspect the precise rotation period measurements for this pair may be somewhat inaccurate as discussed in Section~\ref{subsubsec:rot-notes_RTW-1123+8009}.

We next compare our twin stars' rotation photometric amplitudes to their rotation periods and H$\alpha$ activity in Figure~\ref{fig:amp_prot_ha}, where we find general agreement with the period-amplitude results of \citet{Newton_2016} for field M dwarfs. It is evident that a star with a rotation amplitude $\lesssim$\,50\,mmag does not differentiate if the star is markedly active/inactive or fast/slow in rotation, but a star having a very large amplitude $\gtrsim$\,50\,mmag does indicate it is likely a relatively active star rotating at $\lesssim$\,10\,days, although we caution that this is based on a relatively small number of points. Regarding comparisons between twin stars, however, our twins in panel (a) of Figure~\ref{fig:amp_prot_ha} align in rotation period more often than they align in rotation amplitude, i.e., lines on the plot are negligible or typically more horizontal than they are vertical (regardless of plot scaling) --- this is in agreement with the behavior seen in the equality plots in Figure~\ref{fig:rot-equal}. This all aligns with the general expectation that rotation amplitudes will vary more and show more scatter over human observational timescales compared to rotation periods or H$\alpha$ activity.

%%%%%%%%%%%%%% tab - summary match/mismatch table %%%%%%%%%%%%%%%
\begin{deluxetable}{lc|ccccc|ccccc}[!t]
\tablewidth{0pt}
\tablecaption{A Summary of Matched and Mismatched Behaviors for the Twin Binary Results Sample \label{tab:summary}}
\tablehead{
\colhead{System} & \colhead{Type} & \colhead{$P_{\rm{rot}}$} & \colhead{$\rm{amp}_{\rm{rot}}$} & \colhead{$\rm{MAD}_{long}$} & \colhead{H$\alpha$ EW} & \colhead{$L_{\rm{X}}$} & \colhead{$P_{\rm{cycle}}$} & \colhead{UV} & \colhead{IR} & \colhead{Flare} & \colhead{B Strength} \\[-0.9em]
\colhead{} & \colhead{} & \colhead{} & \colhead{} & \colhead{} & \colhead{} & \colhead{} & \colhead{} & \colhead{Act.} & \colhead{Var.} & \colhead{Rate} & \colhead{\& Topology}\\
\colhead{} & \colhead{} & \colhead{$<$\,25\%} & \colhead{$<$\,25\%} & \colhead{$<$\,25\%} & \colhead{$<$\,25\%/0.2\text{\AA}} & \colhead{$<$\,25\%} & \colhead{} & \colhead{} & \colhead{} & \colhead{} & \colhead{}
}
\startdata
RTW 0143-0151 AB & FC  & \nodata    & \nodata    & \nodata    & \checkmark & \nodata  & \nodata & \nodata & \nodata & \nodata & \nodata \\
RTW 0231-5432 AB & FC  & \checkmark & $\times$   & \checkmark & \checkmark & \nodata  & \nodata & \nodata & \nodata & \nodata & \nodata \\
RTW 0231+4003 AB & FC  & \nodata    & \nodata    & \nodata    & \nodata    & \nodata  & \nodata & \nodata & \nodata & \nodata & \nodata \\
RTW 0409+4623 AB & FC  & \checkmark & \checkmark & \nodata    & \nodata    & \nodata  & \nodata & \nodata & \nodata & \nodata & \nodata \\
KAR 0545+7254 AB & PC  & \nodata    & \nodata    & \nodata    & \checkmark & \nodata  & \nodata & \nodata & \nodata & \nodata & \nodata \\
LP 719-37 AB\tablenotemark{$\rm{\diamond}$}     & FC  & \checkmark & \checkmark & \checkmark & \checkmark & \nodata  & \nodata & \nodata & \nodata & \nodata & \nodata \\
LP 368-99 AB\tablenotemark{$\rm{\diamond}$}     & Gap & $\times$   & $\times$   & \checkmark & \checkmark & \nodata  & \nodata & \nodata & \nodata & \nodata & \nodata \\
L 533-2 AB       & PC  & \nodata    & \nodata    & \checkmark & \checkmark & \nodata  & \nodata & \nodata & \nodata & \nodata & \nodata \\
RTW 0933-4353 AB\tablenotemark{$\rm{\diamond}$} & FC  & \checkmark & $\times$   & $\times$   & \checkmark & \nodata  & \nodata & \nodata & \nodata & \nodata & \nodata \\
LP 551-62 AB     & PC  & \nodata    & \nodata    & \nodata    & \checkmark & \nodata  & \nodata & \nodata & \nodata & \nodata & \nodata \\
RTW 1123+8009 AB\tablenotemark{$\rm{\diamond}$} & FC  & \checkmark & \nodata    & \nodata    & $\times$   & \nodata  & \nodata & \nodata & \nodata & \nodata & \nodata \\
RTW 1133-3447 AB & PC  & \nodata    & \nodata    &($\times$)  & \checkmark & \nodata  & \nodata & \nodata & \nodata & \nodata & \nodata \\
RTW 1336-3212 AB & FC  & \nodata    & \nodata    & \checkmark & \checkmark & \nodata  & \nodata & \nodata & \nodata & \nodata & \nodata \\
L 197-165 AB     & FC  & \nodata    & \nodata    & \checkmark & \checkmark & \nodata  & \nodata & \nodata & \nodata & \nodata & \nodata \\
RTW 1433-6109 AB & PC  & \nodata    & \nodata    & \checkmark & \checkmark & \nodata  & \nodata & \nodata & \nodata & \nodata & \nodata \\
L 1197-68 AB     & PC  & \nodata    & \nodata    & \nodata    & \checkmark & \nodata  & \nodata & \nodata & \nodata & \nodata & \nodata \\
L 1198-23 AB     & PC  & \nodata    & \nodata    & \checkmark & \checkmark & \nodata  & \nodata & \nodata & \nodata & \nodata & \nodata \\
RTW 1512-3941 AB & PC  & \nodata    & \nodata    & \checkmark & \checkmark & \nodata  & \nodata & \nodata & \nodata & \nodata & \nodata \\
RTW 1812-4656 AB & PC  & \nodata    & \nodata    & \checkmark & \checkmark & \nodata  & \nodata & \nodata & \nodata & \nodata & \nodata \\
GJ 745 AB\tablenotemark{$\rm{\diamond}$}        & Gap & \nodata    & \nodata    & \nodata    & \checkmark & \nodata  & \nodata & \nodata & \nodata & \nodata & \nodata \\
RTW 2011-3824 AB & FC  & \nodata    & \nodata    & \nodata    & \checkmark & \nodata  & \nodata & \nodata & \nodata & \nodata & \nodata \\
G 230-39 AB      & FC  & \nodata    & \nodata    & \nodata    & \checkmark & \nodata  & \nodata & \nodata & \nodata & \nodata & \nodata \\
RTW 2202+5537 AB & FC  & \checkmark & \checkmark & \nodata    & \nodata    & \nodata  & \nodata & \nodata & \nodata & \nodata & \nodata \\
RTW 2211+0058 AB & FC  & \nodata    & \nodata    & \nodata    & \checkmark & \nodata  & \nodata & \nodata & \nodata & \nodata & \nodata \\
RTW 2235+0032 AB & FC  & \checkmark & $\times$   & $\times$   & \checkmark & \nodata  & \nodata & \nodata & \nodata & \nodata & \nodata \\
RTW 2236+5923 AB & PC  & \checkmark & \nodata    & \nodata    & \nodata    & \nodata  & \nodata & \nodata & \nodata & \nodata & \nodata \\
RTW 2241-1625 AB\tablenotemark{$\rm{\diamond}$} & PC  & \checkmark & $\times$   & $\times$   & $\times$   & \nodata  & \nodata & \nodata & \nodata & \nodata & \nodata \\
RTW 2244+4030 AB & PC  & \nodata    & \nodata    & \nodata    & \checkmark & \nodata  & \nodata & \nodata & \nodata & \nodata & \nodata \\
L 718-71 AB      & PC  & \nodata    & \nodata    & \checkmark & \checkmark & \nodata  & \nodata & \nodata & \nodata & \nodata & \nodata \\
RTW 2311-5845 AB\tablenotemark{$\rm{\diamond}$} & FC  & \checkmark & $\times$   & $\times$   & $\times$   & \nodata  & \nodata & \nodata & \nodata & \nodata & \nodata \\
\hline
2MA 0201+0117 AB\tablenotemark{$\rm{\diamond}$} & PMS & $\times$   & \checkmark & \checkmark & $\times$   & $\times$ & \nodata & \nodata & \nodata & \nodata & \nodata \\
GJ 1183 AB\tablenotemark{$\rm{\diamond}$}       & FC  & \checkmark & $\times$   & $\times$   & $\times$   & $\times$ & \nodata & \nodata & \nodata & \nodata & \nodata \\
NLTT 44989 AB\tablenotemark{$\rm{\diamond}$}    & FC  & $\times$   & $\times$   & $\times$   & $\times$   & $\times$ & \nodata & \nodata & \nodata & \nodata & \nodata \\
\hline
\hline
All Mismatch Rate:       & & 3/14            & 8/12            & 6/18            & 6/29            & 3/3   & \nodata & \nodata & \nodata & \nodata & \nodata \\
\hspace{37.5pt} Percent:   & & $21\%_{-7\%}^{+14\%}$ & $67\%_{-15\%}^{+10\%}$ & $33\%_{-9\%}^{+12\%}$ & $21\%_{-6\%}^{+9\%}$  & 100\% & \nodata & \nodata & \nodata & \nodata & \nodata \\
                         & & & & & & & & & & & \\%intentional blank row
PC Mismatch Rate:        & & 0/2             & 1/1             & 1/7             & 1/12            & 0/0   & \nodata & \nodata & \nodata & \nodata & \nodata \\
\hspace{38pt} Percent: & & 0\%             & 100\%           & $14\%_{-5\%}^{+21\%}$ & $8\%_{-3\%}^{+15\%}$   & N/A   & \nodata & \nodata & \nodata & \nodata & \nodata \\
                         & & & & & & & & & & & \\%intentional blank row
FC Mismatch Rate:        & & 1/10            & 6/9             & 5/9             & 4/14            & 2/2   & \nodata & \nodata & \nodata & \nodata & \nodata \\
\hspace{38pt} Percent: & & $10\%_{-3\%}^{+17\%}$  & $67\%_{-17\%}^{+11\%}$ & $56\%_{-16\%}^{+14\%}$ & $29\%_{-9\%}^{+14\%}$ & 100\% & \nodata & \nodata & \nodata & \nodata & \nodata \\
\enddata
\tablenotetext{\rm{\diamond}}{These noteworthy systems have individual discussions provided in Section~\ref{sec:sys-notes}.}
\tablecomments{A summary for the Results Sample showing which of the 33 pairs of twin M dwarfs display matching or mismatching properties. The structural type (PC/Gap/FC/PMS) for each system is indicated, as described in Table~\ref{tab:SampleTable-phot}, and the three systems from \citetalias{RTW_P1} are separated at the bottom. The rotation period ($P_{\rm{rot}}$), rotation amplitude ($\rm{amp}_{\rm{rot}}$), multi-year photometric variability ($\rm{MAD}_{long}$), and X-ray luminosity ($L_{\rm{X}}$) are designated as matching with a~\checkmark~symbol if the underlying mean measures for A and B differ by less than 25\%, and are considered mismatching using an~$\times$~symbol if the difference is greater than 25\%. RTW~1133-3447~AB uses parentheses to indicate the unreliable long-term MAD measure for A and subsequently unreliable mismatch flag, as discussed in Section~\ref{subsec:longterm-phot-results}. The same matched/mismatched marking scheme is used for the H$\alpha$ activity, but now with our CHIRON cases designated based on our manual inspection of the H$\alpha$ line profiles and the pairs' positions in panel (a) of Figure~\ref{fig:Ha-equality} relative to the one-to-one line, alongside corresponding numerical EW cutoffs at 25\% relative difference and 0.2\,\text{\AA} absolute difference (see Section~\ref{subsec:chiron-ha-results} for details). For literature H$\alpha$ cases we rely on the externally reported EWs as summarized in Table~\ref{tab:SpecTable}. The summary rows at the bottom give the fractions of mismatched cases, with the PC and FC subcategories only including those corresponding main sequence systems --- asymmetric percent uncertainties are 1-$\sigma$ equivalent errors derived following the small-N binomial statistics method detailed in the appendix of \citet{2003ApJ...586..512B}. Empty columns for several additional activity-related factors are included on the right to emphasize the potential for further study.}
\end{deluxetable}

\section{Discussion: The Results Sample} \label{sec:discussion1}

\subsection{Overall Assessment of Matched and Mismatched Systems} \label{subsec:mismatches}

We now take a complete view of our Results Sample and decide which systems appear ``matched" or ``mismatched" in rotation and activity, as summarized in Table~\ref{tab:summary}. H$\alpha$ is the most stable activity tracer among our observed properties, and our manual inspection of the H$\alpha$ results (Section~\ref{subsec:chiron-ha-results}) indicated a clear set of match/mismatch designations that could be reproduced with 25\% percent difference and 0.2\,$\text{\AA}$ absolute difference cutoffs applied to the EWs. Furthermore, a 25\% threshold cleanly separates cases near versus notably offset from the one-to-one line in the rotation period results in panel (a) of Figure~\ref{fig:rot-equal}, and we also consider a 25\% threshold to constitute a scientifically meaningful difference in the context of M dwarf rotation-activity behaviors. We therefore adopted the 25\% percent difference criterion across our other measures as well to classify a pair of components as being ``matched" or ``mismatched" in a given property. We reiterate here the population-level selection effects and detection biases occurring for our coverage of the parameters in Table~\ref{tab:summary}; the exception is H$\alpha$ activity, for which data were secured for every system observable from CTIO (although in a volume-limited sample, not a volume-complete sample).

Altogether, it is remarkable just how well-matched many of our twin systems appear, particularly in rotation rate where both of the PC systems and nine of the ten FC systems display matching periods. Similarly, for the H$\alpha$ measurements, 11 of the 12 PC systems and 10 of the 14 FC systems host matching behaviors, including both active emission and inactive absorption cases. We see alignment less often in the photometric modulation measures for rotation amplitude and long-term variability, but this is likely because these rely on net spot and faculae contrast balancing, which can change considerably as active regions evolve over time. The extreme congruence of rotation and H$\alpha$ activity, which each match for 79\% of the systems, demonstrates that M dwarf rotation-activity behaviors are, to at least first order, broadly predictable phenomena generally ascribable to a combination of mass, age, and composition. This is a notable result considering the stochastic natures of dynamos and activity, the complexity of M dwarf behaviors across the main sequence, and the \textit{many} subtle factors that could affect activity and rotation throughout a star's formation and lifetime.

However, as shown in the mismatched line at the bottom of Table~\ref{tab:summary} for all types of systems, at least one of five pairs is mismatched in every characteristic measured. This implies that, when randomly selecting field M dwarfs, roughly one out of every five will have their activity and/or rotation behaviors deviated from expectations by at least 25\%, even if mass, age, and composition are constrained; this remains true even after substantial vetting for unresolved companions. Considering only characteristics with at least five measured PC systems, we see a higher rate of mismatches in FC stars than in PC stars. This may be a consequence of FC M dwarfs remaining active for longer than PC M dwarfs, lending more opportunity for activity differences to appear in our FC targets. Indeed, our two cases with very large active/inactive H$\alpha$ mismatches, NLTT~44989~AB and RTW~1123+8009~AB, are both FC pairs. Overall, our mismatched systems are critical targets that inform the predictability of rotation-activity phenomena and constrain the intrinsic scatter possible, and they highlight the need for careful consideration and investigation of higher order factors in M dwarf rotation-activity evolution.

\subsection{Possible Explanations for Mismatches} \label{subsec:explanations}

There are \textit{many} factors possibly relevant to cases with mismatched activity and/or rotation, despite their similar fundamental parameters. We discussed these factors extensively in \citetalias{RTW_P1}, and next highlight the most crucial factors likely relevant to the systems here, with specific systems discussed further in Section~\ref{sec:sys-notes}.

{\it Magnetic Braking} --- Foremost is that FC M dwarfs undergo a phase of fast magnetic braking as they transition from rapidly-rotating active stars to slowly-rotating inactive stars \citep{Newton_2016, Newton_2017, Newton_2018, Medina_2022_spindown, 2023MNRAS.526..870S}, with observed variability to this overall process \citep{Pass2022}, possibly due to the birth environment and/or early disk factors \citep{2021MNRAS.508.3710R, Pass_2024_ApJ, RTW_P1}. The fast magnetic braking phase is likely responsible for the strong active/inactive mismatch case seen in RTW~1123+8009~AB, as we argued for the similar case of NLTT~44989~AB in \citetalias{RTW_P1}. For example, previous minor deviations in rotation-activity trajectories could change when a given star arrives at the rapid transition phase, leading to a short time window in which significantly mismatched behaviors could exist for our twin components.

{\it Dynamo Bistability / Activity Cycles} --- A proposed dynamo bistability in some rapidly-rotating low-mass M dwarfs \citep{Gastine_2013, 2014ApJ...785...10C, 2024A&A...687A..95M} could explain our mismatches as well. The exact areas of parameter space where this dynamo bistability may apply are not precisely constrained, but based on the work of \citet{Gastine_2013} and \citet{2024A&A...687A..95M} a key regime is mid-to-late M dwarfs at $P_{\rm{rot}} \lesssim 1$\,day (see panel (b) of Figure~\ref{fig:mass-Ha-prot}). The dynamo state a star might host from the bistability could be sensitive to the magnetic initial conditions, per \citet{Gastine_2013}, which could explain why some of our low-mass systems with matching rapid rotation periods can sometimes show matching H$\alpha$ activity (e.g., RTW~0933-4353~AB) while other such cases show firmly mismatched H$\alpha$ activity (e.g., RTW~2311-5845~AB). However, \citet{Kitchatinov_2014} argue the hypothesized bistability may instead be linked to long-term magnetic cycles (presumably with corresponding activity cycles), which have been explored for M dwarfs \citep[e.g.,][A.~A.~Couperus et al.~2025 in preparation]{2016A&A...595A..12S, Farrish_2021, 2023ApJ...949...51I, 2024ApJ...977..144W} and could themselves also explain mild mismatches via activity snapshots. We do see candidate activity cycle signals in our long-term photometry for RTW~0231-5432~AB in Figure~\ref{fig:long-term-curves}, but this system is actually matching in rotation period and H$\alpha$ activity, and our other mismatch cases do not (yet) reveal activity cycles, except for GJ~1183~AB in \citetalias{RTW_P1}. The roles of activity cycles and the dynamos causing them therefore remain highly relevant but poorly constrained.

\begin{figure}[!t]
\figurenum{16}
\centering
\gridline{\fig{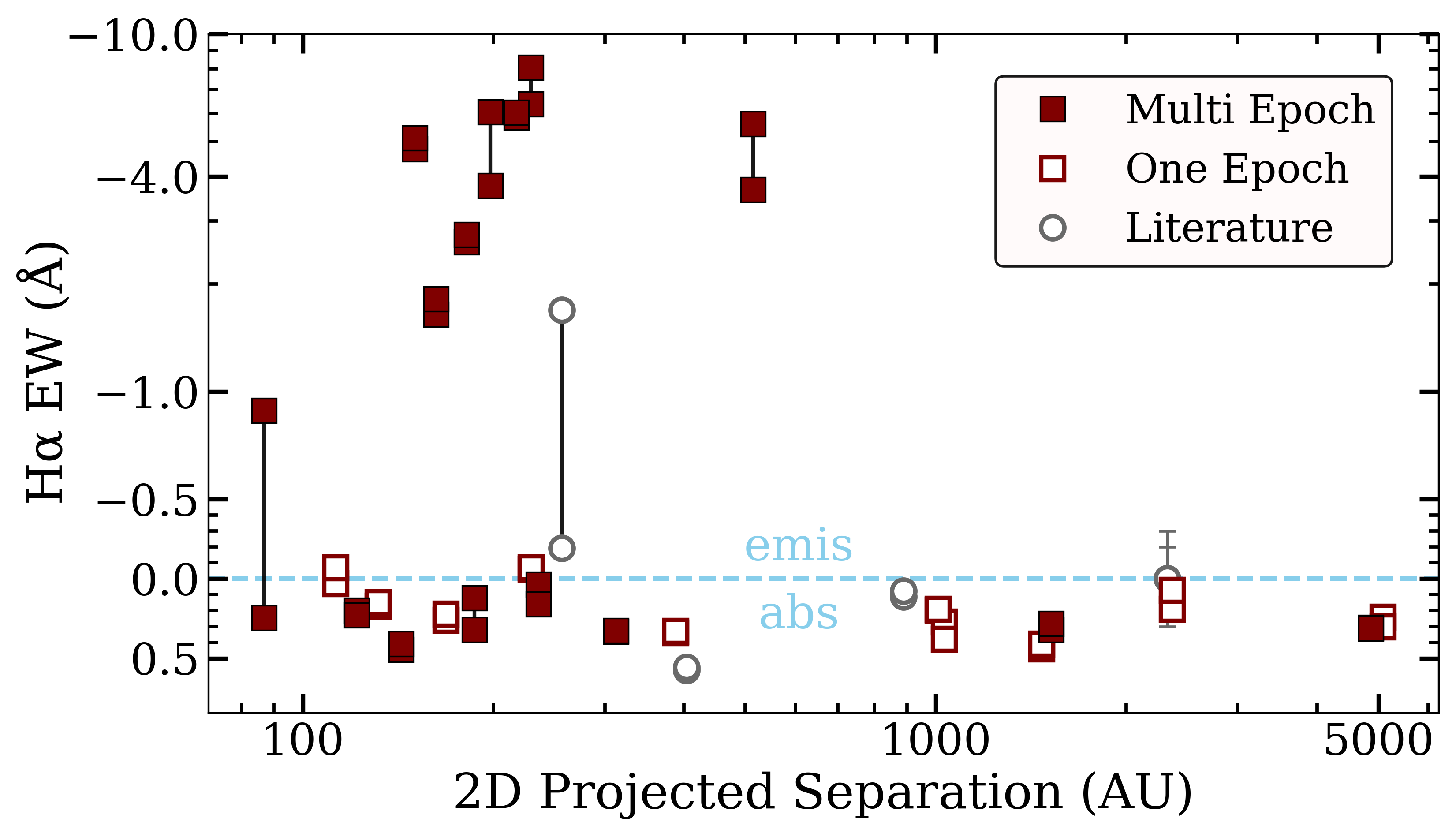}{0.50\textwidth}{(a)}
          \fig{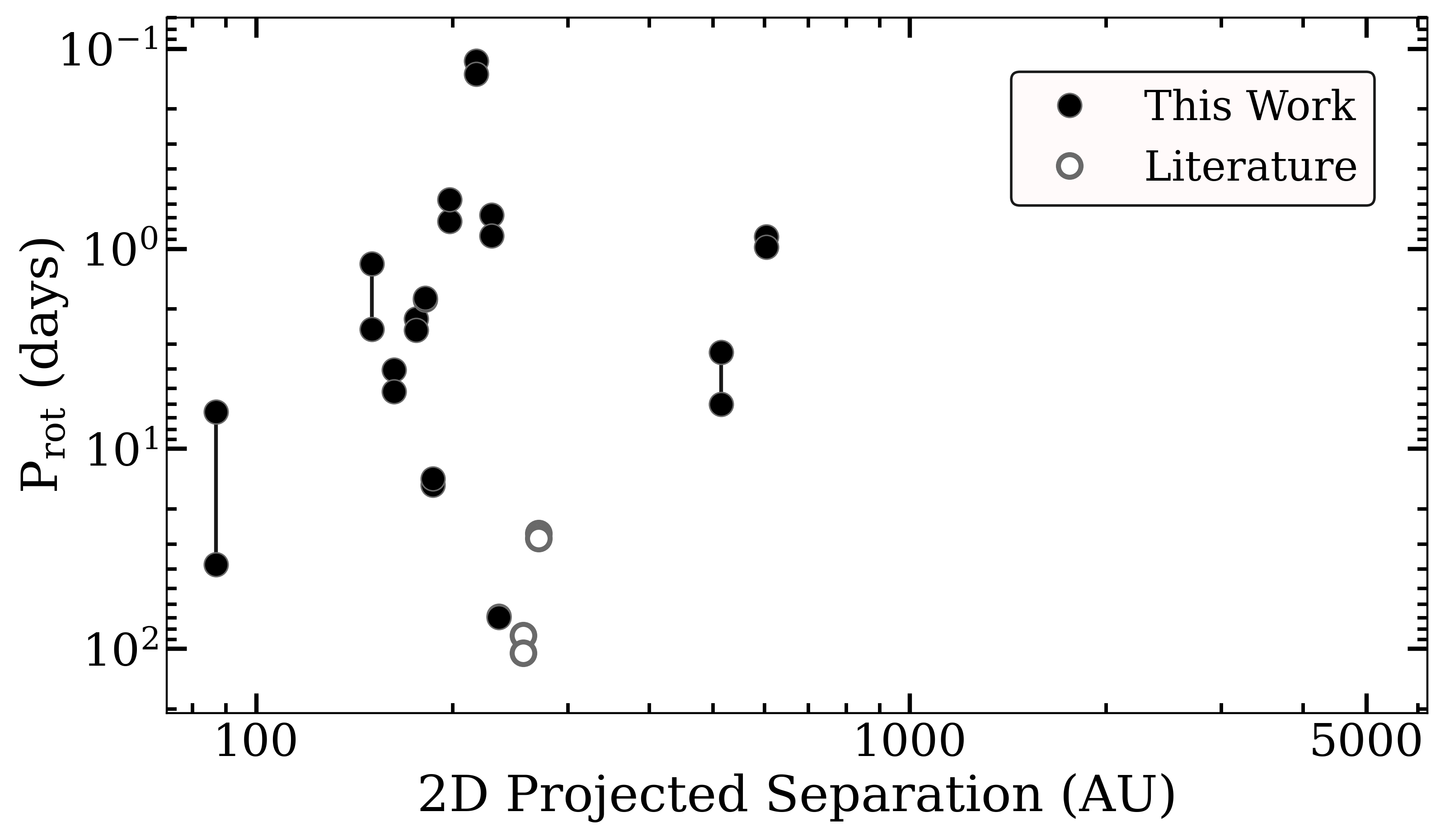}{0.484\textwidth}{(b)}}
\figcaption{The H$\alpha$ activity (a) and rotation rate (b) of our Results Sample binary components shown versus their physical separations, using values from Table~\ref{tab:SampleTable-astr}, Table~\ref{tab:SpecTable}, Table~\ref{tab:RotTable}, and \citetalias{RTW_P1}. A single marker is one star and each twin pair is connected by a vertical black line. (a) --- The y-axis is scaled linearly between $-$0.5 to $+$0.5 and logarithmically outside these bounds, with stronger emission moving upward on the plot. Blue labels indicate the regions of H$\alpha$ abs(orption) and emis(sion). All squares are measurements from CHIRON while circles are from literature sources, and solid squares show those with multiple CHIRON epochs using their mean EWs. The range of observed EWs for multi-epoch cases are not shown for visual clarity. The CHIRON results shown here only include non-flaring epochs with both stars successfully observed back-to-back. (b) --- A faster rotation period moves upward on the plot, to track with stronger emission in panel (a). For RTW~0409+4623~AB, we assign A the 2.55 day period and B the 2.24 day period for plotting purposes. Both panels reveal no obvious trends with separation, indicating dynamical interactions between the wide components are likely not significantly impacting their rotation-activity behaviors. \label{fig:ha-2dsep}}
\end{figure}

{\it Companion Interactions} --- Unresolved, close-in, stellar or brown dwarf companions could disrupt expected rotation-activity patterns, as could possible dynamical interactions between our wide binary components. We have extensively searched for and validated against close companions where possible, so we do not consider close companion interactions to be significant for our 33 systems. We assess dynamical interactions between the wide pairs by considering the physical separation of our pairs versus their activity and rotation in Figure~\ref{fig:ha-2dsep}, but we see no apparent trends between matched or mismatched activity/rotation and pair separation. This is expected in the context of tidal interactions possibly changing rotation and activity, as this only meaningfully occurs below orbital periods of $\sim$100\,days \citep{2019ApJ...881...88F}, significantly shorter than our twin systems' projected orbital periods of $>$\,1000\,years. There is a tentative clustering of active pairs around $\sim$200\,au in Figure~\ref{fig:ha-2dsep}, which we speculate could perhaps relate to twin binary formation, but there are many inactive pairs in this region as well.

%%%%%%%%%%%%%%% fig - Ha EW and Prot versus 2d separation %%%%%%%%%%%%%%%

{\it Disk Interactions} --- Twin components interfering with each other's disks during formation appears unlikely as such impacts are broadly found to cease at separations beyond $\sim$100\,au \cite[e.g.,][]{1996ApJ...458..312J, Meibom2007, 2009ApJ...696L..84C, 2012ApJ...751..115H, 2017A&A...607A...3M, 2019A&A...627A..97M, 2023ASPC..534..275O}, including for low-mass stars. However, this does not rule out possible interactions early in the pairs' lives, as some twin wide binaries may form much closer together and subsequently widen in their birth environments \citep{2019MNRAS.489.5822E, 2022ApJ...933L..32H}. This circumstance could debase our assumption of independent coevality for pair members, especially if they remain on highly eccentric orbits today with close periastron passages \citep[e.g.,][]{2022ApJ...933L..32H}, which could periodically produce short-lived interactions between the stars and hypothetically affect their activity and/or spindown evolution. In contrast, recent work by \citet{2024ApJ...973...42B} found PMS components in the same system hosting twin jets and twin disks, indicating that twin behaviors are possible even for early formation disks.

{\it Star-Planet Interactions} --- Tidal and/or magnetic interactions between the stars and orbiting planets could be partially responsible for the observed differences in our twins \citep[e.g.,][]{2015ApJ...799...27P, 2016A&A...591A..45P, 2022MNRAS.513.4380I, 2024MNRAS.527.3395I}. This is an intriguing possibility because it can be investigated observationally; targeting the discordant pairs NLTT~44989~AB and RTW~1123+8009~AB for higher precision RVs may be particularly revealing. There are no exoplanets yet reported around any of our twin stars based on a crossmatch with the NASA Exoplanet Archive on 2025 March 16, nor did we see evidence for planetary signals in the various photometric and RV time series' we have employed. Another related avenue for future work could be deriving detailed abundance measurements for our twin components, as recent work by \citet{2025MNRAS.538.2408Y} found that abundance differences in otherwise similar co-natal stars can correlate with activity differences and constrain various possible star-planet explanations.

{\it Mass Differences} --- We revisit the possibility that small mass differences could explain our observed mismatched FC twins, hypothesizing that slightly different masses could result in the two stars reaching their active-to-inactive rapid transition phases at sufficiently offset times to explain our observations. \citetalias{Pass_2024_ApJ} used their simple model of mass-dependent spindown in FC M dwarfs to estimate the probability of observing an otherwise twin binary in an active/inactive mismatched state given a component mass difference. The \citet{Benedict_2016} $M_V$ MLR we use to estimate our masses has an rms scatter of 0.023\,$\rm{M}_\odot$, but our twins control for age and composition, so likely have even less underlying scatter between components. The largest difference in estimated mass for an FC pair in our Results Sample is 0.007\,$\rm{M}_\odot$, so we use this value in Equation~(3) from \citetalias{Pass_2024_ApJ} and find a 1.4\% chance of observing a single FC pair as an active/inactive mismatch. We have two FC active/inactive H$\alpha$ mismatch cases, NLTT~44989~AB and RTW~1123+8009~AB, out of our 14 total FC twin systems with H$\alpha$ results. Combining this information and using binomial statistics, we find that the 1.4\% probability yields a 98.5\% chance we would have observed fewer than two such systems if the masses are accurate and spindown is entirely mass dependent (when controlling for composition and environment). This implies (1) the slight mass differences alone are very likely not causing our active/inactive mismatched cases, and (2) FC M dwarf spindown is not entirely determined by mass. Higher order dynamo/disk/planet factors are therefore likely involved in at least some cases, in agreement with the results of \citetalias{Pass_2024_ApJ} and our assessment in \citetalias{RTW_P1}.

\subsection{Trends with Mass} \label{subsec:masstrends}

\begin{figure}[!t]
\figurenum{17}
\centering
\gridline{\fig{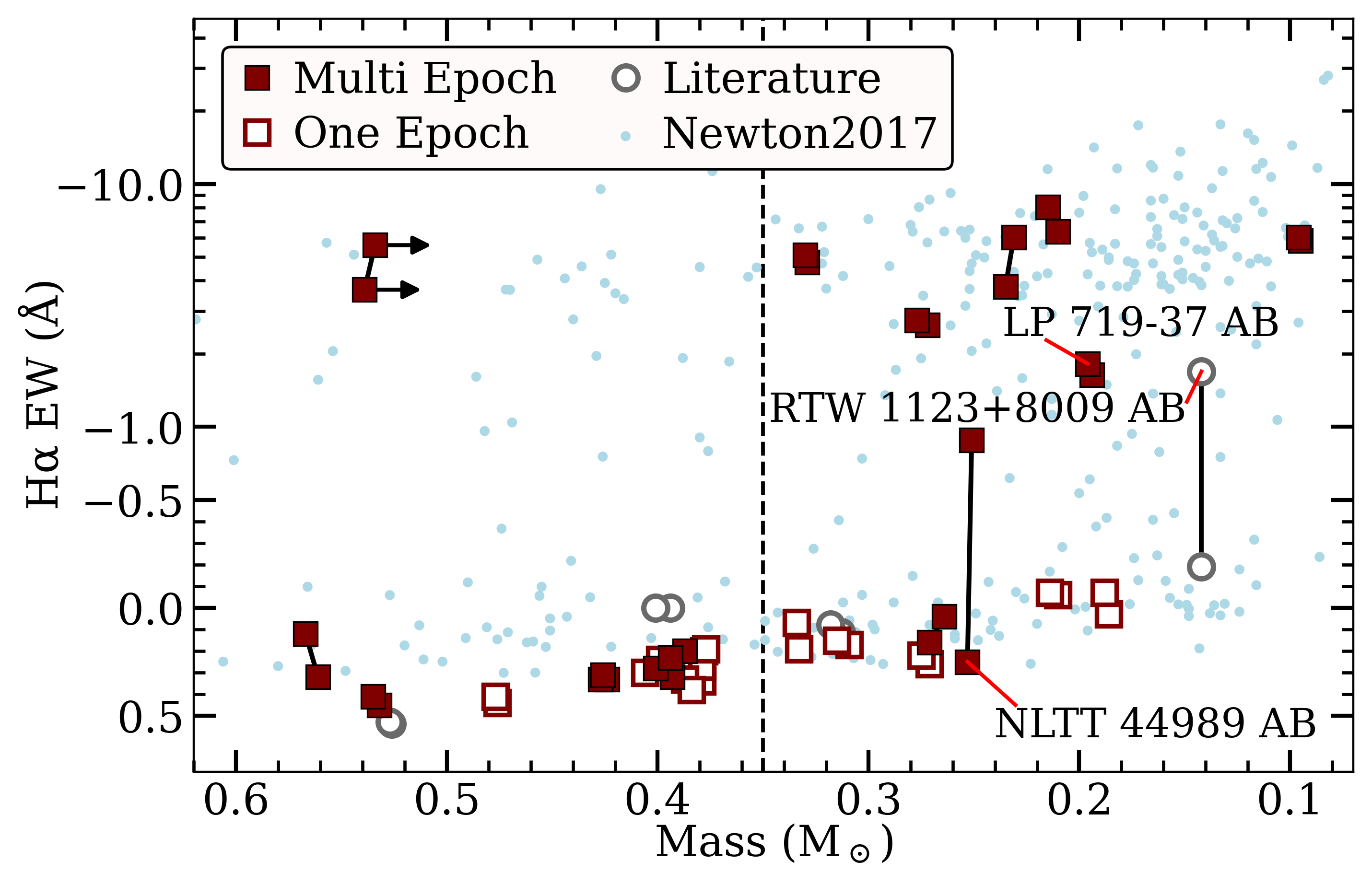}{0.50\textwidth}{(a)}
          \fig{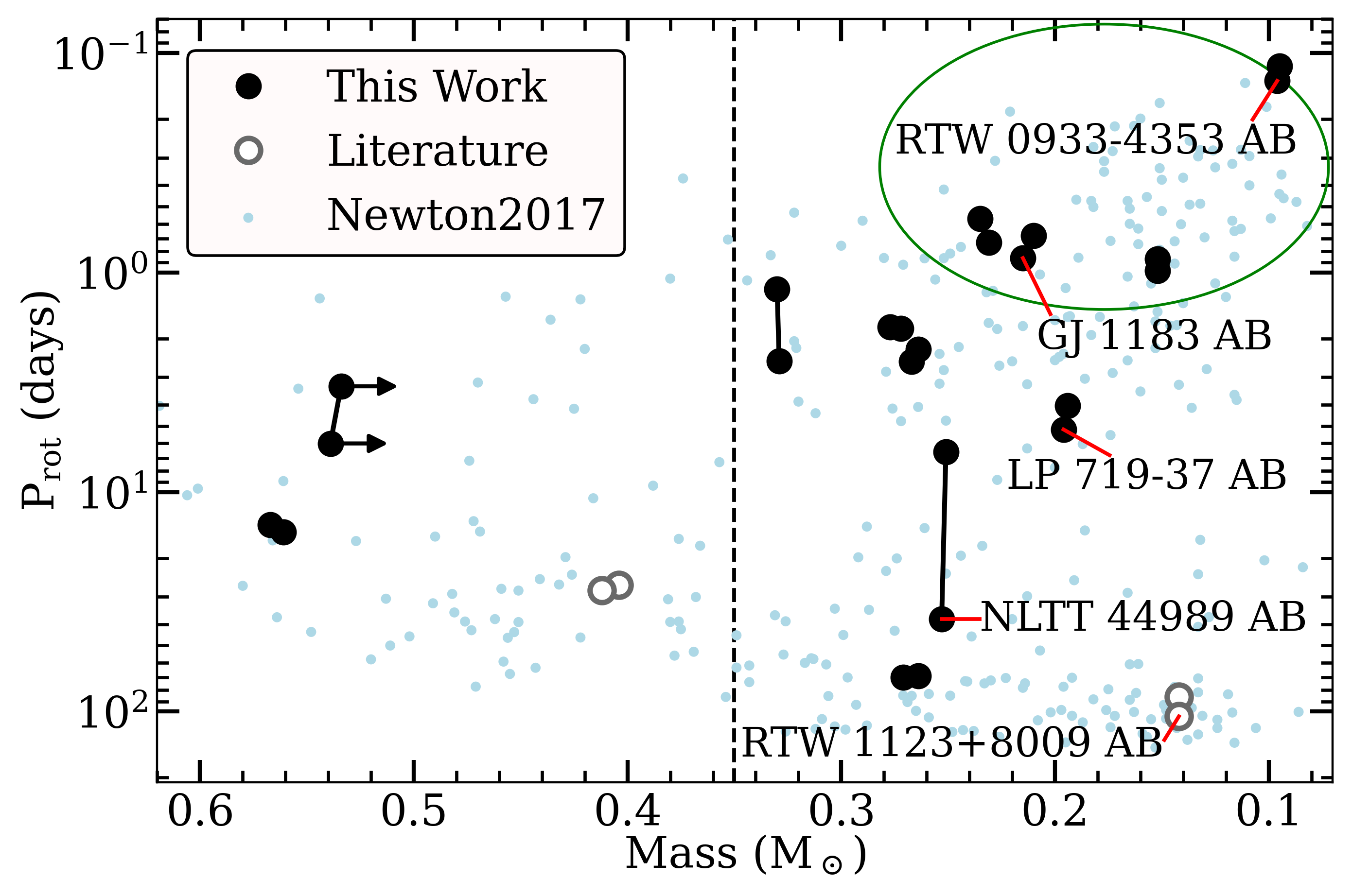}{0.484\textwidth}{(b)}}
\figcaption{The estimated masses of binary components in our Results Sample are plotted against their (a) H$\alpha$ EWs and (b) rotation periods. Values are from Table~\ref{tab:SampleTable-phot}, Table~\ref{tab:SpecTable}, Table~\ref{tab:RotTable}, and \citetalias{RTW_P1}, and we only plot systems with measurements available for both components. Twin pairs are connected by black lines, and field star results from \citetalias{Newton_2017} are underplotted in light blue. The y-axes are set such that a star with more emission or faster rotation moves upward on the plots. A vertical dashed line at 0.35\,$\rm{M_\odot}$ in each plot approximately divides PC and FC stars. Arrows indicate the less reliable mass upper limits for the pre-main-sequence case of 2MA~0201+0117~AB. Various systems discussed in the text are labeled. (a) --- The y-axis is scaled linearly between $-$0.5\,\text{\AA} to $+$0.5\,\text{\AA} and logarithmically outside these bounds. Squares represent CHIRON EWs, only including non-flaring epochs with both stars successfully observed back-to-back, and observed EW ranges in multi-epoch cases are not shown for visual clarity. The lack of active/inactive mismatched cases in the PC regime above 0.35\,$\rm{M_\odot}$ may simply be due to these stars having largely already evolved to inactivity. (b) --- For RTW~0409+4623~AB, we assign A the 2.55\,day period and B the 2.24\,day period for plotting purposes. A green ellipse roughly indicates the approximate region of parameter space in rapidly-rotating low-mass stars where a bistable dynamo is hypothesized to possibly occur, which may be responsible for some of the behaviors we observe. \label{fig:mass-Ha-prot}}
\end{figure}

Finally, we compare H$\alpha$ activity and rotation period to mass in Figure~\ref{fig:mass-Ha-prot}, where several takeaways are evident. Most pairs match well in both H$\alpha$ strength and rotation across the entire range of M dwarf masses, and the general trends in these plots are broadly consistent with our present understanding of spindown across M dwarfs being primarily mass dependent to first order and taking longer in lower mass stars. However, the two standout systems with active/inactive pair mismatches, NLTT~44989~AB and RTW~1123+8009~AB, host FC stars with one component in the middle of the active-inactive transition region despite their twin nature. One system, LP~719-37~AB, is also near the start of the FC active-inactive transition region, but both stars are presently similarly active (see $\S$\ref{subsec:lp719}). We see no active/inactive mismatch cases in the PC regime in Figure~\ref{fig:mass-Ha-prot}, but this may simply be because these stars have largely already spun down to inactivity --- the mismatched PC case of SLW~1022+1733~AB discussed in Section~\ref{subsubsec:gunning-ha-results} serves as a relevant example here despite not being in our twin sample.

%%%%%%%%%%%%%%% fig - mass versus HaEW and mass versus prot %%%%%%%%%%%%%%%

Overall, these results are as expected if higher order factors relating to formation rotation rates and disks, and/or star-planet interactions, are occasionally relevant and result in a few notable large mismatches. These additional factors may be present for only some stars, such as those with specific planetary configurations that could drive suitable star-planet interactions \citep[e.g.,][]{2024A&A...690A.379K, 2025OJAp....8E..59K}, or may be present in all stars, such as stochastic elements intrinsic to their formation and/or dynamo evolution. Finding only a few large mismatch cases does not preclude a factor from existing in all FC stars, as there may only be a short time window for strong mismatches to exist and be observed near the rapid active-inactive transition in FC M dwarfs. Disentangling these various hypotheses requires a much better understanding of how early disk factors, star-planet interaction mechanisms, and dynamos impact the host star activity and rotation across evolutionary timescales.

\section{Systems Worthy of Note} \label{sec:sys-notes}

Here we provide additional information relating to our noteworthy systems, including all of the pairs with mismatches in H$\alpha$ EWs and/or derived rotation periods. We discuss scientific insights offered by these cases along the way. Entries are ordered by ascending RA, except the three systems from \citetalias{RTW_P1} introduced first.

\subsection{2MA 0201+0117 AB, GJ 1183 AB, and NLTT 44989 AB} \label{subsec:p1-stars}

These three systems were examined extensively in \citetalias{RTW_P1}, including via new X-ray luminosity measurements from Chandra, and the results are briefly summarized here. 2MA~0201+0117~AB is a young pre-main-sequence system with a factor of two difference in rotation rate and large mismatches in H$\alpha$ and X-ray activity, implicating early formation factors. GJ~1183~AB is an FC system with matching rotation periods but mismatches in all activity parameters considered, informing the activity scatter in otherwise identical similarly-rotating M dwarfs --- this system also resides in the proposed dynamo bistability region shown in Figure~\ref{fig:mass-Ha-prot}. NLTT~44989~AB is also an FC system and the pair presents a total mismatch in rotation and activity, with roughly a factor of six difference in rotation period, at least a factor of 39 difference in X-ray luminosity, and a significant inactive/active H$\alpha$ mismatch between the components.

\subsection{LP 719-37 AB} \label{subsec:lp719}

A and B are both FC and match in all of their activity and rotation properties examined here, even as active stars showing marked variations in their H$\alpha$ emission. Intriguingly, both components are near the start of the FC active-inactive transition region and near NLTT~44989~B in Figure~\ref{fig:prot-Ha} and Figure~\ref{fig:mass-Ha-prot}. This implies that the components are near the start of their rapid magnetic braking phases, but their H$\alpha$ activity levels and rotation rates suggest that they probably have not progressed far into the fast braking phase, yet. This demonstrates that not all systems in this parameter region appear mismatched in activity. We hypothesize that if we were to observe the LP~719-37~AB system over stellar evolutionary time scales, one of the components would likely progress into its fast braking phase slightly before the other, and quickly shift toward inactivity in Figure~\ref{fig:mass-Ha-prot} while its twin remains more active, akin to the currently observed large mismatches in NLTT~44989~AB and RTW~1123+8009~AB. Our measurements find LP~719-37~A (5.19\,days) to be a slightly slower rotator than B (4.04\,days), suggesting that A would likely be the component to transition first. This system highlights how our largest mismatch cases may have been observed at slightly offset points during their rapid transitions from active to inactive states. It is possible that all FC M dwarfs may go through an analogous `mismatch' phase or time period of less reliable predictability around this transition.

\subsection{G 103-63 A(C)-B} \label{subsec:G103}

The A component likely hosts an unresolved companion, as discussed in Section~\ref{subsec:multiplicity-checks}, making the system a likely hierarchical triple. Despite this, the rotation periods and H$\alpha$ EWs we adopt from \citetalias{2024AJ....167..159L} and \citetalias{Pass_2024_ApJ}, respectively, indicate similar periods and activity levels for the widely separated A and B components. The rotation period for the A component at 95.73\,days may belong to either A or a close unseen C component, but is much longer than the $\sim$7-day tidal circularization timescale of M dwarfs \citep{Vrijmoet_thesis_2023}, indicating the two are likely not tidally synchronized, consistent with no elevated activity in A. This system may be similar to our system KX~Com~A-BC discussed in \citetalias{RTW_P1}, where dynamical interactions in hierarchical triples might deviate the components' rotation spindown evolution compared to that of isolated stars \citep{2023MNRAS.526.6168F}. Alternately, it may be that the rotation period measurements we adopt from \citetalias{2024AJ....167..159L}, or the presence of an unresolved companion, are erroneous results.

\subsection{RTW 0824-3054 A-B(C)} \label{subsec:RTW0824}

This system is very likely a hierarchical triple based on the suspected unresolved companion to B discussed in Section~\ref{subsec:multiplicity-checks}, thus invalidating its twin nature. The single epoch of CHIRON spectral observations we acquired show more H$\alpha$ emission and a larger $v\sin(i)$ in B than A, consistent with the behavior that would be expected if B had a close-in interacting companion causing more rapid rotation and stronger activity compared to A.

\subsection{LP 368-99 AB} \label{subsec:LP368}

%%%%%%%%%%%%%%% fig - LP368-099AB an GJ0745AB Gap Plot %%%%%%%%%%%%%%%
\begin{figure}[!t]
\figurenum{18}
\centering
\includegraphics[width=0.49\textwidth]{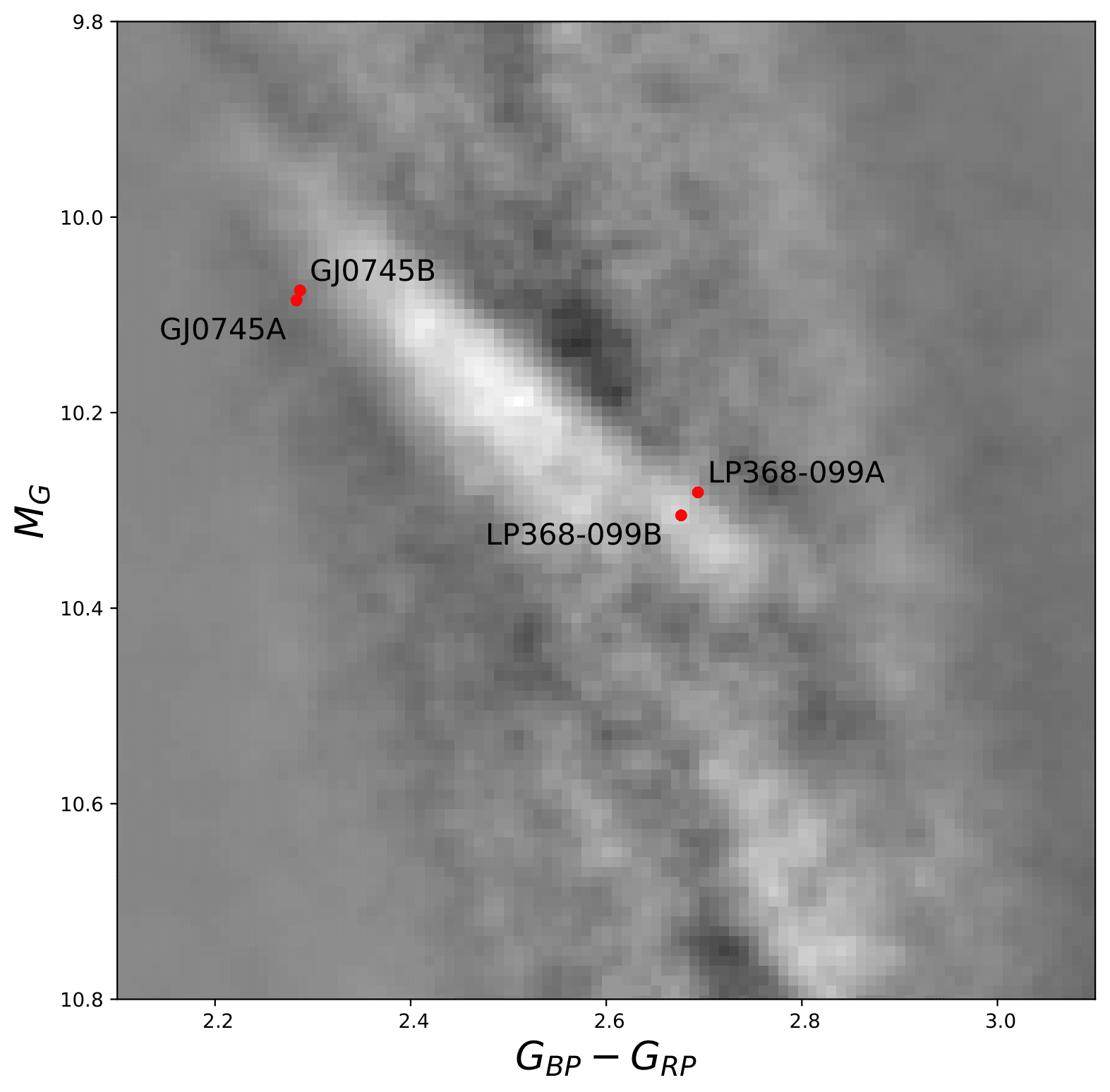}
\figcaption{A visually enhanced plot of the Gaia main sequence from \citet{2020AJ....160..102J} showing the M dwarf PC/FC transition gap, in which white indicates fewer stars. LP~368-99~A ($P_{\rm{rot}}=1.19$\,days) and B ($P_{\rm{rot}}=2.53$\,days) are plotted highlighting their relative positions on the upper edge of the gap and directly within the gap, respectively. Stars in the gap region undergo complex interior structure changes, which we hypothesize might be why the two components have rotation periods differing by roughly a factor of two (see Section~\ref{subsec:LP368} for further details). GJ~745~A and B are also shown to demonstrate their positions either within the gap or just below the gap, as discussed in Section~\ref{subsec:GJ745}. \label{fig:gap}}
\end{figure}

The A component has a rotation period of 1.19\,days while B is rotating every 2.53\,days, in general alignment with their CHIRON mean $v\sin(i)$ measures at $13.94 \pm 0.29$\,$\mathrm{km\,s^{-1}}$ for A and $5.34 \pm 0.42$\,$\mathrm{km\,s^{-1}}$ for B. The rotation difference is perhaps small in absolute terms, but stands out as a clear relative mismatch in panel (a) of Figure~\ref{fig:rot-equal}. The components match in H$\alpha$ activity despite the mismatched rotation rates, the only such case in our results. Both components are also sufficiently rapidly rotating to be in the saturated activity regime (see Figure~\ref{fig:prot-Ha}), presumably leading to the similar H$\alpha$ emission strengths despite their factor of two difference in rotation rate.

Interestingly, the two components are near the Jao Gap PC/FC transition region, with A landing just on the upper edge of the gap while B lands firmly in the middle of the gap, as shown in Figure~\ref{fig:gap}. Stars near the gap undergo complex internal structural changes, with inner and outer convective zones on either side of a middle radiative layer that periodically comes and goes over millions to billions of years \citep{2012ApJ...751...98V, 2018MNRAS.480.1711M, 2018A&A...619A.177B}. This ``convective kissing instability" occurs due to non-equilibrium nuclear burning of $^3$He and results in periodic luminosity and radius changes. Such (very slow) pulsations could hypothetically change the overall angular momentum evolution of a star because the various convective and radiative layers repeatedly couple and decouple \citep{Jao_2023}. Their current positions on the HRD in Figure~\ref{fig:gap} suggest that the A component is at or near the peak of its $^3$He instability pulsation, elevating it just above the gap, while B is not and thus appears in the gap. LP~368-99~A and B therefore likely have periodically mismatched interior structures, which we hypothesize may cause different angular momentum evolution between the stars and the subsequently different rotation periods we observe today.

These gap pulsations are not predicted to start until ages of roughly 0.5--2\,Gyrs depending on the exact stellar mass \citep{2018A&A...619A.177B}, so our hypothesis fails if the stars are too young. The pair's presence in the rapidly-rotating saturated regime in Figure~\ref{fig:prot-Ha} implies they are likely no older than 1--3\,Gyrs \citep{Pass2022, Medina_2022_spindown}, but we otherwise do not see clear indications of youth in either star. Thus, given the lack of evidence for youth, we favor the interior structure hypothesis to explain their rotation mismatch, but of course cannot rule out formation aspects and/or disk effects.

In addition, \citet{2024ApJ...977...15C} argued that a significant increase in the convective turnover time at the PC/FC transition boundary leads to rapid magnetic braking and consequently longer rotation periods compared to stars marginally cooler than the transition region. They map stars in the gap to a range of acceptable gyrochrones spanning ages of roughly 2--8\,Gyr, corresponding to rotation periods of about 50--270\,days for stars in this region. If their results and interpretation are correct, the rapid $<$\,3\,day rotation rates we recover for LP~368-99~A and B indicate they must be relatively young stars, possibly having formed with very different initial rotation rates that might explain their factor of two difference today. However, it may instead be the case that these two rapidly-rotating stars are a counterpoint to the results of \citet{2024ApJ...977...15C}, as \citet{Jao_2023} observationally demonstrated the region of stronger spindown is likely located slightly \textit{below} the transition region gap, instead of \textit{within} the gap where LP~368-99~A and B reside with fast rotation. More precise age information for LP~368-99~A and B would be needed to better disentangled these explanations.

Overall, this system is an excellent candidate for future investigations of astrophysics in the gap region given the components control for mass, age, composition, environment, and current activity level. 

\subsection{RTW 0933-4353 AB} \label{subsec:RTW0933}

This is the lowest mass (0.095\,$\rm{M}_\odot$) system in the sample. This low mass and the pair's extremely rapidly rotation periods of $\sim$3\,hours place them in the parameter regime where M dwarfs may host bistable dynamos (see Figure~\ref{fig:mass-Ha-prot}) \citep{Gastine_2013, 2024A&A...687A..95M}. This makes them similar to the BL+UV Ceti system mentioned in Section~\ref{sec:intro} and the GJ~1183~AB system, although GJ~1183~A and B are twice as massive at 0.21\,$\rm{M}_\odot$. RTW~0933-4353~AB shows matching rotation alongside matching strong H$\alpha$ emission, while BL+UV Ceti and GJ~1183~AB each host matching rotation periods but different activity levels. This suggests RTW~0933-4353~A and B could be in the same dynamo state while the components in GJ~1183~AB and BL+UV Ceti could be in different dynamo states, with correspondingly different magnetic field morphologies. Oscillatory cycling dynamos may also be the culprit instead of a dynamo bistability \citep{Kitchatinov_2014}, and we indeed found candidate activity cycles for GJ~1183~A and B in \citetalias{RTW_P1}, but we do not see cycles appearing in our long-term data for RTW~0933-4353~AB so far (see Figure~\ref{fig:long-term-curves}). For RTW~0933-4353~A and B, the long-term 0.9\,m program data acquired several years before the rotation observations smoothly phase-fold into the $\sim$3\,hour rotation periods, implying the stars' spot/faculae regions have been stable over many years. This disfavors possible activity cycles at these timescales in the two stars, and is akin to the long-term spot stability found in other M stars by \citet{2016A&A...590A..11V} and \citet{2020ApJ...897..125R}. Hypothetically, perhaps the bistable dynamo is indeed involved but one of the dynamo states also manifests activity cycles while the other state does not. The confirmation of cycles appearing in some of these stars but not others would be crucial evidence for disentangling the behaviors of these low-mass rapidly-rotating objects.

\subsection{RTW 1123+8009 AB} \label{subsec:RTW1123}

This system shows significantly mismatched H$\alpha$ activity, with EWs of $-1.691$\,\text{\AA} for A and $-0.192$\,\text{\AA} for B from the \citetalias{Pass_2024_ApJ} measurements. Their three visits over four months on the A component consistently show emission, indicating random flares are not the culprit, and they did not see RV variations in those spectra. We do not have our own time series RVs or speckle observations of the system, but the literature and Gaia parameter checks discussed in $\S$\ref{subsec:multiplicity-checks} do not indicate unresolved companions. Despite the pair's activity mismatch, the stars host similar rotation period measurements of 105\,days in A and 86\,days in B, respectively, from \citetalias{2024AJ....167..159L}. This pair is especially intriguing considering that the more slowly rotating A star is much more active than B, challenging our expectations of rotation-activity behavior and spindown for M dwarfs. However, the rotation periods may be incorrect, a notable possibility considering the nature of their rotation measurements as discussed in Section~\ref{subsubsec:rot-notes_RTW-1123+8009}.

The many possibilities we discussed in \citetalias{RTW_P1} and Section~\ref{sec:discussion1} for explaining strong activity mismatches often instead explain rotation rate mismatches that would presumably result in the activity mismatches. However, if the rotation periods for this system are correct, the fact that the more slowly rotating star is more active indicates that the rapid transition from active to inactive in FC M dwarfs may indeed have a stochastic element, as discussed in \citetalias{Pass_2024_ApJ} and \citetalias{RTW_P1}. The magnetic morphology could be a related factor as well \citep{Garraffo2018}, where different H$\alpha$ activity levels could hypothetically manifest from different magnetic morphologies despite similar rotation rates. This system clearly merits considerable follow up work.

\subsection{GJ 745 AB} \label{subsec:GJ745}

This is our only system that resides within 10\,pc, and it has consequently been thoroughly studied over the years. For example, both components have empirically-derived quantities; a recent set from \citet{2015ApJ...804...64M} includes $\rm{T_A} = 3500 \pm 60$\,K and $\rm{T_B} = 3494 \pm 62$\,K, $\rm{R_A} = 0.31 \pm 0.01 \, R_\odot$, and $\rm{R_B} = 0.32 \pm 0.01 \, R_\odot$, $\rm{[Fe/H]_A} = -0.33 \pm 0.08$ and $\rm{[Fe/H]_B} = -0.35 \pm 0.08$ in log solar units. The congruence of these values further confirms the pair's twin status. The stars are very weak in X-rays based on recent work by \citet{2023A&A...676A..14C}, commensurate with the pair's functionally identical H$\alpha$ absorption we see in our results, and the average magnetic fields for both components are limited to $<$\,220\,G \citep{2022A&A...662A..41R}. This serves as an example of how some of our pairs likely show matching H$\alpha$ behaviors simply because they have quite low magnetic activity levels in the first place. The pair resides in the upper left region of the PC/FC transition gap, as shown in Figure~\ref{fig:gap}, so the stars may currently have (or previously had) the same complex interior pulsations occurring as described earlier for LP~368-99~AB (Section~\ref{subsec:LP368}). Alternately, they may not in fact be in the gap.

\subsection{RTW 2241-1625 AB} \label{subsec:rtw2241}

This is the only PC system in our Results Sample that has a mismatch in H$\alpha$ activity, in which the A star exhibits consistently less absorption than B --- B's mean absorption EW is $2.7 \pm 0.5$ times larger or 0.2\,\text{\AA} larger than A's, as visible in Figure~\ref{fig:Ha-stacked_spec}. The A component is stronger than B in photometric rotation amplitude and long-term variability as well. This is all despite their matching rotation periods near 14\,days for A and 15\,days for B, making the system somewhat of a PC analog to the FC systems GJ~1183~AB and RTW~2311-5845~AB. The fast magnetic braking phase, bistable dynamo hypothesis, and our related explanations employing them are only applicable in the FC regime, however, making this PC mismatch system particularly interesting. A relevant factor unique to PC systems but not FC systems is the intermediate period gap discussed in Section~\ref{sec:intro}, but the periods for this system are shorter than the 20--30\,day period region of this feature. However, some of the other explanations could still apply, such as magnetic star-planet interactions enhancing stellar activity without substantively changing stellar rotation, a particularly strong H$\alpha$ activity cycle contrast between the components, or lurking unseen massive companions, although the latter appears unlikely.

\subsection{RTW 2311-5845 AB} \label{subsec:rtw2311}

The stars in this pair have masses of 0.23\,$\rm{M}_\odot$ and matching periods $<$\,1\,day, with an activity mismatch where B is more active in H$\alpha$, photometric rotation amplitude, and long-term photometric variability --- this makes the pair akin to the GJ~1183~AB and BL+UV Ceti systems. These FC systems with matching rotation period but mismatched activity offer constraints on the activity scatter possible in otherwise similarly-rotating twin stars. For the similar case of GJ~1183~AB in \citetalias{RTW_P1} we reported that A was $26\% \pm 9\%$ stronger than B in H$\alpha$ emission on average, whereas for RTW~2311-5845~AB here we find B's H$\alpha$ emission is $60\% \pm 42\%$ stronger than A's when using the mean and standard deviation of their EWs. This system therefore finds an even greater level of activity scatter is possible in otherwise identical FC M dwarfs, as is also the case for RTW~1123+8009~AB (Section~\ref{subsec:RTW1123}).

In the context of the aforementioned bistable dynamo, the matching rotation rate but mismatched activity suggests that RTW~2311-5845~AB likely hosts a similar configuration of mismatched dynamo states and corresponding field morphologies as GJ~1183~AB, with activity cycles possibly involved as well. This is in contrast to RTW~0933-4353~AB (Section~\ref{subsec:RTW0933}), in which rotation and activity were both matching and did not show evidence for cycles. However, we do not see indications of long-term cycles in the 4\,years of available 0.9\,m data for RTW~2311-5845~A and B shown in Figure~\ref{fig:long-term-curves}. Cycles may therefore not be present in certain dynamo states, the cycles may be much longer than 4\,years, and/or they may be too weak in amplitude to detect here.

\section{Future Prospects} \label{sec:future-work}

To highlight both the need and potential for future study of these twin systems, we intentionally include several additional columns in Table~\ref{tab:summary} relating to other activity behaviors of interest despite our lack of such measurements. For example, we only included X-ray assessments for our systems with Chandra observations reported in \citetalias{RTW_P1} --- more such measurements are clearly needed to investigate what factors drive the large scatter seen in M dwarf X-ray activity behaviors. Similarly, UV activity studies are of crucial value for exoplanet habitability studies because of UV photon effects on planetary atmospheres. Our twin systems can map the array of UV behaviors for otherwise identical exoplanet hosts, a key avenue of work in preparation for the Habitable Worlds Observatory. We note that some of our components have UV measurements available from GALEX \citep{2017ApJS..230...24B} and/or SWIFT \citep{2005SSRv..120...95R}, along with some having X-ray measurements from ROSAT \citep{2016A&A...588A.103B}, SWIFT \citep{2014ApJS..210....8E}, and/or XMM-Newton \citep{2008A&A...480..611S}, but we leave a deeper investigation of the X-ray and UV activity of the full set of twins to future work. 

The metallicities of our twin stars also merit future study, as this would allow us to validate our assumption of matching compositions, identify any discrepant cases that might indicate non-coevality, and look for any trends with metallicity in our cumulative results. This is particularly relevant as metallicity can change the activity level and spindown evolution for a star \citep[e.g.,][]{2019ApJ...872..128V, 2020ApJ...889..108A, 2020MNRAS.499.3481A, 2021ApJ...912..127S}, so it might be a factor in explaining some of our mismatched pairs if they have suitably different compositions. The vast majority of our twin stars do not have reliable photospheric metallicity measurements available in the literature to our knowledge, but we note the four systems we reported coronal abundances for in \citetalias{RTW_P1}.

Altogether, each twin pair is a laboratory controlling for the underlying stellar parameters, while the sample as a whole can offer insights at a broader population level, making our twin systems valuable targets for future investigations.

\section{Summary and Conclusions} \label{sec:conclusions}

We have investigated the rotation and stellar activity properties of 36 M dwarf twin wide binary systems in our full sample, where components in each twin pair have the same mass, age, composition, and environment. 32 of the systems are newly reported here, building upon the 4 systems we previously reported in \citetalias{RTW_P1}. We have vetted for unseen companions using a combination of Gaia parameter assessments, speckle imaging, RV time series, long-term ground-based astrometry, and literature checks, cumulatively revealing three confirmed or suspected non-twin hierarchical triples. We present the remaining 33 systems as robust twin binaries that constitute our Results Sample. We report new radial velocities, $v\sin(i)$ measurements, H$\alpha$ EWs, multi-year photometric variability levels, rotation periods, and rotation amplitudes for both components in observed subsets of the systems. Our key takeaways are:

\begin{itemize}

    \item Roughly 80\% of our observed twin binaries have extremely well-matched rotation rates and activity levels, including pairs across the full range of M dwarfs and across a wide range of rotation-activity parameter space. This broadly supports the general predictability of such behaviors based on fundamental stellar parameters, i.e., mass, age, and composition.
    
    \item In contrast, roughly 20\% of the observed twin pairs {\it do not} match, with rotation periods and/or activity levels that differ by more than 25\% despite the components being effectively identical in fundamental characteristics. These mismatches appear more often in fully convective systems than in higher mass partially convective pairs, but both structural types do show mismatched cases. Of the systems matching in rotation rate, half also match in H$\alpha$ activity while the other half surprisingly do not.
    
    \item Thus, if selecting a random field M dwarf of any structural type, our available data suggest that roughly one out of every five cases will have rotation and/or activity properties deviating from expectations by at least 25\% --- sometimes by much more --- when compared to an otherwise identical comparison case. This includes activity measured via photometric variability, chromospheric activity, and coronal X-ray activity, and is despite the stars' functionally identical masses, ages, compositions, and environments. As a benchmark, consider only H$\alpha$ activity --- where our observational selection effects and detection biases are of least concern among the properties investigated --- where we find that $21\%_{-6\%}^{+9\%}$ of twin systems demonstrate mismatched H$\alpha$ activity levels.

    \item For the data we utilize, and including M dwarfs of any type, we find the frequency of mismatches for different characteristics descends as $L_X > \rm{amp_{rot}} > \rm{MAD_{long}} > EW_{H\alpha} = P_{rot}$. The X-ray results rely on only three systems, however, so are less confident in this order. Photometric rotation amplitudes and multi-year photometric variability appear mismatched beyond 25\% more often than H$\alpha$ activity and rotation periods, presumably due to spot and faculae net brightness inhomogeneities being more variable with time and less consistent tracers of underlying activity levels.

    \item Even for otherwise identical and similarly-rotating M dwarfs, we find that fully convective twin stars can differ by at least $60\% \pm 42\%$ in H$\alpha$ emission EW, with total active/inactive mismatches in some cases. For the one available partially convective pair with matching rotation rates and available activity measurements (RTW~2241-1625~AB), B's absorption EW is $2.7 \pm 0.5$ times larger than A's.
    
    \item We compile three systems with strikingly large active/inactive mismatches between their twin components: (1) NLTT~44989~AB, based on our work from \citetalias{RTW_P1}, and residing in our twin sample as an FC pair. (2) RTW~1123+8009~AB, based on measurements from \citetalias{Pass_2024_ApJ} and \citetalias{2024AJ....167..159L}, and residing in our twin sample as an FC pair. (3) SLW~1022+1733~AB, based on measurements from \citet{Gunning_2014}, and not residing in our twin sample but meeting our equal-magnitude selection criteria and hosting PC stars. NLTT~44989~A and B have very different rotation periods that align with their activity mismatch, RTW~1123+8009~A and B surprisingly have similar rotation periods that do not align with their activity mismatch (although with potentially inaccurate literature rotation rate measurements), and SLW~1022+1733~AB does not yet have rotation rate measurements available.

    \item One twin system residing in the transition region between partially and fully convective M dwarfs, LP~368-99~AB, has component rotation rates differing by a factor of two. We hypothesize that the periodic interior structure changes predicted to occur in this region may alter the stars' angular momentum evolution and explain the mismatch.

    \item Alongside the variety of behaviors we uncover, it is notable that our twin systems broadly align with typical trends seen in the rotation-activity plane across the saturated and unsaturated regimes. The only standout exception case, RTW~1123+8009~AB, may stem from inaccurate rotation period measurements for the pair in the literature.
    
    \item Our results altogether provide evidence in favor of M dwarf rotational spindown being primarily determined by fundamental parameters (i.e., a combination of mass, age, and composition), alongside higher order factors related to some combination of initial rotation rates, early disk properties, star-planet interactions, and dynamo stochasticity. A combination of stellar activity cycles, the fast magnetic braking phase in fully convective M dwarfs, complex magnetic morphology evolution, and the hypothesized bistable dynamo in rapidly rotating low-mass M dwarfs are all also likely relevant for explaining the various types of cases we find with mismatched behaviors.
    
\end{itemize}

Overall, our results demonstrate the ongoing challenges in predicting M dwarf activity and rotation evolution, and throughout this paper we highlight many areas for potential future work. The presence of our large total mismatch cases indicates that such strong deviations in activity and rotation are \textit{possible}, though not necessarily \textit{probable}, requiring they be kept in mind and their root causes carefully disentangled. Meanwhile, this creates a lingering significant uncertainty in accurately reconstructing the activity histories of any planet-hosting M dwarfs studied today, particularly for fully convective M dwarfs which account for roughly half of all stars.

\vskip20pt

\textit{Data Access:} The CHIRON spectra utilized here can be obtained from the NOIRLab Data Archive. The light curve data shown in this work from the CTIO/SMARTS 0.9\,m are available as Data behind the Figure (DbF) products, including for both the long-term program and higher-cadence rotation observations. 

%% Please use the acknowledgment and contribution environments. This will 
%% be anonomyized when the "anonymous" style option is used. 
\begin{acknowledgments}

A.~A.~Couperus thanks the following individuals for conversations that enhanced this work: Benjamin P.~Brown, Gregory A.~Feiden, Emily K.~Pass, and Russel J.~White. We also thank Elliott P.~Horch and Andrei Tokovinin for their assistance in collecting and reducing the speckle data. This work has been supported by the NSF through grants AST-141206, AST-1715551, and AST-2108373, as well as via NASA/Chandra grant GO1-22013B. We have used data from the SMARTS 0.9\,m and 1.5\,m telescopes, which are operated as part of the SMARTS Consortium by RECONS (www.recons.org) members, and with the assistance of staff at Cerro Tololo Inter-American Observatory. This work has made use of data from the European Space Agency (ESA) mission Gaia, processed by the Gaia Data Processing and Analysis Consortium (DPAC). Funding for the DPAC has been provided by national institutions, in particular the institutions participating in the Gaia Multilateral Agreement. This publication makes use of data products from the Two Micron All Sky Survey, which is a joint project of the University of Massachusetts and the Infrared Processing and Analysis Center/California Institute of Technology, funded by NASA and the NSF. This paper includes data collected by the TESS mission, which are publicly available from MAST at \cite{tess_FFI_doi_ref}. This research has made use of NASA’s Astrophysics Data System (ADS), as well as the SIMBAD database \citep{SIMBAD} and VizieR catalog access tool \citep{VizieR} operated at CDS, Strasbourg, France.

\end{acknowledgments}

\begin{contribution}

Andrew~A.~Couperus was responsible for leading the project and its observing campaigns, carrying out nearly all data reduction and analysis (except the speckle observations), and writing and submitting the manuscript.
Todd J. Henry helped develop and guide the project throughout its entirety, maintained funding and observatory resources, aided analysis decisions and interpreting results, and helped line edit and streamline the manuscript.
Aman Kar developed new code adapted for use in various parts of the TESS and TESS-unpopular analysis here, aided the contamination methodology used for the rotation assessment data, provided detailed aid for several rotation period assessments, and helped edit the manuscript.
Wei-Chun Jao previously developed analysis pipelines that were adapted for use in the 0.9\,m differential photometry reductions and CHIRON spectral analyses here; he also helped guide the project and interpret the results, provided details and supporting figures concerning the Jao gap, and helped edit the manuscript.
Eliot Halley Vrijmoet facilitated the SOAR speckle observations through a related program, helped interpret the speckle results, and derived the MLR mass estimates.
Rachel A. Osten supported the overall project, particularly concerning the X-ray results first reported in \citetalias{RTW_P1} and incorporated here, and helped edit the manuscript.
Finally, all authors personally carried out and/or helped facilitate new observations included in this work across the various observing campaigns.

%%This section gives authors the space to recognize author contributions. The text inside this environment is NOT counted towards the total word quanta. At a minimum, manuscripts are expected to include this text:
%All authors contributed equally to the Terra Mater collaboration.

%% But authors are expected to provide more specific details, e.g. 
%%
%%SC was responsible for writing and submitting the manuscript.
%%WWM came up with the initial research concept and edited the manuscript.
%%OTS obtained the funding and edited the manuscript.
%%EBF provided the formal analysis and validation. He also edited the manuscript.
%%GEH Supervised the undergraduates, wrote the software and administers the project github and Zenodo repositories.
%%
%% Authors can use the Contributor Role Taxonomy (CRediT) at
%% for ideas on how write a good statement tailored to their needs.

\end{contribution}

\facilities{CTIO:0.9m, CTIO:1.5m(CHIRON), CTIO:2MASS, FLWO:2MASS, Gaia, TESS, PO:1.2m, SOAR, LDT}

\software{IRAF \citep{10.1117/12.968154,1993ASPC...52..173T}, Source Extractor \citep{1996A&AS..117..393B}, unpopular \citep{unpopular}, BANYAN $\Sigma$ \citep{banyan_2018}, tpfplotter \citep{tpfplotter}, Astropy \citep{2013A&A...558A..33A,Astropy2018, 2022ApJ...935..167A}, Matplotlib \citep{Matplotlib2007}, NumPy \citep{NumPy2020}, and Aladin \citep{2000A&AS..143...33B,2014ASPC..485..277B}.}


\appendix

\section{Additional Rotation Assessment Notes} \label{sec:appendix}

In this appendix we discuss the rotation information for systems that do not have adopted periods (except in a few tentative cases with indications of likely periods), following from Section~\ref{subsec:rot-notes} and in the context of Table~\ref{tab:RotTable}. Throughout, we estimate ranges of likely rotation periods for stars based on their H$\alpha$ EW measurements and estimated masses compared to results in \citetalias{Newton_2017}.

\subsection{RTW 0143-0151 AB} \label{subsec:rot-notes_RTW-0143-0151}

\textit{RTW~0143-0151} A and B show no reliable rotation signals in any of the available datasets. TESS-unpop shows weak variations around a $\sim$5\,day period, but only in one of two sectors, and we consider these to probably be systematics. This is especially likely given the components' flat H$\alpha$ activity and 0.21\,$\rm{M_\odot}$ masses suggest periods much longer than 5\,days.

\subsection{RTW 0231+4003 AB} \label{subsec:rot-notes_RTW-0231+4003}

\textit{RTW~0231+4003} A and B are at least partially blended in all available data, in addition to having moderate extra contamination from two nearby much brighter sources bleeding into the TESS and TESS-unpop apertures. TESS shows questionable signals around 3.5--5\,days and 12--15\,days, but they are inconsistent across the two sectors and show indications of possibly being systematics; they may also be from the contaminants. In contrast, TESS-unpop shows only long-term variations, but with no reliable results and again suspicion of obfuscating systematics. ZTF consistently shows candidate signals around a month in duration across all three filters and for both stars, but with periods very similar to lunar variations, suggesting likely systematics. No H$\alpha$ info is available.

\subsection{KAR 0545+7254 AB} \label{subsec:rot-notes_KAR-0545+7254}

\textit{KAR~0545+7254} A and B were observed by \cite{CARMENES_2015} to have consistent H$\alpha$ absorption, suggesting likely rotation periods of 20--80\,days for their masses of $\sim$0.40\,$\rm{M_\odot}$. The components are almost entirely resolved in TESS and TESS-unpop, with only slight blending on their wings, but A has major contamination from another source. TESS shows no meaningful signals for either star across seven sectors. Merging the seven sectors with TESS-unpop data finds candidate signals at $\sim$42\,days and $\sim$37\,days for A and B respectively, which are congruent with the H$\alpha$ inactivity, but obfuscating systematics are a significant concern in the merged sectors. Given the major contamination in A, we would not adopt its period from TESS-unpop alone, regardless of the systematics concern. The available data therefore do not yield period measurements for both stars that meet our criteria.

\subsection{RTW 0824-3054 A-B(C)} \label{subsec:rot-notes_RTW-0824-3054}

\textit{RTW~0824-3054} A and B show a complex interplay of multiple periodic signals in the blended TESS data across four sectors. We see a strong signal for a period at 1.14\,days, a likely second repeating signal at 1.5--2\,days, and two weaker uncertain signals possibly near roughly 0.9--1\,days and $\sim$2.5\,days. TESS-unpop shows these same intermixed rapid signals, along with evidence for a candidate longer periodic signal around 35--50\,days when considering a merging of all the sectors. However, there are two contaminating background sources in TESS much brighter than the target stars, with both landing close to or directly in the various apertures. Furthermore, the B component has a suspected unresolved companion as well (Section~\ref{subsec:multiplicity-checks}), likely adding a fifth star into the mix. The light curves also show a recurring sharp dip feature at several tens of mmag every $\sim$16.5\,days, with TESS reporting planet candidate Data Validation (DV) products from detected Threshold Crossing Events (TCEs) due to these dips\footnote{Several of our other targets, including LP~368-99~A \& B and RTW~2202+5537~B, have reported planet candidate DV products from TCEs as well, but our manual inspection of these results finds they are all spurious false detections due to stellar variability from flares and/or rotational modulation.}. While intended for planetary transits, our review of the DV products and the strengths of the features might favor a long-period partial stellar eclipse instead, and \citet{2022ApJS..258...16P} report an eclipsing binary signal with period 16.47\,days for the A source from TESS data. However, centroid offset results in the TESS DV products suggest a background star is likely responsible instead of our M dwarf targets, so we henceforth disregard this signal. Both twin components are inactive and nearly flat in H$\alpha$ activity, which, combined with the 0.25\,$\rm{M_\odot}$ mass estimate of A, suggests a rotation period for A of roughly $\gtrsim$\,40\,days. The B component, assuming it is composed of two lower mass stars, would likely host even longer rotation periods to explain the inactivity. The H$\alpha$ inactivity for both components therefore suggests the various rapid 0.9--2.5\,day signals we see in TESS might all belong to the contaminating sources and TESS systematics, while the candidate long-term period in TESS-unpop might belong to one of our components.

\subsection{L 533-2 AB} \label{subsec:rot-notes_L-533-2}

\textit{L~533-2} A and B analyses used identical apertures in a single sector of blended TESS-unpop data, with no SAP pipeline results available to inform polynomial inclusion or exclusion in unpopular for long-term trend validation. Without the polynomial during unpopular processing flat light curves are found, but with a polynomial there is a candidate signal of $\sim$30\,days, suspiciously close to the 27\,day sector baseline. Resolved Gaia light curves, in all three Gaia filters, demonstrate enhanced periodogram power at a range of possible long-term periods spanning roughly 10--80\,days for both stars. Both components show clear H$\alpha$ absorption, implying likely rotation periods of 20--90\,days at their 0.39\,$\rm{M_\odot}$ masses. The long period signals in Gaia and TESS-unpop are congruent with the stellar activity levels for each star, but we are ultimately unable to determine reliable rotation periods.

\subsection{LP 551-62 AB} \label{subsec:rot-notes_LP-551-62}

\textit{LP~551-62} A and B are inactive H$\alpha$ absorbers, suggesting likely periods of roughly 20--90\,days for their masses of 0.38\,$\rm{M_\odot}$. Both components have moderate contamination from a nearby much brighter source bleeding into the TESS and TESS-unpop apertures. TESS shows no signals except very weak likely systematics. TESS-unpop hosts long-term variations in three sectors, with two sectors being consecutive --- when merged these suggest a long period commensurate with the H$\alpha$ inactivity, but do not constrain a precise period measurement.

\subsection{RTW 1133-3447 AB} \label{subsec:rot-notes_RTW-1133-3447}

\textit{RTW~1133-3447} A and B are resolved in TESS, TESS-unpop, and ASAS-SN, but A is heavily contaminated by a close, significantly brighter source, with B also slightly contaminated by the wings of the brighter source in TESS and TESS-unpop. Both components are inactive with H$\alpha$ absorption and have masses of about 0.4\,$\rm{M_\odot}$, suggesting likely rotation periods of roughly 20--80\,days. A has complex variability in TESS from QLP data in sector 63, showing a clear 0.90\,day periodic signal with changing amplitude, a possible second signal at 4.65\,days, and possibly more rapid periodic components in beat patterns --- all are inconsistent with its H$\alpha$ absorption and therefore likely not from A. B does not show variability in TESS beyond evident systematics. In TESS-unpop, A has no SAP flux info available to inform polynomial inclusion or exclusion, but B does show possible long-term signals in its SAP fluxes; we expect long periods for both based on H$\alpha$, so we use polynomial-included results. A in TESS-unpop shows the 0.90\,day signal much weaker than before when using our manually chosen unpopular aperture, indicating that this signal may belong to the brighter contaminating source. No other reliable signals appear for A beyond likely systematics. B in TESS-unpop similarly only shows weak and inconsistent long-term variations at low confidence, which we ascribe to likely systematics.

\subsection{RTW 1336-3212 AB} \label{subsec:rot-notes_RTW-1336-3212}

\textit{RTW~1336-3212} A and B both show inactive flat H$\alpha$ activity, which at their 0.19\,$\rm{M_\odot}$ masses suggests likely slow rotation periods of 60--140\,days. TESS for B shows a weak candidate $\sim$4.5\,day signal, but this only appears in one of three AB-blended sectors, at low confidence, and is inconsistent with the H$\alpha$ inactivity. Blended TESS-unpop sector-merged results suggest long-term variations $\gtrsim$\,30\,days, but with poor period measurements and the possibility of sector-length systematics.

\subsection{L 197-165 AB} \label{subsec:rot-notes_L-197-165}

\textit{L~197-165} A and B are spatially resolved from each other in all available datasets. Both components are inactive with H$\alpha$ absorption, suggesting periods of 20--120\,days at their masses of 0.33\,$\rm{M_\odot}$. TESS-unpop sector-merged results reveal a candidate 30\,day period in A and show weak long-term variations in B at $\gtrsim$\,30\,days, along with $g$-band ASAS-SN data showing weak peaks around $\sim$30\,days for both stars, but both sets of results may also be systematics from sector baselines and lunar variations, respectively. The candidate signals are congruent with the H$\alpha$ activity levels, but we ultimately deem them too unreliable and do not adopt a period for either star.

\subsection{RTW 1433-6109 AB} \label{subsec:rot-notes_RTW-1433-6109}

\textit{RTW~1433-6109} A and B both host prominent H$\alpha$ absorption, indicating possible periods of approximately 20--80\,days at their masses of $\sim$0.43\,$\rm{M_\odot}$. Blended TESS data show an intermixing of rapid signals at about 1.2--1.4\,days and 2.8\,days, which very likely belong to contaminating sources or TESS systematics given the target M dwarfs' H$\alpha$ inactivity. Blended TESS-unpop sector-merged data show a candidate $\sim$40\,day period signal combined with the faster signals mentioned above, but the long-period signal is uncertain given the blend of multiple sources and sector-merging systematics.

\subsection{L 1197-68 AB} \label{subsec:rot-notes_L-1197-68}

\textit{L~1197-68} A and B are both inactive H$\alpha$ absorbers, yielding likely periods of roughly 20--90\,days for their masses of $\sim$0.38\,$\rm{M_\odot}$. We see no rotation signals in any of the available datasets and subsequently do not adopt periods for either star.

\subsection{L 1198-23 AB} \label{subsec:rot-notes_L-1198-23}

\textit{L~1198-23} A and B are both consistently inactive with H$\alpha$ in absorption, suggesting periods of roughly 20--80\,days at their masses of $\sim$0.40\,$\rm{M_\odot}$. Both ASAS-SN and ZTF host resolved data that show very weak signals around $\sim$50\,days. These are congruent with the single sector of mildly-blended TESS-unpop data that shows a candidate $\gtrsim$\,30\,day signal in B, suspiciously close to the 27\,day sector baseline. \cite{Newton_2016} report observations of the B component as part of the MEarth project, but found a 0.46\,day period categorized as an unreliable ``non-detection or undetermined detection" --- this rapid period is also inconsistent with the H$\alpha$ absorption we observe. We deem the various available results too weak in their detection strengths to use as reliable periods.

\subsection{RTW 1512-3941 AB} \label{subsec:rot-notes_RTW-1512-3941}

\textit{RTW~1512-3941} A and B show inactive H$\alpha$ absorption, yielding probable rotation periods of about 10--60\,days at their masses of 0.53\,$\rm{M_\odot}$. SAP fluxes are not available in most cases to inform polynomial inclusion or exclusion for TESS-unpop. With a polynomial we see long-term variations at $\sim$27\,days, making it very likely a result of sector-length systematics. Without a polynomial we see no reliable rotation signals. No other rotation measurements are available.

\subsection{RTW 1812-4656 AB} \label{subsec:rot-notes_RTW-1812-4656}

\textit{RTW~1812-4656} A and B are spatially resolved in the available datasets, with inactive H$\alpha$ absorption and masses that indicate probable rotation rates of 20--80\,days. A is moderately contaminated by a nearby, much brighter source in TESS and TESS-unpop, in addition to the $\Delta G <$ 2\,mag source within most apertures. There are two available sectors of QLP and TESS-SPOC results: A shows candidate uncertain signals at about 7\,days and 12\,days, possibly from contaminating sources given the inconsistency with the H$\alpha$ inactivity, while B shows questionable signals that are likely systematics or QLP processing artifacts. TESS-unpop sector-merged results indicate possible periodic variations of about a month in each star. ASAS-SN $g$-band results also show weak but elevated periodogram peaks at about a month for both stars, but these could possibly be lunar signals, and A and B also have major and mild contamination in ASAS-SN, respectively. Altogether, these roughly month-long signals may be the real periods, but given their weak strength, proximity to a TESS sector length, and the lack of uncontaminated supporting information, we ultimately decide to not adopt the measures into our set of final periods. If they \textit{are} the true rotation rates, the A and B components would have similar rotation periods aligned with their functionally identical H$\alpha$ EWs. Using ASAS-SN data, \cite{2020MNRAS.491...13J} report a rotation period of 0.43\,days for the A component via their automated pipeline, but we consider this result unreliable and do not adopt it based on (1) our own manual inspection of the same ASAS-SN variability data, (2) the lack of this signal appearing in TESS or our other data sources, and (3) this rapid period disagreeing with the prominent H$\alpha$ absorption the star consistently shows in our CHIRON results.

\subsection{GJ 745 AB} \label{subsec:rot-notes_GJ-745}

\textit{GJ~745} A and B are inactive with H$\alpha$ absorption, indicating approximate periods of 20--90\,days for their masses of 0.38\,$\rm{M_\odot}$. None of the available datasets yielded rotation detections. \cite{Newton_2016} report observations for both components as part of the MEarth project, but in each star they find unreliable periods noted as ``non-detection or undetermined detection" cases. Several other literature sources report period measures as follows: $P/\sin(i) = 5.03^{+1.89}_{-1.25}$\,days for A and $P/\sin(i) = 5.55^{+2.22}_{-1.44}$\,days for B from \cite{2016ApJ...822...97H}, $P/\sin(i) > 7.9$\,days for A and $P/\sin(i) > 8.0$\,days for B from \cite{2018A&A...612A..49R}, and $P_{\rm{rot}} = 3.80\pm0.01$\,days for B from \cite{2019A&A...621A.126D} via ASAS data. These results suggest A and B likely have generally similar rotation, depending on their inclinations. However, we note the 3.80\,day rotation period for B from \cite{2019A&A...621A.126D} is inconsistent with the observed H$\alpha$ absorption and has not been confirmed by other studies of the system nor our own TESS analysis.

\subsection{RTW 2011-3824 AB} \label{subsec:rot-notes_RTW-2011-3824}

\textit{RTW~2011-3824} A and B show no evident rotation signals in any of the available data, so we do not adopt any periods. Both components show H$\alpha$ absorption, implying likely periods of about 20--120\,days at their masses.

\subsection{G 230-39 AB} \label{subsec:rot-notes_G-230-39}

\textit{G~230-39} A and B are H$\alpha$ inactive according to the results of \citetalias{Pass_2024_ApJ}, implying likely periods of about 20--120\,days at their masses of $\sim$0.32\,$\rm{M_\odot}$. Despite many sectors of TESS data, neither star shows reliable rotation signals beyond inconsistent weak systematics; this is likely due to the TESS pipeline detrending out the likely longer period signals. Moderately blended TESS-unpop data have nine available sectors, with several consecutive sets at sectors 14/15/16, 55/56/57, and 75/76. They show gradual long-term variations in each sector, with sector-merged results indicating candidate signals of roughly 50-80\,days for A and 40--85\,days for B. In Table~\ref{tab:RotTable} we report the maximum power peak periods from these regions, although systematics are a concern. The signals align across consecutive sectors better for A than B, but this may be coincidental given the simple averaging method we use for merging sectors. Blended ASAS-SN data show a weak $\sim$72\,day signal for A and a slightly stronger $\sim$71\,day signal for B, while resolved ZTF data show a strong candidate signal at 74\,days for A but nothing significant for B. Despite the somewhat uncertain candidate nature of the various detections, their alignment favors A possibly having a period of $\sim$70\,days, which is congruent with the H$\alpha$ inactivity. However, we do not confirm that B has a similar signal because the TESS-unpop and ASAS-SN results for B blend with A. We are thus unable to determine final periods from both stars for our component comparisons, but do find a plausible candidate period of about $\sim$70\,days for A. We note that \cite{Newton_2016} also report observations for both components as part of the MEarth project, but in each star they find unreliable periods noted as ``non-detection or undetermined detection" cases.

\subsection{RTW 2211+0058 AB} \label{subsec:rot-notes_RTW-2211+0058}

\textit{RTW~2211+0058} A and B both show H$\alpha$ absorption, suggesting likely periods of about 20--130\,days for their masses of 0.27\,$\rm{M_\odot}$. TESS-unpop shows weak long-term variations indicative of periods $\gtrsim$\,25\,days when considering the two available sectors merged, but these results are suspect given the possible impact of sector-length systematics. ZTF shows a very weak candidate signal at $\sim$73\,days for A but nothing for B. However, \citetalias{2024AJ....167..159L} report periods of exactly 77.98\,days for both stars from their analysis of improved ZTF data\footnote{The 77.98\,day period results are in the catalog of \citetalias{2024AJ....167..159L} but not \citetalias{Lu_2022}, presumably owing to their stated relaxing of the vetting criteria for \citetalias{2024AJ....167..159L}. These periods therefore have a greater chance of being inaccurate false positives.}. They similarly report a 77.98\,day rotation period in \cite{2024NatAs...8..223L} for just the A component but not the B component, despite \cite{2024NatAs...8..223L} stating they used the dataset from \citetalias{2024AJ....167..159L}. We ourselves find a similar candidate ZTF period result for A but find no indications of such a period for B. Based on the available information and numerically identical periods for A and B in \citetalias{2024AJ....167..159L}, we hypothesize that a crossmatch in their analyses may have erroneously associated the A star's detection with the 5\farcs00 separated twin B star as well in the published catalog from \citetalias{2024AJ....167..159L}. AB blending in their ZTF data may also be a culprit. We adopt the literature period for A (lacking an adopted amplitude as they report none) given the congruence in detected periods and alignment with its H$\alpha$ activity, but do not adopt a rotation result for B, inhibiting our rotation comparisons for the system.

\subsection{RTW 2244+4030 AB} \label{subsec:rot-notes_RTW-2244+4030}

\textit{RTW~2244+4030} A and B are inactive with H$\alpha$ absorption according to the results of both \cite{2013AJ....145..102L} and \cite{2014MNRAS.443.2561G}. This inactivity at the components' 0.53\,$\rm{M_\odot}$ masses implies likely rotation periods of 10--60\,days. The system shows a messy mix of systematics and artifacts in the two sectors of blended TESS data, with only the 30-minute cadence FFI data for sector 16 showing a candidate $\sim$12--14\,day signal that may be rotation or the 13.7\,day TESS systematic. Merging the two sectors of blended TESS-unpop data reveals candidate long-term periods, but with timescales comparable to sector baselines, again suggesting possible systematics. \cite{2020MNRAS.491.5216G} report observations and rotation analyses for each component via the APACHE survey, but find no rotation signals in either star despite several thousand data points collected for each source over several years.

\subsection{L 718-71 AB} \label{subsec:rot-notes_L-718-71}

\textit{L~718-71} A and B are only weakly blended in TESS and TESS-unpop and are resolved in other datasets. Both stars show H$\alpha$ absorption, implying likely periods of roughly 10--80\,days at their masses. ASAS-SN shows weak indications of possible long-term variations between roughly 25--70\,days in A and 10--75\,days in B, but at much too low confidence to be reliable. TESS shows candidate signals near $\sim$6.5\,days and $\sim$11.5\,days (\citet{2023ApJS..268....4F} similarly report a 6.53\,day period for A and 6.47\,day period for B from the same single TESS sector of data), but they are inconsistent across the three sectors of available data for each star, and obfuscating shorter systematic disruptions are plainly evident in the light curves. TESS-unpop shows long-term brightness variations in each sector for both stars, and sector-merged results suggest long periods compatible with the H$\alpha$ activity levels, but the signals phase-fold poorly and could represent a large range of possible periods. The SAP fluxes also show quite clear long-term variations, suggesting these TESS-unpop signals may be legitimate. However, we deem these signals to be too uncertain for reliable periods given the notable proximity to a sector baseline in several sectors' signals and the relatively limited set of only three sectors of data over roughly five years for merging.

%% For this sample we use BibTeX plus aasjournalv7.bst to generate the

\bibliography{main}{}
\bibliographystyle{aasjournalv7}



\end{document}